\documentclass[12pt]{iopart}

\newcommand{\gguide}{{\it Preparing graphics for IOP Publishing journals}}
\usepackage{iopams}
\usepackage{isomath}
\usepackage{dsfont}
\usepackage{graphicx}
\usepackage{bm}
\usepackage{amssymb}
\usepackage{mathrsfs}
\usepackage{color}
\usepackage{cancel}
\usepackage{tikz}
\usepackage{tkz-euclide}
\usepackage{tikz-3dplot}
\usepackage{tabularx}
\usepackage{pgfplots}
\usepgfplotslibrary{groupplots}
\pgfplotsset{compat=1.7}
\usepgfplotslibrary{patchplots}
\usepgfplotslibrary{fillbetween}
\usepackage{bibentry}
\usetikzlibrary{snakes}
\usetikzlibrary{math} 
\usetikzlibrary{decorations.markings}
\usepackage{caption}
\usepackage{subcaption}
\usepackage{xcolor}
\definecolor{OffWhite}{HTML}{FFF7DD}
\definecolor{DarkGreen}{HTML}{013220}
\definecolor{DarkRed}{HTML}{B30900}
\definecolor{DarkBlue}{HTML}{0047AB}
\definecolor{BrightBlue}{HTML}{04D9FF}

\newcommand{\AxisRotator}[1][rotate=0]{%
	\tikz \draw[x=.5em,y=1.25em,line width=.2ex,-stealth,#1] (0,0) arc (-130:170:1 and 1);%
}

\begin{document}

	\title{Beam model of Doppler backscattering}
	
	\author{Valerian H Hall-Chen$^{1,2}$, Felix I Parra$^{1}$, and Jon C Hillesheim$^{2}$} 

	\address{$^1$ Rudolf Peierls Centre for Theoretical Physics, University of Oxford, Oxford OX1 3PU, UK\\
		$^2$ UKAEA/CCFE, Culham Science Centre, Abingdon, Oxon, OX14 3DB, UK}

	\ead{valerian.chen@physics.ox.ac.uk}
	\vspace{10pt}
	\begin{indented}
		\item[]January 2022
	\end{indented}

	\begin{abstract}
		We use beam tracing --- implemented with a newly-written code, Scotty --- and the reciprocity theorem to derive a model for the linear backscattered power of the Doppler Backscattering (DBS) diagnostic. Our model works for both the O-mode and X-mode in tokamak geometry (and certain regimes of stellarators). We present the analytical derivation of our model and its implications on the DBS signal localisation and the wavenumber resolution. To determine these two quantities, we find that it is the curvature of the field lines and the magnetic shear that are important, rather than the curvature of the cut-off surface. We also provide an explicit formula for the hitherto poorly-understood quantitative effect of the mismatch angle. Consequently, one can use this model to correct for the attenuation due to mismatch, avoiding the need for empirical optimisation. This is especially important in spherical tokamaks, since the magnetic pitch angle is large and varies both spatially and temporally. 


	\end{abstract}

	%
	%
	%
	%
	%

	\section{Introduction}
	Turbulent fluctuations in tokamaks are responsible for cross-field transport. The Doppler Backscattering (DBS) microwave diagnostic enables the non-perturbative characterisation of turbulent density fluctuations ($1 \lesssim k_{\perp} \rho_i \lesssim 10$) \cite{Hillesheim:DBS:2009,Hirsch:DBS:2001} and flows \cite{Hillesheim:DBS_rotation:2015, Conway:DBS_Flow:2005, Conway:flows:2010, Schmitz:LH:2017,Tynan:drift_turbulence:2009} with high spatial and temporal resolution, both at the edge and the core of the plasma. Here, $k_{\perp}$ is the wavenumber of turbulent fluctuations perpendicular to the magnetic field and $\rho_i $ is the ion gyroradius. Consequently, it is a widely used diagnostic for both tokamaks and stellarators \cite{Hennequin:DBS:2004, Happel:DBS:2009, Zhou:DBS:2013, Happel:DBS_synthetic:2017, Shi:DBS:2016, Rhodes:DBS:2016, Hu:DBS:2017, Tokuzawa:DBS_LHD:2018, MolinaCabrera:DBS_TCV:2019, Wen:DBS:2021, Tokuzawa:DBS_LHD:2021, Carralero:DBS_JT60SA:2021, Yashin:DBS:2021}. Moreover, since DBS is a microwave diagnostic, it is one of the few diagnostics able to survive the neutron fluxes generated by burning plasmas of future fusion reactors \cite{Volpe:microwave:2017, Costley:new_diagnostics:2010}. 
	
	There is a series of problems we seek to address: mismatch attenuation, wavenumber resolution, and localisation of the signal. We build a model that can account for all of them in realistic geometries and for realistic turbulence spectra. The model that we develop is an extension of previous work on reciprocity by Gusakov and collaborators \cite{Piliya:reciprocity:2002, Gusakov:scattering_slab:2004, Bulanin:spatial_spectral_resolution:2006, Gusakov:1D_RCDR:2011, Krutkin:DBS_synthetic:2019}. We introduce more geometry, do not assume a particular turbulence spectrum, and use beam tracing to make the problem more tractable. Since wavenumber resolution and localisation are already widely studied \cite{Hirsch:DBS:2001, Gusakov:scattering_slab:2004, Bulanin:spatial_spectral_resolution:2006}, we present our insight on them later in the paper with only a brief introduction here. The rest of this section will focus on the mismatch attenuation.

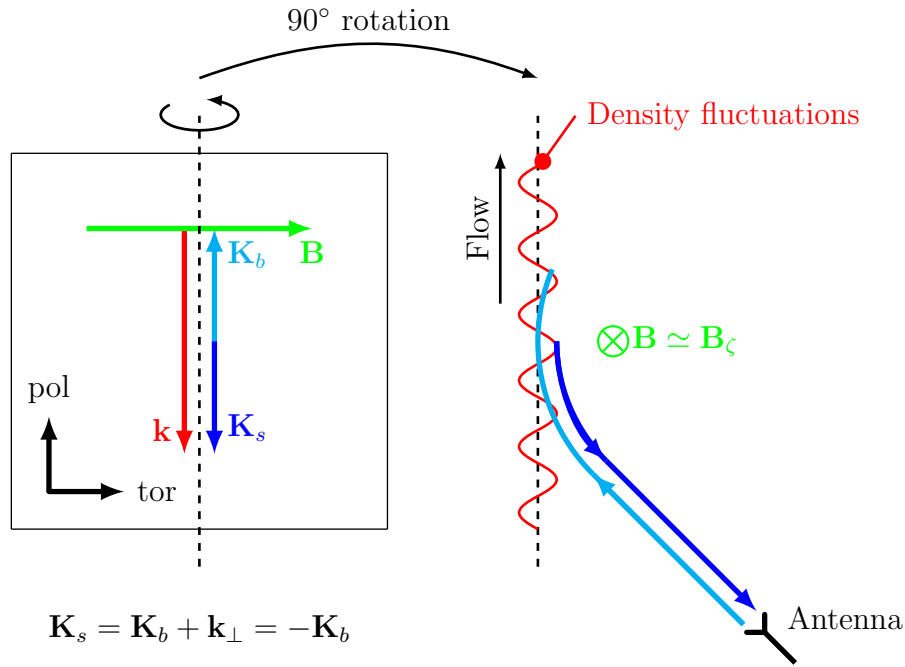
\begin{figure}
		\centering
		\begin{tikzpicture}[>=latex]
		\draw [line width = 1pt, dashed] (0,-3) -- (0,3); 
		\draw [line width = 1pt, segment length = 25pt, segment amplitude=2.5mm, color = red, snake=coil,segment aspect=0] (0,-2.5) -- (0,2.5); 
		\draw[*-,red, line width = 1pt] (0,2.3) -- (0.5,3) node[right,anchor=west] {Density fluctuations};
		
		\draw[green,thick] (1-0.707*0.2,-0.707*0.2) -- (1+0.707*0.2,0.707*0.2); 
		\draw[green,thick] (1+0.707*0.2,-0.707*0.2) -- (1-0.707*0.2,0.707*0.2); 
		\draw[green,thick] (1,0) circle (0.2) node[right, xshift = 1mm] {$\mathbf{B} \simeq \mathbf{B}_{\zeta}$}; 
		
		\draw[->, black, line width = 1pt] (-0.5,0.5) -- (-0.5,2.5) node[midway, above, rotate=90] {Flow}; 
		
		\draw[cyan, line width = 2pt] (0,0) arc (0:45:-2.5);
		\draw[cyan, line width = 2pt] (0,0) arc (0:-22.5:-2.5);
		\draw[cyan, line width = 2pt] ({2.5-(2.5*cos(45)) +2},{-(2.5*sin(45)) -2}) -- ({2.5-(2.5*cos(45))},{-(2.5*sin(45))});
		\draw[->, cyan, line width = 2pt] ({2.5-(2.5*cos(45)) +2},{-(2.5*sin(45)) -2}) -- ({2.5-(2.5*cos(45))},{-(2.5*sin(45))});
		
		\draw[blue, line width = 2pt] (0.25,0) arc (0:45:-2.25);
		\draw[->,blue, line width = 2pt] (0.25,0) arc (0:45:-2.25);
		\draw[<-,blue, line width = 2pt] ({2.5-(2.25*cos(45)) +2},{-(2.25*sin(45)) -2}) -- ({2.5-(2.25*cos(45))},{-(2.25*sin(45))});
		
		\draw[{angle 90 reversed}-, black, line width = 2pt] ({2.5-(2.375*cos(45)) +2.05},{-(2.375*sin(45)) -2.05}) -- ({2.5-(2.375*cos(45)) +2.6},{-(2.375*sin(45)) -2.6}) node[midway, above right, rotate=0] {Antenna}; 
		
		\draw [line width = 0.5pt] (-2,-2.5) -- (-2,2.5); 
		\draw [line width = 0.5pt] (-2,2.5) -- (-7,2.5);
		\draw [line width = 0.5pt] (-7,2.5) -- (-7, -2.5);
		\draw [line width = 0.5pt] (-7, -2.5) -- (-2,-2.5); 
		\draw [line width = 0.5pt] (-4.5, -3.5) node[below] {$\mathbf{K}_s = \mathbf{K}_b + \mathbf{k}_\perp = -\mathbf{K}_b$};
		
		
		\draw[->, red, line width = 2pt] (-4.7,1.5) -- (-4.7,-1.5) node[above left] {$\mathbf{k}$};
		
		\draw[->, cyan, line width = 2pt] (-4.3,0) -- (-4.3,1.5) node[below right] {$\mathbf{K}_b$};
		
		\draw[->, blue, line width = 2pt] (-4.3,0) -- (-4.3,-1.5) node[above right] {
			$\mathbf{K}_s$
		};
		
		\draw[->, green, line width = 2pt] (-6,1.5) -- (-3,1.5) node[below] {$\mathbf{B}$};
		
		\draw [line width = 1pt, dashed] (-4.5,-3.0) -- (-4.5,3); 
		\draw (-4.5,3.0) node {\AxisRotator[rotate=-90,->]}; 
		\draw (-2.25,4.0) node[above] {$90 ^{\circ}$ rotation};
		\draw[line width = 1pt, ->] (-4.5,3.5) to [out=20, in=160] (0,3.5);
		
		\draw[->, black, line width = 2pt] (-6.5,-2) -- (-6.5,-1.0) node[above] {pol};
		\draw[->, black, line width = 2pt] (-6.5,-2) -- (-5.5,-2.0) node[right] {tor};
		\draw[black,fill] (-6.5,-2) circle (0.75pt);
		\end{tikzpicture}
		\caption{A microwave probe beam is launched into the plasma. The emitting antenna also acts as the receiver, hence only the backscattered signal is measured. The Bragg condition determines how the wavevectors of the turbulence $\mathbf{k}$ (red), the probe beam $\mathbf{K}_b$ (light blue), and the scattered beam $\mathbf{K}_s$ (dark blue) must relate to one another for backscattering to occur. The right side of the figure represents a poloidal cut. The left side of the figure is in the plane perpendicular to the page, through the dotted line on the right side; that is, in both cases, the poloidal direction is pointing up, but on the left, the toroidal direction is in the plane of the page, while on the right, the toroidal direction is into the page. In conventional tokamaks, the magnetic field is mostly in the toroidal direction. In this figure, we depict the density fluctuations at the cut-off for illustrative purposes only. Our model accounts for backscattering at every point along the ray, and we later show the extent to which different points contribute to the signal. }
		\label{fig:schematic}
	\end{figure}

	In order for detectable backscattering to occur, the wavevector of the turbulent fluctuations at a particular point has to be twice in magnitude and opposite in direction to the wavevector of the probe beam at that point (Figure \ref{fig:schematic}). This Bragg condition determines the dominant turbulent wavenumber probed by DBS. The extent to which other wavenumbers are also backscattered into the detector is given by the wavenumber resolution, while the contribution of various points along the ray to the backscattered signal is known as the localisation or spatial resolution.

	We now introduce mismatch and the associated attenuation of the backscattered signal. The spatial scale of turbulence perpendicular to the field lines is much shorter than the characteristic length parallel to the field lines \cite{Catto:GK:1978,Frieman:GK:1982}. Hence, one has to launch the probe beam into the plasma such that the beam reaches the scattering location perpendicular to the magnetic field, allowing the Bragg condition for backscattering to be met. When the poloidal field is much smaller than the toroidal field, as in conventional tokamaks, this is achieved by sending a beam that does not propagate toroidally.

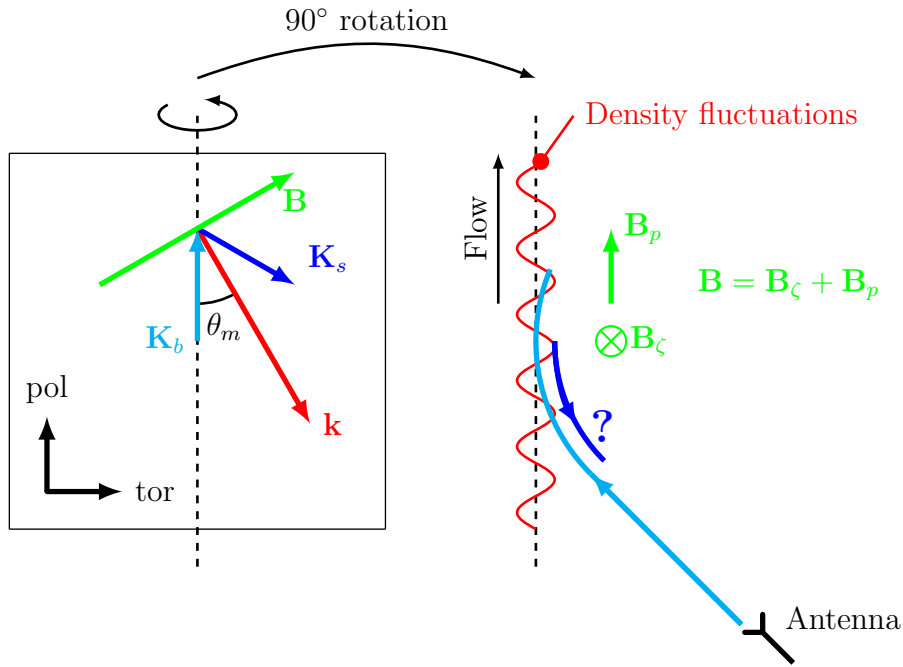
\begin{figure}
	\centering
		\begin{tikzpicture}[>=latex]
		\draw [line width = 1pt, dashed] (0,-3) -- (0,3); 
		\draw [line width = 1pt, segment length = 25pt, segment amplitude=2.5mm, color = red, snake=coil,segment aspect=0] (0,-2.5) -- (0,2.5); 
		\draw[*-,red, line width = 1pt] (0,2.3) -- (0.5,3) node[right,anchor=west] {Density fluctuations};
		
		\draw[green,thick] (1-0.707*0.2,-0.707*0.2) -- (1+0.707*0.2,0.707*0.2); 
		\draw[green,thick] (1+0.707*0.2,-0.707*0.2) -- (1-0.707*0.2,0.707*0.2); 
		\draw[green,thick] (1,0) circle (2.0mm) node[right, xshift = 1mm] {$\mathbf{B}_{\zeta}$}; 
		\draw[->, green, line width = 2pt] (1,0.5) to (1,1.5) node[right] {$\mathbf{B}_{p}$};
		\draw[green,thick] (2.0,0.75) node[right] {$\mathbf{B} = \mathbf{B}_{\zeta} + \mathbf{B}_{p}$};
		
		\draw[->, black, line width = 1pt] (-0.5,0.5) -- (-0.5,2.5) node[midway, above, rotate=90] {Flow}; 
		
		\draw[cyan, line width = 2pt] (0,0) arc (0:45:-2.5);
		\draw[cyan, line width = 2pt] (0,0) arc (0:-22.5:-2.5);
		\draw[cyan, line width = 2pt] ({2.5-(2.5*cos(45)) +2},{-(2.5*sin(45)) -2}) -- ({2.5-(2.5*cos(45))},{-(2.5*sin(45))});
		\draw[->, cyan, line width = 2pt] ({2.5-(2.5*cos(45)) +2},{-(2.5*sin(45)) -2}) -- ({2.5-(2.5*cos(45))},{-(2.5*sin(45))});
		
		\draw[blue, line width = 2pt] (0.25,0) arc (0:45:-2.25);
		\draw[->,blue, line width = 2pt] (0.25,0) arc (0:30:-2.25) node[right] {\Large{\textbf{?}}};
		
		\draw[{angle 90 reversed}-, black, line width = 2pt] ({2.5-(2.375*cos(45)) +2.05},{-(2.375*sin(45)) -2.05}) -- ({2.5-(2.375*cos(45)) +2.6},{-(2.375*sin(45)) -2.6}) node[midway, above right, rotate=0] {Antenna}; 
		
		\draw [line width = 0.5pt] (-2,-2.5) -- (-2,2.5); 
		\draw [line width = 0.5pt] (-2,2.5) -- (-7,2.5);
		\draw [line width = 0.5pt] (-7,2.5) -- (-7, -2.5);
		\draw [line width = 0.5pt] (-7, -2.5) -- (-2,-2.5);
		
		
		\draw [line width = 1pt, dashed] (-4.5,-3) -- (-4.5,3); 
		\draw (-4.5,3.0) node {\AxisRotator[rotate=-90,->]}; 
		\draw (-2.25,4.0) node[above] {$90 ^{\circ}$ rotation};
		\draw[line width = 1pt, ->] (-4.5,3.5) to [out=20, in=160] (0,3.5);
		
		\draw[black, line width = 1pt] (-4.5,0.5) arc (-90:-60:1) node[midway, below, xshift=1mm] {$\theta_m$};
		
		\draw[->, red, line width = 2pt] (-4.5,1.5) -- ({-4.5+3*cos(60)},{1.5-3*sin(60)}) node[right] {$\mathbf{k}$};
		
		\draw[->, cyan, line width = 2pt] (-4.5,0) -- (-4.5,1.5) node[near start, below left] {$\mathbf{K}_b$};
		
		\draw[->, blue, line width = 2pt] (-4.5,1.5) -- ({-4.5+1.5*cos(30)},{1.5-1.5*sin(30)}) node[above right] {
			$\mathbf{K}_s$
		};
		
		\draw[->, green, line width = 2pt] ({-4.5-1.5*cos(30)},{1.5-1.5*sin(30)}) -- ({-4.5+1.5*cos(30)},{1.5+1.5*sin(30)}) node[below, yshift = -0.5mm] {$\mathbf{B}$};
		
		\draw[->, black, line width = 2pt] (-6.5,-2) -- (-6.5,-1.0) node[above] {pol};
		\draw[->, black, line width = 2pt] (-6.5,-2) -- (-5.5,-2.0) node[right] {tor};
		\draw[black,fill] (-6.5,-2) circle (0.75pt);
		\end{tikzpicture}
	\caption{The mismatch angle reduces the backscattered signal. The variables are defined in the caption of Figure \ref{fig:schematic}, except for the poloidal magnetic field $\mathbf{B}_p$, associated with the plasma current and the mismatch angle, which is defined to be $\sin \theta_{m} = \hat{\mathbf{K}} \cdot \hat{\mathbf{b}}$. Here $\hat{\mathbf{b}}$ and $\hat{\mathbf{K}}$ are the unit vectors of the magnetic field and the probe beam's wavevector respectively. Having a significant poloidal field means that a ray propagating entirely in the poloidal plane can no longer meet the Bragg condition.}
	\label{fig:DBS_mismatch}
	\end{figure}

	In spherical tokamaks, the magnetic pitch angle is large (up to $35^{\circ}$ \cite{Hillesheim:DBS_MAST:2015}, compared to $\sim 10^{\circ}$ in conventional tokamaks like JET \cite{Coelho:Pitch:2009} and KSTAR \cite{Ko:Pitch:2016}) and it varies both spatially and temporally. Consequently, the DBS probe beam and the magnetic field are not normal to each other in general. This misalignment decreases the backscattered signal, making interpretation of the signal complicated (Figure \ref{fig:DBS_mismatch}): a decrease in the signal's magnitude could be due to a decrease in the fluctuations or an increase in the mismatch angle, defined to be $\sin \theta_{m} = \hat{\mathbf{K}} \cdot \hat{\mathbf{b}}$. Here $\hat{\mathbf{b}}$ and $\hat{\mathbf{K}}$ are the unit vectors along the external magnetic field $\mathbf{B}$ and the probe beam's wavevector $\mathbf{K}$, respectively.

	The misalignment can be empirically optimised with 2D beam steering \cite{Hillesheim:DBS_MAST:2015, Damba:mismatch:2021}. However, such empirical optimisation is ungainly and expensive, requiring several repeated shots for every measurement. Consequently, a quantitative understanding of the effect of the mismatch angle on the DBS signal would make practical the characterisation of spherical-tokamak plasmas with DBS.

	We seek to develop a quantitative understanding of what affects the backscattered signal and how. We write the electric field due to the microwaves as $\mathbf{E} \textrm{e}^{- \rmi \Omega t}$, where $\Omega$ is the angular frequency of the microwave beam. Hence, the electric field of the probe beam (which is launched into the plasma) and the scattered microwaves satisfy
	\begin{equation} \label{eq:Maxwell_total}
		\frac{c^2}{\Omega^2} \nabla \times (\nabla \times \mathbf{E}) = \bm{\epsilon} \cdot \mathbf{E} .
	\end{equation}
	Here we have already divided away the factor of $\textrm{e}^{- \rmi \Omega t}$, and $\bm{\epsilon}$ is the cold plasma dielectric tensor \cite{Stix:Waves:1992}, given by
	\begin{equation} \label{eq:cold_plasma_dielectric_tensor}
		\bm{\epsilon}
		= \bm{1}
		- \frac{\Omega_{pe}^2}{\Omega^2 - \Omega_{ce}^2} (\bm{1} - \hat{\mathbf{b}}\hat{\mathbf{b}})
		- \frac{\Omega_{pe}^2}{\Omega^2} \hat{\mathbf{b}}\hat{\mathbf{b}}
		+ \frac{\rmi \Omega_{pe}^2 \Omega_{ce}}{\Omega(\Omega^2 - \Omega_{ce}^2)} (\hat{\mathbf{b}} \times \bm{1} ) ,
	\end{equation}
	where $\bm{1}$ is the $3 \times 3$ identity matrix, $\Omega_{pe} = \left( \Sigma n_e e^2 / m_e \epsilon_0 \right)^{ 1 / 2 }$ is the electron plasma frequency, $\Omega_{ce} = e B / m_e$ is the electron cyclotron frequency, $\Sigma n_e = n_e + \delta n_e$ is the electron density, $n_e$ is the equilibrium electron density, $\delta n_e$ is the fluctuating electron density, $m_e$ is the electron mass, $e$ is the absolute value of the electron charge, and $\epsilon_0$ is the permitivity of free space. We split the dielectric tensor into equilibrium and turbulent parts, $\bm{\epsilon} = \bm{\epsilon}_{eq} + \bm{\epsilon}_{tb}$. Assuming the fluctuating part of the electron density is much smaller than its equilibrium part $\delta n_e \ll n_e$, we find that
	\begin{equation} \label{eq:density_tb}
		\bm{\epsilon}_{tb} = \frac{\delta n_e}{n_e} (\bm{\epsilon}_{eq} -\bm{1}).
	\end{equation}
	The inhomogeneity length associated with the equilibrium part of $\bm{\epsilon}$ is long, while the turbulence has a much shorter spatial scale. For $\delta n_e / n_e \ll 1$, we can split the electric field into a large component due to the beam and a small additive term due to the scattered microwaves, $\mathbf{E} \simeq \mathbf{E}_b + \mathbf{E}_s$. The equilibrium part of $\bm{\epsilon}$ is responsible for propagation, refraction, and diffraction of the probe beam,
	\begin{equation} \label{eq:Maxwell_eq}
		\frac{c^2}{\Omega^2} \nabla \times (\nabla \times \mathbf{E}_{b})
		- \bm{\epsilon}_{eq} \cdot \mathbf{E}_{b}
		= 0.
	\end{equation}
	The scattered electric field $\mathbf{E}_{s}$, which is much smaller than the probe beam electric field $\mathbf{E}_{b}$, is associated with the fluctuating part of the dielectric constant, which has a much smaller associated spatial scale,
	\begin{equation} \label{eq:Maxwell_tb}
		\frac{c^2}{\Omega^2} \nabla \times (\nabla \times \mathbf{E}_{s})
		- \bm{\epsilon}_{eq} \cdot \mathbf{E}_{s}
		= \bm{\epsilon}_{tb} \cdot \mathbf{E}_{b} .
	\end{equation}

	We use beam tracing in Section \ref{section_beam} to determine the electric field $\mathbf{E}_{b}$ due to the probe beam. We obtain an integral that determines the backscattered signal using the reciprocity theorem in Section \ref{section_reciprocity}. The main thrust of this paper is the simplifications that we apply to this integral to obtain a manageable result, as presented in Sections \ref{section_turbulence}--\ref{section_backscattered_P}. With the form of the backscattered power, we proceed to discuss localisation in Section \ref{section_localisation}, wavenumber resolution in Section \ref{section_wavenumber_resolution}, and the effect of the mismatch angle in Sections \ref{section_mismatch} and \ref{section_backscattered_E_ST}. We finish with a discussion of the model's insights in Section \ref{section_discussion}. 

	\section{Beam tracing} \label{section_beam}
	To determine the electric field of the probe beam, one could perform full-wave simulations \cite{Happel:DBS_synthetic:2017, Krutkin:DBS_synthetic:2019, Hirsch:2D_fullwave:2001, Silva:2D_fullwave:2004, Hillesheim:2D_fullwave:2012, Williams:3D_fullwave:2014}. However, this method requires the turbulent dielectric tensor to high resolution, which is precisely what we do not know and are trying to determine. A seemingly more sensible approach, especially considering that there are many ray tracing codes in the fusion community \cite{Peysson:C3PO:2012, Marushchenko:TRAVIS:2007, Farina:GRAY:2007, Prater:benchmarking_codes:2008}, would be to trace a bundle of rays to reconstruct the microwave electric field. However, since DBS depends on physics near the cut-off \cite{Hirsch:DBS:2001, Gusakov:scattering_slab:2004}, this method is not suitable as ray tracing breaks down near the cut-off \cite{Honore:quasioptics:2006}.

	One way around this problem is to trace only one ray and perform an expansion around that ray. This method, known as beam tracing, corresponds to tracing the path of a Gaussian beam, where this central ray gives the location of the peak of the Gaussian envelope. The theory of evolving Gaussian beams in isotropic inhomogeneous media is well-studied \cite{Casperson:beam_tracing:1973, Cerveny:beam_tracing:1982, Kravtsov:beam_tracing:2007}. Beam tracing in anisotropic inhomogeneous media, which is relevant for magnetic confinement fusion, was covered briefly by Peeters \cite{Peeters:Beam_Tracing:1996} and more extensively by Pereverzev \cite{Pereverzev:Beam_tracing:1992, Pereverzev:Beam_tracing:1993, Pereverzev:Beam_Tracing:1996, Pereverzev:Beam_tracing:1998} and Poli \cite{Poli:paraxial:1999, Poli:beam_tracing_BC:2001}. Beam tracing has also been implemented numerically \cite{Poli:Torbeam:2001}; it has so far been used to model electron cyclotron resonance heating (ECRH) \cite{Prater:benchmarking_codes:2008}, electron cyclotron current drive (ECCD) \cite{Prater:benchmarking_codes:2008, Ramponi:ITER_EC:2008, Bertelli:ECCD:2010}, lower-hybrid current drive \cite{Rodrigues:LHCD:2002}, synthetic-aperture microwave imaging (SAMI) \cite{Thomas:SAMI:2016}, and conventional reflectometry \cite{Stegmeir:reflectometry:2011, Hillesheim:zonal_flow:2016}. Importantly, it has been shown that the beam tracing method can be applied near the cut-off \cite{Maj:Beam_Tracing:2009}.
	
	The rest of this section is divided into four parts. Section \ref{subsection_beam} lays out the beam-tracing orderings and equations, summarising the results of our beam-tracing derivation in \ref{appendix_beam_tracing}. Next, in Section \ref{subsection_dispersion}, we proceed to describe our choice of the beam-tracing dispersion relation and the subtleties involved in doing this correctly (Section \ref{subsection_dispersion}). Thirdly, Section \ref{subsection_Scotty} describes our new beam-tracing code. Finally, we detail two scenarios that we use to illustrate the findings of this paper in Section \ref{subsubsection_testcases}.
	
	\subsection{Ordering and equations} \label{subsection_beam}
	In beam tracing, we assume that the length scale associated with the inhomogeneity of the density $L$ is long compared to both the width $W$ and wavelength $\lambda$ of the beam, and that the wavelength $\lambda$ is much smaller than the width of the beam $W$, $\lambda \ll W \ll L$. We choose the specific ordering \cite{Pereverzev:Beam_tracing:1992, Pereverzev:Beam_tracing:1993, Pereverzev:Beam_Tracing:1996, Pereverzev:Beam_tracing:1998}
	\begin{equation}
		\frac{W}{L} \sim \frac{\lambda}{W} \ll 1 .
	\end{equation}
	Ordering the width as an intermediate length scale follows from classical optics by taking the Rayleigh length to be the same order as the inhomogeneity length. We then consider a region of space close to the trajectory of the central ray, $\mathbf{r} = \mathbf{q}(\tau)$, where $\tau$ is a parameter that gives the position along the ray. This $\mathbf{q}$ is not to be confused with the safety factor, usually called $q$ in the literature. We will find the equations for $\mathbf{q}(\tau)$ as part of our beam tracing derivation. 
	
	To define a convenient coordinate system, we introduce the effective group velocity (this is not the true group velocity since it is a derivative with respect to the parameter $\tau$, not with respect to time)
	\begin{equation}
		\mathbf{g} = g \hat{\mathbf{g}} = \frac{\rmd \mathbf{q}}{\rmd \tau},
	\end{equation}
	where $g = |\mathbf{g}|$ is the magnitude of $\mathbf{g}$, and $\hat{\mathbf{g}}$ its direction. The group velocity $\mathbf{g}$ is parallel to the central ray. We will describe any arbitrary position as being composed of the position along $\mathbf{q}(\tau)$ and across the ray,
	\begin{equation} \label{eq:coordinates}
		\mathbf{r} = \mathbf{q}(\tau) + \mathbf{w} = \mathbf{q}(\tau) + w_x \hat{\mathbf{x}} (\tau) + w_y \hat{\mathbf{y}} (\tau).
	\end{equation}
	Here $\hat{\mathbf{x}} (\tau)$ and $\hat{\mathbf{y}} (\tau)$ are two mutually perpendicular unit vectors which are also perpendicular to $\mathbf{g}$. In the orthogonal basis ($\hat{\mathbf{x}}$, $\hat{\mathbf{y}}$, $\hat{\mathbf{g}}$), the vector $\mathbf{w}$ is given by
	\begin{eqnarray}
		\mathbf{w} =
		\left( \begin{array}{c}
			w_{x} \\
			w_{y} \\
			0   \\
		\end{array} \right) .
	\end{eqnarray}

	We write the electric field as
	\begin{equation} \label{eq:beam_ansatz}
		\mathbf{E} \left( \mathbf{r} \right) = \mathbf{A} (\mathbf{r}) \exp \left[ \rmi \psi \left( \mathbf{r} \right) \right] .
	\end{equation}
	Using the beam tracing coordinate system, equation (\ref{eq:coordinates}), we propose the following ansatz for $\psi$:
	\begin{equation} \label{eq:phase} 
	\eqalign{
		\psi (\mathbf{r})
		= s (\tau) + \mathbf{K}_{w}(\tau) \cdot \mathbf{w} + \frac{1}{2} \mathbf{w} \cdot \bm{\Psi}_{w}(\tau)  \cdot \mathbf{w} + \ldots ; }
	\end{equation}
	higher order terms of $\psi$ are not required in our derivation. Here 
	\begin{equation} \label{eq:eikonal_s}
		s = \int_0^\tau K_g (\tau') g (\tau') \ \rmd \tau' \sim \frac{L}{\lambda} ,
	\end{equation}
	with $K_g (\tau) = \mathbf{K} (\tau) \cdot \hat{\mathbf{g}} (\tau)$ being the projection of the wavevector along the ray, while
	\begin{equation}
		\mathbf{K}_w \sim \frac{1}{\lambda},
	\end{equation}
	is the projection of the wavevector perpendicular to the ray. Hence, both $K_g$ and $\mathbf{K}_w$ are real. The 2D symmetric matrix $\bm{\Psi}_w (\tau)$ is complex, and
	\begin{equation} \label{eq:size_of_Psi}
		\bm{\Psi}_w \sim \frac{1}{W^2}.
	\end{equation}
	The real part of $\bm{\Psi}_w$ is responsible for the curvature of the Gaussian beam, while its imaginary part gives the characteristic decay width of the Gaussian envelope. In general, the real and imaginary parts are not simultaneously diagonalisable. The eigenvalues of the real part are
	\begin{equation} \label{eq:Psi_real}
		\left[ \textrm{Re} \left( \bm{\Psi}_{w}  \right) \right]_{\alpha \alpha} 
		= \frac{K^3}{K_g^2} \frac{1}{R_{b,\alpha}}
		,
	\end{equation}  
	where $R_{b,\alpha}$ are the radii of curvature of the beam front, while the eigenvalues of the imaginary part are
	\begin{equation} \label{eq:Psi_imag}
		\left[ \textrm{Im} \left( \bm{\Psi}_{w}  \right) \right]_{\alpha \alpha}
		= \frac{2}{W_{\alpha}^2} ,
	\end{equation}	
	where $W_{\alpha}$ are the beam widths. From here on, we use the subscripts $_g$ and $_w$ to indicate projection parallel and perpendicular to the central ray respectively, and we use bold roman and bold italics to denote vectors and matrices respectively. In the basis $(\hat{\mathbf{x}}, \hat{\mathbf{y}}, \hat{\mathbf{g}})$, $\mathbf{K}_w$ and $\bm{\Psi}_w$ are
	\begin{eqnarray}
		\mathbf{K}_w =
		\left( \begin{array}{c}
			K_{x} \\
			K_{y} \\
			0   \\
		\end{array} \right) ,
	\end{eqnarray}
	and
	\begin{eqnarray}
		\bm{\Psi}_w =
		\left( \begin{array}{ccc}
			\Psi_{xx} & \Psi_{xy} & 0 \\
			\Psi_{yx} & \Psi_{yy} & 0 \\
			0		  & 0		 & 0 \\
		\end{array} \right) ,
	\end{eqnarray}
	such that $\mathbf{K}_w \cdot \hat{\mathbf{g}} = 0$ and $\bm{\Psi}_w \cdot \hat{\mathbf{g}} = 0$. 
	
	We expand the amplitude $\mathbf{A}$, given in equation (\ref{eq:beam_ansatz}), in $W^{} / L$ to obtain
	\begin{equation} \label{eq:amplitude}
		\mathbf{A} =  \mathbf{A}^{(0)} (\tau) + \mathbf{A}^{(1)}(\tau, \mathbf{w}) + \mathbf{A}^{(2)}(\tau, \mathbf{w}) + \ldots ,
	\end{equation}
	where $A^{(n)} \sim (W / L)^n A^{(0)}$. Only the lowest order term $\mathbf{A}^{(0)}$ is used in this work; finding the evolution equations for higher order corrections to the amplitude requires going to next order in the beam tracing derivation, as we show in \ref{appendix_beam_tracing}. We split the zeroth order term into a complex amplitude $A^{(0)}$ and a polarisation $\hat{\mathbf{e}}$,
	\begin{equation}
		\mathbf{A}^{(0)} (\tau) = A^{(0)}(\tau) \ \hat{\mathbf{e}} \left( \mathbf{q} (\tau), K_g (\tau), \mathbf{K}_{w} (\tau) \right).
	\end{equation}	
	The polarisation depends on $\mathbf{q} (\tau)$, $K_g (\tau)$, $\mathbf{K}_{w} (\tau)$, and satisfies
	\begin{equation} \label{eq:size_of_e}
		\hat{\mathbf{e}} \cdot \hat{\mathbf{e}}^* = 1 .
	\end{equation} 

	We derive (in \ref{appendix_beam_tracing}) the equations to determine the functions of $\tau$ --- $\mathbf{q}$, $K_g$, $\mathbf{K}_w$, $\bm{\Psi}_w$, $A^{(0)}$, and $\hat{\mathbf{e}}$. These equations are summarised here. First, we define
	\begin{equation} \label{eq:D_definition}
		\bm{D} (\mathbf{q},\mathbf{K}) = \frac{c^2}{\Omega^2} \left( \mathbf{K} \mathbf{K} - K^2 \bm{1} \right) + \bm{\epsilon} (\mathbf{q}).
	\end{equation}
	Since $\bm{D}$ is Hermitian, we can find its eigenvalues and eigenvectors, 
	\begin{equation} \label{eq:H_and_e_definition}
		\bm{D} \cdot \hat{\mathbf{e}} = H \hat{\mathbf{e}}.
	\end{equation}
	This equation defines the dispersion relation $H (\mathbf{q},\mathbf{K})$ corresponding to the polarisation $\hat{\mathbf{e}} (\mathbf{q},\mathbf{K})$. We want $H=0$ at all points along the central ray. One way to ensure that is to choose $\mathbf{K}$ such that $H = 0$ at some arbitrary point $\mathbf{q}$ along the central ray. Then, evolve $\mathbf{q}$ and $\mathbf{K} = K_g \hat{\mathbf{g}} + \mathbf{K}_w$ using
	\begin{equation} \label{eq:dq_dtau_maintext}
		\frac{\rmd \mathbf{q}}{\rmd \tau} = \nabla_K H ,
	\end{equation}
	and
	\begin{equation} \label{eq:dK_dtau_maintext}
		\frac{\rmd \mathbf{K}}{\rmd \tau} = - \nabla H ,
	\end{equation}
	such that $H=0$ along the rest of the ray. The associated polarisation $\hat{\mathbf{e}}$ is then calculated in post-processing. We do not account for mode conversion in this work. That is, either the O-mode or X-mode has $H=0$, and we assume that the dispersion relation of the other mode is always sufficiently different from zero. 
	This assumption has worked well in the cases studied in this paper. However, it is known to fail when the magnetic shear is large, giving rise to mode conversion \cite{Fidone:ModeConversion:1971, Boyd:ModeConversion:1985, Donne:ModeConversion:1997, Minami:ModeConversion:1998, Nagasaki:ModeConversion:1999, Tokuzawa:ModeConversion:2003, Tokuzawa:ModeConversion:2010}.
	
	For convenience, instead of evolving $\bm{\Psi}_w$, we evolve the 3D matrix
	\begin{equation}
		\bm{\Psi} (\tau) = \nabla \nabla \psi \left( \tau, w_x = 0, w_y =0 \right),
	\end{equation}	
	from which $\bm{\Psi}_w$ is subsequently determined. The equation for $\bm{\Psi}$ is
	\begin{equation} \label{eq:dPsi_dtau_maintext}
	\eqalign{
		\frac{\rmd \bm{\Psi}}{\rmd \tau}
		= -\left(
		\bm{\Psi} \cdot \nabla_K \nabla_K H \cdot \bm{\Psi}
		+ \bm{\Psi} \cdot \nabla_K \nabla H
		+ \nabla \nabla_K H \cdot \bm{\Psi}
		+ \nabla \nabla H
		\right) .
	}
	\end{equation}
	The matrix $\bm{\Psi}_{w}$ can be deduced from $\bm{\Psi}$ by projecting $\bm{\Psi}$ on the plane perpendicular to $\hat{\mathbf{g}}$. 
	
	Finally, we split the amplitude into its modulus and its phase,
	\begin{equation}
		A^{(0)} = \left| A^{(0)} \right| \exp \left[ \rmi \left( \phi_G + \phi_P \right) \right] .
	\end{equation}
	Its modulus is given by
	\begin{equation} \label{eq:beamtracingsummary_amplitude} 
		\left| A^{(0)} \right| = C \left[\det \left( \textrm{Im} \left( \bm{\Psi}_w \right) \right) \right]^{\frac{1}{4}} g^{-\frac{1}{2}}  ,
	\end{equation}
	where $C$ is a constant of integration and $\det \left[ \textrm{Im} \left( \bm{\Psi}_w \right) \right] = \textrm{Im} \left( \Psi_{xx} \right) \textrm{Im} \left( \Psi_{yy} \right) - \left[ \textrm{Im} \left( \Psi_{xy} \right) \right]^2$; its phase is $\phi_G + \phi_P$, which is composed of the Gouy phase $\phi_G$, given by
	\begin{equation} \label{eq:beamtracingsummary_Gouy_phase_evolution}
	\eqalign{
		\frac{\rmd \phi_G}{\rmd \tau}
		= - \textrm{Im}(\bm{\Psi}) : \nabla_K \nabla_K H ,
	}
	\end{equation}	
	and the phase associated with the changing polarisation when propagating through a plasma $\phi_P$, given by
	\begin{equation} \label{eq:beamtracingsummary_plasma_phase_evolution}
	\eqalign{
		\frac{\rmd \phi_P}{\rmd \tau}
		= 
		\rmi \frac{\rmd \hat{\mathbf{e}}}{\rmd \tau} \cdot \hat{\mathbf{e}}^* 
		- \frac{1}{2 \rmi} \left( 
			\frac{\partial \hat{\mathbf{e}}^*}{\partial K_\mu} \cdot \bm{D} \cdot \frac{\partial \hat{\mathbf{e}} }{\partial r_\mu} 
			- \frac{\partial \hat{\mathbf{e}}^*}{\partial r_\mu} \cdot \bm{D} \cdot \frac{\partial \hat{\mathbf{e}} }{\partial K_\mu} 
		\right)	
		 .
	}
	\end{equation}
	Our derivation in \ref{appendix_beam_tracing}, an alternative approach to Pereverzev's original work \cite{Pereverzev:Beam_tracing:1992, Pereverzev:Beam_tracing:1993, Pereverzev:Beam_Tracing:1996, Pereverzev:Beam_tracing:1998}, shows that the beam tracing equations for $\mathbf{q}$, $K_g$, $\mathbf{K}_w$, $\bm{\Psi}_w$, $A^{(0)}$, and $\hat{\mathbf{e}}$ are the result of forcing equation (\ref{eq:beam_ansatz}) to be a solution of equation (\ref{eq:Maxwell_eq}), keeping terms up to and including $(W / L)^2$.
	
	Given the properties of the launch beam at the antenna, we can then calculate how it evolves as it propagates into the plasma. This initial condition for a beam launched perpendicular to the surface of the antenna is
	\begin{equation} \label{eq:emit}
		\mathbf{E}_{ant} =
		A_{ant} \hat{\mathbf{e}}_{ant} \exp \left(\frac{\rmi}{2} \mathbf{w} \cdot \bm{\Psi}_{w,ant} \cdot \mathbf{w}  \right) ,
	\end{equation}
	where we have chosen $\rmi s_{ant} = 0$, and noted that due to the vacuum dispersion relation, $ \mathbf{K}_{w,ant} = 0$. The probe beam's electric field is thus
	\begin{equation} \label{eq:beam_field_final}
	\eqalign{
		\mathbf{E}_{b}
		&=
		A_{ant} \exp(\rmi \phi_G + \rmi \phi_P) \left[ \frac{\det (\textrm{Im} [\bm{\Psi}_w])  }{\det (\textrm{Im} [\bm{\Psi}_{w,ant} ])} \right]^{\frac{1}{4}}  \sqrt{\frac{g_{{ant}}}{g}}  \\
		&\times
		\hat{\mathbf{e}} \exp \left(\rmi s + \rmi \mathbf{K}_w \cdot \mathbf{w} + \frac{\rmi}{2} \mathbf{w} \cdot \bm{\Psi}_w \cdot \mathbf{w}  \right)
		.
	}
	\end{equation}
	We then calculate this electric field by finding gradients of the dispersion relation $H$ and using them to evolve the relevant quantities. We now proceed to describe our choice of dispersion relation, and the subtleties involved in doing this correctly.

	\subsection{Dispersion relation} \label{subsection_dispersion}
	In our derivation of beam tracing, the dispersion relation $H$ is defined in equation (\ref{eq:H_and_e_definition}). To calculate $H$, we first express the various components of the cold plasma dielectric tensor in the orthonormal basis $(\hat{\mathbf{u}}_1, \hat{\mathbf{u}}_2, \hat{\mathbf{b}})$. The $\hat{\mathbf{u}}_1$ and $\hat{\mathbf{u}}_2$ directions are perpendicular to $\hat{\mathbf{b}}$, and the $\hat{\mathbf{u}}_1$ direction together with $\hat{\mathbf{b}}$ defines a plane that contains $\mathbf{K}$, while $\hat{\mathbf{u}}_2$ is perpendicular to the plane of $\hat{\mathbf{b}}$ and $\mathbf{K}$. In this basis, we write the cold plasma dielectric tensor, given in equation (\ref{eq:cold_plasma_dielectric_tensor}), as
	\begin{eqnarray}
	\bm{\epsilon} =
	\left(
	\begin{array}{ccc}
		\epsilon_{11} & - \rmi \epsilon_{12} & 0\\
		\rmi \epsilon_{12} & \epsilon_{11} & 0\\
		0 & 0 & \epsilon_{bb}\\
	\end{array}
	\right) ,
	\end{eqnarray}
	where
	\begin{equation}
		\epsilon_{11} = 1 - \frac{\Omega_{pe}^2}{\Omega^2 - \Omega_{ce}^2} ,
		\qquad
		\epsilon_{12} = \frac{\Omega_{pe}^2 \Omega_{ce}}{\Omega \left( \Omega^2 - \Omega_{ce}^2 \right)} ,
		\qquad
		\epsilon_{bb} = 1 - \frac{\Omega_{pe}^2}{\Omega^2} .
	\end{equation}
	Here the subscripts $_1$, $_2$, and $_b$ denote the components in the $\hat{\mathbf{u}}_1$, $\hat{\mathbf{u}}_2$, and $\hat{\mathbf{b}}$ directions, respectively. The components of $\bm{D}$ are
	\begin{equation} \label{eq:D_explicit}
	\bm{D} =
	\left(
	\begin{array}{ccc}
		D_{11} & - \rmi D_{12} & D_{1b}\\
		\rmi D_{12} & D_{22} & 0\\
		D_{1b} & 0 & D_{bb}\\
	\end{array}
	\right) ,
	\end{equation}
	where
	\begin{equation}
		D_{11}
		=
		\epsilon_{11} - \frac{c^2}{\Omega^2} K^2 \sin^2 \theta_m,
	\end{equation}
	\begin{equation}
		D_{22}
		=
		\epsilon_{11} - \frac{c^2}{\Omega^2} K^2,
	\end{equation}
	\begin{equation}
		D_{bb}
		=
		\epsilon_{bb} - \frac{c^2}{\Omega^2} K^2 \cos^2 \theta_m,
	\end{equation}
	\begin{equation}
		D_{12}
		=
		\epsilon_{12},
	\end{equation}
	\begin{equation}
		D_{1b}
		=
		\frac{c^2}{\Omega^2} K^2 \sin \theta_m \cos \theta_m.
	\end{equation}
	Here, the mismatch angle $\theta_m$ is defined in Figure \ref{fig:DBS_mismatch}. We proceed to find the three eigenvalues of $\bm{D}$ in \ref{appendix_Cardano}; this requires solving a cubic equation, which we do with Cardano's formula. To evolve the beam, we would need to choose the eigenvalue that is zero along the entire path of the ray. Unfortunately, it is not immediately obvious which solutions correspond to the O and X modes. Moreover, Cardano's formula is complicated and cumbersome. Hence, we elect to use this form of the dispersion relation only in post-processing, but not to propagate the beam. We now prove that this is indeed a valid approach.
	
	Instead of $H$, we used $\bar{H} \left(H (\mathbf{q}, \mathbf{K}), \mathbf{q}, \mathbf{K}\right)$, with the function $\bar{H}$ satisfying
	\begin{equation} \label{eq:H_bar}
		\bar{H} (H = 0, \mathbf{q}, \mathbf{K}) = 0 ,
	\end{equation}
	such that the derivatives of $\bar{H}$ with respect to $\mathbf{q}$ and $\mathbf{K}$ holding $H$ constant vanish for $H=0$. Bearing this in mind, we now evaluate the gradients of such an alternative dispersion relation, getting
	\begin{equation}
		\frac{\rmd \mathbf{K}}{\rmd \bar{\tau}}
		= 
		- \nabla \bar{H} ,
		\qquad
		\frac{\rmd \mathbf{q}}{\rmd \bar{\tau}}
		= 
		\nabla_K \bar{H} ,
	\end{equation}	
	and
	\begin{equation}\label{eq:Psi_evolution_bar}
	\eqalign{
		\frac{\rmd \bm{\Psi}}{\rmd \bar{\tau}}
		= -\left(
		\bm{\Psi} \cdot \nabla_K \nabla_K \bar{H} \cdot \bm{\Psi}
		+ \bm{\Psi} \cdot \nabla_K \nabla \bar{H}
		+ \nabla \nabla_K \bar{H} \cdot \bm{\Psi}
		+ \nabla \nabla \bar{H}
		\right) ,
	}
	\end{equation}
	where the new parameter $\bar{\tau}$ is defined by
	\begin{equation} \label{eq:taubar}
		\frac{\rmd \tau}{\rmd \bar{\tau}}
		=
		\frac{\partial \bar{H}}{\partial H}.
	\end{equation}
	To obtain this result, we use equation (\ref{eq:constraint}) in the derivation of equation (\ref{eq:Psi_evolution_bar}). Here the derivatives of $\bar{H}$ are evaluated without holding $H$ fixed, that is
	\begin{equation}
		\frac{\partial \bar{H}}{\partial \alpha}
		=
		\frac{\partial \bar{H}}{\partial H} \Bigr|_{\mathbf{q}, \mathbf{K}}
		\frac{\partial H}{\partial \alpha}
		+
		\frac{\partial \mathbf{q}}{\partial \alpha} \cdot
		\nabla \bar{H} \Bigr|_{H, \mathbf{K}}
		+
		\frac{\partial \mathbf{K}}{\partial \alpha} \cdot
		\nabla_K \bar{H} \Bigr|_{H, \mathbf{q}} .
	\end{equation}		
	Consequently, we can calculate the beam parameters $\mathbf{q}$, $\mathbf{K}$, and $\bm{\Psi}$ from $\bar{H}$ as they are unaffected by choosing a new dispersion relation. 

	In our implementation of beam tracing, which we later describe in Section \ref{subsection_Scotty}, we make the following choice for the dispersion relation $\bar{H}$: the solution of the Booker quartic \cite{Booker:propagation:1936, Booker:propagation:1938}. Since at least one of the eigenvalues of $\bm{D}$ is always zero along the central ray, its determinant must also be zero, $\det \left( \bm{D} \right) = 0$. This turns out to be a quartic equation. Fortunately, this quartic is biquadratic, which makes it easier to solve than Cardano's formula. Hence, we have
	\begin{equation} \label{eq:H_Scotty}
		\bar{H}
		=
		K^2 \frac{c^2}{ \Omega^2 }
		+ \frac{
			\beta 
			\pm
			\sqrt{\beta^2 - 4 \alpha \gamma}
		}{
			2 \alpha	
		}
		=
		0 ,
	\end{equation}
	where 
	\begin{equation}
		\alpha = \epsilon_{bb} \sin^2 \theta_m + \epsilon_{11} \cos^2 \theta_m ,
	\end{equation}
	\begin{equation}
		\beta = - \epsilon_{11} \epsilon_{bb} \left( 1 + \sin^2 \theta_m \right) - \left( \epsilon_{11}^2 - \epsilon_{12}^2 \right) \cos^2 \theta_m ,
	\end{equation}
	and
	\begin{equation}
		\gamma = \epsilon_{bb} \left( \epsilon_{11}^2 - \epsilon_{12}^2 \right) .
	\end{equation}
	Recall that $\sin \theta_m$ is the mismatch angle, see Figure \ref{fig:DBS_mismatch}. The sign of the square root in equation (\ref{eq:H_Scotty}) is chosen based on the mode. To figure out which sign corresponds to which mode, we consider the case where $\theta_m = 0$, getting
	\begin{equation}
		\frac{
			\beta 
			\pm
			\sqrt{\beta^2 - 4 \alpha \gamma}
		}{
			2 \alpha	
		}
		=
		\frac{
			- \epsilon_{11}^2
			+ \epsilon_{12}^2
			- \epsilon_{11} \epsilon_{bb} 
			\pm
			\left|
				\epsilon_{11}^2
				- \epsilon_{12}^2
				- \epsilon_{11} \epsilon_{bb}				
			\right|
		}{
			2 \epsilon_{11}	.
		}		
	\end{equation}
	We need to choose the signs such that we recover $K^2 c^2 / \Omega^2 = \epsilon_{bb}$ for the O mode and $K^2 c^2 / \Omega^2 = \epsilon_{11} + \epsilon_{12} / \epsilon_{11}$ for the X mode. Thus, if $\epsilon_{11}^2 - \epsilon_{12}^2 - \epsilon_{11} \epsilon_{bb} > 0$, the $+$ of the $\pm$ corresponds to the O mode. If it is less than zero, then $-$ of the $\pm$ corresponds to the O mode. The other sign, in either case, would correspond to the X mode. One selects the appropriate sign, and uses this $\bar{H} (H, \mathbf{q}, \mathbf{K})$ in place of $H$ to find $\mathbf{q}$, $\mathbf{K}$, and $\bm{\Psi}$.
	
	There are a few quantities that depend on the definition of $H$ given in equation (\ref{eq:H_and_e_definition}). These are equations (\ref{eq:beamtracingsummary_amplitude}), (\ref{eq:beamtracingsummary_Gouy_phase_evolution}), and (\ref{eq:beamtracingsummary_plasma_phase_evolution}). In particular, the exact definition of $g$ in equation (\ref{eq:beamtracingsummary_amplitude}) is crucial for our result on localisation in Section \ref{section_localisation}.	Our calculation of group velocity is dependent on the choice of dispersion relation, since
	\begin{equation}
		\bar{\mathbf{g}} = \frac{\rmd \tau}{\rmd \bar{\tau}} \frac{\rmd \mathbf{q}}{\rmd \tau} ,
	\end{equation}
	where the dependence of $\rmd \tau / \rmd \bar{\tau}$ on our choice of $H$ is given in equation (\ref{eq:taubar}). We use Cardano's solution in equations (\ref{eq:Cardano_H_1}), (\ref{eq:Cardano_H_2}), and (\ref{eq:Cardano_H_3}) in post-processing to determine $g_{ant}$ and $g$ in equation (\ref{eq:beam_field_final}) consistent with our definition of $H$.
	
	Having explained how to handle the subtleties of the dispersion relation, we proceed to describe the code used to solve the beam-tracing equations and determine the probe beam's electric field.
	
	\subsection{Beam tracer: Scotty} \label{subsection_Scotty}
	To simulate the propagation of Gaussian beams in tokamak plasmas, we have developed a new code, Scotty. It is a beam-tracing code written in Python 3, entirely in cylindrical coordinates ($R$, $\zeta$, $Z$), with an option to convert the output to Cartesian coordinates ($X$, $Y$, $Z$). This exploits the toroidal symmetry of tokamak plasmas, simplifying the beam tracing equations. Hence, although the theoretical work presented in this paper is applicable to stellarators (in regions where the appropriate orderings and approximations hold), Scotty cannot be used out-of-the-box for such devices. Nonetheless, the results presented in this paper would still be applicable; one would simply need to run a suitable beam-tracing code and post-process the output appropriately.
	
	
	In cylindrical coordinates, we have $H(R,Z,K_R,K_\zeta,K_Z)$: a very natural choice of variables for a tokamak. The new spatial coordinates $R$, $Z$, $\zeta$ are defined as follows
	\begin{equation}
		R = \sqrt{X^2 + Y^2} ,
	\end{equation}
	\begin{equation}
		\zeta = \tan^{-1} \left( \frac{Y}{X}  \right) ,
	\end{equation}
	\begin{equation}
		Z = Z .
	\end{equation}	
	The components of $\mathbf{K}$ in the new coordinate system are
	\begin{equation} \label{eq:definition_K_R}
		K_R = K_X \cos \zeta + K_Y \sin \zeta ,
	\end{equation}
	\begin{equation} \label{eq:definition_K_zeta}
		K_\zeta = \left( - K_X \sin \zeta + K_Y \cos \zeta \right) \sqrt{X^2 + Y^2},
	\end{equation}	
	and
	\begin{equation} \label{eq:definition_K_Z}
		K_Z = K_Z .
	\end{equation}
	Note that $K_\zeta$ is the toroidal mode number, and is thus dimensionless, instead of having units of inverse length (unlike the other components of the wavevector). Hence, $\mathbf{K}$ becomes
	\begin{equation}
		\mathbf{K} 
		= K_R \nabla R
		+ K_\zeta \nabla \zeta 
		+ K_Z \nabla Z ,
	\end{equation}
	being careful to remember that we have $\nabla R = \hat{\mathbf{R}}$, $\nabla Z = \hat{\mathbf{Z}}$, but $\nabla \zeta = \hat{\boldsymbol{\zeta}} / R$. 
	
	Since Scotty was inspired by TORBEAM \cite{Poli:Torbeam:2001}, it has the option to use the same input files for equilibrium data: specifically, $\mathbf{B}$ and the poloidal flux $\psi_p$ on a grid ($R$ and $Z$), as well as electron density as a function of radial coordinate $\sqrt{\psi_p}$. We split the magnetic field into toroidal and poloidal components, 
	\begin{equation}
		\mathbf{B} = 
		B_\zeta \hat{\boldsymbol{\zeta}}
		+
		B_R \hat{\mathbf{R}}
		+
		B_Z \hat{\mathbf{Z}} ,
	\end{equation}
	where $B_\zeta$ is the toroidal component of the magnetic field, and $B_R$ and $B_Z$ are its poloidal components. Scotty can also calculate $\mathbf{B}$ directly from EFIT \cite{Lao:EFIT:1985, Appel:EFIT:2006} output: this is done by using $R B_\zeta$, given in the EFIT output, and numerically evaluating the gradients of $\psi_p$. The magnetic field is calculated as follows:
	\begin{equation} \label{eq:B_R_Scotty}
		B_R = - \frac{1}{R} \frac{\partial \psi_p}{\partial Z},
	\end{equation}
	\begin{equation} \label{eq:B_zeta_Scotty}
		B_\zeta = \frac{1}{R} I \left( \psi_p (R,Z) \right),
	\end{equation}	
	where $I$ is proportional to the poloidal current, an output of EFIT, and
	\begin{equation} \label{eq:B_Z_Scotty}
		B_Z = \frac{1}{R} \frac{\partial \psi_p}{\partial R}.
	\end{equation}	
	Scotty assumes lossless propagation as it is written specially for DBS. Since we use Mega Ampere Spherical Tokamak (MAST) plasmas as a case study in this work, we do not need to account for the relativistic correction to the electron mass. As such, in its current implementation Scotty does not use temperature profiles, although we expect to add the relativistic correction in the near future.
	
	In its current implementation, Scotty matches boundary conditions at the plasma-vacuum edge, using the generalised Snell's law \cite{Poli:beam_tracing_BC:2001} for $\bm{\Psi}$ (like TORBEAM) but not for $\mathbf{K}$ (unlike TORBEAM). Consequently, a possible discontinuity in $\nabla \nabla H$ arising from a sudden change in the density gradient is properly handled; see \ref{appendix_vacuum_plasma_interface} for details. 
	
	The definitions of $K_R$, $K_\zeta$, $K_Z$, in equations (\ref{eq:definition_K_R}), (\ref{eq:definition_K_zeta}), and (\ref{eq:definition_K_Z}), were chosen such that the electric field has the form:
	\begin{equation} \label{eq:E_b_cylindrical}
	\eqalign{	
		\mathbf{E}_b \propto
		\exp \Big[
			&
			\rmi s
			+ \rmi K_R \Delta R
			+ \rmi K_Z \Delta Z
			+ \rmi K_\zeta \Delta \zeta \\
			&+ \frac{\rmi}{2} \big(
				\Psi_{RR} \left( \Delta R \right)^2
				+ 2 \Psi_{RZ} \Delta R \Delta Z
				+ \Psi_{ZZ} \left( \Delta Z \right)^2 \\
				&+ \Psi_{\zeta \zeta} \left( \Delta \zeta \right)^2
				+ 2 \Psi_{R \zeta} \Delta R \Delta \zeta
				+ 2 \Psi_{\zeta Z} \Delta \zeta \Delta Z
			\big)
		\Big] ,
	}
	\end{equation}
	which gives us the definitions of the components of $\bm{\Psi}$ in cylindrical coordinates. 
	The components $\Psi_{RR}$, $\Psi_{ZZ}$, $\Psi_{\zeta \zeta}$, $\Psi_{RZ}$, $\Psi_{R \zeta}$, and $\Psi_{Z \zeta}$ of the matrix $\bm{\Psi}$ in the new coordinate system must satisfy 	
	\begin{equation}
		\Psi_{\alpha \beta} (\tau) = \frac{
			\partial^2 \psi	
		}{
			\partial r_\alpha \partial r_\beta
		} \left( \tau, w_x=0, w_y = 0 \right).
	\end{equation}
	Hence, they are
	\begin{equation}
		\Psi_{RR} 
		= \Psi_{XX} \cos^2 \zeta 
		+ 2 \Psi_{XY} \sin \zeta \cos \zeta
		+ \Psi_{YY} \sin^2 \zeta,
	\end{equation}
	\begin{equation}
	\Psi_{ZZ} = \Psi_{ZZ} ,
	\end{equation}	
	\begin{equation} \label{eq:Psi_zeta_zeta}
	\eqalign{
		\Psi_{\zeta \zeta} 
		&= \left[
			\Psi_{XX} \sin^2 \zeta
			-2 \Psi_{XY} \sin \zeta \cos \zeta
			+\Psi_{YY} \cos^2 \zeta 
		\right] \left( X^2 + Y^2 \right) \\
		&- \left( K_X \cos \zeta + K_Y \sin \zeta \right) \sqrt{X^2 + Y^2}  ,
	}
	\end{equation}
	\begin{equation}
		\Psi_{RZ} 
		= \Psi_{XZ} \cos \zeta 
		+ \Psi_{YZ} \sin \zeta ,
	\end{equation}
	\begin{equation}
	\fl
	\eqalign{
		\Psi_{R \zeta} 
		&= \left[
			-\Psi_{XX} \sin \zeta \cos \zeta
			+\Psi_{XY} \left(
				\cos^2 \zeta - \sin^2 \zeta
			\right)
			+\Psi_{YY} \sin \zeta \cos \zeta 
		\right] \sqrt{X^2 + Y^2} \\
		&- K_X \sin \zeta + K_Y \cos \zeta ,
	}
	\end{equation}
	\begin{equation}
		\Psi_{Z \zeta}
		= \left(
			- \Psi_{XZ} \sin \zeta 
			+ \Psi_{YZ} \cos \zeta
		\right) \sqrt{X^2 + Y^2} .
	\end{equation}
	The rest of the elements can be found by remembering that $\bm{\Psi}$ is symmetric. 
	
	With these new variables, the gradients become
	\begin{equation}
		\nabla 
		= \nabla R \frac{\partial}{\partial R}
		+ \nabla \zeta \frac{\partial}{\partial \zeta}
		+ \nabla Z \frac{\partial}{\partial Z} ,
	\end{equation}
	and
	\begin{equation}
		\nabla_K
		= \nabla_K K_R \frac{\partial}{\partial K_R}
		+ \nabla_K K_\zeta \frac{\partial}{\partial K_\zeta}
		+ \nabla_K K_Z \frac{\partial}{\partial K_Z} .
	\end{equation}		
	Using all these properties, we can write the evolution equations in a way which makes their Hamiltonian character explicit,
	\begin{equation}
		\frac{\rmd q_\alpha}{\rmd \tau}
		=
		\frac{\partial H}{\partial K_\alpha},
	\end{equation}
	\begin{equation}
		\frac{\rmd K_\alpha}{\rmd \tau}
		=
		- \frac{\partial H}{\partial r_\alpha},
	\end{equation}
	and
	\begin{equation}
	\eqalign{
		\frac{\rmd \Psi_{\alpha \beta}}{\rmd \tau}
		=
		- \frac{\partial^2 H}{\partial r_\alpha \partial r_\beta} 
		- \Psi_{\alpha \gamma} \frac{\partial^2 H}{\partial K_\gamma \partial r_\beta}
		- \frac{\partial^2 H}{\partial r_\alpha \partial K_\eta} \Psi_{\eta \beta}
		- \Psi_{\alpha \gamma} \frac{\partial^2 H}{\partial K_\gamma \partial K_\eta} \Psi_{\eta \beta} .
	}
	\end{equation}
	Here we note that while $\rmd q_\zeta / \rmd \tau = \partial H / \partial K_\zeta$, the corresponding component of group velocity, given by
	\begin{equation}
		\mathbf{g} = \nabla_K H,
	\end{equation}
	is $g_\zeta = R \ \partial H / \partial K_\zeta$; $\rmd q_\zeta / \rmd \tau = \partial H / \partial K_\zeta$ is an angular velocity, while $g_\zeta$ is a linear velocity, and they have different units. 
	
	Moving forward, the most obvious simplification to the above equations is due to toroidal symmetry: spatial gradients of equilibrium properties in the toroidal direction are zero, that is, $\partial H / \partial \zeta = 0$. 
	
	We proceed to solve these equations numerically, evolving $\bm{\Psi}$, from which $\bm{\Psi}_w$ may be obtained. Since $\bm{\Psi}$ is symmetric, it only has six independent components. We can further reduce the number of such components: equation (\ref{eq:constraint}) means we can reduce it by three. Hence, it is in principle possible to solve for only three independent components of $\bm{\Psi}$. However, we deemed the implementation too complex given the unclear benefits, which is why we have made a conscious design decision to solve for all six components
	
	The initial conditions required by Scotty for the beam are as follows: frequency, initial beam widths and curvatures, poloidal $\varphi_p$ and toroidal $\varphi_t$ launch angles, and launch position. Note that only the initial $q_R$ and $q_Z$ need to be specified. The toroidal angle is taken to be zero at launch, $q_\zeta = 0$. The launch angles are defined the same way as TORBEAM \cite{Poli:Torbeam:2001}, and are used to initialise $\mathbf{K}_{ant}$ as follows,
	\begin{equation}
	\eqalign{
		K_{R,ant}     &= - \frac{\Omega}{c}         \cos \varphi_t \cos \varphi_p , \\
		K_{\zeta,ant} &= - \frac{\Omega}{c} R_{ant} \sin \varphi_t \cos \varphi_p , \\
		K_{Z,ant}     &= - \frac{\Omega}{c}       				   \sin \varphi_p .
	}
	\end{equation}
	Scotty uses SciPy's initial value problem solver to evolve the beam tracing ODEs. The solver has the option to easily switch between various integration methods. For this paper, we use an explicit Runge-Kutta method of order 5(4) \cite{SciPy:algorithms:2020, Dormand:RK:1980}.		
	
	Having broadly described the workings of Scotty, we lay out the parameters for two test scenarios that we use throughout the rest of the paper to illustrate our analytical results.

	\subsection{Test scenarios} \label{subsubsection_testcases}
	Throughout the rest of this paper, we use two test scenarios to illustrate our results. One of these scenarios uses equilibrium data from a real shot which was carried out at MAST, while the other is entirely analytical. We detail these two scenarios here, and unless otherwise stated in subsequent sections, the parameters here are what we then use. In both scenarios, the probe beam's frequency was taken to be $55$ GHz with O-mode polarisation.
	
\begin{figure}
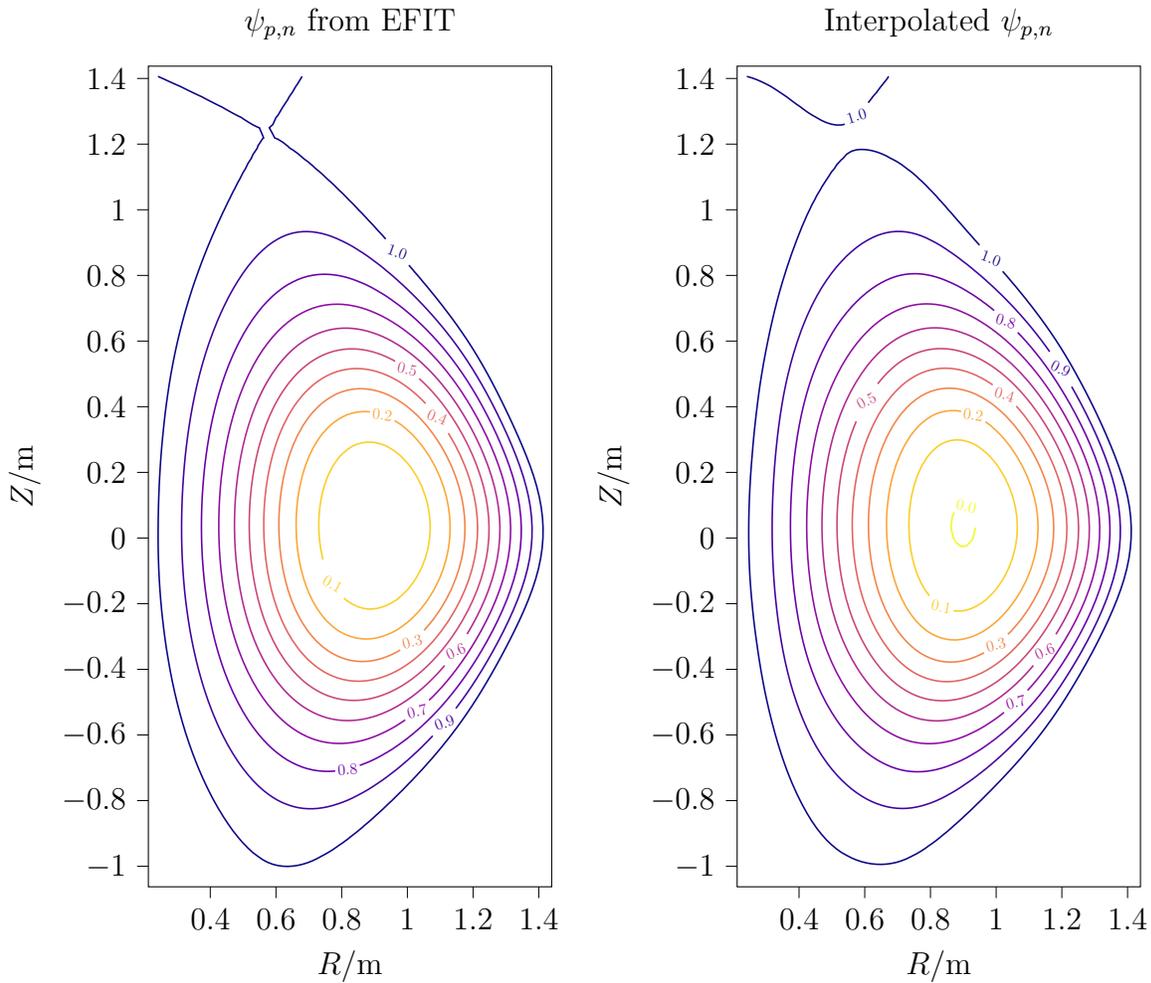

	\centering
	\begin{subfigure}{0.49\textwidth}		
		\centering


		\end{subfigure}		

	\caption{Data from MSE-constrained EFIT (left), after smoothing (right). Here $\psi_{p,n}$ is the normalised $\psi_p$ such that $\psi_{p,n}=1$ on the last closed flux surface and $\psi_{p,n} = 0$ on the magnetic axis (left). The smoothing spline (right) gives $\psi_{p,n} < 0$ in a region of space near the magnetic axis. Thankfully, the beam is never in that region in the simulations used for this paper, so this is not an issue.}
	\label{fig:smooth_flux}
	\end{figure}

	To illustrate our model's ability to deal with real plasmas, we chose MAST shot 29908, at 190ms. This was one of six repeated shots used to study the effect of mismatch in an earlier paper \cite{Hillesheim:DBS_MAST:2015}, in which DBS data was analysed for these shots at 190ms. The $R B_\zeta$ and the normalised poloidal flux density $\psi_p$ were determined by MSE-constrained EFIT, and we used equations (\ref{eq:B_R_Scotty}), (\ref{eq:B_zeta_Scotty}), and (\ref{eq:B_Z_Scotty}) to calculate $\mathbf{B}$. We used SciPy's bivariate spline to interpolate $\psi_p$ \cite{SciPy:algorithms:2020}. To ensure that the second spatial derivatives are smooth, we used a degree of 5 and a smoothing factor of 2. This does not significantly change the poloidal flux profile, as can be seen from Figure \ref{fig:smooth_flux}.
	
	The density profile was acquired by Thomson scattering \cite{Scannell:Thomson:2010}. We chose to fit the density data rather than smooth it because it was noisy, which made the evaluation of $\nabla \nabla n$, and thus $\rmd \bm{\Psi} / \rmd \tau$, challenging. 
	Since data processing is not the main focus of this paper, we instead use the following function to fit the density data
	\begin{equation}
		n_e = C_1 \tanh \left[C_2 \left( \psi_{p,n} - C_3 \right) \right] ,
	\end{equation}
	where $\psi_{p,n}$ is the normalised $\psi_p$ such that $\psi_{p,n}=1$ on the last closed flux surface and $\psi_{p,n} = 0$ on the magnetic axis. The coefficients $C_1 = 3.25 \times 10^{19}\textrm{m}^{-3}$, $C_2 = -2.4$, and $C_3 = 1.22$ were determined via manual fitting by visual inspection. These coefficients were used by Scotty directly to calculate the density. Notice that the fit gives negative densities for $\psi_{p,n} > C_3$; at these values of $\psi$, the density is set to $0$. The fit is not particularly good when $\psi_{p,n} \lesssim 0.4$; the experimental density profile is hollow. Fortunately, this is not a problem for the current work, since the beams studied in this paper do not enter that region.
	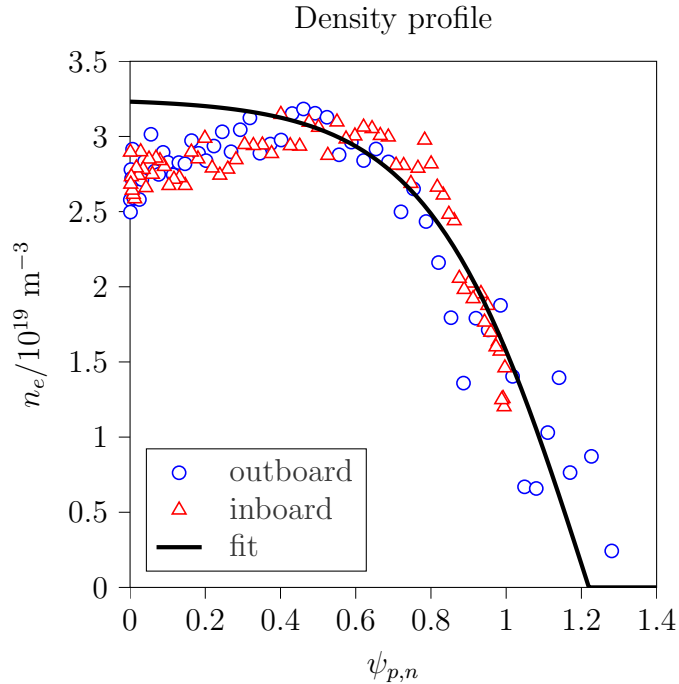
\begin{figure}[htp]
	\centering
		\begin{tikzpicture}
	
		\begin{axis}[
	    clip mode=individual,		
		legend cell align={left},
		legend style={fill opacity=0.8, draw opacity=1, text opacity=1, at={(0.03,0.03)}, anchor=south west},
		tick align=outside,
		tick pos=left,
		title={Density profile},
		x grid style={white!69.0196078431373!black},
		xlabel={$\psi_{p,n}$},
		xmin=0, xmax=1.4,
		xtick style={color=black},
		y grid style={white!69.0196078431373!black},
		ylabel={$n_e / 10^{19} \ \textrm{m}^{-3}$},
		ymin=0, ymax=3.5,
		ytick style={color=black},
		width=7cm,
		height=7cm,
		scale only axis
		]		
		
		\addplot [semithick, blue, mark=*, mark size=2.5, mark options={fill=white}, only marks]
		table {%
		0.000674190541703855 2.57881480279978
		0.00111830352164397 2.49711427188393
		0.00223693958709974 2.77965423555725
		0.00405178732618413 2.72119210279677
		0.00660102343539321 2.915309781169
		0.00993511017038644 2.59010260907286
		0.0140847879043118 2.66755462705898
		0.0190466913503634 2.71831775949944
		0.0247993318406661 2.58059425241817
		0.0313201885419671 2.71335764264422
		0.0385946325727959 2.83453767787164
		0.0466595627018033 2.79665928239244
		0.0555603542239995 3.01429265575256
		0.0653593426626851 2.76987869757702
		0.0761034719637129 2.74770440677501
		0.0878456621824586 2.89409646352967
		0.10066324088452 2.82701965716556
		0.11460384245264 2.74460752232422
		0.129702185437384 2.82626231355635
		0.145981632372369 2.8182299413108
		0.163446535319975 2.97228713342733
		0.18213072203935 2.88794535567924
		0.201986250226751 2.83916354319202
		0.22298423045449 2.93619324532
		0.245142331421566 3.03113431515985
		0.268453370389108 2.89814332602686
		0.292960066885728 3.04472010073498
		0.318582564385274 3.12353617284652
		0.345282615702875 2.88838999818151
		0.372972552189417 2.94945929291377
		0.401589151399695 2.97632893817103
		0.430984625265435 3.15242781998727
		0.461130094740458 3.18274443419761
		0.491975152947361 3.15429918877774
		0.523547693976126 3.12773982570183
		0.55575687887439 2.8780785582339
		0.588293552295689 2.96249070472617
		0.621029967670032 2.83995057361519
		0.653930650355694 2.91669978376883
		0.687038326777582 2.83068982697908
		0.720345049947258 2.49863687558608
		0.753723023088043 2.65148834325159
		0.787003534788673 2.43420857283096
		0.820190773029522 2.16109714126815
		0.853331518919086 1.79428488898161
		0.886533425747921 1.35917988042181
		0.9195060918737 1.7909263207634
		0.952254097996954 1.71294125954014
		0.984853719402011 1.87645886960438
		1.01709248964258 1.40407755813274
		1.04883630315773 0.66916937373424
		1.07999762249445 0.658209221926244
		1.11056990275803 1.0300117176866
		1.14057135702659 1.39514787444776
		1.16991321115013 0.763932542592719
		1.2266891534116 0.871798206690335
		1.28069115975837 0.242736460772829
		};
		\addlegendentry{outboard}
		
		\addplot [semithick, red, mark=triangle*, mark size=3, mark options={fill=white}, only marks]
		table {%
		0.996526741387153 1.46101752699573
		0.994504547089312 1.20208869591959
		0.991700953424645 1.25700116534096
		0.988131140991454 1.24915713943724
		0.983804032165577 1.57239134786851
		0.978728731036578 1.63498027762152
		0.972912783949585 1.60074005616746
		0.966363676003093 1.70847427364997
		0.959086070211288 1.69796404200006
		0.951088822311333 1.87620532222301
		0.942379067622649 1.76521830954089
		0.932959450258031 1.95534882889731
		0.922837808346198 1.95314078964641
		0.912015602462065 1.92106979468279
		0.900496995476962 2.02463323470766
		0.888286736731429 1.98239131738246
		0.875387977655903 2.05660857215966
		0.861800668618842 2.43985874318378
		0.847523515565025 2.482458001788
		0.832549816966477 2.61051196380996
		0.816863348940903 2.66281903047815
		0.800451761798244 2.81782422152015
		0.783300826506 2.97597159689201
		0.765410007799791 2.78873444238408
		0.746782307634858 2.68758662940543
		0.727434471297435 2.80693487826098
		0.707388327730846 2.80902482996306
		0.686674582020632 2.99667540084303
		0.665330416815825 3.00657320451627
		0.643387251289511 3.05244900777127
		0.620868125784508 3.06243675149566
		0.597802272274762 3.0036933636608
		0.574217614140251 2.98363717205853
		0.550163934538731 3.09660077679392
		0.525729163907071 2.87442883933666
		0.500999373337062 3.06087852361678
		0.476061791036489 3.0946154985988
		0.451017977381119 2.93572881160843
		0.425965183833844 2.9388254761569
		0.401002689145618 3.14661470201121
		0.376246154320985 2.88640625930268
		0.351832004701992 2.94348212780285
		0.327875856880447 2.93968551415214
		0.304482167656794 2.9464022107839
		0.281744528418661 2.8457667702238
		0.259762185699132 2.78289977398012
		0.238604420091884 2.74270778613375
		0.218309460901928 2.79035380310947
		0.198912093278524 2.9879325241836
		0.180456361937353 2.85128323996267
		0.162981712654346 2.89632649301312
		0.146514342428792 2.67404460439309
		0.13107949806336 2.72560136432648
		0.1166805364406 2.7152006440347
		0.103308932454028 2.67385768741637
		0.0909309444340689 2.80062939897801
		0.079500618682177 2.83949273697338
		0.0689683819859384 2.85627216402254
		0.0592886586969529 2.74852354293771
		0.0504154559946507 2.849907531014
		0.0423068008807361 2.6597914152599
		0.0349229534786793 2.78097320970593
		0.0282309292432789 2.89846922127333
		0.0222301065172237 2.74188403202222
		0.0169402759458895 2.79074940739314
		0.0123952976938558 2.58556030663619
		0.0086231335470619 2.61516465721406
		0.00562011897113028 2.64279032666658
		0.00335728418515872 2.73244098635823
		0.00178940547338047 2.68303882941062
		0.000896819941831589 2.8960784431899
		};
		\addlegendentry{inboard}

		\addplot [ultra thick, black]
		table {%
		0 3.23144574998695
		0.02 3.22958211364861
		0.04 3.22753193780109
		0.06 3.22527668767502
		0.08 3.22279601567951
		0.1 3.22006759017048
		0.12 3.21706690933247
		0.14 3.21376709915817
		0.16 3.21013869450387
		0.18 3.2061494022121
		0.2 3.20176384532832
		0.22 3.19694328750446
		0.24 3.19164533678434
		0.26 3.18582362811509
		0.28 3.17942748413343
		0.3 3.17240155404967
		0.32 3.16468543080883
		0.34 3.1562132471641
		0.36 3.1469132518714
		0.38 3.13670736792557
		0.4 3.125510735632
		0.42 3.11323124436696
		0.44 3.09976905815169
		0.46 3.08501614167737
		0.48 3.06885579519557
		0.5 3.05116220875814
		0.52 3.03180004867036
		0.54 3.01062409172486
		0.56 2.98747892581286
		0.58 2.96219873884757
		0.6 2.93460722154404
		0.62 2.90451761341259
		0.64 2.87173292523044
		0.66 2.83604637510365
		0.68 2.79724207880461
		0.7 2.75509603808755
		0.72 2.7093774727895
		0.74 2.65985054327993
		0.76 2.60627650871442
		0.78 2.54841636299379
		0.8 2.48603398369911
		0.82 2.41889981892597
		0.84 2.34679512228277
		0.86 2.26951672687137
		0.88 2.18688232457275
		0.9 2.09873618747855
		0.92 2.00495523435046
		0.94 1.90545530762274
		0.96 1.80019748741272
		0.98 1.68919423069071
		1 1.57251508927678
		1.02 1.45029173330853
		1.04 1.32272199118677
		1.06 1.1900726166042
		1.08 1.05268051139276
		1.1 0.910952171798283
		1.12 0.765361186000117
		1.14 0.616443690785747
		1.16 0.464791791539114
		1.18 0.311045056150579
		1.2 0.155880302312179
		1.22 0
		1.4 0
		};
		\addlegendentry{fit}

		\end{axis}

		\end{tikzpicture}

	\caption{Electron density $n_e$ data from the Thomson scattering diagnostic as a function of normalised poloidal flux (blue circles and red triangles, denoting measurements on the outboard and inboard respectively), hyperbolic tangent fit (black line). The fit is less good when $\psi_{p,n} \lesssim 0.4$; the experimental density profile is hollow.	Fortunately, this is not a problem for the current work, since the beams studied in this paper do not enter that region.}
	\label{fig:density_fit}
	\end{figure}
	
\begin{figure}
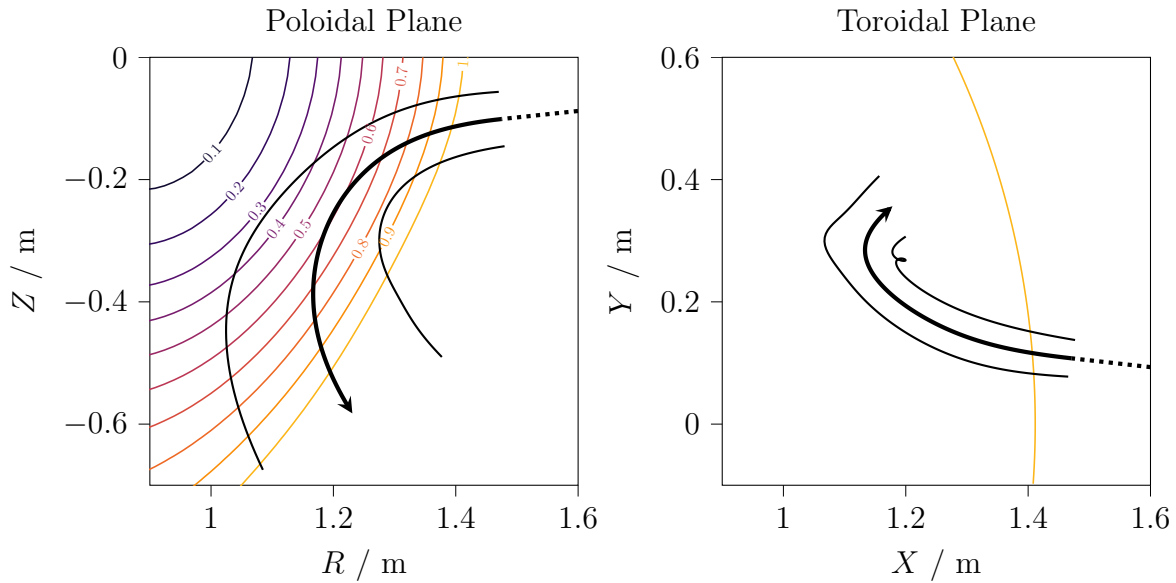
 
	\begin{subfigure}{.5\textwidth} 
		\centering


		\end{subfigure}
			
		\caption{Beam propagation through MAST plasma, shot 29908 at 190ms. Poloidal (left) and toroidal (right) cross-sections. The orange line in the right figure shows the radial position of the last closed flux surface at the midplane. Thin solid black lines give the $1/\textrm{e}$ positions of the Gaussian beam's electric field and the dotted lines show the ray's propagation in vacuum. The beam is launched from the steering mirror, located on the midplane at $R = 2.4 \textrm{m}$, which is out of the graph's boundaries. Thick solid black lines indicate trajectory of the central ray and the arrow the direction of propagation. Notice that the thin solid line loops upon itself in the figure on the right: this is not an artefact but the result of the beam widening as it turns and travels a significant distance in the poloidal plane, while not propagating quite as far in the toroidal plane.}
		\label{fig:ray_in_plasma}
	\end{figure}

	We launch a circular beam, at $q_R = 2.44\textrm{m}$ and $q_Z = 0\textrm{m}$, with $\varphi_p = 6^{\circ}$, $\varphi_t = -6.4^{\circ}$, $R_{b,x} = R_{b,y} = -72.8\textrm{cm}$, and $W_{x} = W_{y} = 3.97\textrm{cm}$. The launch beam's width and curvature were obtained from E-plane measurements of the 50 GHz beam of the MAST DBS; this is a focusing rather than a diverging beam, with the beam waist roughly located at the MAST port window \cite{Hillesheim:DBS_MAST:2015}.
	This beam propagates through the plasma as given in Figure \ref{fig:ray_in_plasma}. 

	For our analytical equilibrium, we use a large aspect ratio, circular flux surface equilibrium. The toroidal component is given by
	\begin{equation}
		B_\zeta = B_{\zeta,a} \frac{ R_a }{ R },
	\end{equation} 
	where $R_a$ is the $R$ of the magnetic axis and $B_{\zeta,a}$ is the corresponding toroidal magnetic field at the said axis. The poloidal field inside the last closed flux surface, $(R - R_a)^2 + Z^2 = a^2$, where $a$ is the minor radius, is
	\begin{equation} 
		B_R = B_p \frac{Z}{\sqrt{\left( R-R_a \right)^2 + Z^2}},
	\end{equation}
	and
	\begin{equation}
		B_Z = B_p \frac{R-R_a}{\sqrt{\left( R-R_a \right)^2 + Z^2}}.
	\end{equation}
	Here, inside the last closed flux surface, we have
	\begin{equation}
		B_p = B_{p,a} \frac{\sqrt{ (R - R_a)^2 + Z^2 }}{a},
	\end{equation}
	where $B_{p,a}$ is the magnitude of the poloidal magnetic field on the last closed flux surface. Unless otherwise stated, we use $B_{p,a} = 0.1\textrm{T}$, $B_{\zeta,a} = 1\textrm{T}$, $R_a = 1.5\textrm{m}$, and $a = 0.5\textrm{m}$. The density profile was linear in $\sqrt{\psi_p}$, going from $n_e = 4 \times 10^{19}\textrm{m}^{-3}$ at $\sqrt{\psi_p} = 0$ to zero density at $\sqrt{\psi_p} = 1$. A circular beam is launched at $q_R = 2.2\textrm{m}$ and $q_Z = 0\textrm{m}$, with $\varphi_p = 6^{\circ}$, $\varphi_t = 0^{\circ}$, $R_{b,x} = R_{b,y} = -4.0\textrm{m}$, and $W_{x} = W_{y} = 4\textrm{cm}$.


	\section{Reciprocity theorem} \label{section_reciprocity}
	The exact expression for the scattered electric field $\mathbf{E}_s$ is complicated. Coherent scattering in fusion plasmas has been studied extensively in cases where the wave frequency is much larger than the plasma frequency \cite{Slusher:scattering:1980, Hutchinson:diagnostics:2002, Sheffield:scattering:2011}. Since the frequency of the DBS beam is close to the plasma frequency, refraction is significant, making analysis even more challenging. Fortunately, we know that the emitted and received patterns of the antenna are the same. To simplify the subsequent equations, we define the antenna surface to be a surface perperpendicular to the beam propagation and close to the physical antenna.	For the antenna to receive a wave, it must be of the same form as the emitted wave, shown in equation (\ref{eq:emit}), but time reversed. Since the received and emitted waves are travelling in opposite directions, their wavevectors and beamfront curvatures, given by $\textrm{Re} (\bm{\Psi})$, have opposite signs. The received and emitted waves must also have the same envelope, given by $\textrm{Im} (\bm{\Psi})$, to pass through the optics of the DBS system. The polarisation of the emitted wave must be the same as that of the received wave. For example, an antenna emitting right circularly polarised light must also receive right circularly polarised light. However, the direction of travel has changed, and thus the direction the polarisation is moving also changes, as one might expect from time reversal symmetry. The received polarisation is thus the complex conjugate of that emitted. Consequently, at the antenna, the received scattered electric field must satisfy
	\begin{equation} \label{eq:scattering_received_form}
		\mathbf{E}_{r,ant} = A_{r} \hat{\mathbf{e}}_{ant}^* \exp \left(
			- \frac{\rmi}{2} \mathbf{w} \cdot \bm{\Psi}_{w,ant}^* \cdot \mathbf{w}  
		\right) ,
	\end{equation}
	where we have assumed that the antenna is in a vacuum and that the probe beam is launched perpendicular to the antenna surface. The scattered wave will be detected by the antenna if and only if it is of the form in equation (\ref{eq:scattering_received_form}).

	To calculate $A_r$ for a given $\mathbf{E}_{s},$ we project the scattered electric field on the Gaussian beam mode, the lowest order mode of the Gauss-Hermite beams, which form an orthogonal basis \cite{Brooker:Optics_Textbook:2003, Goldsmith:Quasioptics_Textbook:1998}, 
	\begin{equation} \label{eq:A_r}
	\fl
		A_{r} = \pi^{-1} \left[ \det \left[\textrm{Im} \left( \bm{\Psi}_{w,ant} \right) \right] \right]^{\frac{1}{2}} \int_{ant} \left[ 
			\mathbf{E}_s \cdot \hat{\mathbf{e}}_{ant} \exp \left( 
				\frac{\rmi}{2} \mathbf{w} \cdot \bm{\Psi}_{w,ant} \cdot \mathbf{w} 
			\right) 
		\right] \rmd S.
	\end{equation}
	The $\pi^{-1} \left[ \det \left[\textrm{Im} \left( \bm{\Psi}_{w,ant} \right) \right] \right]^{1 / 2}$ piece is the reciprocal of what we would get if we evaluated the 2D Gaussian surface integral with standard contour integration methods, taking care to choose the signs of the roots carefully. We have assumed that the DBS optics produces a sufficiently good Gaussian beam, such that the contribution to the signal from higher order modes is negligible.

	We use reciprocity to obtain $A_r$ without calculating $\mathbf{E}_s$ in its entirety. The reciprocity theorem is a standard method of calculating the signal received by an antenna. However, it cannot be used in its usual form when the medium is a magnetised plasma. In order to deal with highly magnetised plasmas in tokamaks, some modifications have to be made. Specifically, the reciprocal beam has to be launched into a plasma which has its dielectric tensor transposed \cite{Villeneuve:reciprocity:1958, Piliya:reciprocity:2002, Gusakov:scattering_slab:2004, Gusakov:Reflectometry:1997}, which is the same as having its magnetic field reversed. The dielectric tensor has to be transposed to maintain time reversal symmetry. Like previous work on reciprocity, we use the superscript $^{(+)}$ to denote solutions in the medium with the transposed dielectric tensor, which is
	\begin{equation} \label{eq:Maxwell_reciprocity}
		\frac{c^2}{\Omega^2} \nabla \times (\nabla \times \mathbf{E}^{(+)}) = \bm{\epsilon}_{eq}^{T} \cdot \mathbf{E}^{(+)}.
	\end{equation}

	To obtain $A_r$ in equation (\ref{eq:A_r}), we contract equation (\ref{eq:Maxwell_tb}) with $\mathbf{E}^{(+)}$ ,
	\begin{equation}
		\frac{c^2}{\Omega^2} \mathbf{E}^{(+)} \cdot \nabla \times (\nabla \times \mathbf{E}_s )
		= \mathbf{E}^{(+)} \cdot \bm{\epsilon}_{eq} \cdot \mathbf{E}_s
		+ \mathbf{E}^{(+)} \cdot \bm{\epsilon}_{tb} \cdot \mathbf{E}_b ,
	\end{equation}
	and integrate by parts, using equation (\ref{eq:Maxwell_reciprocity}) to obtain
	\begin{equation}
	\eqalign{
		\frac{c^2}{\Omega^2}
		\nabla \cdot [(\nabla \times \mathbf{E}_{s}) \times \mathbf{E}^{(+)}
		- (\nabla \times \mathbf{E}^{(+)}) \times \mathbf{E}_{s}]
		= \mathbf{E}^{(+)} \cdot \bm{\epsilon}_{tb} \cdot \mathbf{E}_b .
	}
	\end{equation}
	Choosing the right surface and the right boundary condition for $\mathbf{E}^{(+)}$ on that surface, one can calculate $A_r$ from a volume integral of $\mathbf{E}^{(+)} \cdot \bm{\epsilon}_{tb} \cdot \mathbf{E}_{b}$,
	\begin{equation} \label{eq:reciprocity}
	\fl
	\eqalign{
		\int [(\nabla \times \mathbf{E}_{s}) \times \mathbf{E}^{(+)} + \mathbf{E}_{s} \times (\nabla \times \mathbf{E}^{(+)})] \cdot \rmd \mathbf{S}
		= \int \frac{\Omega^2}{c^2} \mathbf{E}^{(+)} \cdot \bm{\epsilon}_{tb} \cdot \mathbf{E}_{b} \ \rmd V .
	}
	\end{equation}
	We impose that $\mathbf{E}^{(+)}$ at the antenna is
	\begin{equation} \label{eq:R_antenna}
		\mathbf{E}^{(+)}_{ant}
		= \pi^{-1} \left\{
			\det \left[\textrm{Im} \left( \bm{\Psi}_{w,ant} \right)\right]
		\right\}^{\frac{1}{2}} \hat{\mathbf{e}}_{ant} \exp \left( 
			\frac{\rmi}{2} \mathbf{w} \cdot \bm{\Psi}_{w,ant} \cdot \mathbf{w} 
		\right) .
	\end{equation}
	
	The behaviour of $\mathbf{E}^{(+)}$ as it propagates into the transposed plasma is governed by the beam tracing equations. The beam's evolution is governed by the dispersion relation $\bm{D} \cdot \hat{\mathbf{e}} = 0$, where $\bm{D}$ was defined in equation (\ref{eq:D_definition}). When the dielectric tensor of the plasma is transposed, the dispersion relation is $\bm{D}^{T} \cdot \hat{\mathbf{e}}^{(+)} = 0$. The fact that $\bm{D}$ is Hermitian implies that $\bm{D}^{\dagger} \cdot \hat{\mathbf{e}} = 0$. Thus, $\bm{D}^{T} \cdot \hat{\mathbf{e}}^{*} = 0$, and we conclude that $\hat{\mathbf{e}}^{(+)} = \hat{\mathbf{e}}^*$. Intuitively, this makes sense, since transposing the dielectric tensor corresponds to switching the direction of the magnetic field; hence, left-handed polarisation in the transposed plasma behaves the same as right-handed polarisation in the physical plasma and vice versa. As such, the reciprocal beam $\mathbf{E}^{(+)}$ propagates like the probe beam $\mathbf{E}_b$, except for its polarisation being complex conjugated. Since $\mathbf{E}^{(+)}$ at the antenna, equation (\ref{eq:R_antenna}), has the same form as $\mathbf{E}_b$ at the antenna and follows the same set of evolution equations, except for the complex conjugated polarisation, the reciprocal beam is
	\begin{equation} \label{eq:reciprocal_field_final}
	\fl
	\eqalign{
		\mathbf{E}^{(+)}
		&= \pi^{-1} 
		\sum_{i = O, X}
		\alpha_i
		\left\{
			\det \left[ \textrm{Im} \left( \bm{\Psi}_{w,ant} \right) \right]
		\right\}^{\frac{1}{4}} \left\{
			\det \left[ \textrm{Im} \left( \bm{\Psi}_{w,i} \right) \right]
		\right\}^{\frac{1}{4}} \left( \frac{g_{ant}}{g_i} \right)^{\frac{1}{2}} \\
		&\times \hat{\mathbf{e}}^*_i \exp (\rmi \phi_{G,i} - \rmi \phi_{P,i} ) \exp \left(\rmi s_i + \rmi \mathbf{K}_{w,i} \cdot \mathbf{w}_i + \frac{\rmi}{2} \mathbf{w}_i \cdot \bm{\Psi}_{w,i} \cdot \mathbf{w}_i \right) ,
	}
	\end{equation}
	where the coefficients $\alpha_i$ give the relative amplitudes of the O and X modes. We remark that $\phi_G$ does not change sign, unlike $\phi_P$. We now match the expression for the reciprocal electric field, equation (\ref{eq:reciprocal_field_final}), with its initial condition, equation (\ref{eq:R_antenna}), giving
	\begin{equation} \label{eq:R_antenna_matching}
		\hat{\mathbf{e}}_{ant}
		=
		\alpha_{O,ant} \hat{\mathbf{e}}_{O,ant}^*
		+
		\alpha_{X,ant} \hat{\mathbf{e}}_{X,ant}^* .
	\end{equation}
	Since the dispersion relation $\bm{D}$ is Hermitian, its eigenvectors are orthogonal: $\hat{\mathbf{e}}_X \cdot \hat{\mathbf{e}}^*_O  = 0$ and $\hat{\mathbf{e}}_O \cdot \hat{\mathbf{e}}^*_X = 0$. We contract equation (\ref{eq:R_antenna_matching}) with $\hat{\mathbf{e}}^*_O$ and $\hat{\mathbf{e}}^*_X$ in turn, finding
	\begin{equation} \label{eq:alpha_O}
		\alpha_{O} = \hat{\mathbf{e}}_{ant} \cdot \hat{\mathbf{e}}_{O,ant} ,
	\end{equation}
	and
	\begin{equation} \label{eq:alpha_X}
		\alpha_{X} = \hat{\mathbf{e}}_{ant} \cdot \hat{\mathbf{e}}_{X,ant} . 
	\end{equation}
	Having found the coefficients for the O and X modes, we now explicitly have an expression for $\mathbf{E}^{(+)}$. 
	
	We proceed by evaluating the integral on the left side of equation (\ref{eq:reciprocity}) over a surface which both contains the antenna and is far enough from the plasma that we can write $\mathbf{E}_s$ as a summation of plane waves,
	\begin{equation}
		\mathbf{E}_{s} 
		= \sum_{\mathbf{K}_{s}} \mathbf{A}_{s} \exp (\rmi \mathbf{K}_s \cdot \mathbf{r}),
	\end{equation}
	which we can do because the antenna is typically situated away from the plasma. Due to the vacuum dispersion relation, the scattered microwaves at the antenna have almost the same wavenumber as that emitted, $\Omega / c$, at the antenna. The integrand of the surface integral in equation (\ref{eq:reciprocity}) is
	\begin{equation}
	\eqalign{
		&(\nabla \times \mathbf{E}_{s}) \times \mathbf{E}^{(+)}
		+ \mathbf{E}_{s} \times (\nabla \times \mathbf{E}^{(+)}) \\
		&\qquad \simeq 
		\sum_{\mathbf{K}_{s}} \left[
		\rmi (\mathbf{K}_{s} \cdot \mathbf{E}^{(+)} ) \mathbf{A}_{s}
		- \rmi (\mathbf{A}_{s} \cdot \mathbf{E}^{(+)} ) \mathbf{K}_{s}
		\right] \exp (\rmi \mathbf{K}_s \cdot \mathbf{r}) \\
		&\qquad + \sum_{\mathbf{K}_{s}} \left[
		\rmi (\mathbf{A}_{s} \cdot \mathbf{E}^{(+)} ) \mathbf{K}
		- \rmi (\mathbf{K} \cdot \mathbf{A}_{s} ) \mathbf{E}^{(+)}
		\right] \exp (\rmi \mathbf{K}_s \cdot \mathbf{r})
		, }
	\end{equation}
	where we have used $\nabla \mathbf{E}^{(+)} \simeq \rmi \mathbf{K} \mathbf{E}^{(+)}$. 
	At the antenna, $\mathbf{K}$ is normal to the surface, and as a result its electric field is parallel to the surface. Moreover, at the antenna, $\mathbf{E}^{(+)} \propto \exp ( \rmi K_g \hat{\mathbf{g}} \cdot \mathbf{r} )$, and any term that contains an $\mathbf{A}_s$ with $\mathbf{K}_s$ not exactly perpendicular to the surface of the antenna integrates to zero. Therefore, $\int \sum_{\mathbf{K}_{s}} (\mathbf{K}_{s} \cdot \mathbf{E}^{(+)}) [\mathbf{A}_s \exp (\rmi \mathbf{K}_s \cdot \mathbf{r})] \cdot \rmd \mathbf{S} = 0$. Similarly, $\int (\mathbf{K} \cdot \mathbf{E}_{s}) \mathbf{E}^{(+)} \cdot \rmd \mathbf{S} = 0$. The scattered waves are always travelling out of the plasma; the probe beam is also travelling out of the plasma at all points, except at the antenna. Consequently, the terms that contain $\mathbf{A}_s \cdot \mathbf{E}^{(+)}$ cancel at all points other than on the antenna. The surface integral in equation (\ref{eq:reciprocity}) is thus 
	\begin{equation} \label{eq:reciprocity_amplitude}
	\fl
	\eqalign{
		\int [(\nabla \times \mathbf{E}_{s}) \times \mathbf{E}^{(+)} + \mathbf{E}_{s} \times (\nabla \times \mathbf{E}^{(+)})] \cdot \rmd \mathbf{S} 
		& = - \int_{ant} 2 \rmi \frac{\Omega}{c} \mathbf{E}_{s} \cdot \mathbf{E}^{(+)}   \rmd S \\ 
		& = - 2 \rmi \frac{\Omega}{c} A_r .
	}	
	\end{equation}
	Here we have used the fact that the wavenumbers are $\Omega^{} / c$ since the antenna is in vacuum. Hence, we get the reciprocity relation
	\begin{equation} \label{eq:reciprocity_theorem}
		A_r 
		=\frac{\Omega \rmi}{ 2 c } \int \mathbf{E}^{(+)} \cdot \bm{\epsilon}_{tb} \cdot \mathbf{E}_{b} \ \rmd V .
	\end{equation}
	
	Before we conclude this section, we first show that this equation can be further simplified. We use the microwave and reciprocal electric fields, as well as the linearised dielectric tensor, in the reciprocity theorem. Substituting equations (\ref{eq:density_tb}), (\ref{eq:beam_field_final}), (\ref{eq:reciprocal_field_final}), (\ref{eq:alpha_O}), and (\ref{eq:alpha_X}) into the volume integral of the reciprocity theorem, equation (\ref{eq:reciprocity_theorem}), we get
	\begin{equation} \label{eq:reciprocity_theorem_explicit}
		\fl
	\eqalign{
		A_r
		&= \frac{ \rmi \Omega A_{ant} }{ 2 \pi c } \int \sum_{i = O, X}  
		\hat{\mathbf{e}}_{ant} \cdot \hat{\mathbf{e}}_{i,ant}
		\left\{
			\det \left[ \textrm{Im} \left( \bm{\Psi}_{w,i} \right) \right]
		\right\}^{\frac{1}{4}}  \left(
			\frac{g_{ant}}{g_i}
		\right)^{\frac{1}{2}} \\
		&\times \exp (\rmi \phi_{G,i} - \rmi \phi_{P,i} ) \exp \left(\rmi s_i + \rmi \mathbf{K}_{w,i} \cdot \mathbf{w}_i + \frac{\rmi}{2} \mathbf{w}_i \cdot \bm{\Psi}_{w,i} \cdot \mathbf{w}_i \right)  \\
		& \times \frac{\delta n_e}{n_e} 
		\hat{\mathbf{e}}^*_i \cdot (\bm{\epsilon}_{eq} -\bm{1}) \cdot \hat{\mathbf{e}}
		\ \exp(\rmi \phi_G + \rmi \phi_P) \left\{
			\det \left[ \textrm{Im} \left( \bm{\Psi}_{w} \right) \right]
		\right\}^{\frac{1}{4}}
		\left(
			\frac{g_{ant}}{g}
		\right)^{\frac{1}{2}} \\
		&\times
		\exp \left(
			\rmi s 
			+ \rmi \mathbf{K}_w \cdot \mathbf{w} 
			+ \frac{\rmi}{2} \mathbf{w} \cdot \bm{\Psi}_w \cdot \mathbf{w}  
		\right)
		\ \rmd V.
	}
	\end{equation}
	Here we have assumed that the polarisation of the probe beam is appropriately well-matched to the pitch angle at the plasma edge, such that only one mode, either O or X, propagates into the plasma. If this assumption does not hold, then one should calculate $A_r$ for each probe beam mode separately and add the two together. To simplify the above expression, equation (\ref{eq:reciprocity_theorem_explicit}), and thus the subsequent algebra, we first make an argument about the contribution of the reciprocal electric field to the backscattered amplitude. 
	
	In general, the O and X modes take different paths through the plasma, even if they enter the plasma at the same point. For a given polarisation of the probe beam $\mathbf{E}_b$, the reciprocal beam with the same polarisation will follow the same path as the probe beam, and the reciprocal beam with the opposite polarisation follows a different path. As a result, the reciprocal beam with the same polarisation as the probe beam overlaps with the probe beam over a volume of order $W^2 L$, whereas the reciprocal beam with the opposite polarisation in general overlaps only over a small volume $W^3$. Since the contribution to the integral from the reciprocal beam with the opposite polarisation to the probe beam is small by $W / L \ll 1$, equation (\ref{eq:reciprocity_theorem_explicit}) simplifies to
	\begin{equation} \label{eq:mismatch_first}
	\eqalign{
		A_r
		&= \frac{ \rmi \Omega A_{ant} }{ 2 \pi c } \hat{\mathbf{e}}_{ant} \cdot \hat{\mathbf{e}}_{ant} \int [\det[\textrm{Im} (\bm{\Psi}_w) ]]^{\frac{1}{2}} \frac{g_{ant}}{g} \exp (2 \rmi \phi_G ) \\
		& \times \frac{\delta n_e}{n_e} \hat{\mathbf{e}}^*
		\cdot (\bm{\epsilon}_{eq} -\bm{1}) \cdot \hat{\mathbf{e}}
		\ \exp(2 \rmi s + 2 \rmi \mathbf{K}_w \cdot \mathbf{w} + \rmi \mathbf{w} \cdot \bm{\Psi}_w \cdot \mathbf{w} ) \ \rmd V.
	}
	\end{equation}
	This is the form of the backscattered amplitude that we subsequently use.

	\section{Assumptions about turbulent fluctuations} \label{section_turbulence}
	We see that we need to evaluate the three spatial integrals in equation (\ref{eq:mismatch_first}) to determine the backscattered signal. In order to do this, we need to make some assumptions about the nature of the turbulent fluctuations $\delta n_e$.

	In this work, we take the plasma to be in steady state, such that the equilibrium electron density $n_e$ has no time dependence, whereas we assume that the turbulent fluctuations $\delta n_e$ can indeed have a fast time dependence $t$ in addition to a spatial dependence. This time-dependence will be important in Section \ref{subsection:correlation_function}. For now, we will concentrate on the spatial properties $\delta n_e$.

\begin{figure}
		\centering
		\begin{tikzpicture}[>=latex]
		\draw[style=help lines] (0,0) (3,2);
		
		\coordinate (vec1) at (90:2.5);
		\coordinate (vec2) at (0:2.5);
		\coordinate (vec3) at (200:2.5);
		\coordinate (vec4) at (110:2.5);
		\coordinate (vec5) at (140:2.5);
		\coordinate (vec6) at (180:2.5);
		
		\draw[->,thick,black] (0,0) -- (vec1) node[left] {$\hat{\mathbf{g}}$};
		\draw[->,thick,black] (0,0) -- (vec2) node[right] {$\hat{\mathbf{x}}$};
		\draw[->,thick,blue] (0,0) -- (vec3) node [below] {$\hat{\mathbf{b}}$};
		\draw[->,thick,blue] (0,0) -- (vec4) node [left] {$\hat{\mathbf{u}}_{1}$};
		\draw[->,thick,red] (0,0) -- (vec5) node [left] {$\mathbf{K}$};
		\draw[dashed,thick,black] (0,0) -- (vec6) ;
		
		\draw [->,black,thick,domain=90:110] plot ({1.5*cos(\x)}, {1.5*sin(\x)}) node[above right] {$\theta$};
		\draw [->,black,thick,domain=110:140] plot ({1.0*cos(\x)}, {1.0*sin(\x)});
		\node[above] at (-0.75,0.75) {$\theta_m$};
		\draw [->,black,thick,domain=180:200] plot ({1.5*cos(\x)}, {1.5*sin(\x)});
		\node[below] at (-1.65,0.0) {$\theta$};
		
		\draw (0,.3)-|(.3,0);
		\draw [blue] ({0.3*cos(200)}, {0.3*sin(200)}) -- ({0.3*cos(200)-0.3*cos(70)}, {0.3*sin(200)+0.3*sin(70)}) -- ({0.3*cos(110)}, {0.3*sin(110)});
		
		\draw[blue,thick] (0,-0.5) circle (0.2cm) node[below right] {$\hat{\mathbf{u}}_{2} = \hat{\mathbf{y}}$};
		\draw[blue,thick] (-0.14,-0.64) -- (0.14,-0.36);
		\draw[blue,thick] (0.14,-0.64) -- (-0.14,-0.36);
		
		\end{tikzpicture} \\
		
		\caption{Bases for $\mathbf{k}_\perp$ and $\mathbf{w}$, with $\theta > 0$ and $\theta_m > 0$.} 
		\label{fig:basis}
	\end{figure}
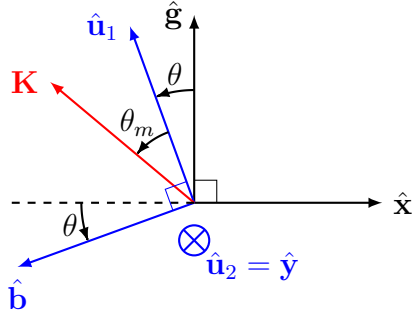

	We consider a tubular region of space around the Gaussian beam as it propagates through the plasma. This region of space is elongated along the ray, and across the ray it is several times the width of the Gaussian beam, such that the beam's electric field is effectively zero on the boundary of this region. We Fourier analyse $\delta n_e$ in this volume. We use the usual assumptions for turbulent fluctuations \cite{Catto:GK:1978,Frieman:GK:1982}, considering electron density fluctuations $\delta n_e$ with very large gradients across the magnetic field and small gradients along it. Hence, we define coordinates aligned with the magnetic field. We use $u_\parallel$, the arc length along magnetic field lines, and two variables $u_1$ and $u_2$ that both satisfy 
	\begin{equation}
		\hat{\mathbf{b}} (\mathbf{q}(\tau) + \mathbf{w}) \cdot \nabla u_i
		= 0 .
	\end{equation}
	With these variables, we get
	\begin{equation} \label{eq:n_FT} 
	\fl
		\delta n_e (\mathbf{r}, t)
		=
		\int
		\delta \tilde{n}_e ( k_{\perp,1}, k_{\perp,2}, u_\parallel, \omega )
		\exp \left( \rmi k_{\perp,1} u_1 + \rmi k_{\perp,2} u_2 - \rmi \omega t \right)
		\rmd k_{\perp,1} \ \rmd k_{\perp,2} \ \rmd \omega,
	\end{equation}
	where $\omega$ has contributions from both the angular frequency of the turbulence in the plasma's frame and the Doppler shift due to the moving plasma, and 
	\begin{equation} \label{eq:size_of_k_perp}
		k_{\perp,\alpha} \sim \frac{1}{\lambda} \gg k_\parallel \sim \frac{1}{L} \sim \frac{\partial}{\partial u_\parallel}
	\end{equation}
	are the components of the turbulence wavevector perpendicular to the magnetic field. We define $u_1$ and $u_2$ such that their gradients are perpendicular to each other at $\mathbf{w} = 0$ (on the central ray). They are only perpendicular to each other along the central ray, and not when we move away from it; this is a consequence of magnetic shear. Any vector perpendicular to the magnetic field is resolved into two directions: $\hat{\mathbf{u}}_{1} = \nabla u_1 (\mathbf{w}=\mathbf{0})$ and $\hat{\mathbf{u}}_{2} = \nabla u_2 (\mathbf{w}=\mathbf{0})$. The subscripts $_1$ and $_2$ indicate projection on these directions, respectively. We choose $\hat{\mathbf{u}}_{1}$ and $\hat{\mathbf{u}}_{2}$ as follows: $\hat{\mathbf{u}}_{1}$ is in the plane of $\hat{\mathbf{g}}$ and $\hat{\mathbf{b}}$, while $\hat{\mathbf{u}}_{2}$ is perpendicular to $\hat{\mathbf{g}}$, and both of them are perpendicular to $\hat{\mathbf{b}}$,
	\begin{equation} \label{eq:kperp1}
		\hat{\mathbf{u}}_{1} 
		= \frac{ (\hat{\mathbf{b}} \times \hat{\mathbf{g}}) \times \hat{\mathbf{b}} }{ |(\hat{\mathbf{b}} \times \hat{\mathbf{g}}) \times \hat{\mathbf{b}}| },
	\end{equation}
	and
	\begin{equation} \label{eq:kperp2}
		\hat{\mathbf{u}}_{2} 
		= \frac{ \hat{\mathbf{b}} \times \hat{\mathbf{g}} }{ | \hat{\mathbf{b}} \times \hat{\mathbf{g}} | }.
	\end{equation}
	Here we take $\hat{\mathbf{b}}$ to be a shorthand for $\hat{\mathbf{b}} (\mathbf{q}(\tau)) = \hat{\mathbf{b}} (\tau)$. That is, $\hat{\mathbf{b}}$ is the unit vector of the magnetic field on the central ray. Note that if $\hat{\mathbf{g}}$ and $\hat{\mathbf{b}}$ are perpendicular to each other (that is, if there is no mismatch, as we later show), $\hat{\mathbf{u}}_{1} = \hat{\mathbf{g}}$ (Figure \ref{fig:basis}). Note that these $\hat{\mathbf{u}}_{1}$ and $\hat{\mathbf{u}}_{2}$ are the same as those used in Section \ref{subsection_dispersion}, as we will prove in equation (\ref{eq:K_y}).
	We align the basis for $\mathbf{w}$, see equation (\ref{eq:coordinates}), with the basis for $\mathbf{k}_\perp$. We choose $\hat{\mathbf{y}} = \hat{\mathbf{u}}_{2}$ and denote projection in that direction with the subscript $_y$. The other basis vector for $\mathbf{w}$, which is perpendicular to both $\hat{\mathbf{g}}$ and $\hat{\mathbf{y}}$, will be
	\begin{equation}
	\hat{\mathbf{x}}
	= \frac{ \hat{\mathbf{y}} \times \hat{\mathbf{g}} }{ |\hat{\mathbf{y}} \times \hat{\mathbf{g}} | }
	= \frac{ \hat{\mathbf{u}}_{2} \times \hat{\mathbf{g}} }{ |\hat{\mathbf{u}}_{2} \times \hat{\mathbf{g}} | }.
	\end{equation}
	
	Based on the bases for $\mathbf{k}_\perp$ and $\mathbf{w}$, we define the angle $\theta$ such that
	\begin{equation} \label{eq:theta}
		\cos \theta = \hat{\mathbf{g}} \cdot \hat{\mathbf{u}}_{1},
	\end{equation}
	and
	\begin{equation}
		\hat{\mathbf{x}} \cdot \hat{\mathbf{u}}_{1} = - \sin \theta .
	\end{equation}
	This angle $\theta$ is not the mismatch angle, but is of the same order as the mismatch angle, as we will later prove. The mismatch angle is
	\begin{equation}
		\sin \theta_m = \hat{\mathbf{K}} \cdot \hat{\mathbf{b}}.
	\end{equation}
	
	Using the definitions of $\hat{\mathbf{u}}_{1} = \nabla u_1 (\mathbf{w}=\mathbf{0})$ and $\hat{\mathbf{u}}_{2} = \nabla u_2 (\mathbf{w}=\mathbf{0})$ above, we derive the coordinates $u_1$ and $u_2$ in \ref{appendix_derivation_u1_u2}. We summarise the results here. They are
	\begin{equation} \label{eq:u1_final}
	\fl
	\eqalign{
		u_1 
		&= \int_{0}^{\tau} g (\tau') \cos \theta (\tau') \ \rmd \tau
		- w_x \sin \theta \\
		&+ \frac{w_x^2}{2} \left(
			\frac{\sin \theta}{g} \frac{\rmd \theta}{\rmd \tau} 
			- \boldsymbol{\kappa} \cdot \hat{\mathbf{x}} \sin \theta
			+ \hat{\mathbf{x}} \cdot \nabla \hat{\mathbf{b}} \cdot \hat{\mathbf{g}}
			- \hat{\mathbf{x}} \cdot \nabla \hat{\mathbf{b}} \cdot \hat{\mathbf{x}} \tan \theta 
		\right) \\
		&+ w_x w_y \left(
			- \boldsymbol{\kappa} \cdot \hat{\mathbf{y}} \sin \theta
			+ \hat{\mathbf{y}} \cdot \nabla \hat{\mathbf{b}} \cdot \hat{\mathbf{g}}
			+ \frac{ \sin \theta \tan \theta}{g} \frac{\rmd \hat{\mathbf{x}}}{\rmd \tau} \cdot \hat{\mathbf{y}}
			- \hat{\mathbf{y}} \cdot \nabla \hat{\mathbf{b}} \cdot \hat{\mathbf{x}} \tan \theta
		\right) ,
	}
	\end{equation}
	and
	\begin{equation} \label{eq:u2_final}
	\eqalign{
		u_2 
		&= w_y 
		+ \frac{w_x^2}{2} \left(
			\frac{
				\tan \theta
			}{
				g
			} \frac{\rmd \hat{\mathbf{x}}}{\rmd \tau} \cdot \hat{\mathbf{y}}
			+
			\frac{
				\hat{\mathbf{x}} \cdot \nabla \hat{\mathbf{b}} \cdot \hat{\mathbf{y}}
			}{
				\cos \theta
			}
		\right)
		+ w_x w_y \frac{
			\hat{\mathbf{y}} \cdot \nabla \hat{\mathbf{b}} \cdot \hat{\mathbf{y}}
		}{
			\cos \theta
		} .
	}
	\end{equation}
	Here,
	\begin{equation}
		\boldsymbol{\kappa} = \frac{1}{g} \frac{\rmd \hat{\mathbf{g}}}{\rmd \tau}
	\end{equation}
	is the curvature of the central ray. This ray curvature should not be confused with the wavefront curvature, $1 / R_b$. Note that we have kept corrections to $u_1$ and $u_2$ to order $\lambda$. Finally, the arc length along the magnetic field is
	\begin{equation} \label{upar_final} 
		u_\parallel
		\simeq - \int_0^\tau g (\tau') \sin \theta(\tau') \rmd \tau'
		- w_x \cos \theta ,
	\end{equation}
	where we have neglected terms that are small in $W / L$. Since $k_\parallel \sim 1 / L$, we have $k_\parallel u_\parallel \sim 1$, and the higher order contributions to $u_\parallel$ are not required.

	\section{Backscattered electric field: general} \label{section_backscattered_E}
	We proceed to evaluate the amplitude of the backscattered electric field, $A_r$. Writing the volume element in equation (\ref{eq:mismatch_first}) as
	\begin{equation}
		\rmd V = g \ \rmd w_x \ \rmd w_y \ \rmd \tau ,
	\end{equation}
	and substituting equations (\ref{eq:n_FT}), (\ref{eq:u1_final}), and (\ref{eq:u2_final}) into equation (\ref{eq:mismatch_first}), we get
	\begin{equation} \label{eq:mismatch_AfterFT}
	\fl
	\eqalign{
		A_r
		&= \frac{\rmi \Omega A_{ant} g_{ant} \hat{\mathbf{e}}_{ant} \cdot \hat{\mathbf{e}}_{ant}}{2 \pi c} \int [\det[\textrm{Im} (\bm{\Psi}_w) ]]^{\frac{1}{2}} \exp (2 \rmi \phi_G ) \exp \left( - \rmi \omega t \right) \\
		& \times \frac{ \delta \tilde{n}_e }{n_e} \ \hat{\mathbf{e}}^*
		\cdot (\bm{\epsilon}_{eq} -\bm{1}) \cdot \hat{\mathbf{e}}
		\ \exp \left(
			2 \rmi s + \rmi k_{\perp,1} \int_0^\tau g (\tau') \cos \theta (\tau') \ \rmd \tau' 
		\right) \\
		& \times \exp 
		\left[ 
			\rmi (2 \mathbf{K}_w + \mathbf{k}_{\perp,w} ) \cdot \mathbf{w} 
			+ \rmi \mathbf{w} \cdot \bm{M}_w \cdot \mathbf{w} 
		\right]
		\ \rmd w_x \ \rmd w_y \ \rmd \tau \ \rmd k_{\perp,1} \ \rmd k_{\perp,2} \ \rmd \omega.
	}
	\end{equation} 
	Here $\mathbf{k}_{\perp} = k_{\perp,1} \hat{\mathbf{u}}_{1} + k_{\perp,2} \hat{\mathbf{u}}_{2}$, $\mathbf{k}_{\perp,w}$ is $\mathbf{k}_{\perp}$ projected on the plane perpendicular to the group velocity, and $\bm{M}_w$ is the symmetric modified $\bm{\Psi}_w$ matrix, given by
	\begin{eqnarray}
	\bm{M}_w =
	\left(
	\begin{array}{ccc}
	M_{xx} & M_{xy} & 0\\
	M_{xy} & M_{yy} & 0\\
	0 & 0 & 0\\

	\end{array}
	\right) ,
	\end{eqnarray}
	where
	\begin{equation}
	\fl
	\eqalign{
		M_{xx} = \Psi_{xx}
		&+ \frac{k_{\perp,1}}{2} \left(
			\frac{\sin \theta}{g} \frac{\rmd \theta}{\rmd \tau}
			- \boldsymbol{\kappa} \cdot \hat{\mathbf{x}} \sin \theta
			+ \hat{\mathbf{x}} \cdot \nabla \hat{\mathbf{b}} \cdot \hat{\mathbf{g}}
			- \hat{\mathbf{x}} \cdot \nabla \hat{\mathbf{b}} \cdot \hat{\mathbf{x}} \tan \theta
		\right) \\
		&+ \frac{k_{\perp,2}}{2} \left(
			\frac{\tan \theta}{g} \frac{\rmd \hat{\mathbf{x}}}{\rmd \tau} \cdot \hat{\mathbf{y}}
			+  \frac{\hat{\mathbf{x}} \cdot \nabla \hat{\mathbf{b}} \cdot \hat{\mathbf{y}} }{\cos \theta}
		\right) ,
	}
	\end{equation}
	\begin{equation}
	\fl
	\eqalign{
		M_{xy} = \Psi_{xy}
		&+ \frac{k_{\perp,1}}{2} \left(
			- \boldsymbol{\kappa} \cdot \hat{\mathbf{y}} \sin \theta
			+ \hat{\mathbf{y}} \cdot \nabla \hat{\mathbf{b}} \cdot \hat{\mathbf{g}}
			+ \frac{\sin \theta \tan \theta}{g} \frac{\rmd \hat{\mathbf{x}}}{\rmd \tau} \cdot \hat{\mathbf{y}}
			- \hat{\mathbf{y}} \cdot \nabla \hat{\mathbf{b}} \cdot \hat{\mathbf{x}} \tan \theta
		\right) \\
		&+ \frac{k_{\perp,2}}{2}
		\frac{\hat{\mathbf{y}} \cdot \nabla \hat{\mathbf{b}} \cdot \hat{\mathbf{y}}}{\cos \theta}
		,
	}
	\end{equation}
	and
	\begin{equation}
	\fl
		M_{yy} = \Psi_{yy}.
	\end{equation}
	It is worth noting a few points about this modified matrix. First, the modifications to $\bm{\Psi}_w$ are only to its real part, that is, the part associated with curvature (as opposed to width). Secondly, that these modifications depend directly on the curvature of the magnetic field and the magnetic shear; the curvature of the cut-off surface does not explicitly enter the corrections. As we will see in Section \ref{section_wavenumber_resolution}, this affects the wavenumber resolution, and our model gives a different result from widely-cited earlier work \cite{Hirsch:DBS:2001, Bulanin:spatial_spectral_resolution:2006}. 
	In Scotty, we calculate the gradients of $\hat{\mathbf{b}}$ in cylindrical coordinates,
	\begin{equation}
	\eqalign{
		\nabla \hat{\mathbf{b}}
		&= \frac{\partial b_R}{\partial R} \hat{\mathbf{R}} \hat{\mathbf{R}}
		+ \frac{\partial b_R}{\partial Z} \hat{\mathbf{Z}} \hat{\mathbf{R}}
		+ b_R \frac{\hat{\boldsymbol{\zeta}} \hat{\boldsymbol{\zeta}}}{R}
		+ \frac{\partial b_Z}{\partial R} \hat{\mathbf{R}} \hat{\mathbf{Z}}
		+ \frac{\partial b_Z}{\partial Z} \hat{\mathbf{Z}} \hat{\mathbf{Z}} \\
		&+ \frac{\partial b_\zeta}{\partial R} \hat{\mathbf{R}} \hat{\boldsymbol{\zeta}}
		+ \frac{\partial b_\zeta}{\partial Z} \hat{\mathbf{Z}} \hat{\boldsymbol{\zeta}}
		- b_\zeta \frac{ \hat{\boldsymbol{\zeta}} \hat{\mathbf{R}} }{R} .		
	}	
	\end{equation}	
	To get this result, one should remember that the $\hat{\mathbf{R}}$ and $\hat{\boldsymbol{\zeta}}$ basis vectors depend on position. That is, we have used
	\begin{equation}
		\nabla \hat{\mathbf{R}}
		= 
		\frac{\hat{\boldsymbol{\zeta}} \hat{\boldsymbol{\zeta}}}{R},
	\end{equation}
	and
	\begin{equation}
	\nabla \hat{\boldsymbol{\zeta}}
	=
	- \frac{\hat{\boldsymbol{\zeta}} \hat{\mathbf{R}}}{R}.
	\end{equation}	
	
	What the DBS community actually uses is not $A_r (t)$ directly, but its Fourier transform $\tilde{A}_r (\omega) = (2 \pi)^{-1} \int A_r (t) \exp (\rmi \omega t) \ \rmd t $, which is
	\begin{equation} \label{eq:tilde_A_r}
	\fl
	\eqalign{
		\tilde{A}_r (\omega)
		&= \frac{\rmi \Omega A_{ant} g_{ant} \hat{\mathbf{e}}_{ant} \cdot \hat{\mathbf{e}}_{ant}}{2 \pi c} \int [\det[\textrm{Im} (\bm{\Psi}_w) ]]^{\frac{1}{2}} \exp (2 \rmi \phi_G )  \\
		& \times \frac{ \delta \tilde{n}_e }{n_e} \ \hat{\mathbf{e}}^*
		\cdot (\bm{\epsilon}_{eq} -\bm{1}) \cdot \hat{\mathbf{e}}
		\ \exp \left(
			2 \rmi s + \rmi k_{\perp,1} \int_0^\tau g (\tau') \cos \theta (\tau') \ \rmd \tau' 
		\right) \\
		& \times \exp\left[ \rmi (2 \mathbf{K}_w + \mathbf{k}_{\perp,w} ) \cdot \mathbf{w} + \rmi \mathbf{w} \cdot \bm{M}_w \cdot \mathbf{w} \right]
		\ \rmd w_x \ \rmd w_y \ \rmd \tau \ \rmd k_{\perp,1} \ \rmd k_{\perp,2}.
	}
	\end{equation}	
	The task ahead of us is to solve the integrals in equation (\ref{eq:tilde_A_r}). We begin by evaluating the Gaussian integrals in $\mathbf{w}$ (Section \ref{subsection:integrals_w}). When calculating Gaussian integrals with complex coefficients, we need to be careful to choose the correct signs of the roots, in accordance with standard contour integration techniques. To solve the integral in $\tau$, we have to make some assumptions about $\theta$, which we do in Section \ref{subsection:ordering_thetam}. Depending on the assumptions made, we can solve this integral with the \emph{small mismatch angle} (Section \ref{section_backscattered_E_CT}) or \emph{large mismatch angle} (Section \ref{section_backscattered_E_ST}) orderings. Moving forward, the form of the backscattered electric field that we will use for the backscattered power will be that of the \emph{small mismatch angle} ordering. We later show in Section \ref{section_backscattered_E_ST} that in the appropriate limit, the \emph{small mismatch angle} and \emph{large mismatch angle} orderings give the same result. This enables us to use the \emph{small mismatch angle} formulation even in cases which are moderately in the \emph{large mismatch angle} regime.

	\subsection{Gaussian integrals in $w_x$ and $w_y$} \label{subsection:integrals_w}
	To solve the spatial integrals perpendicular to the beam, we first define the inverse of $\bm{M}_w$ as
	\begin{eqnarray}
	\bm{M}_w^{-1}
	=
	\left( \begin{array}{ccc}
	M_{xx}^{-1} & M_{xy}^{-1} & 0 \\
	M_{yx}^{-1} & M_{yy}^{-1} & 0 \\
	0		  & 0		 & 0 \\
	\end{array} \right)
	=
	\left(
	\begin{array}{cc}
	&
	\left(
	\begin{array}{cc}
	M_{xx} & M_{xy} \\
	M_{yx} & M_{yy}
	\end{array}
	\right) ^{-1}
	
	\begin{array}{c}
	0 \\
	0 \\
	\end{array} \\
	
	& \begin{array}{cc}
	\ \ \ \ 0 \ \ \ \ & 0~~
	\end{array}

	\begin{array}{c}
	\ \ \ \ \ \ \ 0
	\end{array}
	
	\end{array} \right) ,
	\end{eqnarray}
	which one may recognise as the Moore-Penrose inverse. It is important to bear in mind that $M_{ij}^{-1}$ is the $i j$ component of $\bm{M}_w^{-1}$, and not $1 / M_{ij}$. We then note that
	\begin{equation} 
	\fl
	\eqalign{
		& (2 \mathbf{K}_w + \mathbf{k}_{\perp,w} ) \cdot \mathbf{w} + \mathbf{w} \cdot \bm{M}_w \cdot \mathbf{w} \\
		&\qquad = 
		\left[ \mathbf{w} + \frac{1}{2}  (2 \mathbf{K}_w + \mathbf{k}_{\perp,w} ) \cdot \bm{M}_w^{-1} \right]
		\cdot \bm{M}_w \cdot
		\left[ \mathbf{w} + \frac{1}{2} \bm{M}_w^{-1} \cdot (2 \mathbf{K}_w + \mathbf{k}_{\perp,w} ) \right]\\		
		& \qquad - \frac{1}{4}
		(2 \mathbf{K}_w + \mathbf{k}_{\perp,w} )
		\cdot \bm{M}_w^{-1} \cdot
		(2 \mathbf{K}_w + \mathbf{k}_{\perp,w} ) . 	
	}
	\end{equation}	
	We substitute this expression into equation (\ref{eq:mismatch_AfterFT}). Note that the integral over $w_y$ in equation (\ref{eq:tilde_A_r}) is a Gaussian integral. The integral over $w_x$ in equation (\ref{eq:tilde_A_r}) is not strictly a Gaussian integral because of the dependence of $u_\parallel$ on $w_x$; this dependence is negligible because the turbulent properties change slowly along a field line, allowing us to use the approximate expression 
	\begin{equation}
		\delta \tilde{n}_e \left( k_{\perp,1}, k_{\perp,2}, u_\parallel, \omega \right)
		\simeq
		\delta \tilde{n}_e \left( k_{\perp,1}, k_{\perp,2}, - \int_0^\tau g (\tau') \sin \theta (\tau') \ \rmd \tau', \omega \right)
		,
	\end{equation}	 	
	to treat this integral as Gaussian as well. Thus, we have a 2D complex Gaussian integral; one should be careful when choosing the signs of the roots, as dictated by standard contour integration techniques. Hence, equation (\ref{eq:tilde_A_r}) becomes
	\begin{equation} \label{eq:A_r_after_wx_wy_integration}
	\fl
	\eqalign{
		\tilde{A}_r
		&= - \frac{\Omega A_{ant} g_{ant} \hat{\mathbf{e}}_{ant} \cdot \hat{\mathbf{e}}_{ant}}{2 c} \int \left[
			\frac{\det \left[\textrm{Im} \left( \bm{\Psi}_w \right) \right]}{\det\left( \bm{M}_w \right) }
		\right]^{\frac{1}{2}} \exp (2 \rmi \phi_G ) \\
		& \times \frac{ \delta \tilde{n}_e }{n_e} \ \hat{\mathbf{e}}^*
		\cdot (\bm{\epsilon}_{eq} -\bm{1}) \cdot \hat{\mathbf{e}}
		\ \exp \left(2 \rmi s + \rmi k_{\perp,1} \int_0^\tau g (\tau') \cos \theta (\tau') \ \rmd \tau' \right) \\
		& \times \exp\left[
			- \frac{\rmi}{4} (2 \mathbf{K}_w + \mathbf{k}_{\perp,w} )
			\cdot \bm{M}_w^{-1} \cdot
			(2 \mathbf{K}_w + \mathbf{k}_{\perp,w} )
		\right]
		\ \rmd \tau \ \rmd k_{\perp,1} \ \rmd k_{\perp,2} .
	}
	\end{equation}
	We note that the phase of $\sqrt{\det \left( \bm{M}_{w} \right)}$ is chosen from
	\begin{equation}
		\sqrt{\det \left( \bm{M}_{w} \right)}=
		\sqrt{M_{xx}}
		\sqrt{\frac{\det \left( \bm{M}_{w} \right)}{M_{xx}}} ,
	\end{equation}
	and the fact that these square roots must have
	\begin{equation}
		\textrm{Im} \left( \sqrt{M_{xx}} \right) > 0 ,
	\end{equation}
	and
	\begin{equation}
		\textrm{Im} \left[ \sqrt{\frac{\det \left( \bm{M}_{w} \right)}{M_{xx}}} \right] > 0 .
	\end{equation}	
	A convenient way to remember these rules is to consider a purely imaginary $\bm{M}_{w}$, in which case this result is the most intuitive and sensible one.
	
	To simplify equation (\ref{eq:A_r_after_wx_wy_integration}), we project $\mathbf{K}_w$ onto $\hat{\mathbf{x}}$ and $\hat{\mathbf{y}}$. Note that the dispersion relation for cold plasma depends on $\mathbf{K}$ only via $K^2$ and $(\mathbf{K} \cdot \hat{\mathbf{b}})^2$,
	\begin{equation}
		H(\mathbf{K},\mathbf{q}) = H(K^2, (\mathbf{K} \cdot \hat{\mathbf{b}})^2, \mathbf{q}) .
	\end{equation}
	Thus, we find that
	\begin{equation} \label{eq:g_coldplasma}
		\mathbf{g}
		= \nabla_K H
		= 2 \frac{\partial H}{\partial K^2} \mathbf{K} + 2 K \sin \theta_m \frac{\partial H}{\partial (\mathbf{K} \cdot \hat{\mathbf{b}})^2} \hat{\mathbf{b}}.
	\end{equation}
	and as a consequence
	\begin{equation} \label{eq:K_y}
		K_y = 0.
	\end{equation}
	That is, the beam wavevector $\mathbf{K}$ is always in the plane defined by $\hat{\mathbf{b}}$ and $\hat{\mathbf{g}}$. Using these insights, we remark that there is an exponential decay of the signal with $k_{\perp,2}$ because of the piece
	\begin{equation} \label{eq:kperp2_size_requirement}
	\fl
	\eqalign{
		&\exp\left[
		- \frac{\rmi}{4} (2 \mathbf{K}_w + \mathbf{k}_{\perp,w} )
		\cdot \bm{M}_w^{-1} \cdot
		(2 \mathbf{K}_w + \mathbf{k}_{\perp,w} )
		\right] \\
		&= 
		\exp\left[
		- \frac{\rmi}{4} \left(
			(2 K_x - k_{\perp,1} \sin \theta)^2 M_{xx}^{-1}
			+ 2 k_{\perp,2} (2 K_x - k_{\perp,1} \sin \theta) M_{xy}^{-1}
			+ k_{\perp,2}^2 M_{yy}^{-1}
		\right)
		\right].
	}
	\end{equation}
	Hence, $k_{\perp,2}$ cannot be of order $1 / \lambda$ because otherwise the argument of the exponential will be large and there will be no signal. Instead, by requiring that this argument can at most be of order unity, we find that 
	\begin{equation} \label{eq:kperp2_size_beam}
		k_{\perp,2} \sim \frac{1}{W} .
	\end{equation}
			
	\subsection{Ordering $\theta_m$} \label{subsection:ordering_thetam}
	We consider two orderings: \emph{small mismatch angle} and \emph{large mismatch angle}. These are simply names for two different orderings typical of, but not exclusive to, conventional and spherical tokamaks, respectively. Moreover, these orderings are not only applicable to tokamaks. As long as the particular ordering holds, the results will be applicable. In particular, they are applicable to stellarators.
	
	In the \emph{small mismatch angle} ordering, we take the mismatch angle to be small for the entire length of the beam path,
	\begin{equation}
		\theta_m \sim \frac{\lambda}{W}.
	\end{equation}
	In this situation, the backscattered signal is given by the Bragg condition, which we discuss in more detail in Section \ref{section_backscattered_E_CT}.
			
	In the \emph{large mismatch angle} ordering, we take the mismatch angle to be of order unity $\theta_{m} \sim 1$. In general, this requires more work and is beyond the scope of this paper. However, in the special scenario where there is no mismatch ($\theta_m = 0$) on at least one point along the beam path, the backscattered signal is dominated by this point. We derive the backscattered signal for this particular ordering in Section \ref{section_backscattered_E_ST}. In the right limits, we show that the \emph{large mismatch angle} and \emph{small mismatch angle} orderings coincide, see Section \ref{section_backscattered_E_ST}.

	\section{Backscattered electric field: small mismatch angle ordering} \label{section_backscattered_E_CT}
	In this section, we solve the integral in $\tau$ for the \emph{small mismatch angle} ordering, while the \emph{large mismatch angle} ordering will be handled in Section \ref{section_backscattered_E_ST}. We begin by exploring in detail the orderings involved in the former ordering. Once we do this, we then proceed to evaluate the integral in $\tau$ via the method of stationary phase.
	
	\subsection{Ordering} 
	We contract equation (\ref{eq:g_coldplasma}) with $\hat{\mathbf{b}}$, giving us
	\begin{equation} \label{eq:size_of_theta}
		- g \sin \theta = 2 K \left(
			\frac{\partial H}{\partial K^2 }
			+ \frac{\partial H}{\partial (\mathbf{K} \cdot \hat{\mathbf{b}})^2 }
		\right) \sin \theta_{m} .
	\end{equation}
	When $\theta_m \sim \lambda / W$, equation (\ref{eq:g_coldplasma}) gives us $g \simeq 2 K \ \partial H / \partial K^2$ to leading order. Hence we find that
	\begin{equation} \label{eq:theta_theta_m}
		\sin \theta = - \left[
			1
			+ \frac{\partial H}{\partial (\mathbf{K} \cdot \hat{\mathbf{b}})^2 } \left( \frac{\partial H}{\partial K^2 } \right)^{-1}
		\right] \sin \theta_{m} .
	\end{equation}
	The finer details of the relationship between $\theta$ and $\theta_m$ are discussed in \ref{appendix_theta_thetam}. However, to proceed, all we need is to note that since
	\begin{equation}
		\frac{\partial H}{\partial (\mathbf{K} \cdot \hat{\mathbf{b}})^2 }
		\sim \frac{\partial H}{\partial K^2 } ,
	\end{equation}
	we get
	\begin{equation} \label{eq:ordering_of_theta}
		\theta \sim \theta_m \sim \frac{\lambda}{W} .
	\end{equation}
	Consequently, we have
	\begin{equation} \label{eq:K_x}
		K_x = -K \sin (\theta_m + \theta) \simeq -K (\theta_m + \theta) \sim \frac{1}{W} .
	\end{equation}
	Hence, $K_x / K$ is small in mismatch angle, whereas
	\begin{equation}
		K_g = K \cos (\theta + \theta_{m}) \simeq K .
	\end{equation}
	From equations (\ref{eq:K_y}) and (\ref{eq:K_x}), we find that $K_w / K = K_x / K \sim \theta_{m}$. At this point, we see that our ordering $\theta_{m} \sim \lambda / W$ is necessary for the mismatch angle to be small enough to allow for a backscattered signal to be detected. Indeed, using equations (\ref{eq:ordering_of_theta}) and (\ref{eq:K_x}), the $M_{xx}^{-1}$ and $M_{xy}^{-1}$ terms in the exponential of equation (\ref{eq:kperp2_size_requirement}) are of order unity.
	
	In a conventional tokamak, most of the magnetic field is in the toroidal direction $\hat{\mathbf{b}} \simeq \hat{\boldsymbol{\zeta}}$. Hence, to have $\theta \sim \lambda / W$, we require
	\begin{equation}
		\hat{\mathbf{g}}
		\cdot \hat{\boldsymbol{\zeta}} \sim \frac{\lambda}{W}.
	\end{equation}
	Note that this is a maximal ordering. The consequence of this is that the group velocity of the probe beam can have a small toroidal component and the orderings will still hold, but everything still works perfectly fine if the group velocity is entirely in the poloidal plane. If this were the case, then the mismatch angle is simply the ratio of the poloidal magnetic field to the toroidal magnetic field, leading us to the conclusion that
	\begin{equation}
		\frac{B_p}{B} \sim \frac{\lambda}{W}.
	\end{equation}

	When the mismatch angle is exactly zero, $\hat{\mathbf{e}}$ can be calculated exactly. In the basis of Section \ref{subsection_dispersion}, the polarisations of the O-mode and X mode are 
	\begin{equation}
		\hat{\mathbf{e}}_{O} \propto 
		\left(
		\begin{array}{c}
			0\\
			0\\
			1\\
		\end{array}
		\right) ,
	\end{equation}
	and
	\begin{equation}
		\hat{\mathbf{e}}_{x} \propto 
		\left(
		\begin{array}{c}
			\rmi \epsilon_{12}\\
			\epsilon_{11}\\
			0\\
		\end{array}
	\right) .
	\end{equation}
	Since we assume the antenna surface to be in vacuum, we take the limit $n_e \to 0$ to find the polarisation at the antenna. The polarisation is linear and hence $|\hat{\mathbf{e}}_{ant} \cdot \hat{\mathbf{e}}_{ant} | = 1$. Without loss of generality, we take $\hat{\mathbf{e}}_{ant}$ to be purely real, and in this case, $\hat{\mathbf{e}}_{ant} \cdot \hat{\mathbf{e}}_{ant} = 1$. If, for purely perpendicular propagation, $\hat{\mathbf{e}}_{ant} \cdot \hat{\mathbf{e}}_{ant} = 1$, then for $\theta \sim \theta_m \ll 1$, it will be close to one.

\begin{figure}
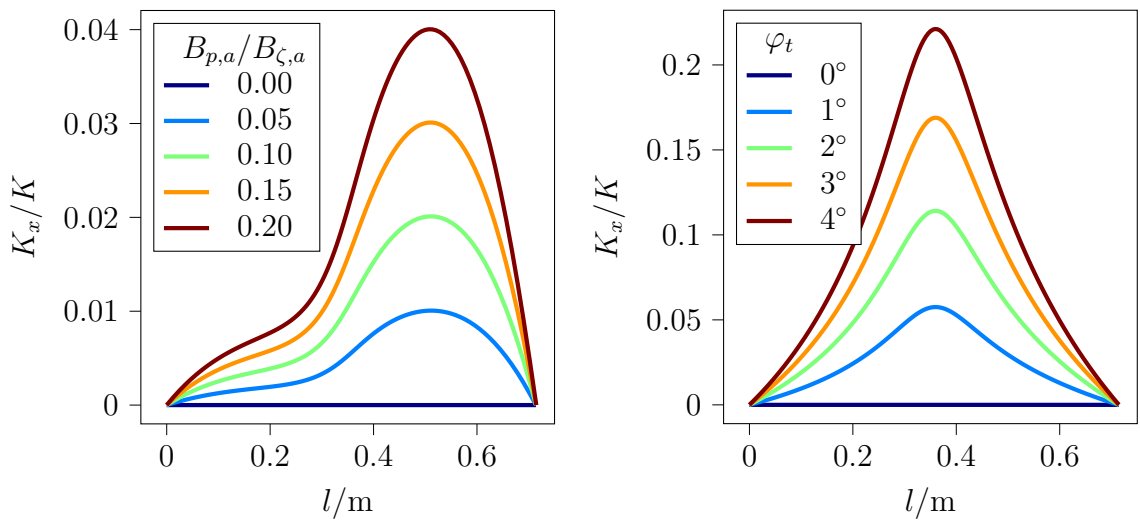

		\centering
		\begin{subfigure}{.45\textwidth}
			\centering


		\end{subfigure}
		\caption{Here $l$ is the arc-length along the central ray, measured from the point of entering the plasma. For a beam in the poloidal plane, increasing the ratio of poloidal field to toroidal field $B_{p,a} / B_{\zeta,a}$ increases the mismatch angle, which increases $K_x / K$ (left). Increasing the toroidal launch angle also increases the mismatch angle, which increases $K_x / K$ (right). We used circular flux surfaces, as described in Section \ref{subsubsection_testcases}.}
		\label{fig:Bp_and_tor_launch_scan}
	\end{figure}

	In order to help develop better intuition of equation (\ref{eq:K_x}) and the various orderings, we launch various probe beams into high-aspect-ratio circular-flux-surface analytic equilibria with no Shafranov shift, described in Section \ref{subsubsection_testcases}. We fix the launch angles, and vary $B_{p,max} / B_{\zeta,max}$, thereby scaling $B_p / B_\zeta \sim B_p / B \sim \theta_m$ everywhere. We see from Figure \ref{fig:Bp_and_tor_launch_scan} (left) that the magnitude of $K_x$ scales accordingly, as expected. We increase the toroidal launch angle for a plasma with $B_p = 0$, starting from a launch angle of $0$, which corresponds to the beam being entirely in the poloidal plane. This also has the effect of increasing the mismatch angle, and we can see in Figure \ref{fig:Bp_and_tor_launch_scan} (right) that $K_x$ also increases.

	\subsection{Stationary phase integral in $\tau$} \label{subsection:stationary_phase}
\begin{figure}
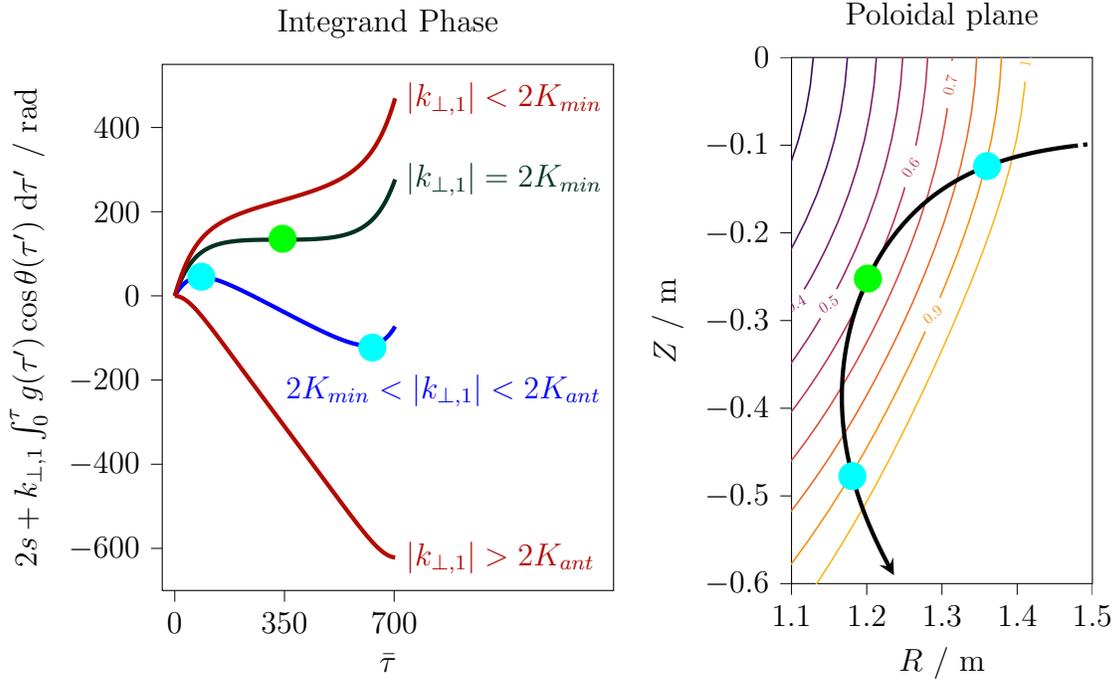

	\centering
		\begin{subfigure}{0.55\textwidth}
		\centering				


		\end{subfigure}		
				
		\caption{The beam is launched from the steering mirror, located on the midplane at $R = 2.4 \textrm{m}$, which is out of the graph's boundaries (right). The thick black line (right) indicates the trajectory of the central ray and the arrow the direction of propagation. Here $K_{min}$ is the minimum wavenumber along the central ray and $K_{ant}$ is the vacuum wavenumber. Every $k_{\perp,1}$ is associated with scattering at different points of the beam path in our MAST test scenario. Every point only backscatters one particular $k_{\perp,1}$. Red lines: these values of $k_{\perp,1}$ are not responsible for backscattering along the ray (either too large or too small). Green line: the $k_{\perp,1}$ for which backscattering occurs at the extremum. Blue line: this value of $k_{\perp,1}$ backscatters from two points along the ray, which is generally the case.  }
		\label{fig:stationary_phase}
	\end{figure}

	We now proceed to evaluate the integral in $\tau$ in equation (\ref{eq:A_r_after_wx_wy_integration}), exploiting the orderings above. Noticing that the function $\exp \left[2 \rmi s + \rmi k_{\perp, 1} \int_0^\tau g (\tau') \cos \theta (\tau') \ \rmd \tau' \right]$ oscillates quickly in $\tau$ since $2 s + k_{\perp, 1} \int_0^\tau g (\tau') \cos \theta (\tau') \ \rmd \tau' \sim L / \lambda \gg 1$, we use the method of stationary phase to evaluate the $\tau$ integral, an approach to DBS that is well-established \cite{Gusakov:scattering_slab:2004}. The method works as follows. Since the exponential fluctuates quickly, the positive fluctuations cancel with the negative fluctuations, integrating to zero to lowest order. This does not happen where the phase is stationary, that is, when
	\begin{equation} \label{eq:first_deriv_phase}
		\frac{\rmd}{\rmd \tau}
		\left[
			2 s + k_{\perp, 1} \int_0^\tau  g (\tau') \cos \theta (\tau') \ \rmd \tau' 
		\right]  \Biggr|_{\tau_\mu}
		= 0 ,
	\end{equation}
	where $\tau_\mu = \tau_a, \tau_b, \ldots$ are various points along the ray that satisfy this equation. Equation (\ref{eq:first_deriv_phase}) is the Bragg condition alluded to in the introduction,
	\begin{equation} \label{eq:Bragg_condition_full}
		k_{\perp, 1} \cos \theta (\tau_\mu) = - 2 K_g (\tau_\mu) .
	\end{equation}	
	Neglecting terms that are small in mismatch, we get
	\begin{equation} \label{eq:Bragg_condition}
		k_{\perp, 1} \simeq - 2 K (\tau_\mu) ,
	\end{equation}	
	which is how the Bragg condition is typically presented in the literature: at every point along the ray, there is a specific $k_{\perp,1}$ that is responsible for backscattering, and its value is determined solely by the wavenumber at that point. We consider the case where the density profile is monotonic. In such a situation, $K$ decreases as we get close to the cut-off, and increases as we go further from it. Hence, for any given $k_{\perp,1}$, there are three possible scenarios, as shown in Figure \ref{fig:stationary_phase}. First, that $k_{\perp,1}$ is either too small or too large, such that at no point of the ray's trajectory is it responsible for backscattering. Second, that it backscatters the beam exactly once, at an extremum value of $K$ along the path. Third, that the same $k_{\perp,1}$ is responsible for backscattering at two points along the path, $\tau_\mu = \tau_a$ and $\tau_\mu = \tau_b$; consequently, we have to add contributions from both these locations when taking the integral in $\tau$. This can be extended to more complicated cases where the density profile is non-monotonic.

	Since the dominant contributions to the integral are due to small intervals centred around $\tau_\mu = \tau_a, \tau_b, \ldots$, we Taylor expand the phase $2 \rmi s + \rmi k_{\perp, 1} \int_0^\tau  g (\tau') \ \rmd \tau'$ around $\tau_a, \tau_b, \ldots$ to find these contributions. Usually, we will have at most two points to expand around ($\tau_a$ and $\tau_b$), but we keep things general and sum over all of them, just in case one were to consider one of the aforementioned complicated cases. We first note that when the mismatch angle is small, $\theta_{m} \sim \lambda / W$, we have
	\begin{equation}
	\fl
	\eqalign{
		\exp \left( \rmi k_{\perp, 1} \int_0^\tau  g (\tau') \cos \theta (\tau') \ \rmd \tau' \right)
		\simeq
		\exp \left( \rmi k_{\perp, 1} \int_0^\tau  g (\tau') \left( 1 - \frac{1}{2} \theta^2 (\tau') \right) \ \rmd \tau' \right) ,
	}
	\end{equation}	
	where we have discarded terms which are small. Hence, expanding the large phase term, we get
	\begin{equation}
	\fl
	\eqalign{
		&\exp \left( 2 \rmi s + \rmi k_{\perp, 1} \int_0^\tau  g (\tau') \left( 1 - \frac{1}{2} \theta^2 (\tau') \right) \ \rmd \tau' \right) \\
		&\simeq
		\exp \left[ 
			2 \rmi s (\tau_\mu) 
			+ \rmi k_{\perp, 1} \int_0^{\tau_\mu}  g (\tau') \left( 1 - \frac{1}{2} \theta^2 (\tau') \right) \ \rmd \tau'
			+ \rmi g (\tau_\mu) \frac{\rmd K}{\rmd \tau} \Bigr|_{\tau_\mu} \left(
				\tau - \tau_\mu
			\right)^2
		\right] .
	}
	\end{equation}
	Here the subscript $_\mu$ on functions of $\tau$  indicates these functions are evaluated at $\tau = \tau_\mu$. We have also used the Bragg condition, equation (\ref{eq:Bragg_condition}), to simplify some of the terms. Thus, the final integral is again a Gaussian integral that gives a significant contribution for sufficiently small values of $| \tau - \tau_\mu |$; hence, slowly-varying functions of $\tau$ such as $\delta \tilde{n}_e$ can be simply evaluated at $\tau_a, \tau_b, \ldots$,
	\begin{equation}
	\eqalign{
		\delta \tilde{n}_{e} \left( 
			k_{\perp,1}, 
			k_{\perp,2}, 
			u_\parallel,
			\omega 
		\right) 
		&  \simeq 
		\delta \tilde{n}_{e,\mu} \left( 
			k_{\perp,1}, 
			k_{\perp,2}, 
			\omega 
		\right) \\ 
		& = \delta \tilde{n}_e \left( 
			k_{\perp,1}, 
			k_{\perp,2}, 
			- \int_0^{\tau_\mu}  g (\tau') \sin \theta (\tau') \ \rmd \tau' , 
			\omega 
		\right) ,
	}
	\end{equation}
	giving
	\begin{equation} \label{eq:A_r_stationary_phase}
	\fl
	\eqalign{
		\tilde{A}_r
		&=
		- \frac{\Omega A_{ant} g_{ant}}{2 c} \int \sum_{\mu = a, b, \ldots} \left[
		\frac{\det \left[\textrm{Im} \left( \bm{\Psi}_{w,\mu} \right) \right]}{\det\left( \bm{M}_{w,\mu} \right) }
		\right]^{\frac{1}{2}}  
		\left[
		\pi \rmi \left(
		g_\mu \frac{\rmd K}{\rmd \tau} \Bigr|_{\tau_\mu}
		\right)^{- 1}
		\right]^{\frac{1}{2}} \\
		& \times
		\exp \left( 2 \rmi \phi_{G,\mu} \right)
		\ \hat{\mathbf{e}}^*_\mu \cdot (\bm{\epsilon}_{eq,\mu} -\bm{1}) \cdot \hat{\mathbf{e}}_\mu
		\frac{\delta \tilde{n}_{e,\mu} }{n_{e,\mu}} \\
		& \times
		\exp\left[
			- \frac{\rmi}{4} (2 \mathbf{K}_{w,\mu} + \mathbf{k}_{\perp} )
			\cdot \bm{M}_{w,\mu}^{-1} \cdot
			(2 \mathbf{K}_{w,\mu} + \mathbf{k}_{\perp} )
		\right]  \\
		& \times 
		\exp \left(
			2 \rmi s_\mu + \rmi k_{\perp, 1} \int_0^{\tau_\mu}  g (\tau') \left(
				1 - \frac{\theta^2 (\tau')}{2}
				\right) \ \rmd \tau' 
		\right)
		\ \rmd k_{\perp, 1} \ \rmd k_{\perp, 2} .
	}
	\end{equation}
	With the mismatch angle being small, $\theta_{m} \sim W / L$, we remark that
	\begin{equation}
	\eqalign{
		&- \frac{\rmi}{4} (2 \mathbf{K}_{w,\mu} + \mathbf{k}_{\perp} )
		\cdot \bm{M}_{w,\mu}^{-1} \cdot
		(2 \mathbf{K}_{w,\mu} + \mathbf{k}_{\perp} ) \\
		& \qquad \simeq 
		- \rmi K_\mu^2 \theta_{m,\mu}^2 M_{xx,\mu}^{-1}
		+ \rmi k_{\perp,2} K_\mu \theta_{m,\mu} M_{xy,\mu}^{-1}
		- \frac{\rmi}{4} k_{\perp,2}^2 M_{yy,\mu}^{-1} ,
	}
	\end{equation}
	and the corrections to $\bm{\Psi}_{w}$ are
	\begin{equation} \label{eq:Mxx_simplified}
	\eqalign{
		M_{xx,\mu} \simeq \Psi_{xx,\mu}
		- \frac{k_{\perp,1}}{2} \left( \hat{\mathbf{b}} \cdot \nabla \hat{\mathbf{b}} \cdot \hat{\mathbf{g}} \right)_\mu ,
	}
	\end{equation}
	and
	\begin{equation} \label{eq:Mxy_simplified}
	\eqalign{
		M_{xy,\mu} \simeq \Psi_{xy,\mu}
		+ \frac{ k_{\perp,1} }{2} \frac{\left[ \left( \hat{\mathbf{b}} \times \hat{\mathbf{g}} \right)  \cdot \nabla \hat{\mathbf{b}} \cdot \hat{\mathbf{g}} \right]_\mu}{\left| \hat{\mathbf{b}}_\mu \times \hat{\mathbf{g}}_\mu \right|}  .
	}
	\end{equation}
	Here we have used, $\hat{\mathbf{x}} \simeq - \hat{\mathbf{b}}$, $\hat{\mathbf{y}} = \hat{\mathbf{b}} \times \hat{\mathbf{g}} / | \hat{\mathbf{b}} \times \hat{\mathbf{g}} |$, and the result that $k_{\perp,2} \sim 1 / W$, as we argued in equation (\ref{eq:kperp2_size_beam}). Hence, the $M_{xx,\mu}$ is modified by the curvature of the magnetic field $\mathbf{B}$, while $M_{xy,\mu}$ is modified by the magnetic shear.
	
	Unfortunately, $\rmd K / \rmd \tau = 0$ at a minimum of one point along the ray, since $K$ decreases as the beam enters the plasma and increases as it leaves. A proper treatment of this divergence requires us to consider the next order terms in the Taylor expansion of the phase $2 \rmi s + \rmi k_{\perp, 1} \int_0^{\tau_\mu}  g (\tau') \left( 1 - \theta^2 (\tau') / 2	\right) \ \rmd \tau' $, which we do in \ref{appendix_airy_function}. However, this is not an important issue; the divergence is integrable, as we see in Section \ref{section_localisation}. 
	
	At this point, we find ourselves in a bit of a difficulty. Consider the inverse Fourier transform of the density fluctuations
	\begin{equation}
	\fl
	\eqalign{
		\delta \tilde{n}_{e,\mu} ( k_{\perp,1}, k_{\perp,2}, \omega )
		=
		\frac{1}{4 \pi^2} \int
		& \delta n_{e} \left(u_1, u_2, - \int_0^{\tau_\mu}  g (\tau') \sin \theta (\tau') \ \rmd \tau', t \right) \\
		& \times
		\exp \left( 
			- \rmi k_{\perp,1} u_1 
			- \rmi k_{\perp,2} u_2 
			+ \rmi \omega t 
		\right)
		\rmd k_{\perp,1} \ \rmd k_{\perp,2} \ \rmd \omega .
	}
	\end{equation} 
	We see from equations (\ref{eq:u1_final}) and (\ref{eq:u2_final}) that $u_1 \sim L$ and $u_2 \sim W$. Consequently, inverse scales as small as $1 / L$ for $k_{\perp,1}$ and $1 / W$ for $k_{\perp,2}$ are large enough to change $\delta \tilde{n}_{e,\mu}$ by order unity. This is in addition to $\delta n_{e}$ changing on the very small scale of $\lambda$. Moreover, in equation (\ref{eq:A_r_stationary_phase}), there is a large phase term $k_{\perp, 1} \int_0^{\tau_\mu}  g (\tau') \rmd \tau' \sim k_{\perp,1} L \sim L / \lambda \gg 1$. These two facts mean that we have order unity change even when $k_{\perp,1}$ changes by as little $1 / L $ or when $k_{\perp,2}$ changes by $1 / W $. Consequently, one would have to consider very small inverse-length scales $1 / L$ which is not only undesirable from a physical point of view, but also prevents us from evaluating either of the remaining integrals. We can get around this by working with the time-averaged backscattered power instead. Using power avoids the need to deal with the phase, and time averaging leads to separation of scales in $\delta \tilde{n}_{e,\mu}$, thereby eliminating the small inverse-length scales.

	\section{Backscattered power} \label{section_backscattered_P}
	For the reasons discussed at the end of Section \ref{subsection:stationary_phase}, it is difficult to evaluate the $k_{\perp,1}$ and $k_{\perp,2}$ integrals when working with the backscattered amplitude. In order to make further analytical progress, we have to eschew the phase of $\tilde{A}_r$, and work with the time-averaged backscattered power instead. We introduce the correlation function (Section \ref{subsection:correlation_function}), and proceed to solve the Gaussian integral in $k_{\perp,2}$ (Section \ref{subsection:k_perp_2}). Unfortunately, the final integral in $k_{\perp,1}$ cannot be solved analytically without making assumptions about the turbulence spectrum. This issue, along with a numerical solution to the $k_{\perp,1}$ integral, is discussed in Section \ref{subsection:power_final}.

	\subsection{Correlation function} \label{subsection:correlation_function}
	We consider the correlation function for two density fluctuations, at $\mathbf{r}, t$ and $\mathbf{r} + \Delta \mathbf{r}, t + \Delta t$,
	\begin{equation}
	\eqalign{
		C (\mathbf{r}, t, \Delta \mathbf{r}, \Delta t)
		=
		\frac{
			\left<
			\delta n_e (\mathbf{r}, t) \delta n_e (\mathbf{r} + \Delta \mathbf{r}, t + \Delta t)
			\right>_t
		}{
			\left<
			\delta n_e^2 (\mathbf{r}, t)
			\right>_t
		}
		.
	}
	\end{equation}	
	The vectors $\mathbf{r}$ and $\Delta \mathbf{r}$ have different scales after time averaging, and this will be important later. Using equation (\ref{eq:n_FT}), we express the density fluctuations in terms of their Fourier transforms
	\begin{equation}
	\fl
	\eqalign{
		& \frac{
			\left<
			\delta n_e (\mathbf{r}, t) \delta n_e (\mathbf{r} + \Delta \mathbf{r}, t + \Delta t)
			\right>_t
		}{
			\left<
			\delta n_e^2 (\mathbf{r}, t)
			\right>_t
		} \\
		& \qquad =
		\left<
			\delta n_e^2 (\mathbf{r}, t)
		\right>_t^{-1}
		\int
		\delta \tilde{n}_e \left( k_{\perp, 1}, k_{\perp, 2}, u_\parallel  + \Delta u_\parallel, \omega \right)
		\delta \tilde{n}_e^* \left( k_{\perp, 1}', k_{\perp, 2}', u_\parallel, \omega' \right) \\
		& \qquad \times \exp \left[
		\rmi (k_{\perp, 1} - k_{\perp, 1}') u_1
		+ \rmi (k_{\perp, 2} - k_{\perp, 2}') u_2
		\right]
		\left< \exp \left[
		- \rmi (\omega - \omega') t
		\right] \right>_t \\
		& \qquad \times \exp \left(
		\rmi k_{\perp, 1} \Delta u_1
		+ \rmi k_{\perp, 2} \Delta u_2
		- \rmi \omega \Delta t
		\right)
		\ \rmd k_{\perp, 1} \ \rmd k_{\perp, 2} \ \rmd \omega
		\ \rmd k_{\perp, 1}' \ \rmd k_{\perp, 2}' \ \rmd \omega'  .
	}
	\end{equation}
	Since we assumed our system was in steady state, we can perform the time average $<>_t$ over a sufficiently large time interval 
	\begin{equation} \label{eq:T_scale}
		T \gg \frac{1}{\omega} ,
	\end{equation}
	such that we get a Dirac delta function as follows
	\begin{equation}
		\int_{-\infty}^{\infty}  \exp \left[
		- \rmi (\omega - \omega') t
		\right] \ \rmd t
		= 2 \pi \delta \left( \omega - \omega' \right) .
	\end{equation}
	However, this is not true of $k_{\perp,i} - k_{\perp,i}'$, since there is a characteristic long length scale $L$. After time averaging, we assume separation of scales for $\mathbf{r}$ and $\Delta \mathbf{r}$, which means we have to order
	\begin{equation}
		k_{\perp, 1} - k_{\perp, 1}' \sim k_{\perp, 2} - k_{\perp, 2}' \sim \frac{1}{L},
	\end{equation}
	but
	\begin{equation}
		k_{\perp,1} \sim k_{\perp,2} \sim \frac{1}{\lambda} .
	\end{equation}
	Note the apparent contradiction with equation (\ref{eq:kperp2_size_beam}): the Fourier transform of the turbulent fluctuations $\delta \tilde{n}_e$ has $k_{\perp,2} \sim 1 / \lambda$, but backscattering requires $k_{\perp,2}$ to be as small as $1 / W$ per equation (\ref{eq:kperp2_size_beam}). We will see shortly that this implies that we eventually should take $k_{\perp,2} = 0$. For now, however, we get
	\begin{equation}
	\fl
	\eqalign{
		& \frac{
			\left<
			\delta n_e (\mathbf{r}, t) \delta n_e (\mathbf{r} + \Delta \mathbf{r}, t + \Delta t)
			\right>_t
		}{
			\left<
			\delta n_e^2 (\mathbf{r}, t)
			\right>_t
		} \\
		& \qquad = \left<
			\delta n_e^2 (\mathbf{r}, t)
		\right>_t^{-1} 
		\int \delta \tilde{n}_e \left( k_{\perp, 1}, k_{\perp, 2}, u_\parallel + \Delta u_\parallel, \omega \right)
		\delta \tilde{n}_e^* \left( k_{\perp, 1}', k_{\perp, 2}', u_\parallel, \omega \right) \\
		& \qquad \times \exp \left[
		\rmi (k_{\perp, 1} - k_{\perp, 1}') u_1
		+ \rmi (k_{\perp, 2} - k_{\perp, 2}') u_2
		\right] \ \rmd k_{\perp, 1} \ \rmd k_{\perp, 2}  \\
		& \qquad \times \exp \left(
		\rmi k_{\perp, 1} \Delta u_1
		+ \rmi k_{\perp, 2} \Delta u_2
		- \rmi \omega \Delta t
		\right)
		\ \rmd \omega \ \rmd k_{\perp, 1}' \ \rmd k_{\perp, 2}'
		,
	}
	\end{equation}
	We compare this to the Fourier transform of the correlation function with respect to $\Delta \mathbf{r}$ and $\Delta t$, which satisfies
	\begin{equation}
	\fl
	\eqalign{
		C (\mathbf{r}, t, \Delta \mathbf{r}, \Delta t)
		&= \int
		\widetilde{C} (\mathbf{r}, t, k_{\perp,1}, k_{\perp,2}, \Delta u_{\parallel}, \omega) \\
		& \times \exp \left(
		\rmi k_{\perp, 1} \Delta u_1
		+ \rmi k_{\perp, 2} \Delta u_2
		- \rmi \omega \Delta t
		\right)
		\ \rmd k_{\perp, 1} \ \rmd k_{\perp, 2} \ \rmd \omega .
	}
	\end{equation}
	Hence, we find that
	\begin{equation} \label{eq:correlation_function_FT}
	\fl
	\eqalign{
		& \widetilde{C} (\mathbf{r}, t, k_{\perp,1}, k_{\perp,2}, \Delta u_{\parallel}, \omega) \\
		& \qquad = \left<
			\delta n_e^2 (\mathbf{r}, t)
		\right>_t^{-1} \int
		\delta \tilde{n}_e \left( k_{\perp, 1}, k_{\perp, 2}, u_{\parallel} + \Delta u_\parallel, \omega \right)
		\delta \tilde{n}_e^* \left( k_{\perp, 1}', k_{\perp, 2}', u_{\parallel}, \omega \right) \\
		& \qquad \times \exp \left[
		\rmi (k_{\perp, 1} - k_{\perp, 1}') u_1
		+ \rmi (k_{\perp, 2} - k_{\perp, 2}') u_2
		\right]
		\ \rmd k_{\perp, 1}' \ \rmd k_{\perp, 2}' .
	}
	\end{equation}
	Since we have assumed a steady-state plasma, there is no slow time dependence, so we drop it from $\widetilde{C}$ from here onwards. The $t$ in $\delta n_e$, is a fast time dependence, which we need to keep. 

	We introduce the backscattered power spectral density $p_r$, such that the total backscattered power $P_r$ is given by
	\begin{equation}
		P_r = \int p_r \ \rmd \omega .
	\end{equation}
	To evaluate $p_r$, we multiply the backscattered amplitude, equation (\ref{eq:A_r_stationary_phase}), by its complex conjugate, and we time average over a time $T$ that satisfies equation (\ref{eq:T_scale}) to find
	\begin{equation} \label{eq:A_r2_start}
	\fl
	\eqalign{
		\frac{p_r}{P_{ant}}
		& =
		\int
		\sum_{\mu = a, b, \ldots}		
		F_\mu ( k_{\perp, 1}, k_{\perp, 2} )
		\exp \left[
			2 \rmi s (\tau_\mu (k_{\perp,1}) )
			+ \rmi k_{\perp, 1} \int_0^{\tau_\mu (k_{\perp,1})}  g (\tau'') \ \rmd \tau'' 
		\right] \\
		& \times \delta \tilde{n}_{e,\mu} ( k_{\perp,1}, k_{\perp,2}, \omega )
		\ \rmd k_{\perp, 1} \ \rmd k_{\perp, 2} \\
		&\times
		\int
		\sum_{\nu = a, b, \ldots}
		F_{\nu}^* (k_{\perp, 1}', k_{\perp, 2}')
		\exp \left[
			- 2 \rmi s (\tau_\nu (k_{\perp,1}') ) 
			- \rmi k_{\perp, 1}' \int_0^{\tau_\nu (k_{\perp,1}')}  g (\tau'') \ \rmd \tau'' 
		\right] \\
		& \times \delta \tilde{n}_{e,\nu}^* ( k_{\perp,1}', k_{\perp,2}', \omega )
		\ \rmd k_{\perp, 1}' \ \rmd k_{\perp, 2}' ,
	}
	\end{equation}
	where $P_{ant}$ is the total power emitted by the antenna and where we have abbreviated the slowly-varying piece of the amplitude (without density fluctuations) as
	\begin{equation}
	\fl
	\eqalign{
		F_\mu (k_{\perp, 1}, k_{\perp, 2})
		&=
		\frac{\Omega g_{ant}}{2 c} 
		\left[
		\frac{\det \left[\textrm{Im} \left( \bm{\Psi}_{w,\mu} \right) \right]}{\det\left( \bm{M}_{w,\mu} \right) }
		\right]^{\frac{1}{2}}  
		\left[
		\pi \rmi \left(
		g_\mu \frac{\rmd K}{\rmd \tau} \Bigr|_{\tau_\mu}
		\right)^{- 1}
		\right]^{\frac{1}{2}} \\
		& \times
		\exp \left( 2 \rmi \phi_{G,\mu} \right)
		\ \hat{\mathbf{e}}^*_\mu \cdot (\bm{\epsilon}_{eq,\mu} -\bm{1}) \cdot \hat{\mathbf{e}}_\mu
		\frac{ 1 }{n_{e,\mu}} \\
		& \times
		\exp\left[
			- \frac{\rmi}{4} (2 \mathbf{K}_{w,\mu} + \mathbf{k}_{\perp} )
			\cdot \bm{M}_{w,\mu}^{-1} \cdot
			(2 \mathbf{K}_{w,\mu} + \mathbf{k}_{\perp} )
		\right]  \\
		& \times 
		\exp \left(
			\frac{\rmi}{2} k_{\perp, 1} \int_0^{\tau_\mu}  g (\tau') \theta^2 (\tau') \ \rmd \tau' 
		\right) .
	}
	\end{equation}
	We remind readers that $\tau_\mu$ is a function of $k_{\perp,1}$. Hence, $F_{\nu} (k_{\perp,1}',k_{\perp,2}')$ would be evaluated at $\tau_\nu (k_{\perp,1}')$, rather than at $\tau_\nu (k_{\perp,1})$. 
	
	Moving forward, we try to match equation (\ref{eq:A_r2_start}) with the Fourier transform of the correlation function $\widetilde{C}$. First, we note that $F_\mu (k_{\perp,1},k_{\perp,2})$ depends slowly on $k_{\perp,1}$ and $k_{\perp,2}$, as we have taken the large phase term out of it. Hence, we can neglect $k_{\perp,i} - k_{\perp,i}' \sim 1 / L \ll 1$, giving us
	\begin{equation} \label{eq:F_and_F_prime}
		F_{\nu} (k_{\perp,1}',k_{\perp,2}') \simeq F_\nu (k_{\perp,1},k_{\perp,2}).
	\end{equation}
	To deal with the exponential term outside $F_\mu (k_{\perp,1},k_{\perp,2})$, we Taylor expand $\tau_\nu (k_{\perp,1}')$ about $\tau_\nu (k_{\perp,1})$
	\begin{equation}
	\eqalign{
		& 2 s \left( \tau_\nu (k_{\perp,1}') \right) + k_{\perp, 1}' \int_0^{\tau_\nu (k_{\perp,1}')}  g (\tau'') \ \rmd \tau'' \\
		& \qquad \simeq
		2 s \left( \tau_\nu (k_{\perp,1}) \right)
		+ k_{\perp, 1}' \int_0^{\tau_\nu (k_{\perp,1})}  g (\tau'') \ \rmd \tau'' \\
		&\qquad + \left[ 2 K_\nu g_\nu
		+ k_{\perp, 1}' g_\nu
		\right] \left[
			\tau_\nu (k_{\perp,1}') - \tau_\nu (k_{\perp,1})
		\right] .
	}
	\end{equation}
	In order to gain the insight we need to proceed further, we consider the following piece of equation (\ref{eq:A_r2_start}), 
	\begin{equation}
		\sum_{\mu,\nu}	
		\delta \tilde{n}_{e,\mu} ( k_{\perp,1}, k_{\perp,2}, \omega )
		\delta \tilde{n}_{e,\nu}^* ( k_{\perp,1}', k_{\perp,2}', \omega )	.
	\end{equation}
	When $\mu \neq \nu$, $\tau_{\mu}$ and $\tau_{\nu}$ generally correspond to positions that, in the perpendicular direction, are many correlation lengths apart. For $\mu \neq \nu$, $\delta \tilde{n}_{e,\mu} \delta \tilde{n}_{e,\nu}^*$ can only become significant when $\tau_\mu$ and $\tau_\nu$ are either close to each other or connected by a length of magnetic field line of the order of the parallel correlation length. Since this only happens on a countable number of flux surfaces, we ignore this possibility, and assume that when $\mu \neq \nu$, $\delta \tilde{n}_{e,\mu} \delta \tilde{n}_{e,\nu}^*$ is small. Hence, we find that
	\begin{equation} \label{eq:phase_Taylor_expansion_for_correlaton}
	\fl
	\eqalign{
		& \sum_{\mu,\nu}
		\delta \tilde{n}_{e,\mu} ( k_{\perp,1}, k_{\perp,2}, \omega )
		\delta \tilde{n}_{e,\nu}^* ( k_{\perp,1}', k_{\perp,2}', \omega ) \\
		& \qquad \times \exp \left[
			2 \rmi s_\mu + \rmi k_{\perp, 1} \int_0^{\tau_\mu}  g (\tau'') \ \rmd \tau'' 
			- 2 \rmi s_{\nu} - \rmi k_{\perp, 1}' \int_0^{\tau_\nu}  g (\tau'') \ \rmd \tau''		
		\right] \\
		& = 
		\sum_{\mu}
		\delta \tilde{n}_{e,\mu} ( k_{\perp,1}, k_{\perp,2}, \omega )
		\delta \tilde{n}_{e,\mu}^* ( k_{\perp,1}', k_{\perp,2}', \omega ) \\
		& \qquad \times \exp \left\{
			\rmi \left( 
				2 K_{\mu} g_\mu + k_{\perp, 1}' g_\mu 
			\right) \left[
				\tau_\mu (k_{\perp,1}) - \tau_\mu (k_{\perp,1}')
			\right] 
		\right\} \\
		& \qquad \times \exp \left[
			\rmi \left( k_{\perp, 1} - k_{\perp, 1}' \right) \int_0^{\tau_\mu}  g (\tau'') \ \rmd \tau''
		\right]		.
	}
	\end{equation}
	Here we can neglect the small term $\left( 2 K_{\mu} g_\mu + k_{\perp, 1}' g_\mu \right) \left[ \tau_\mu (k_{\perp, 1}) - \tau_\mu (k_{\perp, 1}') \right]$. Indeed, using 
	\begin{equation}
	\eqalign{
		\tau_\mu (k_{\perp,1}) - \tau_\mu (k_{\perp,1}') = - \frac{1}{2} \left( \frac{\rmd K}{\rmd \tau} \Bigr|_{\tau_\mu} \right)^{-1} (k_{\perp,1} - k_{\perp,1}' ) ,
	}
	\end{equation}	
	which is a result of the Bragg condition, equation (\ref{eq:Bragg_condition}), we show that term to be small,
	\begin{equation}
	\eqalign{
		& \left( 2 K_{\mu} g_\mu + k_{\perp, 1}' g_\mu \right) \left[
					\tau_\mu (k_{\perp,1}) - \tau_\mu (k_{\perp,1}')
				\right] \\
		& \qquad = \frac{1}{2} \left( \frac{\rmd K}{\rmd \tau} \Bigr|_{\tau_\mu} \right)^{-1} g_\mu (k_{\perp,1} - k_{\perp,1}' )^2
		\sim \frac{\lambda}{L}
		\ll 1 .
	}
	\end{equation}
	Using equations (\ref{eq:phase_Taylor_expansion_for_correlaton}) and (\ref{eq:F_and_F_prime}), we rewrite equation (\ref{eq:A_r2_start}) as
	\begin{equation}
	\fl
	\eqalign{
		\frac{p_r}{P_{ant}} 
		& =
		\int
		\sum_{\mu = a, b, \ldots}
		\delta \tilde{n}_{e,\mu} ( k_{\perp,1}, k_{\perp,2}, \omega )
		\delta \tilde{n}_{e,\mu'}^* ( k_{\perp,1}', k_{\perp,2}', \omega ) \\
		&\times \exp \left[
			\left( k_{\perp, 1} - k_{\perp, 1}' \right) \int_0^{\tau_\mu}  g (\tau'') \ \rmd \tau'' 
		\right] 
		\ \rmd k_{\perp, 1}' \ \rmd k_{\perp, 2}'  \\
		& \times \left| F_\mu (k_{\perp, 1}, k_{\perp, 2}) \right|^2		
		\ \rmd k_{\perp, 1} \ \rmd k_{\perp, 2} ,
	}
	\end{equation}	
	and we match the result to the Fourier transform of the correlation function in equation (\ref{eq:correlation_function_FT}). We now get
	\begin{equation}\label{eq:A_r2_explicit}
	\fl
	\eqalign{
		\frac{p_r}{P_{ant}} 
		& =
		\frac{1}{4} \frac{ \Omega^2 }{c^2} g_{ant}^2 \pi
		\int \sum_{\mu = a, b, \ldots}
		\frac{\det \left[\textrm{Im} \left( \bm{\Psi}_{w,\mu} \right) \right] }{ \left| \det \left[ \bm{M}_{w,\mu} \right] \right| } \left| \hat{\mathbf{e}}^*_\mu \cdot (\bm{\epsilon}_{eq,\mu} -\bm{1}) \cdot \hat{\mathbf{e}}_\mu \right|^2 \\
		& \times
		\left| g_\mu \frac{\rmd K}{\rmd \tau} \Bigr|_{\tau_\mu} \right|^{-1}
		\exp\left[
			2^{} \textrm{Im} \left( M_{xx,\mu}^{-1} \right) K_{\mu}^2 \theta_{m,\mu}^2 
		\right]  \\
		& \times
		\exp\left[
			- 2^{} \textrm{Im} \left( M_{xy,\mu}^{-1} \right) K_{\mu} \theta_{m,\mu} k_{\perp,2}
			+ \frac{1}{2} \textrm{Im} \left( M_{yy,\mu}^{-1} \right) k_{\perp,2}^2
		\right]  \\
		& \times
		\frac{
		\left<
			\delta n_{e,\mu}^2 (t)
		\right>	_t
		}{
			n_{e,\mu}^2
		}
		\widetilde{C}_\mu (k_{\perp,1}, k_{\perp,2}, \omega)
		\ \rmd k_{\perp, 1} \ \rmd k_{\perp, 2} .
	}
	\end{equation}
	Here, $\Delta u_\parallel$ in $\widetilde{C}_\mu$ is zero and the subscript $_\mu$ indicates that it is evaluated at $u_{\parallel,\mu}$, $\delta n_{e,\mu}^2$ is evaluated at $u_{\parallel,\mu} = - \int_{0}^{\tau_\mu} g (\tau') \sin \theta (\tau') \ \rmd \tau' \simeq 0 $, $u_2 = 0$, and $u_1 = \int_0^{\tau_\mu}  g (\tau'') \ \rmd \tau''$. Hence, $\mathbf{r} = \mathbf{q} ( \tau_\mu (k_{\perp,1}) )$, and so, for convenience, we have dropped the $\mathbf{r}$ dependence and denoted it with the subscript $_\mu$ instead.

	\subsection{Gaussian integral in $k_{\perp,2}$} \label{subsection:k_perp_2}
	We begin by manipulating equation (\ref{eq:A_r2_explicit}) into a more wieldly form. Remarking that
	\begin{equation} \label{eq:k_perp_2_gaussian}
	\eqalign{
		& - 2^{} \textrm{Im} \left( M_{xy,\mu}^{-1} \right) K_{\mu} \theta_{m,\mu} k_{\perp,2}
		+ \frac{1}{2} \textrm{Im} \left( M_{yy,\mu}^{-1} \right) k_{\perp,2}^2 \\
		& \qquad = \frac{1}{2} \textrm{Im} \left( M_{yy,\mu}^{-1} \right) \left[
			k_{\perp,2}
			- 2 K_{\mu} \theta_{m,\mu} \frac{ \textrm{Im} \left( M_{xy,\mu}^{-1} \right) }{\textrm{Im} \left( M_{yy,\mu}^{-1} \right)}
		\right]^2  \\
		& \qquad - 2 K_{\mu}^2 \theta_{m,\mu}^2 \frac{ \left[ \textrm{Im} \left( M_{xy,\mu}^{-1} \right) \right]^2 }{\textrm{Im} \left( M_{yy,\mu}^{-1} \right)} ,
	}
	\end{equation}
	we see that the $k_{\perp, 2}$ integral is a Gaussian integral. Thus, we find the $k_{\perp,2}$ selected by the signal,
	\begin{equation}
	\eqalign{
		k_{\perp,2}
		& \simeq k_{\mu,2} (\tau_\mu (k_{\perp,1})) \\
		& =  2 K_{\mu} \theta_{m,\mu} \frac{ \textrm{Im} \left( M_{xy,\mu}^{-1} \right) }{\textrm{Im} \left( M_{yy,\mu}^{-1} \right) } \sim \frac{1}{W}.
	}
	\end{equation}
	Interestingly, this is small in mismatch, and is exactly zero when there is no mismatch. The backscattered power is thus
	\begin{equation}\label{eq:A_r2_explicit_wavenumber_resolution}
	\fl
	\eqalign{
		\frac{p_r}{P_{ant}} 
		& =
		\frac{\pi}{4} \frac{ \Omega^2 }{c^2} g_{ant}^2
		\int \sum_{\mu = a, b, \ldots}
		\frac{\det \left[\textrm{Im} \left( \bm{\Psi}_{w,\mu} \right) \right] }{ \left| \det \left[ \bm{M}_{w,\mu} \right] \right| } \left| \hat{\mathbf{e}}^*_\mu \cdot (\bm{\epsilon}_{eq,\mu} -\bm{1}) \cdot \hat{\mathbf{e}}_\mu \right|^2 \\
		& \times
		\left| g_\mu \frac{\rmd K}{\rmd \tau} \Bigr|_{\tau_\mu} \right|^{-1}
		\exp\left[ 
			- 2 \frac{
			\left(
				k_{\perp,2}
				- k_{\mu,2}
			\right)^2			
			}{
			\left(
				\Delta k_{\mu,2}
			\right)^2
			}
		\right]  
		\exp \left[
			- 2 \frac{\theta_{m,\mu}^2}{\left( \Delta \theta_{m,\mu} \right)^2}
		\right] \\
		& \times
		\frac{
			\left<
			\delta n_{e,\mu}^2 (t)
			\right>	_t
		}{
			n_{e,\mu}^2
		}
		\widetilde{C}_\mu (k_{\perp,1}, k_{\perp,2}, \omega)
		\ \rmd k_{\perp, 1} \ \rmd k_{\perp, 2} .
	}
	\end{equation}		
	Here, we use the notation
	\begin{equation} \label{eq:delta_theta_m}
		\Delta \theta_{m,\mu}
		= 
		\frac{ 1 }{ K_{\mu} } 
		\left(
			\frac{
				\textrm{Im} \left( M_{yy,\mu}^{-1} \right)
			}{
			\left[ \textrm{Im} \left( M_{xy,\mu}^{-1} \right) \right]^2
			-
			\textrm{Im} \left( M_{xx,\mu}^{-1} \right) \textrm{Im} \left( M_{yy,\mu}^{-1} \right)
			}
		\right)^{\frac{1}{2}} ,
	\end{equation} 
	which gives the characteristic $1 / \textrm{e}^2$ width of the mismatch attenuation and
	\begin{equation} \label{eq:delta_k_perp_2}
		\Delta k_{\mu,2} = 2 \left( \frac{- 1}{\textrm{Im} \left( M_{yy,\mu}^{-1} \right) } \right)^{\frac{1}{2}} ,
	\end{equation}
	which gives us the wavenumber resolution (in $1 / \textrm{e}^2$). For those seeking to design a synthetic DBS to study data from gyrokinetic simulations, equation (\ref{eq:A_r2_explicit_wavenumber_resolution}) is the form of the backscattered signal that we recommend using. 
	
	If we take the $k_{\perp,2}$ wavenumber resolution, $\Delta k_{\mu,2}$, to be small, we can evaluate the Gaussian integral by using the approximation $k_{\perp,2} \simeq k_{\mu,2} \simeq 0$ in $\widetilde{C}_\mu$. After doing this, we get
	\begin{equation} \label{eq:A_r2_final}
	\fl
	\eqalign{
		\frac{p_r}{P_{ant}} 
		& =
		\frac{\sqrt{2 \pi^3}}{4} \frac{ \Omega^2 }{c^2} g_{ant}^2
		\int  		
		\sum_{\mu = a, b, \ldots}
		\frac{\det \left[\textrm{Im} \left( \bm{\Psi}_{w,\mu} \right) \right] }{ \left| \det \left[ \bm{M}_{w,\mu} \right] \right| } \left| \hat{\mathbf{e}}^*_\mu \cdot (\bm{\epsilon}_{eq,\mu} -\bm{1}) \cdot \hat{\mathbf{e}}_\mu \right|^2 \\
		& \times
		\left| g_\mu \frac{\rmd K}{\rmd \tau} \Bigr|_{\tau_\mu} \right|^{-1}
		\left[
			 - \textrm{Im} \left( M_{yy,\mu}^{-1} \right)
		\right]^{-\frac{1}{2}}  
		\exp \left(
			- 2 \frac{\theta_{m,\mu}^2}{\left(\Delta \theta_{m,\mu}\right)^2}
		\right) 		
		\\
		& \times
		\frac{
			\left<
			\delta n_{e,\mu}^2 (t)
			\right>	_t
		}{
			n_{e,\mu}^2
		}
		\widetilde{C}_\mu (k_{\perp,1}, k_{\mu,2}, \omega)
		\ \rmd k_{\perp, 1} .
	}
	\end{equation}
	The final integral in $k_{\perp,1}$ cannot be evaluated analytically. However, we can make a few simplifications to the current form of the backscattered power, thereby making clearer the physics involved. 
	
	\subsection{Final simplifications} \label{subsection:power_final}
	We hone in on three pieces of equation (\ref{eq:A_r2_final}), showing how they may be normalised and re-expressed in more explicit forms. We begin by writing equation (\ref{eq:A_r2_final}) as
	\begin{equation} \label{eq:A_r2_final_with_shorthands}
	\fl
	\eqalign{
		\frac{p_r}{P_{ant}} 
		=
		\frac{\sqrt{\pi^3} e^4}{2 c^2 \Omega^2 \epsilon_{0}^2 m_e^2 \bar{W}_{y} }
		\sum_{\mu = a, b, \ldots}
		\int 
		& \varepsilon_{\mu} 
		\frac{
			\bar{W}_{y}
			\det \left[
				\textrm{Im} \left( \bm{\Psi}_{w} \right) 
			\right] 
		}{
			\sqrt{2}
			\left| 
				\det \left[ \bm{M}_{w} \right] 
			\right| 
			\left[
				 - \textrm{Im} \left( M_{yy}^{-1} \right)
			\right]^{\frac{1}{2}}  
		} 
		\exp \left(
			- 2 \frac{\theta_{m}^2}{\left(\Delta \theta_{m}\right)^2}
		\right) \\
		& \times		
		\left<
			\delta n_{e}^2 (t)
		\right>_t
		\widetilde{C}_\mu (k_{\perp,1}, k_{\mu,2}, \omega) \
		g_{ant}^{2} \left|
			g_\mu \frac{\rmd K}{\rmd \tau} \Bigr|_{\tau_\mu} 
		\right|^{-1} 		
		\ \rmd k_{\perp,1},
	}
	\end{equation}
	where $\varepsilon$ and $\bar{W}_y$ will be introduced in the following lines. The first piece we focus on is related to the polarisation; we call this piece $\varepsilon$,
	\begin{equation}
		\varepsilon_\mu
		=
		\frac{\Omega^4 \epsilon_{0}^2 m_e^2}{e^4 n_{e,\mu}^2}
		|\hat{\mathbf{e}}^*_\mu \cdot (\bm{\epsilon}_{eq,\mu} -\bm{1}) \cdot \hat{\mathbf{e}}_\mu|^2 .
	\end{equation}	
	The second pertains to the widths (beam) and curvatures (beam and field lines),
	\begin{equation}
		\frac{\bar{W}_{y}}{\sqrt{2}}
		\frac{
			\det \left[
				\textrm{Im} \left( \bm{\Psi}_{w,\mu} \right) 
			\right] 
		}{
			\left| 
				\det \left[ \bm{M}_{w,\mu} \right] 
			\right| 
			\left[
				 - \textrm{Im} \left( M_{yy,\mu}^{-1} \right)
			\right]^{\frac{1}{2}}  
		} .
	\end{equation}
	Here $\bar{W}_{y}$ is the value that $W_y  = \sqrt{2 / \textrm{Im} (\Psi_{yy}) }$ takes at the beam waist, in vacuum. The final piece contains the integrable divergence
	\begin{equation}
		g_{ant}^2
		\left|
			g_\mu \frac{\rmd K}{\rmd \tau} \Bigr|_{\tau_\mu} 
		\right|^{-1} 
		\ \rmd k_{\perp,1} .
	\end{equation} 

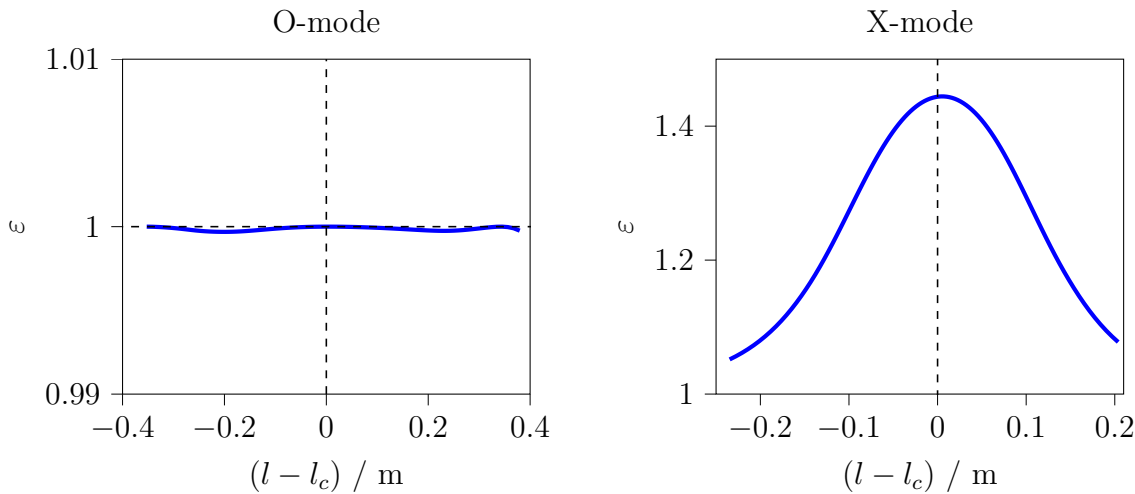
\begin{figure} 
	
		\begin{subfigure}{.5\textwidth}
			\centering		

			\begin{tikzpicture}
			
			\definecolor{color0}{rgb}{0.12156862745098,0.466666666666667,0.705882352941177}
			
			\begin{axis}[
			tick align=outside,
			tick pos=left,
			title={O-mode},
			x grid style={white!69.0196078431373!black},
			xlabel={$(l - l_c)$ / m},
			xmin=-0.4, xmax=0.4,
			xtick style={color=black},
			y grid style={white!69.0196078431373!black},
			ylabel={$ \varepsilon$},
			ymin=0.99, ymax=1.01,
			ytick = {0.99,1.00,1.01},
			ytick style={color=black},
			width = 7cm
			]
			\addplot [ultra thick, blue]
			table {%
			-0.352570424333031 0.999999811013791
			-0.340321159237281 0.999992898116656
			-0.328469754734922 0.999978322942656
			-0.317014938310637 0.999957937470287
			-0.305951373767537 0.999932769233296
			-0.295270262516097 0.999903649235293
			-0.284959452235854 0.999871527136518
			-0.275004354516252 0.999837558655326
			-0.265387174212452 0.999803065424901
			-0.256088337130649 0.999771675051997
			-0.247088509953242 0.999746367110861
			-0.238368636984674 0.999726445370579
			-0.22990997708687 0.9997112626286
			-0.221694142993455 0.999700448401626
			-0.213703142177311 0.999693501241991
			-0.205919418236827 0.999689908323652
			-0.198325891581765 0.999689323560875
			-0.190905998058727 0.999691566264119
			-0.183643724076422 0.999696588453715
			-0.176523935456617 0.999704339385776
			-0.169535418881836 0.999714512303318
			-0.162667909999033 0.999726710317525
			-0.155911471170953 0.99974052801898
			-0.149256802619912 0.999755570172548
			-0.142695230724462 0.999771467962573
			-0.136218693446852 0.999787888302495
			-0.129819723135348 0.999804536434875
			-0.123491426991022 0.999821150769206
			-0.117227465517183 0.999837501512992
			-0.111022029281015 0.999853396745836
			-0.104869814312117 0.999868685541393
			-0.0987659964432693 0.999883257189542
			-0.0927062048671312 0.999897033387416
			-0.0866864951412505 0.999909959863999
			-0.0807033218253898 0.999922000046805
			-0.0747533912066941 0.999933131497012
			-0.0688333611969818 0.99994334416576
			-0.0629402531392075 0.999952635986198
			-0.0570712455573491 0.999961014632377
			-0.0512236483645154 0.999968495634261
			-0.0453948942586195 0.999975100559187
			-0.0395825299952612 0.999980855433808
			-0.033784207567538 0.999985789287384
			-0.0279976753179607 0.999989932952232
			-0.0222207690030891 0.999993318112891
			-0.0164514028270064 0.99999597651242
			-0.0106875604554579 0.999997939341267
			-0.00492728601844827 0.999999236771278
			0 0.999999840463128
			0.000831344945212364 0.999999897607153
			0.00659015429680082 0.999999949050362
			0.0123509906214001 0.999999416608248
			0.018115684968791 0.999998324081259
			0.0238861812752366 0.999996693038973
			0.0296643660793645 0.999994542877769
			0.0354521396041003 0.999991890865839
			0.0412514381891381 0.999988752125375
			0.0470642435584872 0.999985139629482
			0.052892592061175 0.999981064228791
			0.058738583884165 0.999976534668753
			0.0646043922337484 0.999971557621717
			0.0704922724784749 0.99996613783462
			0.076404571243156 0.999960278367029
			0.0823437354396893 0.999953980880637
			0.0883123212164772 0.999947246099296
			0.0943130028041621 0.999940074547024
			0.100348581231362 0.99993246709359
			0.106421992880196 0.999924425281543
			0.112536352643057 0.999915950884899
			0.118695185663749 0.999907044531513
			0.124902044464765 0.999897710218575
			0.131160685683606 0.999887952799351
			0.13747515061653 0.999877781249119
			0.143849789820126 0.999867213334828
			0.15028928738231 0.999856280584498
			0.156798684762134 0.999845033028916
			0.163383404068072 0.999833540123315
			0.170049270618111 0.999821894710841
			0.176802534603124 0.999810219190602
			0.183649891659085 0.999798674176115
			0.190598502144749 0.999787469283954
			0.197656008920432 0.999776876436112
			0.204830550339556 0.999767261797289
			0.212129933573405 0.999759062186777
			0.219563852708665 0.999752881828545
			0.227143721351604 0.999749439380265
			0.234881476660194 0.999749442439512
			0.242789596764062 0.999753531990214
			0.250881115380802 0.999762214487668
			0.259169633580632 0.999775808969867
			0.267669328716497 0.99979445937673
			0.276394960600393 0.999818008348234
			0.285361875066005 0.999845637762706
			0.294586005110029 0.999875824796584
			0.304083869847837 0.999907518830762
			0.31387257155187 0.999940600666037
			0.323969791062703 0.999971366771723
			0.334393275168124 0.999993836049361
			0.345158542573667 0.999999063154587
			0.35627972699155 0.999974296583346
			0.36776879303264 0.999902249690184
			0.379635139549282 0.999760725812132
			};
			\addplot [semithick, black, dashed]
			table {%
			0 0.99
			0 1.01
			};
			\addplot [semithick, black, dashed]
			table {%
			-0.414320018083479 1
			0.459947992827944 1
			};
			\end{axis}
			
			\end{tikzpicture}
		\end{subfigure}	
		\begin{subfigure}{.5\textwidth}
			\centering
			\begin{tikzpicture}
			
			\definecolor{color0}{rgb}{0.12156862745098,0.466666666666667,0.705882352941177}
			
			\begin{axis}[
			tick align=outside,
			tick pos=left,
			title={X-mode},
			x grid style={white!69.0196078431373!black},
			xlabel={$(l - l_c)$ / m},
			xmin=-0.25, xmax=0.21,
			xtick style={color=black},
			y grid style={white!69.0196078431373!black},
			ylabel={$ \varepsilon$},
			ymin=1.0, ymax=1.5,
			ytick style={color=black},
			width = 7cm
			]
			\addplot [ultra thick, blue]
			table {%
			-0.234673502489608 1.05171826025688
			-0.227673632743631 1.05640071470934
			-0.220804845649404 1.06147927769838
			-0.214067797964014 1.06697104870159
			-0.207462695198663 1.07289165551213
			-0.200989319321115 1.07925457152952
			-0.194647113906197 1.08607041848873
			-0.188435187382464 1.09334630161745
			-0.182352316566019 1.10108514562052
			-0.176396950990916 1.10928505549595
			-0.170567218236238 1.11793873028268
			-0.164861087175742 1.12703205154297
			-0.15927619374263 1.13654543833999
			-0.153809912839053 1.14645415012391
			-0.148459564583947 1.15672708738072
			-0.143222414226678 1.16733651559187
			-0.138095672258192 1.17829353630822
			-0.133076494747171 1.18955011394625
			-0.128161983930599 1.20104555849774
			-0.12334918908977 1.21271292311041
			-0.118635107744001 1.2244851955461
			-0.11401668719498 1.23629734199425
			-0.109490826454649 1.24808696267656
			-0.105054378588645 1.25979352924189
			-0.100704153720616 1.27135907787719
			-0.096437115274321 1.28272713314803
			-0.0922503762163327 1.29384537342683
			-0.0881409898750145 1.3046687938383
			-0.0841060312743504 1.31515852467387
			-0.080142595061213 1.32528193397456
			-0.0762477933818524 1.33501256942358
			-0.0724187537135119 1.34433002012143
			-0.0686526166582495 1.3532194187824
			-0.0649465337072617 1.36167089523203
			-0.0612976649852623 1.36967899457716
			-0.0577031769857397 1.37724176249376
			-0.0541602403091568 1.38436005730991
			-0.0506660274173584 1.39103711055449
			-0.0472177104185365 1.39727814229306
			-0.0438124588980487 1.40308996300146
			-0.0404474378111471 1.40848059596461
			-0.0371198054541634 1.41345889757957
			-0.0338267115309066 1.41803424372236
			-0.0305653108664209 1.42221583437674
			-0.0273327488495939 1.42601255828709
			-0.0241261347480264 1.42943336308212
			-0.0209425586071458 1.43248667268574
			-0.0177790917052044 1.43518018971286
			-0.0146327872116972 1.4375207299028
			-0.0115006810832822 1.43951408633808
			-0.00837979318309875 1.44116491762304
			-0.00526712860420572 1.44247665890471
			-0.00215967861416491 1.44345145354157
			0 1.44393121162935
			0.000945652710336004 1.44409009870735
			0.00405175617128939 1.44439200843094
			0.00716174805633438 1.4443551886278
			0.0102786814084741 1.44397622446591
			0.0134056158016942 1.44325028563521
			0.0165456153311557 1.44217115077663
			0.0197017467242884 1.44073125241435
			0.022877080233018 1.43892172619938
			0.0260746882169646 1.43673249276756
			0.0292976447058656 1.43415235978727
			0.0325490255952428 1.43116915419825
			0.0358319084664504 1.42776989490402
			0.039149374835218 1.42394098237539
			0.0425045115764655 1.41966843180446
			0.0459004132538771 1.41493809399341
			0.049340185342625 1.40973592363579
			0.052826948185406 1.4040482626502
			0.0563638416616025 1.3978621808507
			0.0599540306622225 1.39116581746881
			0.0636007113746453 1.3839487617935
			0.0673071180973269 1.37620270972216
			0.0710765301753148 1.36792265711562
			0.0749122783705316 1.35910786709311
			0.0788177523846231 1.34976276569929
			0.0827964070830047 1.33989796382914
			0.0868517695258439 1.32953098005534
			0.0909874477156181 1.31868686205626
			0.0952071367761354 1.30739877467755
			0.0995146255907414 1.2957083446409
			0.10391380581418 1.2836648016543
			0.108408680970405 1.27132361643983
			0.113003370834368 1.25874548918823
			0.117702122432847 1.24599665283157
			0.122509305568548 1.23315218620609
			0.127429402171262 1.22029495250073
			0.132467008100852 1.20751152788426
			0.137626835258604 1.19488167111086
			0.142913713480807 1.18243457086812
			0.148332465669281 1.17017975450283
			0.153888153074875 1.15818365000968
			0.159585943611869 1.14650554373923
			0.165430756244211 1.13520150988574
			0.171427250503216 1.12432277210691
			0.177579818195765 1.11391455306042
			0.183892576496236 1.10401544118801
			0.190369362315784 1.09465725107411
			0.197013671210967 1.0858654233496
			0.203828005765232 1.07766056981389
			};
			\addplot [semithick, black, dashed]
			table {%
			0 0
			0 2
			};
			\end{axis}
			
			\end{tikzpicture}
		
		\end{subfigure}
		\caption{The O-mode polarisation piece (left) and that of the X-mode (right) along the ray, for the MAST test scenario. Here $l - l_c$ is the distance from the cut-off, as measured by the arc-length along the central ray. We define the cut-off to be where $K$ is minimum. Notice that the O-mode polarisation piece is almost exactly $1$, as expected. The X-mode polarisation piece does seem to give more localisation to the cut-off, but this may not be generally true. }
		\label{fig:polarisation}
	\end{figure}

	We first look into the polarisation piece, $\varepsilon$. Using equation (\ref{eq:D_definition}) and recalling that the polarisation $\hat{\mathbf{e}}$ is the eigenvector of $\bm{D}$ corresponding to $H = 0$, see \ref{appendix_beam_tracing_zeroth_order}, we re-express the polarisation piece as
	\begin{equation} \label{eq:polarisation_piece_explicit}
		\varepsilon_\mu
		=
		\frac{\Omega^4 \epsilon_{0}^2 m_e^2}{e^4 n_{e,\mu}^2}
		\left|
		\frac{c^2}{\Omega^2} K_{\mu}^2 
		\left[
			1 - \left(
				\hat{\mathbf{K}}_{\mu} \cdot \hat{\mathbf{e}}_{\mu}
			\right)^2
		\right]
		- 1
		\right|^2
		.
	\end{equation}
	This expression can be made simpler for the O-mode, but is less obvious for the X-mode. Assuming the polarisation is reasonably well-aligned upon entering the plasma, we have $\hat{\mathbf{e}} \cdot \hat{\mathbf{b}} \simeq 1$ for the O-mode. For the cold plasma dispersion relation, $\mathbf{K} = K \cos \theta_m  \hat{\mathbf{u}}_1 + K \sin \theta_m  \hat{\mathbf{b}}$ by definition of the mismatch angle, and consequently $| \hat{\mathbf{K}} \cdot \hat{\mathbf{e}} | \sim \theta_m$ for the O-mode. Thus, the O-mode polarisation piece is
	\begin{equation}
		\varepsilon_\mu
		=
		\frac{\Omega^4 \epsilon_{0}^2 m_e^2}{e^4 n_{e,\mu}^2}
		\left(
			\frac{c^2}{\Omega^2} K_\mu^2 
			- 1
		\right)^2
		.
	\end{equation}	
	Making use of the O-mode dispersion, $\epsilon_{bb,\mu} - K_\mu^2 c^2 / \Omega^2 \simeq 0$, we find that
	\begin{equation} \label{eq:polarisation_piece_O} 
		\varepsilon_\mu
		\simeq
		1
		,
	\end{equation}
	which is constant, see Figure \ref{fig:polarisation} (left). For the X-mode, even when $\theta_m = 0$, the polarisation piece $\varepsilon$ depends on the relative sizes of $\Omega_{ce}$ and $\Omega_{pe}$. 
	Since we do not make any assumptions about them, we will not further simplify the polarisation piece for the X-mode in this work. 
	Should one wish to calculate the polarisation piece of the X-mode, one should use its full expression, given in equation (\ref{eq:polarisation_piece_explicit}). An example is given in Figure \ref{fig:polarisation} (right).
	
	Subsequently, we briefly discuss the piece related to the widths and curvatures,
	\begin{equation}
		\frac{\bar{W}_{y}}{\sqrt{2}}
		\frac{
			\det \left[
				\textrm{Im} \left( \bm{\Psi}_{w,\mu} \right) 
			\right] 
		}{
			\left| 
				\det \left[ \bm{M}_{w,\mu} \right] 
			\right| 
			\left[
				 - \textrm{Im} \left( M_{yy,\mu}^{-1} \right)
			\right]^{\frac{1}{2}}  
		} .
	\end{equation}	
	Due to the complexity of this piece, we do not simplify it any further. Instead, we explain our choice of normalisation. The idea is that we want this piece to be unity under certain conditions which make the problem easier. Consider the waist of a Gaussian beam in vacuum. Take $\bm{M} = \bm{\Psi}$ for simplicity. Assume also that this beam is not astigmatic, that is, the beam widths are minimised at the same point in $\tau$. At this same point, the real parts of $\bm{\Psi}$ are simply zero. If $\bm{\Psi}_w$ is also diagonal in the $(\hat{\mathbf{x}},\hat{\mathbf{y}},\hat{\mathbf{g}})$ basis, then the prefactor $\bar{W}_{y} / \sqrt{2}$ ensures that this piece is unity. 
	
	Finally, we simplify the piece containing the integrable divergence. First, differentiate the Bragg condition with respect to $\tau$
	\begin{equation}
		\frac{\rmd k_{\perp,1}}{\rmd \tau} = - 2 \frac{\rmd K}{\rmd \tau} ,
	\end{equation}
	and use this to find that
	\begin{equation}
		\rmd k_{\perp,1} \left|
			g_\mu \frac{\rmd K}{\rmd \tau} \Bigr|_{\tau_\mu} 
		\right|^{-1}
		= \frac{2}{g} \rmd \tau_\mu .
	\end{equation}
	Since we sum over $\mu$, equation (\ref{eq:A_r2_explicit_wavenumber_resolution}), we can express the integration as being over the beam path,
	\begin{equation}
		\frac{2}{g_\mu} \rmd \tau_\mu
		=
		\frac{2}{g^2} \rmd l .
	\end{equation}
	This piece is thus
	\begin{equation}
		g_{ant}^2
		\left|
			g_\mu \frac{\rmd K}{\rmd \tau} \Bigr|_{\tau_\mu} 
		\right|^{-1} 
		\ \rmd k_{\perp,1} 
		=
		\frac{g_{ant}^2}{g^2} \rmd l ,
	\end{equation}
	where one has to be careful to use the form of the dispersion relation used in our beam tracing derivation, equation (\ref{eq:H_and_e_definition}). Using the definition of $H$, we find that equation (\ref{eq:g_coldplasma}) gives us $g_{ant} = 2 c / \Omega$ (since we take the antenna to be in vacuum).

	In summary, after all these simplifications are applied, equation (\ref{eq:A_r2_final_with_shorthands}) is now
	\begin{equation} \label{eq:A_r2_final_cleaned}
	\fl
	\eqalign{
		\frac{p_r}{P_{ant}} 
		=
		\frac{\sqrt{\pi^3} e^4}{2 c^2 \Omega^2 \epsilon_{0}^2 m_e^2 \bar{W}_{y} }
		\int 
		& \varepsilon_\mu \ \frac{ g_{ant}^{2} }{ g^{2} }
		\frac{
			\bar{W}_{y}
			\det \left[
				\textrm{Im} \left( \bm{\Psi}_{w} \right) 
			\right] 
		}{
			\sqrt{2}
			\left| 
				\det \left[ \bm{M}_{w} \right] 
			\right| 
			\left[
				 - \textrm{Im} \left( M_{yy}^{-1} \right)
			\right]^{\frac{1}{2}}  
		} \\
		& \times		
		\exp \left(
			- 2 \frac{\theta_{m}^2}{\left(\Delta \theta_{m}\right)^2}
		\right) 
		\left<
			\delta n_{e}^2 (t)
		\right>_t
		\widetilde{C}_l (\omega)
		\ \rmd l .
	}
	\end{equation}
	We remind readers that the variables in this equation are now functions of arc-length along the ray, that is $\tau (l)$ and $k_{l,1} = k_{\perp,1} (l)$.	As such, we have dropped the subscript $_\mu$. The notation $\widetilde{C}_l$ indicates that the correlation function is evaluated at $\mathbf{q} (\tau (l))$, $k_{l,1} (l)$, $k_{\perp,2} = 0$, and $\Delta u_{\parallel} = 0$, which are all functions of arc-length as well. Note that $\varepsilon$ is given by equation (\ref{eq:polarisation_piece_explicit}); for the O-mode, this can be simplified to equation (\ref{eq:polarisation_piece_O}), while we do not further simplify it for the X-mode.
	
	Unlike typical papers on DBS \cite{Hirsch:DBS:2001, Holzhauer:DBS:1998, Conway:DBS:2004, Pinzon:DBS_tilt_2019}, our model does not require one to assume that the signal comes entirely from the cut-off. We will later assess the validity of this assumption. Literature on reciprocity does not \emph{a priori} make such an assumption \cite{Gusakov:scattering_slab:2004, Bulanin:spatial_spectral_resolution:2006}, and we now compare our work with theirs. The beam model is applicable in general tokamak geometry (and some regimes of stellarators), can easily account for a wide variety of initial beam conditions and equilibrium density profiles, and does not rely on assuming the density fluctuations have an exponential spectrum. Consequently, we better understand the effect of geometry on wavenumber resolution. We can also perform a more realistic assessment of localisation. Finally, for the first time, we present a quantitative description of the mismatch attenuation as a function of beam properties. The subsequent sections explore localisation, wavenumber resolution, and mismatch attenuation, which are various pieces of equations (\ref{eq:A_r2_explicit_wavenumber_resolution}) and (\ref{eq:A_r2_final_cleaned}).

	\section{Localisation} \label{section_localisation}
	In the previous section, we expressed the backscattered signal as a line integral along the central ray, equation (\ref{eq:A_r2_final_cleaned}). In this section, we study the integrand of this equation to understand where, along the line integral, does most of the signal come from. Traditionally, the DBS signal is thought to come from the region around the cut-off \cite{Hirsch:DBS:2001, Gusakov:scattering_slab:2004}; we now use the beam model to evaluate and understand this insight. The integrand in equation (\ref{eq:A_r2_final_cleaned}) consists of two parts, the turbulence that one seeks to measure $\left< \delta n_{e}^2 (t) \right>_t \widetilde{C}_l (\omega)$, and the prefactor
	\begin{equation} \label{eq:localisation}
	\eqalign{
		\frac{ g_{ant}^{2} }{ g^{2} }
		\frac{
			\bar{W}_{y}
			\det \left[
				\textrm{Im} \left( \bm{\Psi}_{w} \right) 
			\right] 
		}{
			\sqrt{2}
			\left| 
				\det \left[ \bm{M}_{w} \right] 
			\right| 
			\left[
				 - \textrm{Im} \left( M_{yy}^{-1} \right)
			\right]^{\frac{1}{2}}  
		} 
		,
	}
	\end{equation}
	which we call the localisation.	Other authors refer to analogous quantities as the spatial resolution \cite{Gusakov:scattering_slab:2004}, instrumentation response function \cite{Lechte:2D_fullwave:2012, Lechte:2D_fullwave:2017}, weighting function \cite{Bulanin:spatial_spectral_resolution:2006}, or filter function \cite{RuizRuiz:RCDR:2022}. In our expression of the localisation, we have deliberately omitted the pieces associated with polarisation, equation (\ref{eq:polarisation_piece_explicit}), and mismatch
	\begin{equation} \label{eq:localisation_ray_piece}
		\exp\left[
			- 2 \frac{\theta_{m}^2}{\left( \Delta \theta_{m} \right)^2} 
		\right] .
	\end{equation}
	The mismatch may indeed affect localisation; in fact, it is the dominant mechanism of localisation in the \emph{large mismatch angle} ordering, see Section \ref{section_backscattered_E_ST}. However, assuming we are not dealing with such extreme situations, we deem it more physically insightful to discuss the effect of mismatch later (Section \ref{section_mismatch}). 

	We analyse two contributions to the localisation, equation (\ref{eq:localisation}), which we call the \emph{ray} piece
	\begin{equation}
		\frac{g_{ant}^2}{g^2},
	\end{equation} 
	and the \emph{beam} piece 
	\begin{equation} \label{eq:localisation_beam_piece}
		\frac{
			\bar{W}_{y}
			\det \left[
				\textrm{Im} \left( \bm{\Psi}_{w} \right) 
			\right] 
		}{
			\sqrt{2}
			\left| 
				\det \left[ \bm{M}_{w} \right] 
			\right| 
			\left[
				 - \textrm{Im} \left( M_{yy}^{-1} \right)
			\right]^{\frac{1}{2}}  
		} 
		.
	\end{equation} 
	They are named as such because the former can be determined with ray tracing alone, while the latter requires beam tracing. The \emph{ray} piece requires one to use the appropriate dispersion relation, equation (\ref{eq:H_and_e_definition}), to calculate the group velocity. We now briefly discuss the physical interpretations of these two pieces. It seems that the ray piece appears loosely related to scattering efficiency. The beam piece is more complicated. The beam area is proportional to $\det \left[\textrm{Im} \left( \bm{\Psi}_{w} \right) \right] $; hence, that part decreases as the beam expands. However, the interpretation of $\left| \det \left[ \bm{M}_{w} \right] \right|$ is less straightforward. For a circular beam in vacuum, it turns out that the overall beam piece also decreases as the beam expands.
		
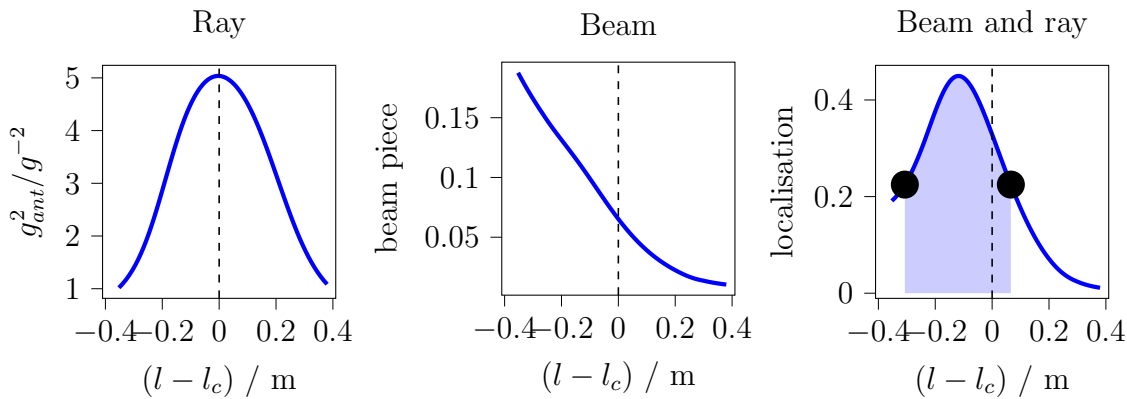
\begin{figure}
	\centering
		\begin{subfigure}{.3\textwidth}
		\centering
		\begin{tikzpicture}
		
		\begin{axis}[
		tick align=outside,
		tick pos=left,
		title={Ray},
		x grid style={white!69.0196078431373!black},
		xlabel={$ (l - l_c)$ / m},
		xmin=-0.41, xmax=0.41,
		xtick style={color=black},
		y grid style={white!69.0196078431373!black},
		ylabel={$g_{ant}^2 / g^{-2}$},
		ymin=0.817429415494464, ymax=5.23792722913036,
		ytick style={color=black},
		width = 3.1cm,
		height = 3.1cm,
		scale only axis		
		]
		\addplot [ultra thick, blue]
		table {%
		-0.352570424333031 1.0183611342961
		-0.340321159237281 1.08794410767026
		-0.328469754734922 1.1651494828826
		-0.317014938310637 1.25018686985741
		-0.305951373767537 1.3431321437022
		-0.295270262516097 1.44391972643272
		-0.284959452235854 1.55230717892193
		-0.275004354516252 1.66774321205897
		-0.265387174212452 1.78961043032079
		-0.256088337130649 1.91732107114041
		-0.247088509953242 2.05010904827075
		-0.238368636984674 2.18705038851473
		-0.22990997708687 2.32709103442269
		-0.221694142993455 2.4690908708525
		-0.213703142177311 2.61188289211926
		-0.205919418236827 2.75434119075612
		-0.198325891581765 2.89545279478611
		-0.190905998058727 3.03438906051207
		-0.183643724076422 3.17057103282462
		-0.176523935456617 3.3036715130753
		-0.169535418881836 3.43304228797262
		-0.162667909999033 3.55808703645198
		-0.155911471170953 3.67835811982067
		-0.149256802619912 3.79350314372668
		-0.142695230724462 3.90326296427049
		-0.136218693446852 4.00746640032153
		-0.129819723135348 4.10602236310434
		-0.123491426991022 4.19891023677839
		-0.117227465517183 4.28616890897791
		-0.111022029281015 4.36788486732229
		-0.104869814312117 4.44418021864753
		-0.0987659964432693 4.51520119081308
		-0.0927062048671312 4.58110769547325
		-0.0866864951412505 4.64206400531041
		-0.0807033218253898 4.69823057398947
		-0.0747533912066941 4.74972487197942
		-0.0688333611969818 4.79657722964266
		-0.0629402531392075 4.83889104550836
		-0.0570712455573491 4.87678129597383
		-0.0512236483645154 4.91036580679095
		-0.0453948942586195 4.9397620636238
		-0.0395825299952612 4.96508446639527
		-0.033784207567538 4.98644203434243
		-0.0279976753179607 5.00393654194348
		-0.0222207690030891 5.01766106792428
		-0.0164514028270064 5.0276989401128
		-0.0106875604554579 5.03412304956631
		-0.00492728601844827 5.03699551032873
		0 5.03667280819582
		0.000831344945212364 5.03636764110562
		0.00659015429680082 5.0322802440193
		0.0123509906214001 5.02476415425557
		0.018115684968791 5.01384309557541
		0.0238861812752366 4.99954292086563
		0.0296643660793645 4.98187953817424
		0.0354521396041003 4.96086302067584
		0.0412514381891381 4.93649806681076
		0.0470642435584872 4.90878370577066
		0.052892592061175 4.87771322376676
		0.058738583884165 4.84327432382283
		0.0646043922337484 4.80544952640779
		0.0704922724784749 4.76421681278942
		0.076404571243156 4.71955051965871
		0.0823437354396893 4.6714224915522
		0.0883123212164772 4.61980348632531
		0.0943130028041621 4.56466482677968
		0.100348581231362 4.50598031577409
		0.106421992880196 4.44372839723739
		0.112536352643057 4.37789273672834
		0.118695185663749 4.30845207553445
		0.124902044464765 4.23540164111126
		0.131160685683606 4.15874838175942
		0.13747515061653 4.07850928317751
		0.143849789820126 3.99471278199563
		0.15028928738231 3.9074006898133
		0.156798684762134 3.81663058994782
		0.163383404068072 3.72247870338927
		0.170049270618111 3.62504290917531
		0.176802534603124 3.52444584698683
		0.183649891659085 3.42083796674106
		0.190598502144749 3.31440037533717
		0.197656008920432 3.2053472916728
		0.204830550339556 3.09392836859647
		0.212129933573405 2.98060672779795
		0.219563852708665 2.86564865119474
		0.227143721351604 2.74921418048685
		0.234881476660194 2.63159954308154
		0.242789596764062 2.51321755231688
		0.250881115380802 2.39457428629192
		0.259169633580632 2.2762430554642
		0.267669328716497 2.15883702717165
		0.276394960600393 2.04298264603234
		0.285361875066005 1.92929480610936
		0.294586005110029 1.81835394436272
		0.304083869847837 1.7106865079917
		0.31387257155187 1.60675229598816
		0.323969791062703 1.50693773497678
		0.334393275168124 1.41162757896776
		0.345158542573667 1.32140363274477
		0.35627972699155 1.23664685909149
		0.36776879303264 1.15764540155376
		0.379635139549282 1.08460194486862
		};
		\addplot [semithick, black, dashed]
		table {%
		0 0.817429415494464
		0 5.23792722913036
		};
		\end{axis}
		
		\end{tikzpicture}

		\end{subfigure}
		\begin{subfigure}{.3\textwidth}
		\centering
		\begin{tikzpicture}
		\begin{axis}[
		tick align=outside,
		tick pos=left,
		title={Beam},
		x grid style={white!69.0196078431373!black},
		xlabel={$ (l - l_c)$ / m},
		xmin=-0.41, xmax=0.41,
		xtick style={color=black},
		y grid style={white!69.0196078431373!black},
		ylabel={beam piece},
		ymin=0.0015287682545884, ymax=0.196466693868004,
		ytick style={color=black},
		yticklabel style={
		        /pgf/number format/fixed,
		        /pgf/number format/precision=2
		},		
		scaled y ticks=false,
		width = 3.1cm,
		height = 3.1cm,
		scale only axis
		]
		\addplot [ultra thick, blue]
		table {%
		-0.352570424333031 0.187605879067394
		-0.340321159237281 0.182186000710253
		-0.328469754734922 0.177155865147291
		-0.317014938310637 0.172469841470032
		-0.305951373767537 0.168090318398959
		-0.295270262516097 0.163985605237763
		-0.284959452235854 0.160129209025693
		-0.275004354516252 0.156496975597793
		-0.265387174212452 0.153074981219758
		-0.256088337130649 0.149718488652108
		-0.247088509953242 0.146584891016021
		-0.238368636984674 0.143655165315208
		-0.22990997708687 0.140884405876343
		-0.221694142993455 0.138247020073389
		-0.213703142177311 0.135727342220874
		-0.205919418236827 0.133303298467118
		-0.198325891581765 0.130952770042553
		-0.190905998058727 0.128655500331387
		-0.183643724076422 0.126395773857154
		-0.176523935456617 0.124170226077219
		-0.169535418881836 0.121972688127069
		-0.162667909999033 0.119799558004086
		-0.155911471170953 0.117646353224116
		-0.149256802619912 0.115508670965228
		-0.142695230724462 0.113382019910539
		-0.136218693446852 0.111262352441519
		-0.129819723135348 0.109146674138107
		-0.123491426991022 0.107034579303558
		-0.117227465517183 0.104926358894702
		-0.111022029281015 0.102822427923518
		-0.104869814312117 0.100721885499884
		-0.0987659964432693 0.0986234837756906
		-0.0927062048671312 0.096527276921084
		-0.0866864951412505 0.094434976984122
		-0.0807033218253898 0.09235035830091
		-0.0747533912066941 0.0902778184662723
		-0.0688333611969818 0.0882179056118786
		-0.0629402531392075 0.0861720414167927
		-0.0570712455573491 0.0841417445134863
		-0.0512236483645154 0.0821285949692244
		-0.0453948942586195 0.0801339535001382
		-0.0395825299952612 0.0781590754166471
		-0.033784207567538 0.076205293126009
		-0.0279976753179607 0.0742738669206781
		-0.0222207690030891 0.072365970231573
		-0.0164514028270064 0.0704828309810105
		-0.0106875604554579 0.0686255907670019
		-0.00492728601844827 0.0667952930675304
		0 0.0652513374633073
		0.000831344945212364 0.0649928773897668
		0.00659015429680082 0.0632191209843408
		0.0123509906214001 0.0614746713647207
		0.018115684968791 0.0597605592201685
		0.0238861812752366 0.0580788052378253
		0.0296643660793645 0.0564290172557922
		0.0354521396041003 0.0548106263329351
		0.0412514381891381 0.0532232826194818
		0.0470642435584872 0.0516666705678469
		0.052892592061175 0.0501405364121642
		0.058738583884165 0.0486447839673663
		0.0646043922337484 0.0471792907911056
		0.0704922724784749 0.0457437739559231
		0.076404571243156 0.0443380059234889
		0.0823437354396893 0.0429616649198634
		0.0883123212164772 0.041614322140503
		0.0943130028041621 0.0402954557866497
		0.100348581231362 0.0390048060053673
		0.106421992880196 0.0377425648877168
		0.112536352643057 0.0365089341804818
		0.118695185663749 0.0353029842182688
		0.124902044464765 0.0341252244157551
		0.131160685683606 0.0329755681601637
		0.13747515061653 0.0318524597809329
		0.143849789820126 0.0307540092248257
		0.15028928738231 0.0296780617414144
		0.156798684762134 0.0286235480608639
		0.163383404068072 0.027590372263531
		0.170049270618111 0.0265784752675874
		0.176802534603124 0.0255878019822754
		0.183649891659085 0.024617995196289
		0.190598502144749 0.0236687409063296
		0.197656008920432 0.0227386983770099
		0.204830550339556 0.0218242737926663
		0.212129933573405 0.0209218249848407
		0.219563852708665 0.0200335845692373
		0.227143721351604 0.0191680788934881
		0.234881476660194 0.0183344479418745
		0.242789596764062 0.0175406006597914
		0.250881115380802 0.0167939292117922
		0.259169633580632 0.0160963598007539
		0.267669328716497 0.0154460217371522
		0.276394960600393 0.0148499894919347
		0.285361875066005 0.0143258261837844
		0.294586005110029 0.0138728394211137
		0.304083869847837 0.0133728551873584
		0.31387257155187 0.0128638846239707
		0.323969791062703 0.0123869431313811
		0.334393275168124 0.0119402653284833
		0.345158542573667 0.0115202572790356
		0.35627972699155 0.0111240559662635
		0.36776879303264 0.0107484285282331
		0.379635139549282 0.0103895830551982
		};
		\addplot [semithick, black, dashed]
		table {%
		0 0.0015287682545884
		0 0.196466693868004
		};
		\end{axis}
		
		\end{tikzpicture}
		
		\end{subfigure}
		\quad
		\begin{subfigure}{.3\textwidth}
		\centering		
		\begin{tikzpicture}
		
		\begin{axis}[
		tick align=outside,
		tick pos=left,
		title={Beam and ray},
		x grid style={white!69.0196078431373!black},
		xlabel={$ (l - l_c)$ / m},
		xmin=-0.41, xmax=0.41,
		xtick style={color=black},
		y grid style={white!69.0196078431373!black},
		ylabel={localisation},
		ymin=-0.0106546147738924, ymax=0.471655273988664,
		ytick style={color=black},
		width = 3.1cm,
		height = 3.1cm,
		scale only axis
		]
		\addplot [ultra thick, blue, name path=f]
		table {%
		-0.352570424333031 0.191050535807687
		-0.340321159237281 0.198208185972729
		-0.328469754734922 0.206413064665986
		-0.317014938310637 0.215619531252224
		-0.305951373767537 0.225767509686779
		-0.295270262516097 0.236782050253814
		-0.284959452235854 0.248569720725673
		-0.275004354516252 0.260996768760976
		-0.265387174212452 0.273944583012038
		-0.256088337130649 0.287058413031983
		-0.247088509953242 0.300515011411726
		-0.238368636984674 0.314181085114773
		-0.22990997708687 0.327850837804807
		-0.221694142993455 0.341344455185768
		-0.213703142177311 0.354503923139518
		-0.205919418236827 0.367162765831641
		-0.198325891581765 0.379167564004693
		-0.190905998058727 0.390390842780267
		-0.183643724076422 0.400746779262944
		-0.176523935456617 0.410217638663427
		-0.169535418881836 0.418737396317924
		-0.162667909999033 0.426257254307016
		-0.155911471170953 0.43274541864922
		-0.149256802619912 0.438182506434284
		-0.142695230724462 0.442559839130987
		-0.136218693446852 0.445880139030121
		-0.129819723135348 0.448158684869529
		-0.123491426991022 0.44942859072698
		-0.117227465517183 0.449732097226729
		-0.111022029281015 0.449116526948469
		-0.104869814312117 0.447626211123467
		-0.0987659964432693 0.445304871386133
		-0.0927062048671312 0.442201851126255
		-0.0866864951412505 0.43837320750031
		-0.0807033218253898 0.433883276888218
		-0.0747533912066941 0.428794799757297
		-0.0688333611969818 0.423143997304702
		-0.0629402531392075 0.416977119584894
		-0.0570712455573491 0.410340885853978
		-0.0512236483645154 0.403281444496663
		-0.0453948942586195 0.395842663508176
		-0.0395825299952612 0.388066411259011
		-0.033784207567538 0.379993276882917
		-0.0279976753179607 0.371661716795828
		-0.0222207690030891 0.363107911473531
		-0.0164514028270064 0.354366454619376
		-0.0106875604554579 0.34546966827027
		-0.00492728601844827 0.336447591292243
		0 0.328649637099849
		0.000831344945212364 0.327328024588167
		0.00659015429680082 0.318136333573764
		0.0123509906214001 0.30889572506809
		0.018115684968791 0.299630067233768
		0.0238861812752366 0.290367479579103
		0.0296643660793645 0.281122566425912
		0.0354521396041003 0.271908009315139
		0.0412514381891381 0.262736631760394
		0.0470642435584872 0.253620510614867
		0.052892592061175 0.244571157504372
		0.058738583884165 0.235600033177054
		0.0646043922337484 0.226717700588374
		0.0704922724784749 0.217933256961248
		0.076404571243156 0.209255458896833
		0.0823437354396893 0.200692087781179
		0.0883123212164772 0.19224999050576
		0.0943130028041621 0.183935249708376
		0.100348581231362 0.175754888080772
		0.106421992880196 0.167717707376122
		0.112536352643057 0.159832197774424
		0.118695185663749 0.15210121562776
		0.124902044464765 0.144534031493779
		0.131160685683606 0.137137090723678
		0.13747515061653 0.129910552908573
		0.143849789820126 0.122853433748023
		0.15028928738231 0.115964078920724
		0.156798684762134 0.109245509121935
		0.163383404068072 0.102704573169576
		0.170049270618111 0.0963481133054588
		0.176802534603124 0.0901828224299519
		0.183649891659085 0.0842141726325145
		0.190598502144749 0.0784476837436972
		0.197656008920432 0.0728854252589134
		0.204830550339556 0.0675227398111466
		0.212129933573405 0.0623597323076273
		0.219563852708665 0.0574092145994307
		0.227143721351604 0.0526971543066681
		0.234881476660194 0.0482489248264891
		0.242789596764062 0.0440833454563689
		0.250881115380802 0.0402143110563642
		0.259169633580632 0.0366392272147191
		0.267669328716497 0.0333454436486623
		0.276394960600393 0.0303382708257852
		0.285361875066005 0.0276387420496006
		0.294586005110029 0.0252257322808928
		0.304083869847837 0.0228767629423409
		0.31387257155187 0.0206690761548916
		0.323969791062703 0.0186663520256897
		0.334393275168124 0.0168552078378796
		0.345158542573667 0.015222909818672
		0.35627972699155 0.0137565288710377
		0.36776879303264 0.0124428688596383
		0.379635139549282 0.011268561988042
		};
    	\path[name path=axis] (axis cs:-0.306934167811999,0) -- (axis cs:0.0658454845039634,0);
		\addplot [semithick, black, mark=*, mark size=5, mark options={solid}, only marks]
		table {%
		-0.306934167811999 0.224866048613365
		0.0658454845039634 0.224866048613365
		};
		\addplot [
		    thick,
		    color=blue,
		    fill=blue, 
		    fill opacity=0.2
		]
		fill between[
		    of=f and axis,
		    soft clip={domain=-0.306934167811999:0.0658454845039634},
		];		
		\addplot [semithick, black, dashed]
		table {%
		0 -0.0106546147738924
		0 0.471655273988664
		};
		\end{axis}
		
		\end{tikzpicture}
		
		\end{subfigure}
		
		\caption{Ray (left) and beam (middle) contributions to localisation as a function of distance along the central ray from the cut-off location for the MAST test scenario. The beam piece is given explicitly in equation (\ref{eq:localisation_beam_piece}). Product of the ray and beam pieces (right); the shaded area is $80\%$ of the total area under the curve. Black points mark the start and end of the shaded area; note that we have chosen these points such that they have the same ordinate value, that is, the same value of localisation. The corresponding half width is $19\textrm{cm}$. }
		\label{fig:localisation_MAST_ray_beam}
	\end{figure}

	We now apply our model to our MAST test scenario. In Figure \ref{fig:localisation_MAST_ray_beam}, we show the localisation due to the \emph{ray} and \emph{beam} pieces. We see that there is some localisation to the cut-off due to the \emph{ray} piece. Here, we define the cut-off to be where $K$ is minimum. Since the localisation associated with the \emph{ray} and \emph{beam} pieces is complicated, we use the following method to determine the localisation length. First, we integrate the localisation with respect to arc length along the central ray, from the point the beam enters the plasma until the point it leaves. Secondly, we choose two points along the ray, integrate localisation with respect to arc length $l$ from the first point to the second, and make sure that this result is $80\%$ that of the first integral. These two points show the start and end of the region where most of the signal is coming from. In this work, we choose the first and second points such that the value of the localisation piece is the same at these two points. Thirdly, we take the localisation length to be half of the arc length between the first and second points. The localisation length associated with the combined \emph{beam} and \emph{ray} pieces calculated by this method is $19\textrm{cm}$, which is large, indicating that, in experiments, there are probably other mechanisms of localisation. 

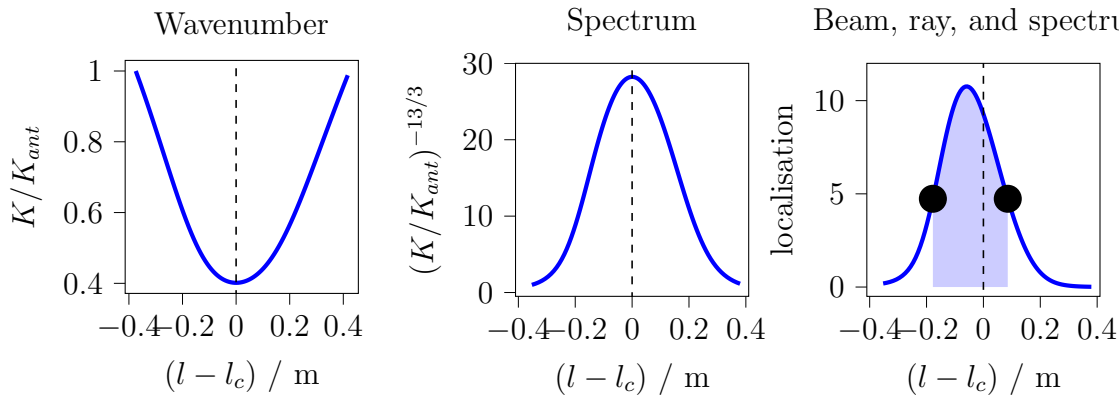
\begin{figure}
	\begin{subfigure}{.3\textwidth}
		\centering
		\begin{tikzpicture}
			\begin{axis}[
				tick align=outside,
				tick pos=left,
				title={Wavenumber},
				x grid style={white!69.0196078431373!black},
				xlabel={$ (l - l_c)$ / m},
				xmin=-0.41396289110446, xmax=0.456185483528433,
				xtick style={color=black},
				y grid style={white!69.0196078431373!black},
				ylabel={$K / K_{ant}$},
				ymin=0.371512886749067, ymax=1.02992795777385,
				ytick style={color=black},
				width = 3.1cm,
				height=3.1cm,
				scale only axis
		]
			\addplot [ultra thick, blue]
			table {%
			-0.37441069225751 1
			-0.356533478829015 0.962558511020698
			-0.340141554295917 0.92698040361795
			-0.324364076943073 0.891740256240884
			-0.309192909438258 0.857080052970303
			-0.294615867342869 0.823223700859437
			-0.28061708266782 0.790372962759241
			-0.26717742419877 0.758704958540095
			-0.254274963291678 0.728369037493594
			-0.241885488069664 0.699484226089045
			-0.229983040453507 0.672138395087072
			-0.218540446740328 0.646389403792068
			-0.207529817370456 0.622266899910868
			-0.196922999118495 0.59977582821625
			-0.186691966384159 0.578899419759459
			-0.176809153924176 0.559602994782834
			-0.167247729081306 0.541837246498091
			-0.157981807018855 0.525541727831222
			-0.148986610419824 0.510647695592287
			-0.140238585663037 0.497080797028535
			-0.131715477237115 0.484763431138467
			-0.123396364113873 0.473616964050509
			-0.11526166146011 0.463563304613795
			-0.107293102884643 0.454526267085791
			-0.0994737009815763 0.446432396194646
			-0.0917876938833679 0.439211978101006
			-0.0842204798873064 0.432799332422724
			-0.0767585449800989 0.427133394385738
			-0.0693893846287116 0.422157838347959
			-0.0621014223812901 0.417821264473961
			-0.0548839265469933 0.414077143892263
			-0.0477269272089867 0.4108839254063
			-0.0406211333318961 0.408204831216864
			-0.0335578518725752 0.406007805029367
			-0.0265289087095061 0.404265329652341
			-0.0195265722216378 0.402954283121877
			-0.0125434793117807 0.402055762838959
			-0.00557256423558111 0.401554936903209
			0.0013930096497673 0.401440844522921
			0.00835991602444414 0.401706281948599
			0.0153347287710237 0.402347630460331
			0.0223239831645058 0.403364727028455
			0.0293342348667785 0.40476076056439
			0.036372117260227 0.406542159015217
			0.0434443972091213 0.408718499772426
			0.0505580294824728 0.411302439698282
			0.0577202101073697 0.414309654842343
			0.064938428865829 0.417758770833896
			0.072220520994256 0.421671364757438
			0.0795747184567905 0.426071848597261
			0.08700970024349 0.43098749514475
			0.0945346421523113 0.436448343860496
			0.102159265450833 0.442487131087768
			0.109893883862767 0.449139240991052
			0.117749449631391 0.456442553829972
			0.125737596268365 0.464437299835942
			0.133870679181969 0.473165895543813
			0.142161810654327 0.482672565719631
			0.150624890453878 0.493003147585761
			0.159274630225181 0.504204600104409
			0.168126566952688 0.51632442013467
			0.177197067460228 0.52940997330198
			0.186503318605279 0.543507726653099
			0.196063301087078 0.55866212968667
			0.205895745675785 0.574914636238814
			0.216020067390196 0.592302385232862
			0.226456279261423 0.610857011134387
			0.23722488398876 0.630603414057532
			0.248346742417281 0.651558476165466
			0.259842921773144 0.673729772703928
			0.271734519676982 0.697114508151669
			0.284042469150095 0.721698105261971
			0.296787322522273 0.747453435269083
			0.309989023280102 0.774339961307263
			0.323666671513911 0.80230352239149
			0.337838287405532 0.831275965282943
			0.352520580459157 0.861175442101083
			0.367728733936241 0.891906701259618
			0.383476203399927 0.923362150677152
			0.399774544456424 0.955423897638054
			0.416633284681483 0.987967479695346
			};
			\addplot [semithick, black, dashed]
			table {%
			0 0
			0 2
		};
		\end{axis}
		\end{tikzpicture}
	\end{subfigure}
	\quad
	\begin{subfigure}{.3\textwidth}
		\centering
		\begin{tikzpicture}
			\begin{axis}[
				tick align=outside,
				tick pos=left,
				title={Spectrum},
				x grid style={white!69.0196078431373!black},
				xlabel={$ (l - l_c)$ / m},
				xmin=-0.41, xmax=0.41,
				xtick style={color=black},
				y grid style={white!69.0196078431373!black},
				ylabel={$(K / K_{ant})^{- 13 / 3}$},
				ymin=-0.5, ymax=30,
				ytick style={color=black},
				width = 3.1cm,
				height=3.1cm,
				scale only axis
				]
				\addplot [ultra thick, blue]
				table {%
					-0.352570424333031 1
					-0.340321159237281 1.15078669847587
					-0.328469754734922 1.33102646448438
					-0.317014938310637 1.54534873470029
					-0.305951373767537 1.79859432432298
					-0.295270262516097 2.0956690527152
					-0.284959452235854 2.44125400189318
					-0.275004354516252 2.83907310321753
					-0.265387174212452 3.29234113031974
					-0.256088337130649 3.80398104271354
					-0.247088509953242 4.37565778463055
					-0.238368636984674 5.00749793960915
					-0.22990997708687 5.69790916256145
					-0.221694142993455 6.44354709275618
					-0.213703142177311 7.23946730719902
					-0.205919418236827 8.07947927196831
					-0.198325891581765 8.95669265949456
					-0.190905998058727 9.86421937836289
					-0.183643724076422 10.7959735064561
					-0.176523935456617 11.7471080299458
					-0.169535418881836 12.7100619243236
					-0.162667909999033 13.6770561927188
					-0.155911471170953 14.6409979036072
					-0.149256802619912 15.5952707207746
					-0.142695230724462 16.533869465632
					-0.136218693446852 17.4514897474273
					-0.129819723135348 18.3435733179165
					-0.123491426991022 19.206312777755
					-0.117227465517183 20.0366213617753
					-0.111022029281015 20.8320747434415
					-0.104869814312117 21.590832220329
					-0.0987659964432693 22.3115444352925
					-0.0927062048671312 22.9932541413249
					-0.0866864951412505 23.6352956152834
					-0.0807033218253898 24.2371973226915
					-0.0747533912066941 24.798254740167
					-0.0688333611969818 25.3170261031161
					-0.0629402531392075 25.7929915778573
					-0.0570712455573491 26.2259261183402
					-0.0512236483645154 26.6158023903436
					-0.0453948942586195 26.9627485984463
					-0.0395825299952612 27.2670094530164
					-0.033784207567538 27.5289110625033
					-0.0279976753179607 27.7488303052784
					-0.0222207690030891 27.9271690364258
					-0.0164514028270064 28.0643333187733
					-0.0106875604554579 28.1607177320261
					-0.00492728601844827 28.2166947052834
					0 28.2327733007253
					0.000831344945212364 28.2326087316241
					0.00659015429680082 28.208775253574
					0.0123509906214001 28.1454839499008
					0.018115684968791 28.0430290067239
					0.0238861812752366 27.9018116187683
					0.0296643660793645 27.7222053017749
					0.0354521396041003 27.5046014782951
					0.0412514381891381 27.2494154167068
					0.0470642435584872 26.9570845073691
					0.052892592061175 26.6280697364898
					0.058738583884165 26.2628604165487
					0.0646043922337484 25.8619821627154
					0.0704922724784749 25.4260080296481
					0.076404571243156 24.9555726404816
					0.0823437354396893 24.4513890480824
					0.0883123212164772 23.9142679666986
					0.0943130028041621 23.3451388998263
					0.100348581231362 22.745072568589
					0.106421992880196 22.1153039169512
					0.112536352643057 21.45723586916
					0.118695185663749 20.7723331727206
					0.124902044464765 20.0623460346826
					0.131160685683606 19.3292457793414
					0.13747515061653 18.5751984184068
					0.143849789820126 17.8025679547452
					0.15028928738231 17.0139204574222
					0.156798684762134 16.2120275180422
					0.163383404068072 15.3998674967105
					0.170049270618111 14.580622833665
					0.176802534603124 13.7576716625651
					0.183649891659085 12.9345720372216
					0.190598502144749 12.1150372969665
					0.197656008920432 11.3029014630092
					0.204830550339556 10.5020786043357
					0.212129933573405 9.71770094611203
					0.219563852708665 8.95314884578083
					0.227143721351604 8.21085891485714
					0.234881476660194 7.49389888911144
					0.242789596764062 6.80566210512942
					0.250881115380802 6.14957189064943
					0.259169633580632 5.52881302405847
					0.267669328716497 4.94610588706372
					0.276394960600393 4.40353533948279
					0.285361875066005 3.90244121105738
					0.294586005110029 3.44337145155273
					0.304083869847837 3.02609333189147
					0.31387257155187 2.64965347024782
					0.323969791062703 2.31247442792801
					0.334393275168124 2.01271514241273
					0.345158542573667 1.74887575532806
					0.35627972699155 1.51861045160446
					0.36776879303264 1.31925590847362
					0.379635139549282 1.14798311915535
				};
				\addplot [semithick, black, dashed]
				table {%
					0 -1
					0 32
				};
			\end{axis}
			
		\end{tikzpicture}
	\end{subfigure}
	\begin{subfigure}{.3\textwidth}
		\centering		
		\begin{tikzpicture}
			\begin{axis}[
				tick align=outside,
				tick pos=left,
				title={Beam, ray, and spectrum},
				x grid style={white!69.0196078431373!black},
				xlabel={$ (l - l_c)$ / m},
				xmin=-0.41, xmax=0.41,
				xtick style={color=black},
				y grid style={white!69.0196078431373!black},
				ylabel={localisation},
				ymin=-0.5, ymax=12,
				ytick style={color=black},
				width = 3.1cm,
				height=3.1cm,
				scale only axis
				]
				\addplot [ultra thick, blue, name path=f]
				table {%
					-0.352570424333031 0.191050535807687
					-0.340321159237281 0.228095343946448
					-0.328469754734922 0.274741251685754
					-0.317014938310637 0.333207369797294
					-0.305951373767537 0.406064161539173
					-0.295270262516097 0.496216814955373
					-0.284959452235854 0.60682182547102
					-0.275004354516252 0.740988906215973
					-0.265387174212452 0.901919018078822
					-0.256088337130649 1.0919647613251
					-0.247088509953242 1.31495084908206
					-0.238368636984674 1.57326113637639
					-0.22990997708687 1.86806429268146
					-0.221694142993455 2.1994690718407
					-0.213703142177311 2.56641956184233
					-0.205919418236827 2.9664839559753
					-0.198325891581765 3.39608733723927
					-0.190905998058727 3.85090091648853
					-0.183643724076422 4.32645161172035
					-0.176523935456617 4.81887091716855
					-0.169535418881836 5.32217823723084
					-0.162667909999033 5.82994441971111
					-0.155911471170953 6.33582476723882
					-0.149256802619912 6.83357481295022
					-0.142695230724462 7.31722661092283
					-0.136218693446852 7.78127267486561
					-0.129819723135348 8.22083169396524
					-0.123491426991022 8.631866084768
					-0.117227465517183 9.01111174636909
					-0.111022029281015 9.35602905790536
					-0.104869814312117 9.66462242178836
					-0.0987659964432693 9.9354394251839
					-0.0927062048671312 10.1676595447103
					-0.0866864951412505 10.3610803490898
					-0.0807033218253898 10.5161145969557
					-0.0747533912066941 10.6333626756403
					-0.0688333611969818 10.71274762514
					-0.0629402531392075 10.7550873336124
					-0.0570712455573491 10.7615697557407
					-0.0512236483645154 10.7336592344155
					-0.0453948942586195 10.6730062207104
					-0.0395825299952612 10.5814105041976
					-0.033784207567538 10.460801123659
					-0.0279976753179607 10.3131779103359
					-0.0222207690030891 10.1405760221848
					-0.0164514028270064 9.9450582994301
					-0.0106875604554579 9.72867381313577
					-0.00492728601844827 9.49343896782116
					0 9.27869069960569
					0.000831344945212364 9.24132404509314
					0.00659015429680082 8.97423633377836
					0.0123509906214001 8.69401967209691
					0.018115684968791 8.40253466672317
					0.0238861812752366 8.10177871543268
					0.0296643660793645 7.79333750142098
					0.0354521396041003 7.47872143496947
					0.0412514381891381 7.15941962402531
					0.0470642435584872 6.83686953744709
					0.052892592061175 6.51245783756045
					0.058738583884165 6.18753078546319
					0.0646043922337484 5.86336912858838
					0.0704922724784749 5.54117274142404
					0.076404571243156 5.22208980491721
					0.0823437354396893 4.90720031720952
					0.0883123212164772 4.59751778955001
					0.0943130028041621 4.29399395301626
					0.100348581231362 3.9975576836814
					0.106421992880196 3.70912807087723
					0.112536352643057 3.42955716713205
					0.118695185663749 3.15949712699565
					0.124902044464765 2.89969175361591
					0.131160685683606 2.65075653206181
					0.13747515061653 2.41311429692168
					0.143849789820126 2.18710660277297
					0.15028928738231 1.97300361467543
					0.156798684762134 1.77109120010734
					0.163383404068072 1.58163681811769
					0.170049270618111 1.40481550084211
					0.176802534603124 1.24070566059469
					0.183649891659085 1.08927428247028
					0.190598502144749 0.950396614415522
					0.197656008920432 0.823816779791023
					0.204830550339556 0.709129121076769
					0.212129933573405 0.605993229645123
					0.219563852708665 0.513993243428077
					0.227143721351604 0.432688899226508
					0.234881476660194 0.361572564158048
					0.242789596764062 0.300016353639739
					0.250881115380802 0.24730079687405
					0.259169633580632 0.202571436616177
					0.267669328716497 0.1649300951374
					0.276394960600393 0.133595647720145
					0.285361875066005 0.107858565996146
					0.294586005110029 0.0868615663805384
					0.304083869847837 0.0692272197950798
					0.31387257155187 0.054765889360625
					0.323969791062703 0.0431654617221096
					0.334393275168124 0.033924732043814
					0.345158542573667 0.0266229779074209
					0.35627972699155 0.0208908085213564
					0.36776879303264 0.0164153282614402
					0.379635139549282 0.0129361189394279
				};
				\path[name path=axis] (axis cs:-0.177735296350424,0) -- (axis cs:0.0856608455966574,0);
				\addplot [semithick, black, mark=*, mark size=5, mark options={solid}, only marks]
				table {
					-0.177735296350424 4.73509069252591
					0.0856608455966574 4.73509069252591
				};
				\addplot [
				thick,
				color=blue,
				fill=blue, 
				fill opacity=0.2
				]
				fill between[
				of=f and axis,
				soft clip={domain=-0.177735296350424:0.0856608455966574},
				];		
				\addplot [semithick, black, dashed]
				table {%
					0 -1
					0 15
				};
			\end{axis}
			
		\end{tikzpicture}
			
		\end{subfigure}
		
		\caption{Normalised wavenumber along the central ray (left). Localisation due to the power-law turbulence spectrum (middle) and overall localisation (right). Here $l_c$ is the distance along the central ray, from launch to cut-off location. The overall localisation is the product of equations (\ref{eq:localisation_ray_piece}), (\ref{eq:localisation_beam_piece}), and (\ref{eq:localisation_spectrum_piece}). The shaded area is $80\%$ of the total area under the curve, and the black points mark the start and end of the shaded area. The corresponding half width is $13\textrm{cm}$. Note that we have chosen the black points such that they have the same ordinate value, that is, the same value of localisation. 
		}
		
		\label{fig:localisation_MAST_spectrum}
	\end{figure}

	We now consider the turbulence spectrum, which may further contribute to localisation. For electrostatic turbulece, references \cite{Schekochihin:spectrum:2008, Schekochihin:spectrum:2009, Barnes:spectrum:2011} suggest that the spectrum is of the form $\widetilde{C}_\mu (k_{\perp,1}, k_{\mu,2}, \omega) \propto k_\perp^{- 10 / 3}$ for $k_{\perp} \rho_i \ll 1$ and $\widetilde{C}_\mu (k_{\perp,1}, k_{\mu,2}, \omega) \propto k_\perp^{- 13 / 3}$ for $k_{\perp} \rho_i \gg 1$. The spectrum piece associated with the backscattered power is thus
	\begin{equation} \label{eq:localisation_spectrum_piece}
		\left(  
			\frac{K}{K_{ant}}
		\right)^{-\frac{10}{3}} 
		\qquad
		\textrm{or}
		\qquad
		\left(  
			\frac{K}{K_{ant}}
		\right)^{-\frac{13}{3}} 
		,
	\end{equation}	
	where we have used the Bragg condition to express $k_{\perp,1}$ in terms of the beam's wavenumber. Since the magnitude of the wavevector is minimum at the cut-off, there is significantly more turbulence with the appropriate $k_{\perp,1}$ for backscattering. We now multiply the piece $(K / K_{ant})^{-13 / 3}$, Figure \ref{fig:localisation_MAST_spectrum} (middle), together with the beam and ray pieces, and see what this overall localisation gives, Figure \ref{fig:localisation_MAST_spectrum} (right). We find the localisation length, after taking the spectrum into account, to be around $13\textrm{cm}$ for our MAST test scenario, Figure \ref{fig:localisation_MAST_spectrum}. One more consideration is the effect of the Doppler shift on localisation, which we do not assess in this work because it requires assuming particular realisations of turbulence. Investigating this effect using gyrokinetic simulations of turbulence together with a synthetic DBS based on the beam model would be interesting further work.
	
\begin{figure}
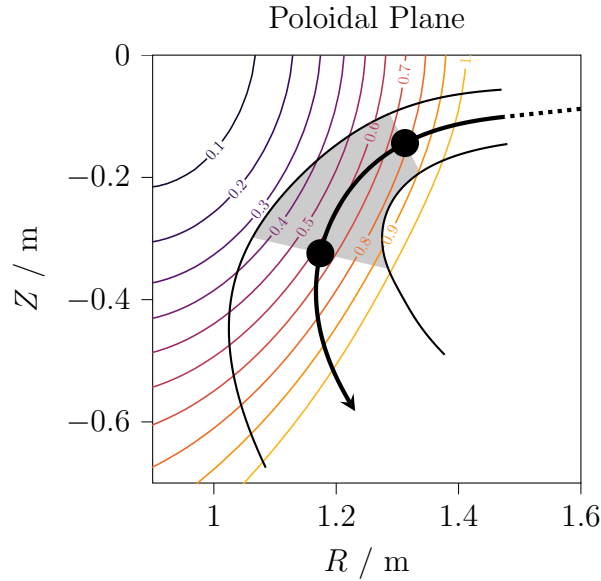
 
		\centering


		\caption{Most of the signal ($80\%$) comes from the region between the two black points. The thick solid line shows the path of the central ray. Thin solid lines give the $1/\textrm{e}$ positions of the Gaussian beam's electric field, and the dotted lines show the ray's propagation in vacuum.}
		\label{fig:localisation_poloidal_cut}
	\end{figure}

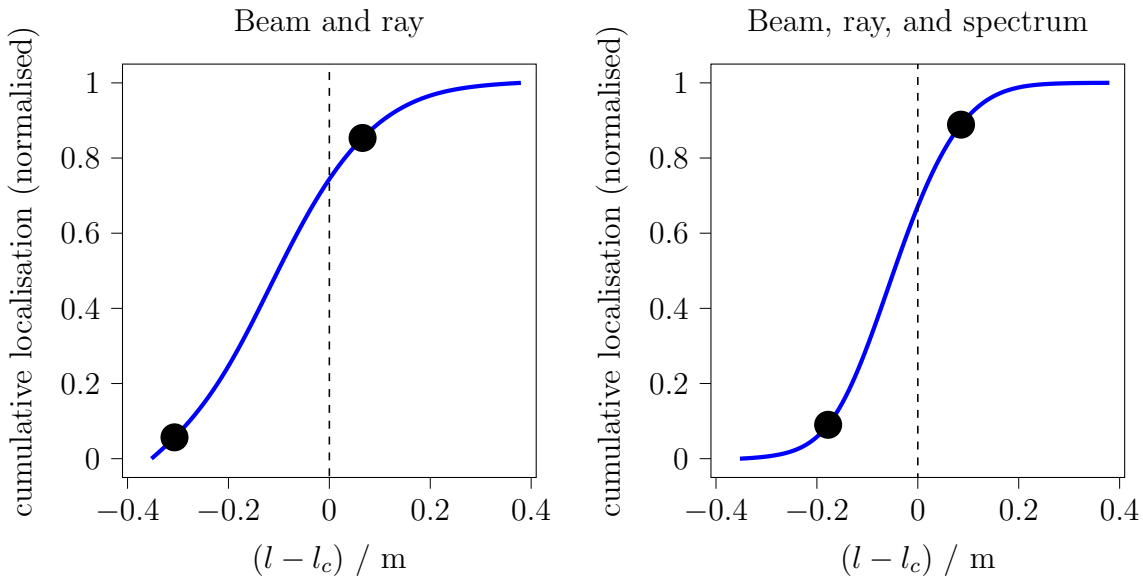
\begin{figure}
		
		\begin{subfigure}{0.49\textwidth}
		\centering
		\begin{tikzpicture}
		\begin{axis}[
		tick align=outside,
		tick pos=left,
		title={Beam and ray},
		x grid style={white!69.0196078431373!black},
		xlabel={$ (l - l_c)$ / m},
		xmin=-0.41, xmax=0.41,
		xtick style={color=black},
		y grid style={white!69.0196078431373!black},
		ylabel={cumulative localisation (normalised)},
		ymin=-0.05, ymax=1.05,
		ytick style={color=black},
		yticklabel style={
		        /pgf/number format/fixed,
		        /pgf/number format/precision=2
		},		
		scaled y ticks=false,
		width = 5.5cm,
		height=5.5cm,		
		scale only axis
		]
		\addplot [ultra thick, blue]
		table {%
		-0.352570424333031 0
		-0.340321159237281 0.0143656446522231
		-0.328469754734922 0.028813229150268
		-0.317014938310637 0.0433782416766208
		-0.305951373767537 0.0580909080069783
		-0.295270262516097 0.0729759973048443
		-0.284959452235854 0.088053386594446
		-0.275004354516252 0.103336897140784
		-0.265387174212452 0.118836860189182
		-0.256088337130649 0.134553892758678
		-0.247088509953242 0.150485993659907
		-0.238368636984674 0.16663505734106
		-0.22990997708687 0.182997002004986
		-0.221694142993455 0.199561614169513
		-0.213703142177311 0.216314611842478
		-0.205919418236827 0.233238526156367
		-0.198325891581765 0.25031315792896
		-0.190905998058727 0.267516625520334
		-0.183643724076422 0.28482679821859
		-0.176523935456617 0.302222647374458
		-0.169535418881836 0.319676556359586
		-0.162667909999033 0.337160118566955
		-0.155911471170953 0.354646063162818
		-0.149256802619912 0.372107717181912
		-0.142695230724462 0.389519109566912
		-0.136218693446852 0.406855061575178
		-0.129819723135348 0.42409132835893
		-0.123491426991022 0.441204882595845
		-0.117227465517183 0.458174151140136
		-0.111022029281015 0.474979038626669
		-0.104869814312117 0.49160076406065
		-0.0987659964432693 0.508021637005287
		-0.0927062048671312 0.52422503397931
		-0.0866864951412505 0.540195539522709
		-0.0807033218253898 0.555919158951953
		-0.0747533912066941 0.571383712852129
		-0.0688333611969818 0.586579004438656
		-0.0629402531392075 0.601495370722467
		-0.0570712455573491 0.616124345142234
		-0.0512236483645154 0.630458663384249
		-0.0453948942586195 0.644492186061193
		-0.0395825299952612 0.658219806086836
		-0.033784207567538 0.671637382178026
		-0.0279976753179607 0.68474167710014
		-0.0222207690030891 0.697530284488622
		-0.0164514028270064 0.710001570147539
		-0.0106875604554579 0.722154617340255
		-0.00492728601844827 0.733989163051021
		0 0.743862629269801
		0.000831344945212364 0.745505665138836
		0.00659015429680082 0.756704728477902
		0.0123509906214001 0.76758781334548
		0.018115684968791 0.778156767389014
		0.0238861812752366 0.788414233881207
		0.0296643660793645 0.798363174343676
		0.0354521396041003 0.808006734759441
		0.0412514381891381 0.817348252036901
		0.0470642435584872 0.826391255643288
		0.052892592061175 0.835139455697958
		0.058738583884165 0.843596742079141
		0.0646043922337484 0.851767176634946
		0.0704922724784749 0.859654959262782
		0.076404571243156 0.867264402435222
		0.0823437354396893 0.874599913179682
		0.0883123212164772 0.8816659632603
		0.0943130028041621 0.888467061812007
		0.100348581231362 0.895007760781019
		0.106421992880196 0.901292706680487
		0.112536352643057 0.907326705214171
		0.118695185663749 0.913114819010271
		0.124902044464765 0.918661987450473
		0.131160685683606 0.923973266430932
		0.13747515061653 0.929053714771101
		0.143849789820126 0.933908248686185
		0.15028928738231 0.938541596337412
		0.156798684762134 0.942958362374633
		0.163383404068072 0.947163184122378
		0.170049270618111 0.951160803042644
		0.176802534603124 0.954956061259425
		0.183649891659085 0.958553875248146
		0.190598502144749 0.961959214822135
		0.197656008920432 0.965177036638586
		0.204830550339556 0.968212068430274
		0.212129933573405 0.97106843173781
		0.219563852708665 0.973750925972622
		0.227143721351604 0.97626542142132
		0.234881476660194 0.978618742477631
		0.242789596764062 0.980818645338019
		0.250881115380802 0.982873694769508
		0.259169633580632 0.984792882665599
		0.267669328716497 0.986585068362706
		0.276394960600393 0.988259247814555
		0.285361875066005 0.989825550153345
		0.294586005110029 0.991294699914579
		0.304083869847837 0.992671181225588
		0.31387257155187 0.993955428023203
		0.323969791062703 0.995152064985717
		0.334393275168124 0.996267596055973
		0.345158542573667 0.997308020223538
		0.35627972699155 0.998279016904642
		0.36776879303264 0.999185903050603
		0.379635139549282 1.00003362127914
		};
		\addplot [semithick, black, mark=*, mark size=5, mark options={solid}, only marks]
		table {%
		-0.306934167811999 0.0567839584437246
		0.0658454845039634 0.853429823570966
		};
		\addplot [semithick, black, dashed]
		table {%
		0.0 -10.0
		0.0 10.0
		};
		\end{axis}
		
		\end{tikzpicture}

		\end{subfigure}
		\begin{subfigure}{0.49\textwidth}
		\centering
		\begin{tikzpicture}
		\begin{axis}[
		tick align=outside,
		tick pos=left,
		title={Beam, ray, and spectrum},
		x grid style={white!69.0196078431373!black},
		xlabel={$ (l - l_c)$ / m},
		xmin=-0.41, xmax=0.41,
		xtick style={color=black},
		y grid style={white!69.0196078431373!black},
		ylabel={cumulative localisation (normalised)},
		ymin=-0.05, ymax=1.05,
		ytick style={color=black},
		width = 5.5cm,
		height=5.5cm,	
		scale only axis
		]
		\addplot [ultra thick, blue]
		table {%
		-0.352570424333031 0
		-0.340321159237281 0.000923731366097019
		-0.328469754734922 0.00199591001485462
		-0.317014938310637 0.00324883614681934
		-0.305951373767537 0.00472036747265631
		-0.295270262516097 0.00645428603200198
		-0.284959452235854 0.00850051381945531
		-0.275004354516252 0.0109145562164192
		-0.265387174212452 0.0137572588928534
		-0.256088337130649 0.0170930506781199
		-0.247088509953242 0.0209903654998454
		-0.238368636984674 0.0255215284411983
		-0.22990997708687 0.030758711601038
		-0.221694142993455 0.0367711860038852
		-0.213703142177311 0.0436231570723049
		-0.205919418236827 0.051371538251436
		-0.198325891581765 0.0600640770771233
		-0.190905998058727 0.069738513751604
		-0.183643724076422 0.0804230531754698
		-0.176523935456617 0.0921378991282815
		-0.169535418881836 0.104888725823918
		-0.162667909999033 0.118668027959873
		-0.155911471170953 0.13345665571397
		-0.149256802619912 0.149224156459805
		-0.142695230724462 0.165929652820621
		-0.136218693446852 0.183522944708075
		-0.129819723135348 0.201945830066188
		-0.123491426991022 0.221133696332359
		-0.117227465517183 0.241017131707325
		-0.111022029281015 0.261523291084369
		-0.104869814312117 0.28257693850804
		-0.0987659964432693 0.304101259682927
		-0.0927062048671312 0.326018766661895
		-0.0866864951412505 0.348252290357549
		-0.0807033218253898 0.370725969990278
		-0.0747533912066941 0.39336626049445
		-0.0688333611969818 0.41610221089582
		-0.0629402531392075 0.438863827722755
		-0.0570712455573491 0.461583911311687
		-0.0512236483645154 0.484198567040387
		-0.0453948942586195 0.506647474622347
		-0.0395825299952612 0.528874045815394
		-0.033784207567538 0.550825545783283
		-0.0279976753179607 0.572453155435516
		-0.0222207690030891 0.593711958421975
		-0.0164514028270064 0.614560904757271
		-0.0106875604554579 0.63496274321739
		-0.00492728601844827 0.654883909097382
		0 0.671525382453577
		0.000831344945212364 0.674295467126675
		0.00659015429680082 0.693168673692225
		0.0123509906214001 0.711481259615916
		0.018115684968791 0.72921316200715
		0.0238861812752366 0.746348042684898
		0.0296643660793645 0.762872438571116
		0.0354521396041003 0.778775457502774
		0.0412514381891381 0.794048721310432
		0.0470642435584872 0.808686296662388
		0.052892592061175 0.822684598433828
		0.058738583884165 0.836042307837542
		0.0646043922337484 0.848760278084972
		0.0704922724784749 0.860841400025371
		0.076404571243156 0.872290484555093
		0.0823437354396893 0.883114160921976
		0.0883123212164772 0.893320761894738
		0.0943130028041621 0.902920218387895
		0.100348581231362 0.911924004362352
		0.106421992880196 0.920345143141122
		0.112536352643057 0.928198216919422
		0.118695185663749 0.935499378695646
		0.124902044464765 0.942265765366725
		0.131160685683606 0.948515744571763
		0.13747515061653 0.954268681735528
		0.143849789820126 0.959544681308636
		0.15028928738231 0.964364461980571
		0.156798684762134 0.968749345676494
		0.163383404068072 0.972721316466783
		0.170049270618111 0.976302962439715
		0.176802534603124 0.979517329153579
		0.183649891659085 0.982387750248532
		0.190598502144749 0.984937681797463
		0.197656008920432 0.987190507607359
		0.204830550339556 0.989169259901189
		0.212129933573405 0.990896380962229
		0.219563852708665 0.992394343891161
		0.227143721351604 0.993685373849755
		0.234881476660194 0.994791103458801
		0.242789596764062 0.995732411858286
		0.250881115380802 0.99652919392469
		0.259169633580632 0.997200061095657
		0.267669328716497 0.997762056751005
		0.276394960600393 0.998230707300422
		0.285361875066005 0.99862024390256
		0.294586005110029 0.998943395813891
		0.304083869847837 0.999210123525785
		0.323969791062703 0.999606400983785
		0.334393275168124 0.999750972422005
		0.345158542573667 0.999868244062249
		0.35627972699155 0.999963313546848
		0.36776879303264 1.00004042794258
		0.379635139549282 1.00010309178211
		};
		\addplot [semithick, black, mark=*, mark size=5, mark options={solid}, only marks]
		table {%
		-0.177735296350424 0.0901447351423829
		0.0856608455966574 0.888786596741955
		};
		\addplot [semithick, black, dashed]
		table {%
		0 -2.0
		0 2.0
		};
		\end{axis}
		
		\end{tikzpicture}

		\end{subfigure}

		\caption{Here we show the cumulative integrals of localisation for the MAST test scenario, normalised such that it is $0$ upon entering the plasma, and $1$ when leaving. On the left, we look at the \emph{beam} and \emph{ray} pieces, and on the right, we include the \emph{spectrum} piece as well. These pieces are shown as a function of arc length from the cut-off, $l - l_c$. The vertical dashed line shows the location of the cut-off. The two black points in each figure correspond to that of Figure \ref{fig:localisation_MAST_ray_beam} (right) and Figure \ref{fig:localisation_MAST_spectrum} (right), indicating the start and end of $80\%$ range, which gives where most of the backscattered signal comes from.}
			
		\label{fig:cumulative_localisation_MAST}
	\end{figure}

	It is important to note that this distance $l - l_c$, where $l_c$ is the arc-length of the cut-off position, is measured along the ray. For an O-mode beam, the ray's radial component of group velocity is small near the cut-off, hence more of the signal is coming from fairly similar flux surfaces, unlike what it might ostensibly seem given the fairly large half-width, see Figure \ref{fig:localisation_poloidal_cut}. 
	Apart from that, it is interesting to note that the peak of localisation is shifted away from the cut-off, by around $l - l_c = 6\textrm{cm}$ or so, due to the changing \emph{beam} piece near the cut-off. The physical intuition as to why the \emph{beam} piece decreases as the beam propagates is as follows: the MAST DBS was designed to have the beam waist before the plasma, hence the beam is always getting wider while in the plasma. Thus far, we have shown plots of the integrand of equation (\ref{eq:A_r2_final_cleaned}), and calculated the associated localisation lengths. Since these lengths are calculated from the cumulative integrals of the localisation pieces, we plot the cumulative integrals themselves in Figure \ref{fig:cumulative_localisation_MAST}.
			
	Since every point along the ray is associated with a particular $k_{\perp,1}$ as a result of the stationary phase integral in Section \ref{subsection:stationary_phase}, it is impossible to divorce the localisation and $k_{\perp,1}$ resolution. Instead of $k_{\perp,1}$ resolution, it is more physically suitable to think of the spread of backscattered $k_{\perp,1}$ as localisation along the ray. Nonetheless, we show how to calculate the associated $k_{\perp,1}$ resolution from the spatial localisation in Section \ref{subsection_kperp1_resolution}.  

	\section{Wavenumber resolution} \label{section_wavenumber_resolution}
	The wavenumber resolution is different for $k_{\perp,1}$ and $k_{\perp,2}$. The calculation of the former follows from the previous section on localisation, while that of the latter is simply given by equation (\ref{eq:delta_k_perp_2}). Consequently, this section is split into two parts, discussing the $k_{\perp,1}$ and $k_{\perp,2}$ resolutions in turn.
	
	\subsection{$k_{\perp,1}$ resolution} \label{subsection_kperp1_resolution}
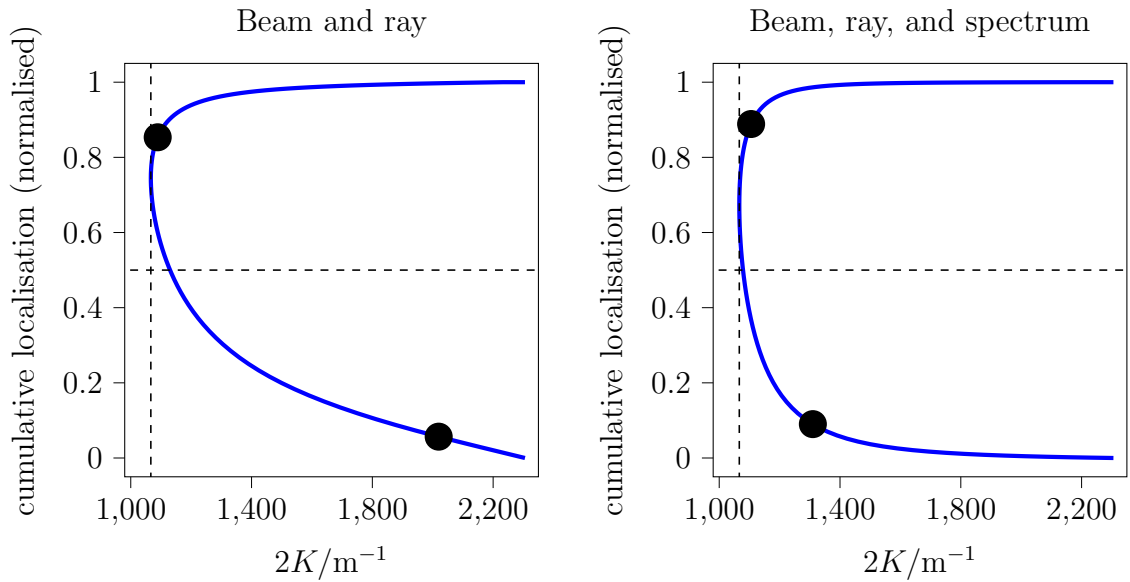
\begin{figure}
		\begin{subfigure}{0.49\textwidth}
		\centering
		\begin{tikzpicture}
		\begin{axis}[
		tick align=outside,
		tick pos=left,
		title={Beam and ray},
		x grid style={white!69.0196078431373!black},
		xlabel={$2 K / \textrm{m}^{-1}$},
		xmin=980, xmax=2350,
		xtick={1000,1400,1800,2200},
		xtick style={color=black},
		y grid style={white!69.0196078431373!black},
		ylabel={cumulative localisation (normalised)},
		ymin=-0.05, ymax=1.05,
		ytick style={color=black},
		yticklabel style={
		        /pgf/number format/fixed,
		        /pgf/number format/precision=2
		},		
		scaled y ticks=false,
		width = 5.5cm,
		height=5.5cm,	
		scale only axis
		]
		\addplot [ultra thick, blue]
		table {%
		2305.42952414685 0
		2305.42952414685 0
		2231.90713114961 0.0143651616771124
		2158.20842490403 0.028812260445217
		2085.11662202933 0.0433767832936817
		2013.35847813338 0.0580889549820082
		1943.57121445674 0.0729735438409573
		1876.30183833652 0.0880504262264879
		1812.06028033474 0.103333422938928
		1751.16738335324 0.118832864876261
		1693.75540570351 0.134549369036784
		1639.90530224109 0.15048093429841
		1589.64775675959 0.16662945504564
		1542.96459582388 0.182990849618551
		1499.79083425441 0.199554904878354
		1460.01735171987 0.216307339313044
		1423.49415432724 0.233230684642416
		1390.0341330621 0.250304742363348
		1359.41720093361 0.267507631571583
		1331.39468068455 0.284817222299257
		1305.70391217444 0.302212486604088
		1282.17865566999 0.319665808786199
		1260.66502696174 0.337148783193605
		1241.0063443046 0.354634139909409
		1223.05445122295 0.372095206865097
		1206.669966272 0.389506013876495
		1191.72244327118 0.406841383047475
		1178.09044517762 0.424077070345373
		1165.66153758274 0.441190049222048
		1154.33220985229 0.458158747257006
		1144.00773318621 0.474963069760722
		1134.60196540749 0.491584236369821
		1126.03711219995 0.508004557242266
		1118.24345392595 0.524207409455672
		1111.15904620086 0.540177378068299
		1104.72940120647 0.555900468867118
		1098.91059868879 0.571364502846691
		1093.67273101024 0.586559283565254
		1088.98195989616 0.601475148358611
		1084.80687151621 0.616103630950081
		1081.11897490986 0.630437467270185
		1077.8926373144 0.644470518138008
		1075.10501336753 0.658197676639023
		1072.73596952201 0.671614801629309
		1070.76800481009 0.684718655983078
		1069.18616889011 0.697506833416676
		1067.97797810294 0.709977699789114
		1067.13333006929 0.722130338394576
		1066.64441717414 0.733964486226154
		1066.50420449655 0.74383762049753
		1066.50563911625 0.745480601127455
		1066.71351455122 0.756679287952346
		1067.26659172878 0.767562006928991
		1068.16515659065 0.778130605642718
		1069.41032578749 0.788387727277361
		1071.00523765786 0.798336333254972
		1072.95469296739 0.8079795694528
		1075.26516422249 0.81732077266706
		1077.94489806245 0.82636347224632
		1081.00401688514 0.835111378185203
		1084.45461973671 0.84356838023116
		1088.31088240338 0.851738540095739
		1092.58915653172 0.85962605753515
		1097.30806747509 0.867235244876978
		1102.48861041944 0.874570509000475
		1108.15424418562 0.881636321519438
		1114.3309819402 0.8884371914172
		1121.04747787873 0.894977670486939
		1128.33510877869 0.901262405085592
		1136.22828099056 0.907296200755347
		1144.76616819004 0.9130841199542
		1153.99047352285 0.918631101897774
		1163.94651839206 0.923942202312243
		1174.68399557781 0.929022479846982
		1186.25738970227 0.933876850551914
		1198.72638993069 0.938510042429304
		1212.1562918082 0.942926659974194
		1226.61838465693 0.947131340355206
		1242.19032044573 0.95112882487493
		1258.95645953647 0.954923955494565
		1277.00818825537 0.958521648524243
		1296.44420289167 0.96192687361021
		1317.3707545474 0.965144587243012
		1339.90171624214 0.96817951699648
		1364.11999401197 0.971035784272656
		1390.16108295574 0.973718188321613
		1418.20528380416 0.9762325992326
		1448.42561876205 0.978585841169907
		1480.98613604429 0.980785670069234
		1516.04022221205 0.982840650409655
		1553.72894906656 0.984759773782364
		1594.17949080186 0.98655189922592
		1637.50365069353 0.988226022391608
		1683.79653519707 0.98979227207108
		1733.13540776279 0.99126137243928
		1785.57874609833 0.992637807472783
		1841.1655163563 0.99392201109383
		1899.91466712572 0.995118607825231
		1961.77068114362 0.996234101391166
		2026.42508433852 0.997274490579516
		2093.53202151359 0.998245454615567
		2162.63689948208 0.99915231027188
		2233.16380862569 1
		2305.42952414685 1
		};
		\addplot [semithick, black, mark=*, mark size=5, mark options={solid}, only marks]
		table {%
		2019.73286824859 0.0567820493585933
		1089.21268962939 0.853401131133321
		};
		\addplot [semithick, black, dashed]
		table {%
		1066.50420449655 -2
		1066.50420449655 2
		};
		\addplot [semithick, black, dashed]
		table {%
		0 0.5
		3000 0.5
		};
		\end{axis}
		
		\end{tikzpicture}

		\end{subfigure}
		\begin{subfigure}{0.49\textwidth}
		\centering
		\begin{tikzpicture}
		\begin{axis}[
		tick align=outside,
		tick pos=left,
		title={Beam, ray, and spectrum},
		x grid style={white!69.0196078431373!black},
		xlabel={$2 K / \textrm{m}^{-1}$},
		xmin=980, xmax=2350,
		xtick={1000,1400,1800,2200},
		xtick style={color=black},
		y grid style={white!69.0196078431373!black},
		ylabel={cumulative localisation (normalised)},
		ymin=-0.05, ymax=1.05,
		ytick style={color=black},
		width = 5.5cm,
		height=5.5cm,	
		scale only axis
		]
		\addplot [ultra thick, blue]
		table {%
		2305.42952414685 0
		2305.42952414685 0
		2231.90713114961 0.000923636146800613
		2158.20842490403 0.00199570427414439
		2085.11662202933 0.003248501253036
		2013.35847813338 0.00471988089172383
		1943.57121445674 0.00645362071674119
		1876.30183833652 0.00849963757667025
		1812.06028033474 0.0109134311313549
		1751.16738335324 0.0137558407787131
		1693.75540570351 0.0170912887067085
		1639.90530224109 0.0209882017887197
		1589.64775675959 0.0255188976525618
		1542.96459582388 0.0307555409575108
		1499.79083425441 0.0367673955875504
		1460.01735171987 0.0436186603468764
		1423.49415432724 0.0513662428139239
		1390.0341330621 0.0600578856026667
		1359.41720093361 0.0697313250250381
		1331.39468068455 0.0804147630742364
		1305.70391217444 0.0921284014471932
		1282.17865566999 0.104877913772883
		1260.66502696174 0.118655795522454
		1241.0063443046 0.133442898847717
		1223.05445122295 0.14920877426136
		1206.669966272 0.165912548600311
		1191.72244327118 0.183504026950912
		1178.09044517762 0.201925013256719
		1165.66153758274 0.221110901615468
		1154.33220985229 0.240992287382944
		1144.00773318621 0.261496332961388
		1134.60196540749 0.28254781015076
		1126.03711219995 0.304069912573753
		1118.24345392595 0.325985160270781
		1111.15904620086 0.348216392109126
		1104.72940120647 0.370687755128995
		1098.91059868879 0.393325711845865
		1093.67273101024 0.416059318599201
		1088.98195989616 0.438818589132377
		1084.80687151621 0.461536330708845
		1081.11897490986 0.484148655292706
		1077.8926373144 0.506595248815337
		1075.10501336753 0.528819528867748
		1072.73596952201 0.55076876604966
		1070.76800481009 0.572394146302903
		1069.18616889011 0.59365075790739
		1067.97797810294 0.614497555109212
		1067.13333006929 0.634897290524252
		1066.64441717414 0.65481640290745
		1066.50420449655 0.671456160841346
		1066.50563911625 0.674225959970915
		1066.71351455122 0.693097221064529
		1067.26659172878 0.711407919305706
		1068.16515659065 0.729137993871956
		1069.41032578749 0.746271108266408
		1071.00523765786 0.762793800798807
		1072.95469296739 0.778695180428902
		1075.26516422249 0.793966869850881
		1077.94489806245 0.80860293634466
		1081.00401688514 0.822599795154981
		1084.45461973671 0.835956127630577
		1088.31088240338 0.848672786894941
		1092.58915653172 0.860752663499333
		1097.30806747509 0.872200567844195
		1102.48861041944 0.883023128494014
		1108.15424418562 0.893228677358556
		1114.3309819402 0.902827144328647
		1121.04747787873 0.911830002182444
		1128.33510877869 0.9202502729005
		1136.22828099056 0.928102537174882
		1144.76616819004 0.935402946338915
		1153.99047352285 0.942168635523039
		1163.94651839206 0.948417970473
		1174.68399557781 0.954170314617356
		1186.25738970227 0.959445770334332
		1198.72638993069 0.964265054177707
		1212.1562918082 0.968649485874754
		1226.61838465693 0.972621047229705
		1242.19032044573 0.976202324002432
		1258.95645953647 0.979416359375661
		1277.00818825537 0.982286484584292
		1296.44420289167 0.984836153283332
		1317.3707545474 0.987088746869341
		1339.90171624214 0.989067295191099
		1364.11999401197 0.990794238218505
		1390.16108295574 0.992292046735688
		1418.20528380416 0.993582943613421
		1448.42561876205 0.994688559242582
		1480.98613604429 0.99562977061091
		1516.04022221205 0.996426470544098
		1553.72894906656 0.997097268561302
		1594.17949080186 0.997659206285488
		1637.50365069353 0.998127808525865
		1683.79653519707 0.998517304974121
		1733.13540776279 0.998840423574579
		1785.57874609833 0.999107123791872
		1841.1655163563 0.999325471345401
		1899.91466712572 0.999503360401134
		1961.77068114362 0.999647916936763
		2026.42508433852 0.999765176488511
		2093.53202151359 0.999860236173238
		2162.63689948208 0.999937342619927
		2233.16380862569 1
		2305.42952414685 1
		};
		\addplot [semithick, black, mark=*, mark size=5, mark options={solid}, only marks]
		table {%
		1310.07493973816 0.0901354429189409
		1105.63735151395 0.888694979592755
		};
		\addplot [semithick, black, dashed]
		table {%
		1066.50420449655 -2
		1066.50420449655 2
		};
		\addplot [semithick, black, dashed]
		table {%
		0 0.5
		3000 0.5
		};
		\end{axis}
		
		\end{tikzpicture}

		\end{subfigure}

		\caption{Here we show the normalised cumulative integrals of localisation as a function of backscattered $k_{\perp,1} = - 2 K$ (given by the Bragg condition) for the MAST test scenario. On the left, we look at the \emph{beam} and \emph{ray} pieces, and on the right, we include the \emph{spectrum} piece as well. These two plots are analogous to those in Figure \ref{fig:cumulative_localisation_MAST}; the difference being that here we plot the cumulative localisation with respect to the backscattered wavenumber rather than arc-length along the ray. The start and end of the $80\%$ range are marked on the graphs. The vertical dashed line indicates the backscattered $k_{\perp,1}$ at the cut-off. Note that the derivatives are infinite at the cut-off, but the plots themselves do not diverge, showing that the localisation piece is indeed integrable. The backscattered $|k_{\perp,1}|$ at the cutoff is $1070\textrm{m}^{-1}$, while the values of $|k_{\perp,1}|$ at the ends of the $80\%$ range are $2020\textrm{m}^{-1}$ and $1090\textrm{m}^{-1}$ (left) and $1310\textrm{m}^{-1}$ and $1110 \textrm{m}^{-1}$ (right). The intersection of the horizontal dashed lines and the curves give the median backscattered $k_{\perp,1}$.}
			
		\label{fig:kperp1_resolution_MAST}
	\end{figure}

	The $k_{\perp,1}$ resolution, $\Delta k_{\perp,1}$, cannot be divorced from the localisation, Section \ref{section_localisation}. Nonetheless, one may consider it insightful to calculate it. Since the \emph{ray} piece as a function of backscattered $k_{\perp,1}$ diverges at cut-off, namely at $k_{l,1} = - 2 K (l_c)$, one might wrongly suspect that the dominant backscattered $k_{l,1}$ comes from the cut-off as well. To properly deal with this divergence, which is integrable, we instead consider the cumulative integral of the \emph{ray} piece. We show this in Figure \ref{fig:kperp1_resolution_MAST}. As we can see from the figure, the median backscattered $k_{\perp,1}$ is not exactly that of the cut-off, regardless of whether we use the spectrum piece or not. The divergence means that it is not particularly insightful to calculate and plot the localisation, the integrand of equation (\ref{eq:A_r2_final_with_shorthands}), to find the backscattered $k_{\perp,1}$ resolution. Instead, we directly take a cumulative integral of the localisation piece, and calculate the resolution from there, see Figure \ref{fig:kperp1_resolution_MAST}. 

	\subsection{$k_{\perp,2}$ resolution}
\begin{figure}
		\centering
		\begin{tikzpicture}

		\begin{axis}[
		legend cell align={left},
		legend style={
			fill opacity=1.0,
			draw opacity=1,
			text opacity=1,
			at={(0.03,0.97)},
			anchor=north west,
			draw=black
		},
		tick align=outside,
		tick pos=left,
		x grid style={white!69.0196078431373!black},
		xlabel={$(l - l_c)$ / m},
		ymin=-25, ymax=625,
		xtick style={color=black},
		y grid style={white!69.0196078431373!black},
		ylabel={$\Delta k_{\mu,2}$ / m$^{-1}$},
		xmin=-0.41, xmax=0.41,
		ytick style={color=black},
		width=6cm,
		height=6cm,
		scale only axis
		]
		\addplot [ultra thick, blue]
		table {%
		-0.352570424333031 78.4695926831428
		-0.340321159237281 79.3960036649419
		-0.328469754734922 80.2354874399316
		-0.317014938310637 80.9368545718394
		-0.305951373767537 81.4600183808329
		-0.295270262516097 81.7814708138768
		-0.284959452235854 81.8955348383461
		-0.275004354516252 81.8156504130133
		-0.265387174212452 81.5772628686819
		-0.256088337130649 81.0130971005844
		-0.247088509953242 80.4245179067591
		-0.238368636984674 79.8790855551273
		-0.22990997708687 79.3989284461781
		-0.221694142993455 79.0272137168136
		-0.213703142177311 78.8143821151145
		-0.205919418236827 78.788791866283
		-0.198325891581765 78.9624006116434
		-0.190905998058727 79.3300883031854
		-0.183643724076422 79.8704966190009
		-0.176523935456617 80.5653329281197
		-0.169535418881836 81.4449717283653
		-0.162667909999033 82.5301866575276
		-0.155911471170953 83.8267413582151
		-0.149256802619912 85.3365036988843
		-0.142695230724462 87.0570508716095
		-0.136218693446852 88.9829133692261
		-0.129819723135348 91.1074242617987
		-0.123491426991022 93.4275613241712
		-0.117227465517183 95.9401038933521
		-0.111022029281015 98.6400772064865
		-0.104869814312117 101.517075407817
		-0.0987659964432693 104.55739061653
		-0.0927062048671312 107.748906952624
		-0.0866864951412505 111.082677880187
		-0.0807033218253898 114.554812404156
		-0.0747533912066941 118.163900349307
		-0.0688333611969818 121.903837250196
		-0.0629402531392075 125.772352460239
		-0.0570712455573491 129.767163412193
		-0.0512236483645154 133.886221866883
		-0.0453948942586195 138.126581623062
		-0.0395825299952612 142.484896040676
		-0.033784207567538 146.958393270626
		-0.0279976753179607 151.544300513336
		-0.0222207690030891 156.239832813146
		-0.0164514028270064 161.043068802838
		-0.0106875604554579 165.952313708781
		-0.00492728601844827 170.966074730242
		0 175.337994948125
		0.000831344945212364 176.083061595189
		0.00659015429680082 181.30204090552
		0.0123509906214001 186.621883669715
		0.018115684968791 192.042265666195
		0.0238861812752366 197.563525054727
		0.0296643660793645 203.181468102151
		0.0354521396041003 208.89062324146
		0.0412514381891381 214.685555699079
		0.0470642435584872 220.55936845222
		0.052892592061175 226.503855788505
		0.058738583884165 232.510422563584
		0.0646043922337484 238.568426361584
		0.0704922724784749 244.663665991414
		0.076404571243156 250.780448082215
		0.0823437354396893 256.900063507363
		0.0883123212164772 263.000576820977
		0.0943130028041621 269.056988437694
		0.100348581231362 275.046000765584
		0.106421992880196 280.949815483308
		0.112536352643057 286.755511189566
		0.118695185663749 292.458711251296
		0.124902044464765 298.050757860666
		0.131160685683606 303.506149943187
		0.13747515061653 308.772195283044
		0.143849789820126 313.784090706594
		0.15028928738231 318.465117242337
		0.156798684762134 322.75148296659
		0.163383404068072 326.595190504948
		0.170049270618111 329.949773208786
		0.176802534603124 332.770996067324
		0.183649891659085 335.011360793384
		0.190598502144749 336.62807327877
		0.197656008920432 337.558640959232
		0.204830550339556 337.685840249674
		0.212129933573405 336.876016074841
		0.219563852708665 335.120810731758
		0.227143721351604 332.589701579101
		0.234881476660194 329.504847643127
		0.242789596764062 326.094860358206
		0.250881115380802 322.615925237792
		0.259169633580632 319.190806675522
		0.267669328716497 315.834143371513
		0.276394960600393 312.876915246719
		0.285361875066005 311.078234539024
		0.294586005110029 310.615279750254
		0.304083869847837 307.187114034465
		0.31387257155187 302.023535811348
		0.323969791062703 296.701692651014
		0.334393275168124 291.217180313372
		0.345158542573667 285.454222518639
		0.35627972699155 279.327954980258
		0.36776879303264 272.730917866174
		0.379635139549282 265.508818896758
		};
		\addlegendentry{Full model}
		\addplot [ultra thick, black, dash dot]
		table {%
		-0.352570424333031 85.5767957102298
		-0.340321159237281 86.4906226823385
		-0.328469754734922 87.1620753984335
		-0.317014938310637 87.5629451327348
		-0.305951373767537 87.6840824742122
		-0.295270262516097 87.5368531090057
		-0.284959452235854 87.1503446255962
		-0.275004354516252 86.5698213755221
		-0.265387174212452 85.8501311435597
		-0.256088337130649 85.0502106518184
		-0.247088509953242 84.2316386858043
		-0.238368636984674 83.4529466834784
		-0.22990997708687 82.7706363476812
		-0.221694142993455 82.2370669322235
		-0.213703142177311 81.8976550473092
		-0.205919418236827 81.7886820800148
		-0.198325891581765 81.9351994818327
		-0.190905998058727 82.3485722946023
		-0.183643724076422 83.0237114198205
		-0.176523935456617 83.94177469346
		-0.169535418881836 85.1275847407589
		-0.162667909999033 86.5990599945393
		-0.155911471170953 88.3631375424479
		-0.149256802619912 90.4226359101081
		-0.142695230724462 92.7764070342151
		-0.136218693446852 95.419672168195
		-0.129819723135348 98.3445392764772
		-0.123491426991022 101.540661162379
		-0.117227465517183 104.996023469467
		-0.111022029281015 108.697839315533
		-0.104869814312117 112.633457274029
		-0.0987659964432693 116.791219957148
		-0.0927062048671312 121.161198692175
		-0.0866864951412505 125.735789910932
		-0.0807033218253898 130.510161926096
		-0.0747533912066941 135.480290944976
		-0.0688333611969818 140.637923200755
		-0.0629402531392075 145.978711281945
		-0.0570712455573491 151.498706553343
		-0.0512236483645154 157.194122368429
		-0.0453948942586195 163.061410423939
		-0.0395825299952612 169.097330738299
		-0.033784207567538 175.299014448149
		-0.0279976753179607 181.664023144955
		-0.0222207690030891 188.190407328464
		-0.0164514028270064 194.876763716175
		-0.0106875604554579 201.72229367112
		-0.00492728601844827 208.726863047327
		0 214.846866906143
		0.000831344945212364 215.89106270016
		0.00659015429680082 223.21627040772
		0.0123509906214001 230.704715423862
		0.018115684968791 238.359627365411
		0.0238861812752366 246.183893065974
		0.0296643660793645 254.178151956503
		0.0354521396041003 262.3425742103
		0.0412514381891381 270.676890400821
		0.0470642435584872 279.180263877479
		0.052892592061175 287.851159810932
		0.058738583884165 296.687216226925
		0.0646043922337484 305.685121035367
		0.0704922724784749 314.840499771604
		0.076404571243156 324.147823549999
		0.0823437354396893 333.600347478781
		0.0883123212164772 343.190088692066
		0.0943130028041621 352.907856039238
		0.100348581231362 362.74335081916
		0.106421992880196 372.685349803646
		0.112536352643057 382.723174878994
		0.118695185663749 392.852319916619
		0.124902044464765 403.056432792655
		0.131160685683606 413.312910659412
		0.13747515061653 423.594930181269
		0.143849789820126 433.871609939476
		0.15028928738231 444.108328758681
		0.156798684762134 454.267175846146
		0.163383404068072 464.307529352795
		0.170049270618111 474.186868084113
		0.176802534603124 483.861802727161
		0.183649891659085 493.289318015396
		0.190598502144749 502.428219888517
		0.197656008920432 511.240767550757
		0.204830550339556 519.694513275156
		0.212129933573405 527.760731163134
		0.219563852708665 535.390976119259
		0.227143721351604 542.517387336306
		0.234881476660194 549.05965814821
		0.242789596764062 554.928891281107
		0.250881115380802 560.031671594068
		0.259169633580632 564.274327969533
		0.267669328716497 567.567572061762
		0.276394960600393 569.830112042604
		0.285361875066005 570.989950058866
		0.294586005110029 570.986552698661
		0.304083869847837 569.782991828916
		0.31387257155187 567.378863221372
		0.323969791062703 563.776297847971
		0.334393275168124 558.969581745303
		0.345158542573667 552.880233556903
		0.35627972699155 545.501054949778
		0.36776879303264 536.858724466974
		0.379635139549282 526.992020931591
		};
		\addlegendentry{Beam only}
		\addplot [semithick, black, dashed]
		table {%
		0 -100
		0 1000
		};
		\end{axis}
		
		\end{tikzpicture}

		\caption{Wavenumber resolution for $k_{\perp,2}$ as a function of distance from the cut-off along the central ray $l - l_c$ for our MAST test scenario. The solid blue line includes the corrections arising from curvature and shear of $\hat{\mathbf{b}}$, while the dash-dot black line does not have these corrections, and only uses the beam properties. }
		\label{fig:wavenumber_resolution_MAST}
	\end{figure}
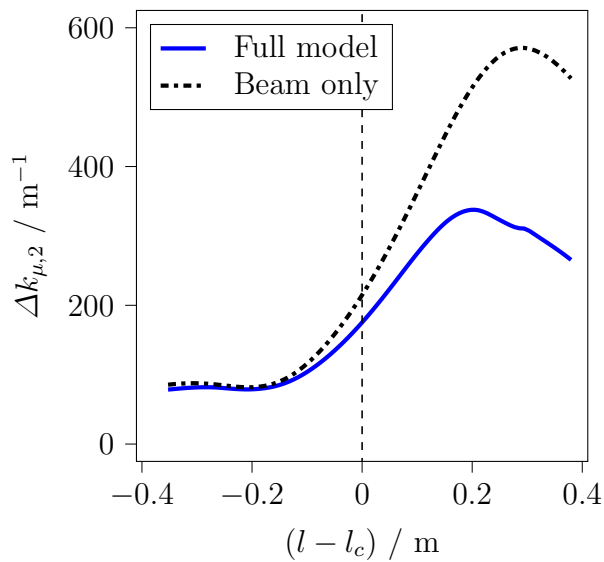

	To understand the width of the Gaussian, $\Delta k_{\mu,2}$, we first consider a simple case with $M_{xy,\mu}=0$. Then, $M_{yy,\mu}^{-1} = 1 / M_{yy,\mu}$. Remembering that $M_{yy,\mu} = \Psi_{yy,\mu}$ exactly, and using the definitions given by equations (\ref{eq:Psi_real}) and (\ref{eq:Psi_imag}), we find that equation (\ref{eq:delta_k_perp_2}) gives 
	\begin{equation}\label{eq:wavenumber_resolution_simplified}
		\Delta k_{\mu,2} 
		= \frac{2 \sqrt{2}}{W_{y,\mu}} \left[ 1 + \frac{1}{4} K_\mu^2 W_{y,\mu}^4 \left( \frac{1}{R_{b,y,\mu}} \right)^2 \right]^{1/2}
		.
	\end{equation}
	This recovers the widely-used expression for the wavenumber resolution for a circular beam in a slab. We do not see the corrections due to the curvature of the magnetic field lines or the magnetic shear in equation (\ref{eq:wavenumber_resolution_simplified}) because we have taken $M_{xy,\mu} = 0$. As useful as this might have been to gain some physical insight, in general, we cannot neglect $M_{xy,\mu}$. The corrections due to curvature and shear of the magnetic field are therefore important. These corrections affect the wavenumber resolution in a way that in general cannot be easily further simplified. Hence, one has to use equation (\ref{eq:delta_k_perp_2}) in its presented form to determine the wavenumber resolution. We now apply our full model, equation (\ref{eq:delta_k_perp_2}), to our MAST test scenario, Figure \ref{fig:wavenumber_resolution_MAST}. By noticing the difference between the solid and dash-dot lines in the figure, one sees that the corrections indeed significantly affect wavenumber resolution.
	
	In our model, the curvature of the cut-off surface does not affect the wavenumber resolution at all. One can understand this as follows. Physically, it is the curvature of the field lines and the magnetic shear that are important, since the beam is scattered off the turbulent fluctuations perpendicular to the field lines. The beam is not scattered from the cut-off surface per se --- hence, strictly speaking, it is not the curvature of the cut-off surface that is important, as previously argued \cite{Hirsch:DBS:2001,Bulanin:spatial_spectral_resolution:2006}.   

	\section{Mismatch} \label{section_mismatch}
	The backscattered spectral density decays exponentially with mismatch, $p_r \propto \exp [ - 2 \theta_{m,\mu}^2 / (\Delta \theta_{m,\mu})^2 ]$, where $\Delta \theta_{m,\mu}$ is given by equation (\ref{eq:delta_theta_m}). This is consistent with our choice to take the mismatch angle to be small, such that the backscattered signal is large enough to detect. Like we did with the wavenumber resolution, we take $M_{xy,\mu} = 0$ to simplify the expression in order to gain some physical insight, getting
	\begin{equation} \label{eq:delta_theta_m_reduced}
		\Delta \theta_{m,\mu} 
		= \frac{ 1 }{ K_{\mu} }  \left[ - \textrm{Im} \left( \frac{ 1 }{ M_{xx,\mu}  } \right) \right]^{ - 1 / 2 } .
	\end{equation}
	We write this out in full, 
	\begin{equation} \label{eq:delta_theta_m_reduced_explicit}
	\Delta \theta_{m,\mu} 
	=
	\frac{\sqrt{2}}{W_{x,\mu} K_\mu}
	\left[
		1 + \frac{W_{x,\mu}^4 K_\mu^2}{4 R_{b,M,\mu}^2} 
	\right]^{\frac{1}{2}} ,
	\end{equation}
	where we use the shorthand
	\begin{equation}
		\frac{1}{R_{b,M,\mu}}
		=
		\frac{1}{R_{b,x,\mu}}
		+
		\left(
			\hat{\mathbf{b}} \cdot \nabla \hat{\mathbf{b}} \cdot \hat{\mathbf{g}} 
		\right)_\mu.
	\end{equation}
	
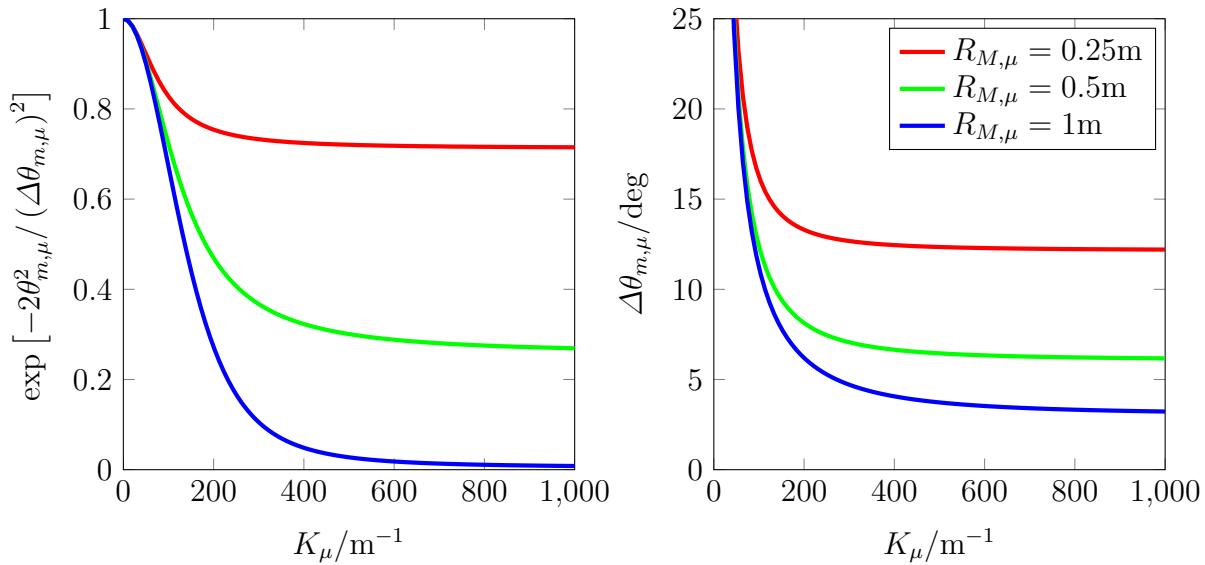
\begin{figure} 
	
	\begin{subfigure}{.45\textwidth}
		\centering
		\begin{tikzpicture}
		\begin{axis}[
		    xlabel = {$K_\mu / \textrm{m}^{-1}$},
		    ylabel = {$\exp \left[ - 2 \theta_{m,\mu}^2 / \left( \Delta \theta_{m,\mu} \right)^2 \right]$},
		    xmin=0, xmax=1000, 
		    ymin=0, ymax=1, 
		    domain=0:1000,
		    samples=100,
			legend cell align={left},
			width=6cm,
			height=6cm,
			scale only axis
		]
			
			\def\Wx{0.075} 
			\def\thetam{5} 
			
			\addplot[mark = none, red, ultra thick]{ 
				e^(
					- (\thetam / 57.3)^2 * \Wx^2 * x^2 /
					((
						1 + 0.25 * \Wx^4 * x^2 * (1/0.25)^2
					) ) 
				)
			};
			\addplot[mark = none, green, ultra thick]{ 
				e^(
					- (\thetam / 57.3)^2 * \Wx^2 * x^2 /
					((
						1 + 0.25 * \Wx^4 * x^2 * (1/0.5)^2
					) ) 
				)
			};
			\addplot[mark = none, blue, ultra thick]{ 
				e^(
					- (\thetam / 57.3)^2 * \Wx^2 * x^2 /
					((
						1 + 0.25 * \Wx^4 * x^2 * (1/1.0)^2
					) ) 
				)
			};
					
		\end{axis}
		\end{tikzpicture}
	\end{subfigure}
	\qquad
	\begin{subfigure}{.45\textwidth}
		\begin{tikzpicture}
		\begin{axis}[
		    xlabel = {$K_\mu / \textrm{m}^{-1}$},
		    ylabel = {$\Delta \theta_{m,\mu} / \textrm{deg}$},
		    xmin=0, xmax=1000, 
		    ymin=0, ymax=25, 
		    domain=25:1000,
		    samples=100,
			legend cell align={left},
			width=6cm,
			height=6cm,
			scale only axis
		]
			
			\def\Wx{0.075} 
			
			\addplot[mark = none, red, ultra thick]{ 
				sqrt(2) * ((
					1 + 0.25 * \Wx^4 * x^2 * (1/0.25)^2
				)^(0.5)) * 57.3 / (x * \Wx)
			};
			\addplot[mark = none, green, ultra thick]{ 
				sqrt(2) * ((
					1 + 0.25 * \Wx^4 * x^2 * (1/0.5)^2
				)^(0.5)) * 57.3 / (x * \Wx)
			};
			\addplot[mark = none, blue, ultra thick]{ 
				sqrt(2) * ((
					1 + 0.25 * \Wx^4 * x^2 * (1/1.0)^2
				)^(0.5)) * 57.3 / (x * \Wx)
			};
			
			\legend{$R_{M,\mu} = 0.25 \textrm{m}$,$R_{M,\mu} = 0.5 \textrm{m}$,$R_{M,\mu} = 1 \textrm{m}$}
		\end{axis}
		\end{tikzpicture}
	\end{subfigure}
	\caption{How beam parameters at the scattering location influence the attenuation due to mismatch. We do not use beam tracing for these graphs; instead, we vary the beam properties by hand and see what happens. Here, we have $1 / R_{M} = 1 / R_x + \hat{\mathbf{b}} \cdot \nabla \hat{\mathbf{b}} \cdot \hat{\mathbf{g}}$, $M_{xy} = 0$, $W_x = 7.5 \textrm{cm}$, and $\theta_m = 5^\circ$. The line colours correspond to the same $R_M$ for both graphs.}
	\label{fig:mismatch_largeK}
	\end{figure}

	It has been reported \cite{Hillesheim:DBS_MAST:2015, Damba:mismatch:2021} that for high $K$, the backscattered signal's amplitude is especially sensitive to the toroidal steering angle. This can be explained by mismatch. At a given finite mismatch angle, the attenuation due to mismatch is larger at larger $K_\mu$, as seen from equation (\ref{eq:delta_theta_m_reduced_explicit}) and from Figure \ref{fig:mismatch_largeK}. Indeed, for
	\begin{equation}
		K_\mu \gg \frac{ 2 R_{b,M,\mu} }{ W_{x,\mu}^{2} } ,
	\end{equation}
	equation (\ref{eq:delta_theta_m_reduced_explicit}) reduces to
	\begin{equation}
		\exp \left[
		- 2 \frac{\theta_{m,\mu}^2}{\left( \Delta \theta_{m,\mu} \right)^2}
	\right]
	\approx \exp \left( -\theta_{m,\mu}^2 \frac{ 4 R_{b,M,\mu}^{2} }{ W_{x,\mu}^{2} } \right) ,
	\end{equation}
	at which point further increasing $K_\mu$ has no effect on the mismatch attenuation. These considerations are especially important when designing DBS systems with no toroidal optimisation; there will be a finite mismatch over many of the channels, and this will be more problematic at larger wavevectors. Understanding mismatch attenuation will help one to mitigate this issue.

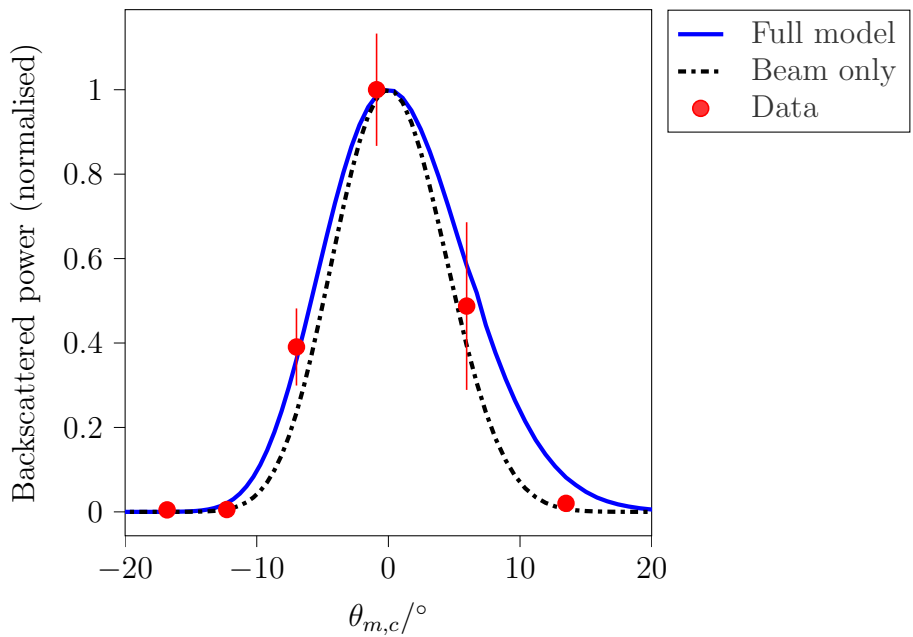
\begin{figure}
	\centering
	\begin{tikzpicture} 	
	\begin{axis}[
	legend cell align={left},
	legend style={fill opacity=0.8, draw opacity=1, text opacity=1, draw=black},
    legend pos=outer north east,
	tick align=outside,
	tick pos=left,
	xlabel=$\theta_{m,c} / ^{\circ}$,
	ylabel=$\textrm{Backscattered power (normalised)}$,
	x grid style={white!69.0196078431373!black},
	xmin=-20.0, xmax=20.0,
	xtick style={color=black},
	y grid style={white!69.0196078431373!black},
	ymin=-0.0566298510126379, ymax=1.18968690857077,
	ytick style={color=black},
	clip mode=individual,	
	width=7cm,
	height=7cm,
	scale only axis		
	]
	\addplot [ultra thick, blue]
	table {%
	29.3321007590341 3.64045292776103e-05
	28.4105649859415 5.88891162148755e-05
	27.5626977012114 9.49719838444684e-05
	26.7047043989058 0.000155635768533665
	25.7890209403511 0.000265150580849538
	24.9457737306792 0.000421161107491566
	24.1188558748486 0.000666620612081822
	23.2328002940721 0.00114127004870988
	22.405258469302 0.00175820327682703
	21.5593036235711 0.00259904280036635
	20.6907628831982 0.00401998337590055
	19.8867858161985 0.0059086434637798
	19.0954288831927 0.00859820835541496
	18.2359452390249 0.0127623153650437
	17.4412218330883 0.0178235794177551
	16.6249934514793 0.02503471428689
	15.8474913311352 0.0340170468438913
	15.0130422905345 0.0473780921218284
	14.2404681932683 0.0635628638957787
	13.4897883222156 0.0815476957061582
	12.6723621203874 0.107772972573098
	11.934157721246 0.13617988994899
	11.1901725776573 0.17026602268711
	10.3903436859841 0.214645183252507
	9.61998902370897 0.263523093879979
	8.91315310308776 0.314287774739618
	8.1328817450157 0.376984048531855
	7.40693444895965 0.441976727148973
	6.71337590757465 0.518671068092831
	5.93486229314117 0.58513081333817
	5.24829799726002 0.653228362985522
	4.53782450289336 0.725028020266096
	3.81385302186327 0.794574216927193
	3.13269946817282 0.8544874991241
	2.45509641002577 0.906627514670231
	1.75243846576039 0.950635716843213
	1.08050620695721 0.980678498142311
	0.425956241786194 0.996925549973572
	-0.262547729033368 0.998813994677669
	-0.898070278227627 0.985984143663364
	-1.53432772993433 0.958970413257339
	-2.19559983259328 0.916640977975401
	-2.81291488319379 0.86475079787801
	-3.42023347796766 0.803579060857359
	-4.06020354626425 0.730989214389566
	-4.65646221246562 0.657042760935795
	-5.25253148126871 0.580170686186496
	-5.84971094070441 0.502752737869188
	-6.41858439561814 0.429519562758621
	-6.98413842082148 0.360562393446395
	-7.56783462975541 0.294745662569393
	-8.11423014474117 0.238566888989506
	-8.66250181610178 0.188869372736914
	-9.20696087703705 0.146201516625662
	-9.7339169592242 0.111412371223395
	-10.2561060904426 0.0830735270582685
	-10.78166737237 0.0602805245681353
	-11.2792747907173 0.043239837779333
	-11.7702464975796 0.0303169466947965
	-12.2738116897361 0.0205956389932773
	-12.75857019114 0.0137167553217755
	-13.2267507095642 0.00898998820708123
	-13.6881200582644 0.00576094850310632
	-14.1678707412096 0.00354267619121959
	-14.6173697407003 0.00215271642571159
	-15.0637813199752 0.00127895301144634
	-15.5160825888171 0.000731456270796149
	-15.9492049751481 0.000412973832257923
	-16.3751349335059 0.000228500163331914
	-16.8092878633116 0.000120773769975242
	-17.2241841110326 6.32536929929739e-05
	-17.6321056224456 3.23746703838915e-05
	-18.0492032036162 1.58811490103964e-05
	-18.4458077360552 7.74180824107122e-06
	-18.8378946514559 3.68505380803259e-06
	-19.2375746146845 1.67648677110534e-06
	-19.6191658152329 7.6377725468743e-07
	-19.994998090383 3.38329796726196e-07
	-20.3788591310365 1.4404562378872e-07
	-20.7455397690534 6.10835425478535e-08
	-21.1079554838783 2.5520696914516e-08
	};
	\addlegendentry{Full model}
	\addplot [ultra thick, black, dash dot]
	table {%
	29.3321007590341 2.73245824623738e-13
	28.4105649859415 2.45854491359326e-12
	27.5626977012114 1.74064425672832e-11
	26.7047043989058 1.10366322765402e-10
	25.7890209403511 6.92262900836508e-10
	24.9457737306792 3.46431301639475e-09
	24.1188558748486 1.57973809775292e-08
	23.2328002940721 6.84684485081114e-08
	22.405258469302 2.60827558732601e-07
	21.5593036235711 9.63511583707835e-07
	20.6907628831982 3.23600093458222e-06
	19.8867858161985 9.59298596164063e-06
	19.0954288831927 2.65304484779802e-05
	18.2359452390249 7.25948698989955e-05
	17.4412218330883 0.000178409198486797
	16.6249934514793 0.000418796742493979
	15.8474913311352 0.000912376875204789
	15.0130422905345 0.00194198714358532
	14.2404681932683 0.00378298822976258
	13.4897883222156 0.00700172248464839
	12.6723621203874 0.0128363590228005
	11.934157721246 0.0215693088479363
	11.1901725776573 0.0349663000126804
	10.3903436859841 0.0559919632503173
	9.61998902370897 0.0852837696460883
	8.91315310308776 0.122016077630716
	8.1328817450157 0.174146295345675
	7.40693444895965 0.235604342883697
	6.71337590757465 0.305455778895158
	5.93486229314117 0.395828478293334
	5.24829799726002 0.485789597192102
	4.53782450289336 0.582789463833132
	3.81385302186327 0.682340908715726
	3.13269946817282 0.772470888723161
	2.45509641002577 0.853359160367755
	1.75243846576039 0.922053087293667
	1.08050620695721 0.969532126230978
	0.425956241786194 0.995185034729118
	-0.262547729033368 0.998155850450494
	-0.898070278227627 0.978527059951587
	-1.53432772993433 0.938176915486522
	-2.19559983259328 0.876558822186293
	-2.81291488319379 0.804108224534582
	-3.42023347796766 0.722375753625106
	-4.06020354626425 0.629429817602249
	-4.65646221246562 0.540275186280738
	-5.25253148126871 0.453369048614743
	-5.84971094070441 0.370283070336218
	-6.41858439561814 0.298227667626229
	-6.98413842082148 0.234457140627352
	-7.56783462975541 0.178136175396736
	-8.11423014474117 0.134093266321516
	-8.66250181610178 0.0985662773111079
	-9.20696087703705 0.0703901222319287
	-9.7339169592242 0.0495714440040782
	-10.2561060904426 0.0341751110666767
	-10.78166737237 0.0226405359667597
	-11.2792747907173 0.0151066364924855
	-11.7702464975796 0.00978649718766445
	-12.2738116897361 0.00612821885416048
	-12.75857019114 0.00379927152503006
	-13.2267507095642 0.00232322416264205
	-13.6881200582644 0.00139751607151986
	-14.1678707412096 0.000802212159560228
	-14.6173697407003 0.000462258438151336
	-15.0637813199752 0.000262457671190705
	-15.5160825888171 0.000142424933360078
	-15.9492049751481 7.77916820964065e-05
	-16.3751349335059 4.18191896733083e-05
	-16.8092878633116 2.15793571134415e-05
	-17.2241841110326 1.11865532345508e-05
	-17.6321056224456 5.70750487895141e-06
	-18.0492032036162 2.79362732329323e-06
	-18.4458077360552 1.37816853007817e-06
	-18.8378946514559 6.68838674835783e-07
	-19.2375746146845 3.10808923407684e-07
	-19.6191658152329 1.46199808537233e-07
	-19.994998090383 6.73415030109699e-08
	-20.3788591310365 2.96983867655111e-08
	-20.7455397690534 1.32409646552501e-08
	-21.1079554838783 5.82992857935411e-09
	};
	\addlegendentry{Beam only}
	\addplot [semithick, red, mark=*, mark size=3, mark options={solid}, only marks]
	table {%
	-6.98316048031629 0.390600855989362
	-12.2755241153475 0.005534323792466
	-16.8115423442074 0.00470783687449848
	5.94506879425094 0.487522663003621
	13.4926939700493 0.0201913892617379
	-0.896694703845822 1
	};
	\addlegendentry{Data}
	\path [draw=red, semithick]
	(axis cs:-6.98316048031629,0.299268820624092)
	--(axis cs:-6.98316048031629,0.481932891354633);
	
	\path [draw=red, semithick]
	(axis cs:-12.2755241153475,0.00434167178425971)
	--(axis cs:-12.2755241153475,0.00672697580067228);
	
	\path [draw=red, semithick]
	(axis cs:-16.8115423442074,0.0025631556316714)
	--(axis cs:-16.8115423442074,0.00685251811732557);
	
	\path [draw=red, semithick]
	(axis cs:5.94506879425094,0.288945057451303)
	--(axis cs:5.94506879425094,0.686100268555939);
	
	\path [draw=red, semithick]
	(axis cs:13.4926939700493,0.0123240707514649)
	--(axis cs:13.4926939700493,0.028058707772011);
	
	\path [draw=red, semithick]
	(axis cs:-0.896694703845822,0.866963853228478)
	--(axis cs:-0.896694703845822,1.13303614677152);
		
	\end{axis}
	
	\end{tikzpicture}
		\caption{Red points indicate DBS data from repeated shots, where only the toroidal launch angle was varied \cite{Hillesheim:DBS_MAST:2015}. We have taken there to be a $1.1^{\circ}$ systematic error in the steering mirror's rotation angle, which affects the launch angles of the beam. The data points are from similar flux surfaces, but at different mismatch angles. Hence, the attenuation that we see is mainly from the mismatch, since we expect the turbulent fluctuations to be largely the same. The solid blue line indicates the model's predicted attenuation due to mismatch, $\exp \left[-2 \theta_{m,c}^2 / \left( \Delta \theta_{m,c} \right)^2 \right] $, given the beam properties at the cut-off when we launch the beam at various toroidal angles. The dash-dot black line shows the attenuation but with $\Delta \theta_{m,c}$ calculated from $\bm{\Psi}_{w}$ instead of $\bm{M}_{w}$, that is, calculated only from the beam properties without the corrections arising from the curvature and shear of $\hat{\mathbf{b}}$.}
	\label{fig:mismatch_MAST}
	\end{figure}

	We proceed to apply the unsimplified expression for mismatch attenuation, equation (\ref{eq:delta_theta_m}), to a real tokamak, Figure \ref{fig:mismatch_MAST}. The beam and equilibrium plasma properties account for the experimentally-observed mismatch attenuation, and not backscattering from some $k_\parallel$ (which we neglect in our model). This preliminary analysis of MAST data is a good proof of concept; we now understand, and are able to calculate, how mismatch attenuates the DBS signal. A more detailed analysis of the other channels (different frequencies) and other times will be performed in a future paper, which will enable us to better evaluate our model for use in real plasmas. This advancement means that we can now operate DBS in regimes where the mismatch is small, but not so small as to be negligible (as was required previously). New insights can thus be gained from existing data, and new experiments can be performed with less strict tolerances.

	\section{Backscattered electric field and power: large mismatch angle ordering}	\label{section_backscattered_E_ST}
	At the end of Section \ref{section_backscattered_E}, we introduced the \emph{small mismatch angle} and \emph{large mismatch angle} orderings. Thus far, we have focused on the former. That is, we have taken $\theta_{m} \sim \lambda / W \ll 1$ along the beam.	When the mismatch angle is not small at every point along the beam (\emph{large mismatch angle} ordering), we consider only the simplified case in which one point along the path has no mismatch, $\tau = \tau_0$, and that the mismatch angle is large elsewhere. 
	
	To begin, we remind readers of the exponentially decaying piece in equation (\ref{eq:A_r_after_wx_wy_integration})
	\begin{equation} \label{eq:mismatch_decay}
	\tilde{A}_r \propto
	\exp \left[
		- \frac{\rmi}{4} \left(
		2 K \sin (\theta + \theta_m)
		+ k_{\perp,1} \sin \theta
		\right)^2 M_{xx}^{-1}
	\right] ,
	\end{equation}
	where we have used equation (\ref{eq:K_x}). Since $k_{\perp,1} \sim K$ and $K^2 M_{xx}^{-1} \sim W^2 / \lambda^2 \gg 1$, the signal is localised to the points where the mismatch is zero, $\theta_{m} = 0$, regardless of whether those are near the cut-off or not; due to equation (\ref{eq:size_of_theta}), $\theta = 0$ at these points as well. At the location with zero mismatch, the dominant scattered wavevector is given by the Bragg condition at this point. In a case in which the mismatch angle does not vanish at any point along the ray, we would need to use a steepest descent method to optimise for both the real (mismatch attenuation) and imaginary (stationary phase, conventionally called the Bragg condition) parts of the exponential. However, we will not cover this here. 
	
	We perform a Taylor expansion around the point of zero mismatch, for $g \left( \tau - \tau_0 \right) \sim W$, which is a sensible ordering because the equilibrium properties such as the mismatch vary on length scales of $L$. Hence, a distance of $W$ away from zero mismatch, we expect the mismatch to be $\sim W / L$. Thus, the argument of equation (\ref{eq:mismatch_decay}) is of order unity in the region $g \left( \tau - \tau_0 \right) \sim W$, which means the backscattered signal is significantly attenuated away from this region. The Bragg condition is also not exactly met away from $\tau_0$, hence
	\begin{equation} \label{eq:delta_kperp1_assertion}
		2 K_{0} + k_{\perp,1} \sim \frac{1}{W} ,
	\end{equation}
	which we will later prove in equation (\ref{eq:delta_kperp1_size}). 
	
	We begin by introducing a few more orderings. From earlier, we had $\theta_0 = \theta_{m,0} = 0$. However, the derivatives of both $\theta$ and $\theta_{m}$ change on the length scale $L$,
	\begin{equation}
		\frac{\rmd \theta}{\rmd \tau}
		\sim \frac{\rmd \theta_{m}}{\rmd \tau}
		\sim \frac{1}{\tau_L},
	\end{equation}
	where $\tau_L \sim L / g$; at a distance of $L$ away from $\tau_0$, the mismatch is large $\theta \sim \theta_m \sim 1$. Using these orderings, we find that
	\begin{equation}
		\left(\tau -\tau_0 \right) \frac{\rmd }{\rmd \tau} \left( 2 K_x - k_{\perp,1} \sin \theta \right) \bigr|_{\tau_0}
		\simeq - 2 K_0 \frac{\rmd \theta_m}{\rmd \tau} \Bigr|_{\tau_0} \left(\tau -\tau_0 \right) 
		\sim \frac{1}{W},
	\end{equation}
	and
	\begin{equation}
	\eqalign{
		& \left(\tau -\tau_0 \right)^2 \frac{\rmd^2 }{\rmd \tau^2} \left( 2 K_x - k_{\perp,1} \sin \theta \right) \bigr|_{\tau_0} \\
		& \qquad \simeq \left[
			- 4 \frac{\rmd K}{\rmd \tau} \Bigr|_{\tau_0} \left( 
			\frac{\rmd \theta_m}{\rmd \tau} \Bigr|_{\tau_0}
			+ \frac{\rmd \theta}{\rmd \tau} \Bigr|_{\tau_0} 
			\right)
			- 2 K_{0} \frac{\rmd^2 \theta_m}{\rmd \tau^2} \Bigr|_{\tau_0}
		\right] \left(\tau - \tau_0 \right)^2 \\
		& 
		\qquad  \sim \frac{1}{L} ,
	}
	\end{equation}
	where we have used equation (\ref{eq:K_x}), and equation (\ref{eq:delta_kperp1_assertion}) to write $k_{\perp,1} \simeq - 2 K_0$. Taking into account all these considerations, and recalling that $k_{\perp,2} \sim 1 / W$, we now Taylor expand the exponentially decaying piece, as well as the large phase piece, keeping terms of order unity
	\begin{equation}
	\fl
	\eqalign{
		&
		- \frac{1}{4}(2 K_x - k_{\perp,1} \sin \theta)^2 M_{xx}^{-1}
		- \frac{1}{2} k_{\perp,2} (2 K_x - k_{\perp,1} \sin \theta) M_{xy}^{-1}
		- \frac{1}{4} k_{\perp,2}^2 M_{yy}^{-1} \\
		& \qquad + 2 s
		+ k_{\perp,1} \int_0^{\tau}  g (\tau') \cos \theta (\tau') \ \rmd \tau' \\
		&  \simeq
		- \frac{1}{4} k_{\perp,2}^2 M_{yy,0}^{-1}
		+ 2 s_0
		+ k_{\perp,1} \int_0^{\tau_0}  g (\tau') \cos \theta (\tau') \ \rmd \tau'\\
		&\qquad  + \left(
			k_{\perp,2} K_0 \frac{\rmd \theta_m}{\rmd \tau} \Bigr|_{\tau_0} M_{xy,0}^{-1}
			+ 2 K_{0} g_0
			+ k_{\perp, 1} g_0
		\right)
		\left(\tau -\tau_0 \right) \\
		&\qquad  + \left[
			g_0 \frac{\rmd K}{\rmd \tau} \Bigr|_{\tau_0} 
			- K_0^2 \left( \frac{\rmd \theta_m}{\rmd \tau} \Bigr|_{\tau_0} \right)^2 M_{xx,0}^{-1}
		\right] \left(\tau -\tau_0 \right)^2 .
	}
	\end{equation}
	To make the equations more manageable, we use the shorthand
	\begin{equation} \label{eq:G0}
		G_0 \left( k_{\perp,1}, k_{\perp,2} \right)
		=
		\left[
			g_0 \frac{\rmd K}{\rmd \tau} \Bigr|_{\tau_0} 
			- K_0^2 \left( \frac{\rmd \theta_m}{\rmd \tau} \Bigr|_{\tau_0} \right)^2 M_{xx,0}^{-1}
		\right]^{-1} .
	\end{equation} 
	We now proceed to solve the Gaussian integral in $\tau$ to get
	\begin{equation} \label{eq:A_r_spherical}
	\fl
	\eqalign{
		\tilde{A}_r
		&= - \frac{\Omega A_{ant} g_{ant} \hat{\mathbf{e}}_{ant} \cdot \hat{\mathbf{e}}_{ant}}{2 c}
		\int
		\left[ \frac{\det \left[\textrm{Im} \left( \bm{\Psi}_{w,0} \right) \right]}{\det \left( \bm{M}_{w,0} \right)} \right]^{\frac{1}{2}}
		\exp (2 \rmi \phi_{G,0} ) \\
		& \times \frac{ \delta \tilde{n}_{e,0} }{n_{e,0}} \ \hat{\mathbf{e}}^*_{0}
		\cdot (\bm{\epsilon}_{eq,0} -\bm{1}) \cdot \hat{\mathbf{e}}_{0}
		\ \exp \left(
			2 \rmi s_0 + \rmi k_{\perp,1} \int_0^{\tau_0} g (\tau') \cos \theta (\tau') \ \rmd \tau'
		\right) \\
		& \times \exp \left[
		- \frac{\rmi}{4}
		\left(
			k_{\perp,2} K_0 \frac{\rmd \theta_m}{\rmd \tau} \Bigr|_{\tau_0} M_{xy,0}^{-1}
			+ 2 K_{0} g_0
			+ k_{\perp, 1} g_0
		\right)^2
		G_0
		\right] \\
		& \times
		\left(
			\pi \rmi G_0
		\right)^{\frac{1}{2}}
		\exp\left[
		-\frac{\rmi}{4} k_{\perp,2}^2 M_{yy,0}^{-1}
		\right]
		\ \rmd k_{\perp,1} \ \rmd k_{\perp,2} .
	}
	\end{equation}
	Here $\textrm{Re} \left( \sqrt{\pi \rmi G_0} \right) \geq 0$ as per standard contour integration methods, and the density fluctuations are evaluated at
	\begin{equation}
		\delta \tilde{n}_{e,0} \left( 
			k_{\perp,1}, 
			k_{\perp,2}, 
			\omega 
		\right) 
		=
		\delta \tilde{n}_e \left(
			k_{\perp,1},
			k_{\perp,2}, 
			- \int_0^{\tau_0} g (\tau') \sin \theta (\tau') \ \rmd \tau', 
			\omega 
		\right) .
	\end{equation}
	In principle, the mismatch could be zero at more than one point along the ray. In that case, the total backscattered signal is the sum of the backscattered signal of each of those points. In the interests of simplicity, we will not discuss this further in this paper.
	
	Before we continue further, we first show that in the appropriate limit, the \emph{large mismatch angle} and \emph{small mismatch angle} orderings give the same result. Showing that the backscattered amplitudes in these orderings match is challenging due to the large phase term $k_{\perp,1} \int_{0}^{\tau} g (\tau') \cos \theta (\tau') \rmd \tau'$. Instead, it is easier to show that the backscattered powers match. To do this, we perform subsidiary expansions of both the \emph{small mismatch angle} and \emph{large mismatch angle} expressions. For the former, we take the mismatch to be large and perform a Taylor expansion of $k_{\perp,1}$ about $- 2 K_0$. 	
	For the \emph{large mismatch angle} ordering, we make the mismatch less than order unity $\theta_{m} \ll 1$ throughout the beam path. In this intermediate range, 
	\begin{equation} \label{eq:subsidiary_expansion_ordering}
		1 \gg \theta_{m} \gg \frac{\lambda}{W} ,
	\end{equation}
	both models should be applicable and must coincide. 
	
	In such an endeavour, the backscattered power of the \emph{small mismatch angle} ordering is given by equation (\ref{eq:A_r2_explicit}); that of the \emph{large mismatch angle} ordering can be determined by applying the same methods used in Section \ref{section_backscattered_P}, giving
	\begin{equation} \label{eq:A_r2_spherical}
	\fl 
	\eqalign{
		\frac{p_r}{P_{ant}} 
		&=
		\frac{1}{4} \frac{ \Omega^2 }{c^2} g_{ant}^2 \left| \hat{\mathbf{e}}_{ant} \cdot \hat{\mathbf{e}}_{ant} \right|^2 \pi
		\int
		\frac{\det \left[\textrm{Im} \left( \bm{\Psi}_{w,0} \right) \right]}{\left| \det \left( \bm{M}_{w,0} \right) \right|} 
		\left| \hat{\mathbf{e}}^*_{0} \cdot (\bm{\epsilon}_{eq,0} -\bm{1}) \cdot \hat{\mathbf{e}}_{0} \right|^2 \\
		& \times
		\frac{1}{\left| G_0 \right|}
		\exp \left\{
		\frac{1}{2} \ \textrm{Im} 
		\left[
		\left(
			k_{\perp,2} K_0 \frac{\rmd \theta_m}{\rmd \tau} \Bigr|_{\tau_0} M_{xy,0}^{-1}
			+ 2 K_{0} g_0
			+ k_{\perp, 1} g_0
		\right)^2
		G_0
		\right]
		\right\} \\
		& \times
		\exp\left[
		\frac{1}{2} \textrm{Im} \left( M_{yy,0}^{-1} \right) k_{\perp,2}^2
		\right] 
		\frac{
		\left<
			\delta n_{e,0}^2 (t)
		\right>	_t
		}{
			n_{e,0}^2
		}
		\widetilde{C}_0 (k_{\perp,1}, k_{\perp,2}, \omega)
		\ \rmd k_{\perp, 1} \ \rmd k_{\perp, 2} .
	}
	\end{equation}
	We now perform the subsidiary expansion of the \emph{large mismatch angle} ordering. We begin by expanding $G_0$ --- see equation (\ref{eq:G0}) --- to get
	\begin{equation} \label{eq:G0_spherical_expansion}
	\eqalign{
		G_0 
		& = \left(
		\frac{\rmd K}{\rmd \tau} \Bigr|_{\tau_0} g_0
		- K_0^2 \left( \frac{\rmd \theta_m}{\rmd \tau} \Bigr|_{\tau_0} \right)^2 M_{xx,0}^{-1}
		\right)^{-1} \\
		& \simeq
		\left(
		\frac{\rmd K}{\rmd \tau} \Bigr|_{\tau_0} g_0
		\right)^{-1}
		+ K_0^2 \left( \frac{\rmd \theta_m}{\rmd \tau} \Bigr|_{\tau_0} \right)^2 M_{xx,0}^{-1} \left(
		\frac{\rmd K}{\rmd \tau} \Bigr|_{\tau_0} g_0
		\right)^{-2} ,
	}
	\end{equation}
	where we have kept terms up to order unity for the exponential. When it comes to the $|G_0|^{-1}$ outside the exponential, see equation (\ref{eq:A_r2_spherical}), we only need to keep the zeroth order term, ignoring the first order correction. It is worth noting that we looked at several MAST shots, and the difference between the full expression and the expansion in equation (\ref{eq:G0_spherical_expansion}) was negligible, that is, the MAST DBS seems to be in the intermediate regime where the \emph{small mismatch angle} and \emph{large mismatch angle} orderings are both valid. We introduce the notation
	\begin{equation} \label{eq:delta_kperp1}
		\Delta k_{\perp,1} = k_{\perp,1} - k_{\perp,1,0} = k_{\perp,1} + 2 K_{0},
	\end{equation}
	where we have used the Bragg condition at the zero-mismatch point. We see that the signal decays exponentially with $\Delta k_{\perp,1}$ and hence we order the most quickly decaying term to be of order one
	\begin{equation}
		- \textrm{Im} \left[ \left( \Delta k_{\perp,1} \right)^2 K_0^2 \left( \frac{\rmd \theta_m}{\rmd \tau} \Bigr|_{\tau_0} \right)^2 M_{xx,0}^{-1} \left(
		\frac{\rmd K}{\rmd \tau} \Bigr|_{\tau_0} 
		\right)^{-2} \right]
		\sim 1 .
	\end{equation}
	This gives us the size of $\Delta k_{\perp,1}$
	\begin{equation} \label{eq:delta_kperp1_size}
		\Delta k_{\perp,1} \sim \frac{1}{\theta_m W} ,
	\end{equation}
	which makes physical sense, since one would expect better localisation at $\tau_0$ with larger mismatch along the ray. In the \emph{large mismatch angle} subsection, we had $\theta_{m} \sim 1$, which gives $\Delta k_{\perp,1} \sim 1 / W$, justifying equation (\ref{eq:delta_kperp1_assertion}). We now take $\theta_{m} \ll 1$ and expand the exponentially decaying piece from equation (\ref{eq:A_r_spherical}), using equations (\ref{eq:G0_spherical_expansion}), (\ref{eq:delta_kperp1}), and (\ref{eq:delta_kperp1_size}), getting
	\begin{equation} \label{eq:A_r2_spherical_subsidiary}
	\fl
	\eqalign{
		& \frac{1}{2} \textrm{Im} \left[
			\left(
				k_{\perp,2} K_0 \frac{\rmd \theta_m}{\rmd \tau} \Bigr|_{\tau_0} M_{xy,0}^{-1}
				+ g_0 \Delta k_{\perp,1}
			\right)^2
			\left(
				g_0 \frac{\rmd K}{\rmd \tau} \Bigr|_{\tau_0}		
				- K_0^2 \left( \frac{\rmd \theta_m}{\rmd \tau} \Bigr|_{\tau_0} \right)^2 M_{xx,0}^{-1}
			\right)^{-1} 
		\right]\\
		& \quad
		+ \frac{1}{2} \textrm{Im} \left( M_{yy,0}^{-1} \right) k_{\perp,2}^2 \\
		& \simeq 
		\frac{1}{2}
		\Delta k_{\perp,1}^2 K_0^2 \left( \frac{\rmd \theta_m}{\rmd \tau} \Bigr|_{\tau_0} \right)^2 \left( \frac{\rmd K}{\rmd \tau} \Bigr|_{\tau_0} \right)^{-2} \textrm{Im} \left( M_{xx,0}^{-1} \right) \\
		& \quad + k_{\perp,2} \Delta k_{\perp,1} K_0 \frac{\rmd \theta_m}{\rmd \tau} \Bigr|_{\tau_0} \left( \frac{\rmd K}{\rmd \tau} \Bigr|_{\tau_0} \right)^{-1} \textrm{Im} \left( M_{xy,0}^{-1} \right) 
		+ \frac{1}{2} k_{\perp,2}^2 \textrm{Im} \left( M_{yy,0}^{-1} \right) ,
	}
	\end{equation}
	where we have neglected terms that are small. 
	
	We proceed to perform a subsidiary expansion of the \emph{small mismatch angle} ordering, equation (\ref{eq:A_r_stationary_phase}), in the limit $\theta_m \gg \lambda / W $. To determine the relationship between $\tau_{\mu}$ and $k_{\perp,1}$ around $\tau_0$, we differentiate equation (\ref{eq:Bragg_condition_full}) with respect to $\tau$, at $\tau = \tau_0$, 
	\begin{equation} 
	\eqalign{
		\frac{\rmd \tau_\mu}{\rmd k_{\perp,1}} \Bigr|_{\perp,1,0}
		= - \left( 2 \frac{\rmd K}{\rmd \tau} \Bigr|_{\tau_0} \right)^{-1} ,
	}
	\end{equation} 
	which we then use in our expansion in $k_{\perp, 1}$ of the exponential decaying piece in equation (\ref{eq:A_r2_explicit}). Together with equations (\ref{eq:delta_kperp1}), and (\ref{eq:delta_kperp1_size}), this gives us
	\begin{equation} \label{eq:A_r2_subsidiary} 
	\fl
	\eqalign{
		& 2^{} \textrm{Im} \left( M_{xx,\mu}^{-1} \right) K_{\mu}^2 \theta_{m,\mu}^2 
		- 2^{} \textrm{Im} \left( M_{xy,\mu}^{-1} \right) K_{\mu} \theta_{m,\mu} k_{\perp,2}
		+ \frac{1}{2} \textrm{Im} \left( M_{yy,\mu}^{-1} \right) k_{\perp,2}^2 \\
		&  = 
		\frac{1}{2}  \Delta k_{\perp,1}^2 K_0^2 \left(
			\frac{\rmd \theta_m}{\rmd \tau} \Bigr|_{\tau_0}
		\right)^2
		\left( \frac{\rmd K}{\rmd \tau} \Bigr|_{\tau_0} \right)^{-2} \textrm{Im} \left( M_{xx,0}^{-1} \right) \\
		& \quad 
		+ k_{\perp,2} \Delta k_{\perp,1} K_0 \frac{\rmd \theta_m}{\rmd \tau} \Bigr|_{\tau_0} \left( \frac{\rmd K}{\rmd \tau} \Bigr|_{\tau_0} \right)^{-1} \textrm{Im} \left( M_{xy,0}^{-1} \right) 
		+\frac{1}{2} k_{\perp,2}^2 \textrm{Im} \left( M_{yy,0}^{-1} \right)
		,
	}
	\end{equation}
	where we have neglected terms which are small and recalled that the Bragg condition is exactly met at $\tau_0$. We note that equations (\ref{eq:A_r2_spherical_subsidiary}) and (\ref{eq:A_r2_subsidiary}) match, showing that the exponential decay is indeed the same in both orderings. 
	As such, we can use the \emph{small mismatch angle} formulation and the associated results, such as equations (\ref{eq:A_r2_explicit_wavenumber_resolution}) and (\ref{eq:A_r2_final_cleaned}), even for moderate \emph{large mismatch angle} situations. Recall that the MAST DBS seems to be in this regime. As we explain below equation (\ref{eq:G0_spherical_expansion}), for the few MAST shots that we have reviewed, the difference between the exact and approximate expressions of $G_0$ was negligble.
	
	\section{Discussion} \label{section_discussion}
	Having presented the physical and quantitative insight our model sheds on localisation, wavenumber resolution, and mismatch attenuation, we now discuss how the various aspects of our model might come together and enable us to better understand DBS as a whole.
	
	A point of considerable interest is that the optimisations of both $k_{\perp,2}$ wavenumber resolution and mismatch attenuation have opposite requirements. To gain intuition, we consider the simplified case for $M_{xy,\mu} = 0$, where the associated mismatch attenuation is given in equation (\ref{eq:delta_theta_m_reduced_explicit}) and the $k_{\perp,2}$ wavenumber resolution in equation (\ref{eq:wavenumber_resolution_simplified}). When designing a DBS system, one would want to maximise resolution (minimise $\Delta k_{\mu,2}$) and minimise mismatch attenuation (maximise $\Delta \theta_{m,\mu}$). Hence, to optimise both simultaneously, one would need an elliptical beam.
	
	We examine how the $- \textrm{Im} \left( 1 / \Psi_{\alpha \alpha} \right)$ piece behaves in vacuum and answer the question of whether the wavenumber resolution $\Delta k_{\mu,2}$ and the mismatch attenuation $\Delta \theta_{m,\mu}$ are larger at or far from the beam waist. The answer is not obvious. Consider the following example. The intuitive reason why the backscattered power is nonzero when there is a mismatch is because there is a spread in $\mathbf{K}$ due to the width and curvature of the beam. Specifically, we see that only the width and curvature in one direction matters, $\Psi_{xx,\mu}$. To maximise the aforementioned spread, we want very curved wavefronts (small $R_{x,\mu}$) and narrow beams (small $W_{x,\mu}$). The dependence on beam curvature means we want to be far from the waist, while the dependence on width means we want to be at the waist. Conversely, the wavenumber resolution, which depends on $\Psi_{yy,\mu}$, benefits from the low curvature near the waist, and from the larger beam widths far from the waist. 
	
	In fact, in vacuum, it can be shown that a elliptical beam satisfies
	\begin{equation}
	- \textrm{Im} \left( \frac{1}{\Psi_{\alpha \alpha}} \right) 
	=
	\frac{2}{W^2_{\alpha,waist}}	
	\end{equation}
	at all points along the beam. Here, $\Psi_{\alpha \alpha}$ are the eigenvalues of $\bm{\Psi}_w$. Hence, it does not matter where we are along the beam nor where is the beam waist, the only important thing is the beam waist's width. We would want a narrower waist to optimise wavenumber resolution, and a wider waist to optimise mismatch attenuation. In order to reconcile these seemingly contradictory requirements, one could imagine using an elliptical Gaussian beam to simultaneously optimise widths in the relevant directions.
	
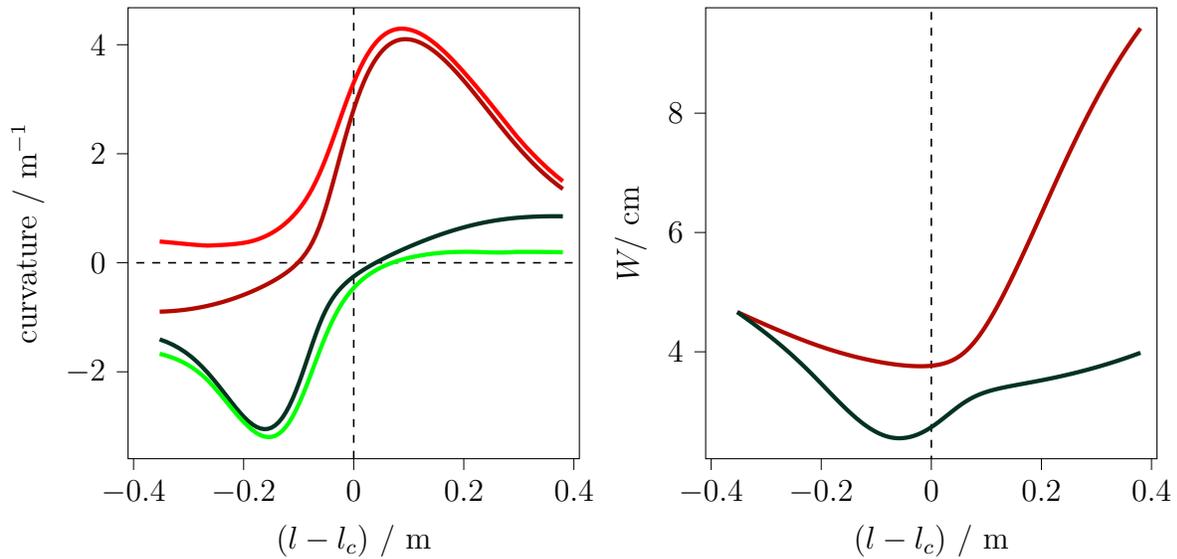
\begin{figure} 
		\begin{subfigure}{.5\textwidth}
		\centering
		\begin{tikzpicture}
		\begin{axis}[
		tick align=outside,
		tick pos=left,
		x grid style={white!69.0196078431373!black},
		xlabel={$(l - l_c)$ / m},
		xmin=-0.41, xmax=0.41,
		xtick style={color=black},
		y grid style={white!69.0196078431373!black},
		ylabel={ curvature / m$^{-1}$},
		ymin=-3.59418044084233, ymax=4.67878894808322,
		ytick style={color=black},
		width=6cm,
		height=6cm,
		scale only axis,
		clip mode=individual
		]
		\addplot [semithick, black, dashed]
		table {%
		0 -3.59418044084233
		0 4.67878894808322
		};
		\addplot [semithick, black, dashed]
		table {%
		-1 0
		1 0
		};	
		\addplot [ultra thick, DarkRed]
		table {%
		-0.352554126400679 -0.897046393753908
		-0.340304861359129 -0.892306808986977
		-0.328453454564625 -0.884594392306891
		-0.316999266512588 -0.874167396815586
		-0.305936113573163 -0.861284600092043
		-0.295254588438315 -0.846194297555151
		-0.284942846301458 -0.829120638669502
		-0.274986681766248 -0.810252811574927
		-0.265369627325126 -0.789743321041771
		-0.25607307521478 -0.767763691774941
		-0.247076585928555 -0.744471104682316
		-0.238360273668691 -0.719975501239681
		-0.22990442155989 -0.694356101735906
		-0.221689885330229 -0.667681601670845
		-0.213698400191981 -0.640009726289272
		-0.205912595840857 -0.611388548376027
		-0.198316007708519 -0.581862533402627
		-0.190893083845035 -0.551471501963444
		-0.183629186936977 -0.520246619603208
		-0.176510591124369 -0.488197862391148
		-0.169524473454506 -0.455297488695797
		-0.162658899990588 -0.421473147087716
		-0.155903142549375 -0.386603582514837
		-0.149248336286477 -0.350523031438152
		-0.142686136152776 -0.313012856908805
		-0.136208740355325 -0.273779504342562
		-0.129808888928435 -0.232438222946841
		-0.123479846504725 -0.188490582193123
		-0.117215383287622 -0.141295345121022
		-0.111009754494758 -0.0900324173639486
		-0.104857678541444 -0.0336582118493545
		-0.0987543142212113 0.0291453491112985
		-0.0926952371180463 0.100020069870876
		-0.0866764154542638 0.180955381729502
		-0.0806941855409071 0.274242082189097
		-0.0747452044922307 0.382313866154534
		-0.0688261115053266 0.507726540340855
		-0.0629338695262489 0.652165294737296
		-0.0570656672872623 0.815885205646271
		-0.0512188229005763 0.997472442862547
		-0.0453907753346585 1.19397723089484
		-0.0395790757682557 1.40143844100284
		-0.0337813788513414 1.61552148249648
		-0.0279954338978096 1.83202587405276
		-0.0222190760303 2.04718035343219
		-0.0164502172931969 2.25776235293182
		-0.0106868377456718 2.46111185419928
		-0.0049269765427426 2.65509564623198
		0 2.81237353412451
		0.000831297092852901 2.83805475719014
		0.00658981239704753 3.00875056937036
		0.0123504266147555 3.16631540467422
		0.0181149659807234 3.31022158038399
		0.0238853663577394 3.44047342253076
		0.0296635314709709 3.55742150609481
		0.0354513689497203 3.6614746307001
		0.0412508218229409 3.75308083326718
		0.0470638774664113 3.83271702183059
		0.0528925765229316 3.90088215666317
		0.058739021793362 3.95809293396476
		0.0646053870937403 4.00488126455184
		0.0704939260707614 4.04179305163499
		0.0764069809646697 4.06938787747287
		0.0823469913051393 4.08823930611587
		0.0883165025221053 4.09893558236888
		0.0943181795914769 4.10207807274244
		0.100354927085528 4.09815818589862
		0.10642974438242 4.0875138315834
		0.112545776629706 4.07045154349713
		0.118706384311982 4.04724913476797
		0.124915162638833 4.01815752779065
		0.131175960734128 3.98340362457693
		0.137492900561555 3.94319440274307
		0.143870395501539 3.89772171226809
		0.15031316847636 3.84716767420497
		0.156826269504143 3.7917104436497
		0.163415092549276 3.73152995699844
		0.170085391527333 3.6668142615475
		0.176843294225396 3.59776588157642
		0.183695243742033 3.52458544273781
		0.190648504989686 3.44745003658496
		0.197711061061226 3.36649586006435
		0.204891465135448 3.2818288554279
		0.212198878363125 3.19354990387035
		0.219643105952836 3.10177631169991
		0.227234631116475 3.00665801564338
		0.234984646526918 2.9083931715889
		0.242905082951709 2.80724181959078
		0.251008634757101 2.70353569381136
		0.259308782025353 2.59768449254747
		0.26781980909268 2.49017806648475
		0.276556819392387 2.38157998195066
		0.28553574657311 2.27251568025393
		0.294773232298513 2.16356755467868
		0.304286602726894 2.05495520891865
		0.314093763722956 1.94718347678689
		0.324211868607484 1.84091133459446
		0.334657161855222 1.73684638047485
		0.345444903823858 1.63566434900921
		0.356589305587382 1.53795660064791
		0.368088385087482 1.4434530687533
		0.380002056326798 1.35244945485057
		};
		\addplot [ultra thick, DarkGreen]
		table {%
		-0.352554126400679 -1.40563987590398
		-0.340304861359129 -1.45054072280748
		-0.328453454564625 -1.50468305387275
		-0.316999266512588 -1.56826774960869
		-0.305936113573163 -1.64128714806167
		-0.295254588438315 -1.72343996990461
		-0.284942846301458 -1.81406831081989
		-0.274986681766248 -1.91213944239764
		-0.265369627325126 -2.01625974154143
		-0.25607307521478 -2.12474222790553
		-0.247076585928555 -2.23574473544966
		-0.238360273668691 -2.34732794231057
		-0.22990442155989 -2.45739361852864
		-0.221689885330229 -2.56377017858877
		-0.213698400191981 -2.66428581785151
		-0.205912595840857 -2.75684231591756
		-0.198316007708519 -2.8394855015834
		-0.190893083845035 -2.91046430679519
		-0.183629186936977 -2.96827204233075
		-0.176510591124369 -3.01166568110125
		-0.169524473454506 -3.03966273851885
		-0.162658899990588 -3.05151762750633
		-0.155903142549375 -3.04664977907978
		-0.149248336286477 -3.0241855012654
		-0.142686136152776 -2.98321442468865
		-0.136208740355325 -2.92316850762216
		-0.129808888928435 -2.84386633283524
		-0.123479846504725 -2.7455527717221
		-0.117215383287622 -2.62894000670678
		-0.111009754494758 -2.49525119850565
		-0.104857678541444 -2.34626784202989
		-0.0987543142212113 -2.18437899778139
		-0.0926952371180463 -2.0126246034722
		-0.0866764154542638 -1.83471053834367
		-0.0806941855409071 -1.65495126649294
		-0.0747452044922307 -1.47802473354561
		-0.0688261115053266 -1.30820362307185
		-0.0629338695262489 -1.14994834223552
		-0.0570656672872623 -1.00654070422811
		-0.0512188229005763 -0.879447372904271
		-0.0453907753346585 -0.768423554879671
		-0.0395790757682557 -0.672013757287545
		-0.0337813788513414 -0.588170985239484
		-0.0279954338978096 -0.514750401811739
		-0.0222190760303 -0.449794912486387
		-0.0164502172931969 -0.39164910665471
		-0.0106868377456718 -0.338971188881073
		-0.0049269765427426 -0.290698574905307
		0 -0.252258550206868
		0.000831297092852901 -0.24599961372528
		0.00658981239704753 -0.204226675930319
		0.0123504266147555 -0.164876158709064
		0.0181149659807234 -0.12755347616486
		0.0238853663577394 -0.0919126031279719
		0.0296635314709709 -0.057672288904573
		0.0354513689497203 -0.0246163002188779
		0.0412508218229409 0.00742557288448741
		0.0470638774664113 0.0385890806706703
		0.0528925765229316 0.0689838473663118
		0.058739021793362 0.09869946157095
		0.0646053870937403 0.127809980437099
		0.0704939260707614 0.156377298098971
		0.0764069809646697 0.184453692426136
		0.0823469913051393 0.212083762957405
		0.0883165025221053 0.239305908548024
		0.0943181795914769 0.26615276585203
		0.100354927085528 0.292633952998761
		0.10642974438242 0.318754174313269
		0.112545776629706 0.344519393907885
		0.118706384311982 0.369932155171976
		0.124915162638833 0.394991649073738
		0.131175960734128 0.419693712071854
		0.137492900561555 0.444030855088587
		0.143870395501539 0.467992264098732
		0.15031316847636 0.491563775024188
		0.156826269504143 0.514727800289891
		0.163415092549276 0.537463131107317
		0.170085391527333 0.559744830497982
		0.176843294225396 0.581544247905568
		0.183695243742033 0.60283606797206
		0.190648504989686 0.623587664085844
		0.197711061061226 0.643758105338181
		0.204891465135448 0.663305050393623
		0.212198878363125 0.682185525716736
		0.219643105952836 0.700355891177094
		0.227234631116475 0.717770184470209
		0.234984646526918 0.734379220301331
		0.242905082951709 0.750130465344008
		0.251008634757101 0.764968145435936
		0.259308782025353 0.778834198534711
		0.26781980909268 0.791670372622464
		0.276556819392387 0.803417751293694
		0.28553574657311 0.814013588578207
		0.294773232298513 0.823398227546956
		0.304286602726894 0.831555024523703
		0.314093763722956 0.838489646809036
		0.324211868607484 0.84418045916923
		0.334657161855222 0.84860857893651
		0.345444903823858 0.85175534469781
		0.356589305587382 0.85359359798887
		0.368088385087482 0.854006670505419
		0.380002056326798 0.852713193762398
		};
		\addplot [ultra thick, red]
		table {%
		-0.352554126400679 0.389811694830858
		-0.340304861359129 0.378993430788326
		-0.328453454564625 0.367768874388136
		-0.316999266512588 0.356513920935823
		-0.305936113573163 0.345667942659127
		-0.295254588438315 0.335678719819866
		-0.284942846301458 0.326968778073835
		-0.274986681766248 0.319914734682283
		-0.265369627325126 0.31456487647733
		-0.25607307521478 0.318436317452267
		-0.247076585928555 0.322727844964322
		-0.238360273668691 0.326970049765522
		-0.22990442155989 0.332329827715245
		-0.221689885330229 0.339098958115571
		-0.213698400191981 0.347072433553491
		-0.205912595840857 0.35658482663518
		-0.198316007708519 0.368215197169155
		-0.190893083845035 0.382707306485378
		-0.183629186936977 0.400808551352278
		-0.176510591124369 0.422587811333246
		-0.169524473454506 0.44779953337093
		-0.162658899990588 0.476356730604342
		-0.155903142549375 0.508392529479214
		-0.149248336286477 0.54409375392235
		-0.142686136152776 0.583701997177286
		-0.136208740355325 0.627502798747945
		-0.129808888928435 0.675784434183333
		-0.123479846504725 0.728694294778129
		-0.117215383287622 0.786415524649092
		-0.111009754494758 0.849247132209007
		-0.104857678541444 0.917763422854071
		-0.0987543142212113 0.992746210640416
		-0.0926952371180463 1.07501280165827
		-0.0866764154542638 1.16534771526998
		-0.0806941855409071 1.26440391788198
		-0.0747452044922307 1.37265430614944
		-0.0688261115053266 1.49060119846014
		-0.0629338695262489 1.61833964536878
		-0.0570656672872623 1.75552533020492
		-0.0512188229005763 1.90138651220923
		-0.0453907753346585 2.05474249444498
		-0.0395790757682557 2.21402488716329
		-0.0337813788513414 2.37733631458548
		-0.0279954338978096 2.54257855917965
		-0.0222190760303 2.70757893828613
		-0.0164502172931969 2.87019052210361
		-0.0106868377456718 3.0284038435473
		-0.0049269765427426 3.18041890948455
		0 3.30439221471522
		0.000831297092852901 3.32469110485566
		0.00658981239704753 3.45995772375611
		0.0123504266147555 3.58524312744331
		0.0181149659807234 3.6998597125142
		0.0238853663577394 3.80357055958597
		0.0296635314709709 3.89646552113287
		0.0354513689497203 3.9787243214499
		0.0412508218229409 4.05058682206795
		0.0470638774664113 4.11235067009456
		0.0528925765229316 4.16435586447028
		0.058739021793362 4.20697212777407
		0.0646053870937403 4.24060314842791
		0.0704939260707614 4.26568682319373
		0.0764069809646697 4.28267942654026
		0.0823469913051393 4.29206145707199
		0.0883165025221053 4.29433672822771
		0.0943181795914769 4.2900249351605
		0.100354927085528 4.27955243640621
		0.10642974438242 4.26318105247402
		0.112545776629706 4.24112880214907
		0.118706384311982 4.21357825227471
		0.124915162638833 4.18068897191113
		0.131175960734128 4.14264608250768
		0.137492900561555 4.09967702626428
		0.143870395501539 4.05200808646892
		0.15031316847636 3.9998680166219
		0.156826269504143 3.94343463850685
		0.163415092549276 3.88285320704571
		0.170085391527333 3.81827825334907
		0.176843294225396 3.74988107150902
		0.183695243742033 3.67782970016518
		0.190648504989686 3.60227807647509
		0.197711061061226 3.52340991256501
		0.204891465135448 3.44147953491348
		0.212198878363125 3.3567624975821
		0.219643105952836 3.26928371734089
		0.227234631116475 3.17880648025223
		0.234984646526918 3.08507630664704
		0.242905082951709 2.98790944709763
		0.251008634757101 2.88717506769671
		0.259308782025353 2.78308840232545
		0.26781980909268 2.67610900391315
		0.276556819392387 2.56623441242918
		0.28553574657311 2.45300981091851
		0.294773232298513 2.33712144699948
		0.304286602726894 2.22569727788652
		0.314093763722956 2.11617791030338
		0.324211868607484 2.00744672656908
		0.334657161855222 1.900239476252
		0.345444903823858 1.79527430367029
		0.356589305587382 1.69319314561037
		0.368088385087482 1.59357105272311
		0.380002056326798 1.49664024074119
		};
		\addplot [ultra thick, green]
		table {%
		-0.352554126400679 -1.66954030604003
		-0.340304861359129 -1.70328096002526
		-0.328453454564625 -1.74351125877177
		-0.316999266512588 -1.79107611801856
		-0.305936113573163 -1.84673075803329
		-0.295254588438315 -1.91096407235757
		-0.284942846301458 -1.98386781975759
		-0.274986681766248 -2.0650725323921
		-0.265369627325126 -2.15347248104949
		-0.25607307521478 -2.2549876534746
		-0.247076585928555 -2.35938071891729
		-0.238360273668691 -2.46417946649079
		-0.22990442155989 -2.56835571362654
		-0.221689885330229 -2.669985613347
		-0.213698400191981 -2.76665853103747
		-0.205912595840857 -2.85655818583599
		-0.198316007708519 -2.93825798414339
		-0.190893083845035 -3.01070109895242
		-0.183629186936977 -3.07308071777574
		-0.176510591124369 -3.12417624409116
		-0.169524473454506 -3.16268544349416
		-0.162658899990588 -3.18766209127879
		-0.155903142549375 -3.19849551818591
		-0.149248336286477 -3.19426877692506
		-0.142686136152776 -3.17404171035832
		-0.136208740355325 -3.1371916651202
		-0.129808888928435 -3.08340069017307
		-0.123479846504725 -3.01253015029481
		-0.117215383287622 -2.92481235773
		-0.111009754494758 -2.8209445574084
		-0.104857678541444 -2.70223838171857
		-0.0987543142212113 -2.57054875591085
		-0.0926952371180463 -2.42809622619062
		-0.0866764154542638 -2.2773891598638
		-0.0806941855409071 -2.12111316513927
		-0.0747452044922307 -1.96199496505699
		-0.0688261115053266 -1.80225985560384
		-0.0629338695262489 -1.64478239906746
		-0.0570656672872623 -1.49225113349932
		-0.0512188229005763 -1.34678251446992
		-0.0453907753346585 -1.20990989286497
		-0.0395790757682557 -1.08258066635235
		-0.0337813788513414 -0.965196093295766
		-0.0279954338978096 -0.857725227387621
		-0.0222190760303 -0.759821462942165
		-0.0164502172931969 -0.670916980400803
		-0.0106868377456718 -0.590332224962259
		-0.0049269765427426 -0.517348965751424
		0 -0.460401457341229
		0.000831297092852901 -0.45126003324172
		0.00658981239704753 -0.391399406122513
		0.0123504266147555 -0.337157430676598
		0.0181149659807234 -0.28798062292178
		0.0238853663577394 -0.243283245056384
		0.0296635314709709 -0.202523236994242
		0.0354513689497203 -0.165255290575052
		0.0412508218229409 -0.131101448785658
		0.0470638774664113 -0.0997473712886641
		0.0528925765229316 -0.070925233212253
		0.058739021793362 -0.0443990905836683
		0.0646053870937403 -0.0199669232100719
		0.0704939260707614 0.00254154901220283
		0.0764069809646697 0.0232831382009476
		0.0823469913051393 0.0423978786242549
		0.0883165025221053 0.0600123684301529
		0.0943181795914769 0.0762460841297063
		0.100354927085528 0.0911764630990631
		0.10642974438242 0.1048925536951
		0.112545776629706 0.117499491888916
		0.118706384311982 0.129106213269574
		0.124915162638833 0.139815485300266
		0.131175960734128 0.149679029281233
		0.137492900561555 0.158685193618294
		0.143870395501539 0.16680691886715
		0.15031316847636 0.17400331973641
		0.156826269504143 0.180278086409536
		0.163415092549276 0.185669131727035
		0.170085391527333 0.190214224661862
		0.176843294225396 0.193949617715705
		0.183695243742033 0.19692154619687
		0.190648504989686 0.199154971891912
		0.197711061061226 0.200596935295355
		0.204891465135448 0.201095830584475
		0.212198878363125 0.200472669833586
		0.219643105952836 0.198806964140279
		0.227234631116475 0.196464532214475
		0.234984646526918 0.19387468899531
		0.242905082951709 0.191457217786928
		0.251008634757101 0.189649779972109
		0.259308782025353 0.188626858762189
		0.26781980909268 0.188397451317593
		0.276556819392387 0.189492330979286
		0.28553574657311 0.192959295244283
		0.294773232298513 0.198707751389887
		0.304286602726894 0.199928668686962
		0.314093763722956 0.19946900151596
		0.324211868607484 0.199013731352754
		0.334657161855222 0.198471299158185
		0.345444903823858 0.197720135529217
		0.356589305587382 0.196601729501931
		0.368088385087482 0.194899961442204
		0.380002056326798 0.192212833058876
		};	
		\end{axis}
		
		\end{tikzpicture}

		\end{subfigure}
		\begin{subfigure}{.5\textwidth}
		\centering
		\begin{tikzpicture}

		\begin{axis}[
		tick align=outside,
		tick pos=left,
		x grid style={white!69.0196078431373!black},
		xlabel={$(l - l_c)$ / m},
		xmin=-0.41, xmax=0.41,
		xtick style={color=black},
		y grid style={white!69.0196078431373!black},
		ylabel={ $W /$ cm},
		ymin=2.20747789824422, ymax=9.76715201620405,
		ytick style={color=black},
		width=6cm,
		height=6cm,
		scale only axis,
		clip mode=individual
		]
		\addplot [semithick, black, dashed]
		table {%
		0 0
		0 12
		};
		\addplot [ultra thick, DarkRed]
		table {%
		-0.352554126400679 4.66585521033228
		-0.340304861359129 4.61474308691741
		-0.328453454564625 4.56564392060262
		-0.316999266512588 4.5185960512512
		-0.305936113573163 4.47360508428893
		-0.295254588438315 4.43066403401647
		-0.284942846301458 4.38975629407585
		-0.274986681766248 4.3508526975921
		-0.265369627325126 4.31390811187551
		-0.25607307521478 4.27885830376173
		-0.247076585928555 4.24561998422551
		-0.238360273668691 4.2141210258669
		-0.22990442155989 4.18428194611981
		-0.221689885330229 4.15601993836299
		-0.213698400191981 4.12925374729072
		-0.205912595840857 4.10390379433787
		-0.198316007708519 4.07989228598513
		-0.190893083845035 4.05714329229675
		-0.183629186936977 4.03558279527271
		-0.176510591124369 4.01513870946384
		-0.169524473454506 3.99574088191926
		-0.162658899990588 3.97732107776452
		-0.155903142549375 3.9598138598452
		-0.149248336286477 3.94316826552758
		-0.142686136152776 3.92734466262104
		-0.136208740355325 3.91230712320045
		-0.129808888928435 3.89802299886456
		-0.123479846504725 3.88446278661904
		-0.117215383287622 3.87160016420698
		-0.111009754494758 3.85941219000349
		-0.104857678541444 3.84787968274531
		-0.0987543142212113 3.83698781240918
		-0.0926952371180463 3.826726951343
		-0.0866764154542638 3.81709386062721
		-0.0806941855409071 3.80809332531958
		-0.0747452044922307 3.79974142656666
		-0.0688261115053266 3.79207006117706
		-0.0629338695262489 3.78509148359959
		-0.0570656672872623 3.7788197628798
		-0.0512188229005763 3.7732785098187
		-0.0453907753346585 3.76850213997889
		-0.0395790757682557 3.76453744478425
		-0.0337813788513414 3.76144550142936
		-0.0279954338978096 3.75930395848437
		-0.0222190760303 3.75820972093893
		-0.0164502172931969 3.75828201290272
		-0.0106868377456718 3.7596656937379
		-0.0049269765427426 3.76253450227908
		0 3.76632224402123
		0.000831297092852901 3.76709353766717
		0.00658981239704753 3.77357965949136
		0.0123504266147555 3.78225749330954
		0.0181149659807234 3.79341236787659
		0.0238853663577394 3.80741959870637
		0.0296635314709709 3.82474216655448
		0.0354513689497203 3.84589294987013
		0.0412508218229409 3.87142372476899
		0.0470638774664113 3.90189908235612
		0.0528925765229316 3.93785806430841
		0.058739021793362 3.97976805924963
		0.0646053870937403 4.0279808930317
		0.0704939260707614 4.0827039592501
		0.0764069809646697 4.14399669855108
		0.0823469913051393 4.21179503978194
		0.0883165025221053 4.28595776313373
		0.0943181795914769 4.36631268653196
		0.100354927085528 4.45243267498884
		0.10642974438242 4.54395251623145
		0.112545776629706 4.64060360771994
		0.118706384311982 4.74213088773565
		0.124915162638833 4.84829449414975
		0.131175960734128 4.95886833245672
		0.137492900561555 5.07363728523688
		0.143870395501539 5.19239431522969
		0.15031316847636 5.31493832128693
		0.156826269504143 5.44107332563578
		0.163415092549276 5.57060938253093
		0.170085391527333 5.70336547100905
		0.176843294225396 5.83917630433508
		0.183695243742033 5.97815698947696
		0.190648504989686 6.12028515158919
		0.197711061061226 6.26532086752525
		0.204891465135448 6.41302734668187
		0.212198878363125 6.56317187870655
		0.219643105952836 6.7155269614863
		0.227234631116475 6.86987128734322
		0.234984646526918 7.02599029164314
		0.242905082951709 7.18367600711563
		0.251008634757101 7.3427260369538
		0.259308782025353 7.50294155851484
		0.26781980909268 7.66412438542263
		0.276556819392387 7.82607324344293
		0.28553574657311 7.98857955719461
		0.294773232298513 8.15139073231049
		0.304286602726894 8.31411973919874
		0.314093763722956 8.47644137287535
		0.324211868607484 8.63803602522937
		0.334657161855222 8.79857680393306
		0.345444903823858 8.9577260253074
		0.356589305587382 9.1151320970258
		0.368088385087482 9.27024567723759
		0.380002056326798 9.42353046538769
		};
		\addplot [ultra thick, DarkGreen]
		table {%
		-0.352554126400679 4.6658547977855
		-0.340304861359129 4.58501539332213
		-0.328453454564625 4.50558735868723
		-0.316999266512588 4.42732870217269
		-0.305936113573163 4.34997310061683
		-0.295254588438315 4.27327152241942
		-0.284942846301458 4.19701266274978
		-0.274986681766248 4.12103262967941
		-0.265369627325126 4.0452226019194
		-0.25607307521478 3.96953324388002
		-0.247076585928555 3.89397975688266
		-0.238360273668691 3.81865190167675
		-0.22990442155989 3.74362369984351
		-0.221689885330229 3.66901378544214
		-0.213698400191981 3.59499140810126
		-0.205912595840857 3.52176552959653
		-0.198316007708519 3.44957357546144
		-0.190893083845035 3.37867043622625
		-0.183629186936977 3.30931828957558
		-0.176510591124369 3.24177772500937
		-0.169524473454506 3.17630051782166
		-0.162658899990588 3.11312424776639
		-0.155903142549375 3.05236323994152
		-0.149248336286477 2.9941364659759
		-0.142686136152776 2.93881607320373
		-0.136208740355325 2.88667845685041
		-0.129808888928435 2.83791951338438
		-0.123479846504725 2.79267656382531
		-0.117215383287622 2.75104709297023
		-0.111009754494758 2.71310462716017
		-0.104857678541444 2.67891222415416
		-0.0987543142212113 2.64853411982586
		-0.0926952371180463 2.62204610395932
		-0.0866764154542638 2.59954520961736
		-0.0806941855409071 2.58115931808366
		-0.0747452044922307 2.56706794851718
		-0.0688261115053266 2.55747965173569
		-0.0629338695262489 2.5522241108013
		-0.0570656672872623 2.55109944906057
		-0.0512188229005763 2.55394791265971
		-0.0453907753346585 2.56064583479797
		-0.0395790757682557 2.57109455907137
		-0.0337813788513414 2.58521167664508
		-0.0279954338978096 2.60292198422549
		-0.0222190760303 2.62414758959071
		-0.0164502172931969 2.64879659999516
		-0.0106868377456718 2.67674985607349
		-0.0049269765427426 2.70784526741363
		0 2.73677926038934
		0.000831297092852901 2.74185954197154
		0.00658981239704753 2.77848759419333
		0.0123504266147555 2.81732081470074
		0.0181149659807234 2.85783519333638
		0.0238853663577394 2.89952841282066
		0.0296635314709709 2.94191160018186
		0.0354513689497203 2.98443458924336
		0.0412508218229409 3.02650253967752
		0.0470638774664113 3.06750238223838
		0.0528925765229316 3.1068427134149
		0.058739021793362 3.14400284981032
		0.0646053870937403 3.17858137671745
		0.0704939260707614 3.21033161081503
		0.0764069809646697 3.23917391521028
		0.0823469913051393 3.26518246349254
		0.0883165025221053 3.28855265238298
		0.0943181795914769 3.30955717405592
		0.100354927085528 3.32844772095109
		0.10642974438242 3.34552285768418
		0.112545776629706 3.3610901912574
		0.118706384311982 3.37542963359479
		0.124915162638833 3.38879044019685
		0.131175960734128 3.40139173111066
		0.137492900561555 3.41342492752855
		0.143870395501539 3.42505704315661
		0.15031316847636 3.43643418241231
		0.156826269504143 3.44768488991001
		0.163415092549276 3.45892318284202
		0.170085391527333 3.47025120655282
		0.176843294225396 3.48176152680155
		0.183695243742033 3.49354110151094
		0.190648504989686 3.50567046499199
		0.197711061061226 3.51822331954107
		0.204891465135448 3.53126966851487
		0.212198878363125 3.54487758045674
		0.219643105952836 3.55911469972653
		0.227234631116475 3.5740495634328
		0.234984646526918 3.58975276426283
		0.242905082951709 3.60629799889959
		0.251008634757101 3.62376303546524
		0.259308782025353 3.64223062280842
		0.26781980909268 3.66178935931389
		0.276556819392387 3.68253454452066
		0.28553574657311 3.70456901908374
		0.294773232298513 3.72800562136673
		0.304286602726894 3.75297849977292
		0.314093763722956 3.77963147511002
		0.324211868607484 3.8081123973554
		0.334657161855222 3.83857438188195
		0.345444903823858 3.87117697401784
		0.356589305587382 3.90608719292457
		0.368088385087482 3.94342207900122
		0.380002056326798 3.98349479662373
		};
		\end{axis}
		
		\end{tikzpicture}

		\end{subfigure}

		\caption{MAST test scenario, with launch parameters $R_{b,x} = R_{b,y} = -1.79\textrm{m}$, $W_{x} = W_{y} = 9.47\textrm{cm}$. 
		Evolution of curvatures $1 / R_b$ (left, dark green and dark red, the two colours corresponding to two different directions) and widths (right) along the beam, calculated from the non-zero eigenvalues of the real and imaginary parts of $\bm{\Psi}_w$ respectively. The effective curvatures $1 / R_{b,M}$ (left, bright green and bright red) are calculated from the non-zero eigenvalues of the real part of $\bm{M}_w$. We plot the curvatures $1 / R_b$ and $1 / R_{b,M}$ rather than the radii of curvature $R_b$ and $R_{b,M}$ to avoid a divergence. Note that the directions of the eigenvectors corresponding to $W$, $1 / R_b$, and $1 / R_{b, M}$ are not aligned with one another (nor with $\hat{\mathbf{x}}$ or $\hat{\mathbf{y}}$) in general. 
		Even at the point the beam enters the plasma (when these plots start), the principal curvatures are not the equal to each other, despite us launching a circular beam, because we have applied the vacuum-plasma boundary conditions, as described in \ref{appendix_vacuum_plasma_interface}. Recall that $l-l_c$ is the distance along the central ray from the cut-off location.}
		\label{fig:widths_curvatures}
	\end{figure}

	Despite the intuition that considering beams in vacuum may give us, the evolution of $\bm{\Psi}_{w}$ is ultimately more complicated in a plasma. This is one reason why using beam tracing is important. In Figure \ref{fig:widths_curvatures}, we see that the beam in the plasma can indeed be quite different from what one would expect from vacuum propagation; the beam curvature goes to zero at two points, and neither of them are where the beam widths are at a minimum. Thus, the notion of a beam waist, where the widths are at a minimum and the wavefront curvatures are zero simultaneously, is not properly achieved to begin with. This insight is a demonstration of the strength of our model.

	\section{Conclusion}
	We have successfully derived the full analytical form of the linear backscattered signal from DBS measurements by combining the beam-tracing equations with the reciprocity theorem in general geometry. This is the first analytical model to self-consistently account for signal localisation (spatial resolution), wavenumber resolution, and mismatch attenuation, given in equations (\ref{eq:localisation}), (\ref{eq:delta_k_perp_2}), and (\ref{eq:delta_theta_m}), respectively. In particular, we find that it is the curvature of the field lines and the magnetic shear, rather than the curvature of the cut-off surface, that is important for the calculation of these quantities. This is a result of the magnetic curvature and shear modifying the effective beam curvature, as shown in equations (\ref{eq:Mxx_simplified}) and (\ref{eq:Mxy_simplified}). We also show that the localisation and one of the two components of wavenumber resolution are inextricably intertwined, via the Bragg condition.
	
	To calculate these quantities, one needs to solve the beam tracing equations (\ref{eq:dq_dtau_maintext}), (\ref{eq:dK_dtau_maintext}), and (\ref{eq:dPsi_dtau_maintext}), evolving the probe beam as it propagates through the plasma. Beam tracing requires the following additional input parameters: the equilibrium density, magnetic field, and potentially temperature. Beam-tracing simulations are swift, which makes the beam model suitable for intershot analysis of DBS data and for large parameter sweeps. Since beam tracing solves for the next-order corrections to ray tracing, it should not be too complicated to upgrade existing ray-tracing codes to solve for the beam properties and use them in post-processing for our model.

	Using our model, one can now correct for the effect of mismatch, since we have found, for the first time, the quantitative analytical dependence of attenuation on the mismatch angle. This is vital for DBS measurements of spherical tokamak plasmas, where the pitch angle is large and varies both spatially and temporally, making optimisation for all channels at all times impossible. Furthermore, we find analytical evidence that mismatch attenuation is also important in conventional tokamaks, especially for high wavenumber measurements. Being able to calculate the quantitative effect of mismatch enables one to relax the criterion that the probe wavevector has to be exactly perpendicular to the magnetic field. As long as the mismatch angle is not so large that the signal is completely attenuated, the mismatch attenuation can be quantitatively corrected with our model. This allows one to run DBS studies with fewer repeated shots by optimising the launch angles such that there is useful data on as many channels as possible, to glean new physical insights from legacy data, and to potentially design simpler DBS systems with smaller ranges of launch angles or without 2D steering.
	
	We derived our model in two limits: the small-mismatch and large-mismatch orderings. Conventionally, DBS is operated in the small-mismatch regime, where the signal does indeed come from around the cut-off region. In the large-mismatch regime, we focus on the specific case where the mismatch angle is zero at at least one point along the beam path, and find that the signal comes from that point, which in general will not be at the cut-off. This provides an alternative method of getting a localised signal, via a different physical mechanism. We show that in cases where the mismatch angle is moderately large, the small-mismatch and large-mismatch formulations give the same results, see Section \ref{section_backscattered_E_ST}. In such situations, the same formulation can be used for both cases.

	Finally, our model provides the basis for synthetic diagnostics, a quick and realistic physical method to calculate the expected DBS signal given particular realisations of turbulence from gyrokinetic simulations. Depending on how much physics one intends to keep, one could use equation (\ref{eq:A_r2_final_cleaned}), (\ref{eq:A_r2_explicit_wavenumber_resolution}), or even (\ref{eq:A_r_after_wx_wy_integration}).
	
	\ack {This work has been funded by the RCUK Energy Programme [EP/T012250/1] and the Engineering and Physical Sciences Research Council (EPSRC) [EP/R034737/1]. In order to obtain further information on the data and models underlying this work, please contact PublicationsManager@ukaea.uk. VH Hall-Chen's DPhil is funded by a National Science Scholarship from A*STAR, Singapore. VH Hall-Chen would like to thank E Poli and NA Crocker for assistance with the TORBEAM code, E Lock, PG Ivanov, MR Hardman, NE Bricknell, and S Dash for general help with Python and Fortran, AM Hall-Chen for proofreading, and J Ruiz Ruiz, SJ Freethy, M Barnes, DC Speirs, RGL Vann, NA Crocker, K Ronald, P Shi, D Farina, and P Hennequin for helpful feedback. VH Hall-Chen is grateful to PG Ivanov for wittily suggesting that the beam tracing code be christened `Scotty'.} 


	

	\appendix

%
%
%
%
%

	\section{Derivation of beam tracing} \setcounter{section}{1} \label{appendix_beam_tracing}

	In beam tracing, we seek to find the electric field of the Gaussian probe beam as it propagates through the plasma. Our derivation is an alternative to that of Pereverzev's \cite{Pereverzev:Beam_tracing:1992, Pereverzev:Beam_tracing:1993, Pereverzev:Beam_Tracing:1996, Pereverzev:Beam_tracing:1998}; the results are equivalent, but ours is specific to our choice of coordinates, and is thus more convenient for our work on Doppler backscattering. Unlike Pereverzev, our derivation does not require some of the intermediate steps in Section III of his work \cite{Pereverzev:Beam_tracing:1998}.	
	
	By directly using the ansatz for the probe beam's electric field, equation (\ref{eq:beam_ansatz}), and the expansion of its amplitude in equation (\ref{eq:amplitude}), we get
	\begin{equation}
	\eqalign{
		\mathbf{E}
		= \left[ \mathbf{A}^{(0)}(\tau) + \mathbf{A}^{(1)}(\tau, \mathbf{w}) + \mathbf{A}^{(2)}(\tau, \mathbf{w}) \right]
		\exp\left( \rmi \psi \right) .
	}
	\end{equation}
	We take $\mathbf{A}^{(1)}$ and $\mathbf{A}^{(2)}$, the higher order amplitudes, to be linear and bilinear in $\mathbf{w}$, that is
	\begin{equation}
		\mathbf{A}^{(1)}(\tau,\mathbf{w}) = \mathbf{w} \cdot \nabla_w \mathbf{A}^{(1)} (\tau),
	\end{equation}
	and
	\begin{equation}
		\mathbf{A}^{(2)}(\tau,\mathbf{w}) = \frac{1}{2} \mathbf{w} \mathbf{w} : \nabla_w \nabla_w \mathbf{A}^{(2)}(\tau) + \mathbf{A}^{(2)} (\tau,\mathbf{w} = 0) .
	\end{equation}
	This simplified dependence on the intermediate length scale $\mathbf{w}$ allows us to neglect $\nabla \nabla \mathbf{A}^{(1)}$ and $\nabla \nabla \nabla \mathbf{A}^{(2)}$ compared to $\nabla \nabla \mathbf{A}^{(0)}(\tau)$ and $\nabla \nabla \nabla \mathbf{A}^{(0)}(\tau)$, respectively.

	Since we are looking for a solution to equation (\ref{eq:Maxwell_eq}), we need to evaluate the derivatives of $\mathbf{E}$,
	\begin{equation} \label{eq:Maxwell_index}
	\eqalign{
		\fl
		\frac{ \partial^2 \mathbf{E} }{ \partial r_\mu \partial r_\nu}
		=  \Bigg[ 
			- \underbrace{ \frac{ \partial \psi }{ \partial r_\mu} \frac{ \partial \psi }{ \partial r_\nu} \mathbf{A}}_{\sim \lambda^{-2} \mathbf{A}}
			+ \underbrace{\rmi \frac{ \partial \psi }{ \partial r_\mu} \frac{ \partial \mathbf{A} }{ \partial r_\nu} 
			+ \rmi \frac{ \partial^2 \psi }{ \partial r_\mu \partial r_\nu}  \mathbf{A} 
			+ \rmi \frac{ \partial \mathbf{A} }{ \partial r_\mu} \frac{ \partial \psi }{ \partial r_\nu}}_{\sim \lambda^{-1} L^{-1} \mathbf{A}}
			+ \underbrace{\frac{ \partial^2 \mathbf{A} }{ \partial r_\mu \partial r_\nu}}_{\sim L^{-2} \mathbf{A}} 
		\Bigg] \exp(\rmi \psi)  . }
	\end{equation}
	The first order derivatives of $\psi$, where $\psi$ is given in equation (\ref{eq:phase}), are
	\begin{equation} \label{eq:grad_psi}
		\nabla \psi 
		= \frac{\rmd s}{\rmd \tau}\nabla \tau+ \mathbf{K}_w 
		+ \frac{\rmd \mathbf{K}_w}{\rmd \tau} \cdot \mathbf{w} \nabla \tau 
		+ \bm{\Psi}_{w}  \cdot \mathbf{w} 
		+ \frac{1}{2}  \mathbf{w} \cdot \frac{\rmd \bm{\Psi}_{w}}{\rmd \tau} \cdot \mathbf{w} \nabla \tau .
	\end{equation}
	Here we have used $\mathbf{w} = [\mathbf{r} - \mathbf{q}(\tau)]_w$, and noted that since $\rmd \mathbf{q} / \rmd \tau = \mathbf{g}$, we get $\mathbf{K}_w \cdot \rmd \mathbf{q} / \rmd \tau = 0$ and $\rmd \mathbf{q} / \rmd \tau \cdot \bm{\Psi}_w \cdot \mathbf{w} = 0$. In order to proceed, we need to know the form of $\nabla \tau$. We do this by using the reciprocal vector
	\begin{equation}
	\eqalign{
		\nabla \tau
		&=
		\left(
		\frac{\partial \mathbf{r}}{\partial w_x}
		\times \frac{\partial \mathbf{r}}{\partial w_y}
		\right)
		\left[
		\frac{\partial \mathbf{r}}{\partial \tau} \cdot
		\left(\frac{\partial \mathbf{r}}{\partial w_x}
		\times \frac{\partial \mathbf{r}}{\partial w_y} \right)
		\right]^{-1}  .
	}
	\end{equation}
	Using $\rmd \hat{\mathbf{x}} / \rmd \tau \cdot \hat{\mathbf{g}} = - \hat{\mathbf{x}} \cdot \rmd \hat{\mathbf{g}} / \rmd \tau$ and $\rmd \hat{\mathbf{y}} / \rmd \tau \cdot \hat{\mathbf{g}} = - \hat{\mathbf{y}} \cdot \rmd \hat{\mathbf{g}} / \rmd \tau$, we find that
	\begin{equation} \label{eq:grad_tau_intermediate}
		\eqalign{\nabla \tau &= \hat{\mathbf{g}} \left[g - \mathbf{w} \cdot \frac{\rmd \hat{\mathbf{g}}}{\rmd \tau} \right]^{-1} .}
	\end{equation}
	Note that $g^{-1} \rmd \hat{\mathbf{g}} / \rmd \tau = \rmd \hat{\mathbf{g}} / \rmd l = \boldsymbol{\kappa}$, where $\boldsymbol{\kappa} \sim L^{-1}$ is the curvature of the central ray and $l$ is arc length along the ray. This ray curvature should not be confused with the wavefront curvature. Using the ray curvature $\boldsymbol{\kappa}$ and equation (\ref{eq:grad_tau_intermediate}), 
	\begin{equation} \label{eq:grad_tau}
		\nabla \tau
		= \frac{ \hat{\mathbf{g}} }{ g (1 - \boldsymbol{\kappa} \cdot \mathbf{w}) } .
	\end{equation}
	We note that $\boldsymbol{\kappa} \cdot \mathbf{w} \sim W^{} / L \ll 1$ and that we can re-express $\nabla \tau$ as
	\begin{equation} \label{eq:grad_tau_expanded}
		\nabla \tau = \frac{\hat{\mathbf{g}}}{g} [1 + \boldsymbol{\kappa} \cdot \mathbf{w} + (\boldsymbol{\kappa} \cdot \mathbf{w})^2 + \ldots] .
	\end{equation}			
	Substituting equation (\ref{eq:grad_tau_expanded}) into the equation for $\nabla \psi$, equation (\ref{eq:grad_psi}), and separating the terms by order, we get 
	\begin{equation} \label{eq:grad_psi_ordering}
		\nabla \psi =
		\underbrace{(\nabla \psi)^{(0)}}_{\sim 1^{} / \lambda} +
		\underbrace{(\nabla \psi)^{(1)}}_{\sim 1^{} / W} +
		\underbrace{(\nabla \psi)^{(2)}}_{\sim 1^{} / L} +
		\ldots ,
	\end{equation}
	where
	\begin{equation} \label{eq:grad_psi_ordered}
	\fl
	\eqalign{
		(\nabla \psi)^{(0)} &= K_g \hat{\mathbf{g}} + \mathbf{K}_w = \mathbf{K} , \\
		(\nabla \psi)^{(1)} &= K_g \hat{\mathbf{g}} (\boldsymbol{\kappa} \cdot \mathbf{w}) + \frac{\hat{\mathbf{g}}}{g} \frac{\rmd \mathbf{K}_w}{\rmd \tau} \cdot \mathbf{w} + \bm{\Psi}_w \cdot \mathbf{w} , \\
		\textrm{and} \\
		(\nabla \psi)^{(2)} &= K_g \hat{\mathbf{g}} (\boldsymbol{\kappa} \cdot \mathbf{w})^2 + \frac{\hat{\mathbf{g}}}{g} (\boldsymbol{\kappa} \cdot \mathbf{w}) \frac{\rmd \mathbf{K}_w}{\rmd \tau} \cdot \mathbf{w} + \frac{1}{2} \frac{\hat{\mathbf{g}}}{g}  \mathbf{w} \cdot \frac{\rmd \bm{\Psi}_w}{\rmd \tau} \cdot \mathbf{w} .}
	\end{equation}
	Here we have defined the wavenumber parallel to the central ray to be $K_g = g^{-1} \rmd s / \rmd \tau = \rmd s / \rmd l$, and the total wavevector to be $\mathbf{K} = K_g \hat{\mathbf{g}} + \mathbf{K}_w$. 
	
	We will need $\nabla \nabla \psi$ only to lowest order. Note that the gradient of $\psi$ can only have two possible length scales, $W$ and $L$. Consequently, to evaluate $\nabla \nabla \psi$ to leading order, we need to find the gradients of both $(\nabla \psi)^{(0)}$ and $(\nabla \psi)^{(1)}$, but not $(\nabla \psi)^{(2)}$. Hence, we have
	\begin{equation} \label{eq:grad_grad_psi}
	\eqalign{\nabla \nabla \psi
		& \simeq \nabla \left[\mathbf{K} + K_g \hat{\mathbf{g}} (\boldsymbol{\kappa} \cdot \mathbf{w}) + \frac{\hat{\mathbf{g}}}{g} \frac{\rmd \mathbf{K}_w}{\rmd \tau} \cdot \mathbf{w} + \bm{\Psi}_w \cdot \mathbf{w} \right] \\
		& \simeq \bm{\Psi}_w + \frac{\hat{\mathbf{g}}}{g} \frac{\rmd \mathbf{K}}{\rmd \tau} + K_g \boldsymbol{\kappa} \hat{\mathbf{g}} + \left( \frac{\rmd \mathbf{K}_w}{\rmd \tau} \right)_w \frac{\hat{\mathbf{g}}}{g} ,}
	\end{equation}
	where all the terms are of order $(\lambda L)^{-1} \sim W^{-2}$.

	We substitute equation (\ref{eq:beam_ansatz}) into equation (\ref{eq:Maxwell_eq}), and remember the sizes of various terms as shown in equation (\ref{eq:Maxwell_index}). We perform a Taylor expansion of the dielectric tensor,
	\begin{equation} \label{eq:dielectric_tensor_expansion}
		\bm{\epsilon} (\mathbf{r})
		= \bm{\epsilon} (\mathbf{q} + \mathbf{w})
		\simeq \bm{\epsilon}(\mathbf{q})
		+ \mathbf{w} \cdot \nabla \bm{\epsilon}(\mathbf{q})
		+ \frac{1}{2} \mathbf{w} \mathbf{w} : \nabla \nabla \bm{\epsilon}(\mathbf{q}) .
	\end{equation}

	\subsection{Zeroth order} \label{appendix_beam_tracing_zeroth_order}
	To lowest order $\nabla \psi \simeq K_g \hat{\mathbf{g}} + \mathbf{K}_w = \mathbf{K} \sim \lambda^{-1}$ and $\mathbf{A} \simeq \mathbf{A}^{(0)}$, giving
	\begin{equation}
		\left[\frac{c^2}{\Omega^2} (\mathbf{K}\mathbf{K} - K^2 \bm{1}) +\bm{\epsilon} \right] \cdot \mathbf{A}^{(0)}
		= 0 .
	\end{equation}
	We do not need to keep higher order terms because $(c^2 / \Omega^{2}) \mathbf{K} \times (\mathbf{K} \times \mathbf{E}) \sim (c^2 K^2 / \Omega^2) E \sim N^2 E$ and $\bm{\epsilon} \cdot \mathbf{E} \sim N^2 E$, where $N$ is the refractive index. For convenience, we use the notation
	\begin{equation} \label{eq:D}
		\bm{D} (\mathbf{q}, \mathbf{K})
		= \frac{c^2}{\Omega^2} (\mathbf{K}\mathbf{K} - K^2 \bm{1}) +\bm{\epsilon} (\mathbf{q}),
	\end{equation}
	leading to
	\begin{equation} 
		\bm{D} \cdot \mathbf{A}^{(0)} 
		= 0 .
	\end{equation} 
	Since $\bm{D}$ is Hermitian, it can be diagonalised --- there exist three vectors $\hat{\mathbf{e}} (\mathbf{r},\mathbf{K})$ such that $\bm{D} \cdot \hat{\mathbf{e}} = H \hat{\mathbf{e}}$. To solve equation (\ref{eq:D}), $K_g (\tau)$ must be such that one of the three eigenvalues $H ( \mathbf{q} (\tau), \mathbf{K} (\tau) )$ vanishes, that is, $H = 0$ will give $K_g (\tau)$ once $\mathbf{K}_w (\tau)$ and $\mathbf{q} (\tau)$ are known. To obtain the equations for $\mathbf{K}_w (\tau)$ and $\mathbf{q} (\tau)$, we need to go to first order in the expansion $\lambda / W \sim W / L$.

	In general, only one of the eigenvalues $H$ goes to zero. The vector $\mathbf{A}^{(0)}$ has to be parallel to the $\hat{\mathbf{e}}$ that corresponds to $H=0$, $\mathbf{A}^{(0)} = A^{(0)} \hat{\mathbf{e}}$.

	\subsection{First order} \label{appendix_beam_tracing_first_order}
	In this subsection, we get contributions to equation (\ref{eq:Maxwell_eq}) that are of first order in $\lambda^{} / W \sim W^{} / L \ll 1$. The terms come from $\nabla \psi$ in equation (\ref{eq:grad_psi}), the next order correction to the amplitude $\mathbf{A}^{(1)}$, and the expansion of $\bm{\epsilon}$ in equation (\ref{eq:dielectric_tensor_expansion}),
	\begin{equation} \label{eq:K_firstorder} 
		(\nabla \psi)^{(1)} \cdot \nabla_K \bm{D} \cdot \hat{\mathbf{e}} A^{(0)}
		+ \bm{D} \cdot \mathbf{A}^{(1)}
		+ \mathbf{w} \cdot \nabla \bm{\epsilon} (\mathbf{q}) \cdot \hat{\mathbf{e}} A^{(0)}
		= 0.
	\end{equation}
	Here we have used
	\begin{equation} \label{eq:d_D_d_K}
	\eqalign{
		\frac{\partial D_{\alpha \beta} }{\partial K_\mu}
		&= \frac{c^2}{\Omega^2} \frac{\partial}{\partial K_\mu} (K_\alpha K_\beta - K^2 \delta_{\alpha \beta}) \\
		&= 
		\frac{c^2}{\Omega^2} \left[
			K_\beta \delta_{\alpha \mu} 
			+ K_\alpha \delta_{\beta \mu} 
			- 2 K_\mu \delta_{\alpha \beta}
		\right]  ,
	}
	\end{equation}
	where $\delta_{\cdot \cdot}$ are Kronecker deltas. Substituting the expression for $(\nabla \psi)^{(1)}$ given in equation (\ref{eq:grad_psi_ordered}) and realising that $\nabla \bm{\epsilon} = \nabla \bm{D}$, we get
	\begin{equation} \label{eq:firstorder_expanded}
	\eqalign{
		& \left(K_g \hat{\mathbf{g}} (\boldsymbol{\kappa} \cdot \mathbf{w})
		+ \frac{\hat{\mathbf{g}}}{g} \frac{d \mathbf{K}_w}{d \tau} \cdot \mathbf{w}
		+ \bm{\Psi}_w \cdot \mathbf{w} \right) \cdot \nabla_K \bm{D} \cdot \hat{\mathbf{e}} A^{(0)} \\
		& \qquad 
		+ \bm{D} \cdot \mathbf{A}^{(1)}
		+ \mathbf{w} \cdot \nabla \bm{D} \cdot \hat{\mathbf{e}} A^{(0)} = 0 .
	}
	\end{equation} 
	Since $\bm{D}$ is Hermitian, $\hat{\mathbf{e}}^* \cdot \bm{D} = 0$. Thus, we multiply equation (\ref{eq:firstorder_expanded}) by $\hat{\mathbf{e}}^*$ to eliminate $\bm{D} \cdot \mathbf{A}^{(1)}$,
	\begin{equation} \label{eq:firstorder_estar}
	\fl
		\hat{\mathbf{e}}^* \cdot
		\left[
			\left(
				K_g \hat{\mathbf{g}} (\boldsymbol{\kappa} \cdot \mathbf{w})
				+ \frac{\hat{\mathbf{g}}}{g} \frac{\rmd \mathbf{K}_w}{\rmd \tau} \cdot \mathbf{w}
				+ \bm{\Psi}_w \cdot \mathbf{w}
			\right)
			\cdot \nabla_K \bm{D}
			+ \mathbf{w} \cdot \nabla \bm{\epsilon} (\mathbf{q})
		\right]
		\cdot \hat{\mathbf{e}}= 0 .
	\end{equation}
	This equation can be simplified by using derivatives of the dispersion relation $H = \hat{\mathbf{e}}^* \cdot \bm{D} \cdot \hat{\mathbf{e}}$. Since $\bm{D} \cdot \hat{\mathbf{e}} = 0 = \hat{\mathbf{e}}^* \cdot \bm{D}$, we find that $\partial_\mu H = \hat{\mathbf{e}}^* \cdot \partial_\mu \bm{D} \cdot \hat{\mathbf{e}}$, where $\partial_\mu$ is shorthand for either $\partial / \partial K_\mu$ or $\partial / \partial r_\mu$. 
	Hence, equation (\ref{eq:firstorder_expanded}) can be rewritten as
	\begin{equation}
	\left( K_g \hat{\mathbf{g}} (\boldsymbol{\kappa} \cdot \mathbf{w} ) + \frac{\hat{\mathbf{g}}}{g} \frac{\rmd \mathbf{K}_w}{\rmd \tau} \cdot \mathbf{w} + \bm{\Psi}_w \cdot \mathbf{w} \right) \cdot \nabla_K H + \mathbf{w} \cdot \nabla H= 0 .
	\end{equation}
	We note that the curvature $\boldsymbol{\kappa}$ is always in the same plane as $\mathbf{w}$ (since it is always normal to the group velocity), and so is the matrix $\bm{\Psi}_w$ (since we have chosen it to be so). Since equation (\ref{eq:firstorder_estar}) is satisfied for all possible values of $\mathbf{w}$, we obtain
	\begin{equation}
		\left[
			K_g \boldsymbol{\kappa} + \frac{1}{g} \left(\frac{\rmd \mathbf{K}_w}{\rmd \tau} \right)_{w}
		\right] \hat{\mathbf{g}} \cdot \nabla_K H
		+ \bm{\Psi}_w \cdot \nabla_K H
		+ \nabla_w H = 0 .
	\end{equation} 
	Note that $H$ is real and $\bm{\Psi}_w$ is complex. There are four equations at play here; there are two directions 
	in the plane perpendicular to $\hat{\mathbf{g}}$,
	and since $\bm{\Psi}_{w}$ is complex, both the real and imaginary parts in either of these directions must be satisfied. Two of these equations are used to determine $\rmd \mathbf{K}_{w} / \rmd \tau$, and the other two are used to determine $\hat{\mathbf{g}}$, the direction of $\mathbf{g} = \rmd \mathbf{q} / \rmd \tau$. The magnitude of $\mathbf{g}$ is not determined by these equations, since we do not have enough of them; instead we choose it shortly. Since we want $\mathbf{K}$ to be real, we have to enforce $\textrm{Im}(\bm{\Psi}_w) \cdot \nabla_K H = 0$, which will lead to $\nabla_K H \propto \hat{\mathbf{g}}$, since $\bm{\Psi}_w$ is perpendicular to the group velocity. This is the key step --- we select the group velocity such that it is in the direction of the central ray, the only direction in which there is no decay of the electric field. Since $\hat{\mathbf{g}}$ is now a unit vector in the direction of $\nabla_K H$, we get
	\begin{equation} \label{eq:firstorder_semifinal}
		\left[
			K_g \boldsymbol{\kappa}
			+ \frac{1}{g} \left(\frac{\rmd \mathbf{K}_w}{\rmd \tau} \right)_w
		\right] |\nabla_K H|
		= - \nabla_w H
		= - \nabla H
		+ \hat{\mathbf{g}} \hat{\mathbf{g}} \cdot \nabla H.
	\end{equation}
	We obtain $\hat{\mathbf{g}} \cdot \nabla H$ from taking the derivative of the equation for $K_g$, $H = 0$, with respect to $\tau$. Since $\rmd H / \rmd \tau = 0$, we obtain
	\begin{equation}
	\eqalign{
		\frac{\rmd \mathbf{K}}{\rmd \tau} \cdot \nabla_K H + \frac{\rmd \mathbf{q}}{\rmd \tau} \cdot \nabla H = 0 , }
	\end{equation}
	which can be rewritten as
	\begin{equation} \label{eq:gradg_H}
	\eqalign{
		\frac{1}{g} \frac{\rmd \mathbf{K}}{\rmd \tau} \cdot \nabla_K H = - \hat{\mathbf{g}} \cdot \nabla H . }
	\end{equation}
	Substituting equation (\ref{eq:gradg_H}) into equation (\ref{eq:firstorder_semifinal}), we get
	\begin{equation} \label{eq:firstorder_absgradH}
	\left[K_g \boldsymbol{\kappa} + \frac{1}{g} \left(\frac{\rmd \mathbf{K}_w}{\rmd \tau} \right)_w \right] |\nabla_K H| + \frac{\hat{\mathbf{g}}}{g} \frac{\rmd \mathbf{K}}{\rmd \tau} \cdot \hat{\mathbf{g}} |\nabla_K H| = - \nabla H .
	\end{equation}
	We then note that
	\begin{equation} \label{eq:dK_dtau_w}
	\eqalign{
		\left(\frac{\rmd \mathbf{K}}{\rmd \tau} \right)_w
		= \left(\frac{\rmd \mathbf{K}_w}{\rmd \tau} \right)_w +  K_g \left(\frac{\rmd \hat{\mathbf{g}}}{\rmd \tau} \right)_w
		= \left(\frac{\rmd \mathbf{K}_w}{\rmd \tau} \right)_w +  K_g g \boldsymbol{\kappa} . }
	\end{equation}
	Consequently, we get
	\begin{equation}
		\frac{|\nabla_K H|}{g} \frac{\rmd \mathbf{K}}{\rmd \tau}  = - \nabla H .
	\end{equation}
	We then choose $g = |\nabla_K H|$. This gives us
	\begin{equation} \label{eq:group_velocity}
		\frac{\rmd \mathbf{q}}{\rmd \tau} = g \hat{\mathbf{g}} = |\nabla_K H| \hat{\mathbf{g}} = \nabla_K H .
	\end{equation}
	Hence, we have
	\begin{equation} \label{eq:dK_dtau}
		\frac{\rmd \mathbf{K}}{\rmd \tau} = - \nabla H.
	\end{equation}
	Note that we could have chosen any other prescription for $g$, and it would only have modified the definition of the free parameter $\tau$. The choice $g = |\nabla_K H|$ emphasises the Hamiltonian character of the equations.

	Finally, we can use equations (\ref{eq:firstorder_expanded}) and (\ref{eq:firstorder_semifinal}) to solve for $\mathbf{A}^{(1)}$,
	\begin{equation} \label{eq:firstorder_simplified}
	\fl
	\eqalign{
		\bm{D} \cdot \mathbf{A}^{(1)}
		= - \left[\left(
				-\frac{\hat{\mathbf{g}} \mathbf{w} \cdot \nabla H}{g}  + \mathbf{w} \cdot  \bm{\Psi}_w 
			\right)\cdot \nabla_K \bm{D} \cdot \hat{\mathbf{e}} + \mathbf{w} \cdot \nabla \bm{D} \cdot \hat{\mathbf{e}} 
		\right] A^{(0)} . 
	}
	\end{equation}
	We can write the solution to equation (\ref{eq:firstorder_simplified}) in terms of the derivatives of $\hat{\mathbf{e}} (\mathbf{r},\mathbf{K})$ Taking any derivative of equation (\ref{eq:H_and_e_definition}), we get
	\begin{equation} \label{eq:deriv_eigenvalue_equation}
	\eqalign{
		\partial_\mu \bm{D} \cdot \hat{\mathbf{e}} + \bm{D} \cdot \partial_\mu \hat{\mathbf{e}}
		&= (\partial_\mu H) \hat{\mathbf{e}},
	}
	\end{equation}
	where we have used $H=0$. The projection of equation (\ref{eq:deriv_eigenvalue_equation}) on $\hat{\mathbf{e}}^*$ gives
	\begin{equation} \label{eq:partial_H}
		\partial_\mu H = \hat{\mathbf{e}}^* \cdot \partial_\mu \bm{D} \cdot \hat{\mathbf{e}} .
	\end{equation}
	The projection perpendicular to $\hat{\mathbf{e}}^*$ gives
	\begin{equation} \label{eq:shift_derivative}
		(\bm{1} - \hat{\mathbf{e}} \hat{\mathbf{e}}^*) \cdot \partial_\mu \bm{D} \cdot \hat{\mathbf{e}} =- \bm{D} \cdot \partial_\mu \hat{\mathbf{e}}.
	\end{equation}
	This equation only gives the component of $\partial_\mu \hat{\mathbf{e}}$ perpendicular to $\hat{\mathbf{e}}$. The component of $\partial_\mu \hat{\mathbf{e}}$ parallel to $\hat{\mathbf{e}}$ is a free choice that is partially constrained by the condition $\hat{\mathbf{e}} \cdot \hat{\mathbf{e}}^* = 1$, equation (\ref{eq:size_of_e}). Indeed, even with this condition, $\hat{\mathbf{e}}$ is defined only up to a phase factor $\alpha$,
	\begin{equation}
		\hat{\mathbf{e}} \rightarrow \hat{\mathbf{e}} \exp(\rmi \alpha).
	\end{equation}
	Hence, one can always add a vector parallel to $\hat{\mathbf{e}}$ to $\partial_\mu \hat{\mathbf{e}}$,
	\begin{equation}
		\partial_\mu \hat{\mathbf{e}} \rightarrow \partial_\mu \hat{\mathbf{e}} + \rmi \hat{\mathbf{e}}^{} \partial_\mu \alpha.
	\end{equation}
	This result shows that the condition $\hat{\mathbf{e}} \cdot \hat{\mathbf{e}}^* = 1$ constrains the component of $\partial_\mu \hat{\mathbf{e}}$ along $\hat{\mathbf{e}}$ to be purely imaginary. To summarise, while $\partial_\mu \hat{\mathbf{e}} \cdot (\bm{1} - \hat{\mathbf{e}}^* \hat{\mathbf{e}})$ is uniquely determined by the beam tracing equations and is thus physical, $\partial_\mu \hat{\mathbf{e}} \cdot \hat{\mathbf{e}}^* \hat{\mathbf{e}}$ can be chosen at will and it  does not have any particular physical meaning.
	Using the result in equation (\ref{eq:firstorder_simplified}), and noticing that equation (\ref{eq:firstorder_simplified}) is perpendicular to $\hat{\mathbf{e}}^*$ when equations (\ref{eq:group_velocity}) and (\ref{eq:dK_dtau}) are satisfied, we find
	\begin{equation} \label{eq:firstorder_final}
	\eqalign{
		\mathbf{A}^{(1)}
		= \left[ \left(-\mathbf{w} \cdot \nabla_w H \frac{\hat{\mathbf{g}} }{g}  + \mathbf{w} \cdot  \bm{\Psi}_w \right)\cdot \nabla_K \hat{\mathbf{e}} + \mathbf{w} \cdot \nabla \hat{\mathbf{e}} \right]  A^{(0)} . }
	\end{equation}

	\subsection{Second order}
	To get the contributions to equation (\ref{eq:Maxwell_eq}) that are second order in $\lambda^{} / W \sim W^{} / L \ll 1$, we need to evaluate $\nabla \mathbf{A}$,
	\begin{equation} \label{eq:grad_A}
	\eqalign{
		\nabla \mathbf{A}
		\simeq \frac{d}{d \tau} (A^{(0)} \hat{\mathbf{e}})  \nabla \tau
		+ \nabla_w \mathbf{A}^{(1)} ,}
	\end{equation}
	where we have noted that $\nabla_w \mathbf{A}^{(1)} \sim \nabla (A^{(0)} \hat{\mathbf{e}})$. To lowest order, $\nabla \tau \simeq \hat{\mathbf{g}}^{} / {g}$. To find $\nabla_w \mathbf{A}^{(1)}$, we take the derivative of equation (\ref{eq:firstorder_final}),
	\begin{equation}
		\nabla_w \mathbf{A}^{(1)} = \left[- \nabla_w H \frac{\hat{\mathbf{g}}}{g} \cdot \nabla_K \hat{\mathbf{e}} + \bm{\Psi}_w \cdot \nabla_K \hat{\mathbf{e}} + \nabla_w \hat{\mathbf{e}} \right] A^{(0)} .
	\end{equation}

	It is convenient to define a new tensor $\bm{\Psi}$ that contains $\bm{\Psi}_w$. The tensor $\bm{\Psi}$ is the lowest order result for $\nabla \nabla \psi$, given in equation (\ref{eq:grad_grad_psi}),
	\begin{equation} \label{eq:Psi_definition}
		\bm{\Psi} = \nabla \nabla \psi .
	\end{equation}
	With this definition of $\bm{\Psi}$, we find
	\begin{equation} \label{eq:constraint}
		\bm{\Psi} \cdot \nabla_K H
		= \frac{\rmd \mathbf{K}}{\rmd \tau}
		= - \nabla H ,
	\end{equation}
	\begin{equation} \label{eq:grad_A1}
		\nabla_w \mathbf{A}^{(1)}
		= A^{(0)} (\bm{\Psi} \cdot \nabla_K + \nabla )_w \hat{\mathbf{e}} ,
	\end{equation}
	and
	\begin{equation} \label{eq:grad_psi1}
		(\nabla \psi)^{(1)}
		= \bm{\Psi} \cdot \mathbf{w} .
	\end{equation}

	Using equations (\ref{eq:Maxwell_index}), (\ref{eq:grad_psi_ordering}), (\ref{eq:dielectric_tensor_expansion}), (\ref{eq:grad_A}), and (\ref{eq:Psi_definition}), we find the second order contribution to (\ref{eq:Maxwell_eq}),
	\begin{equation} \label{eq:second_order}
	\fl
	\eqalign{
		&\bm{D} \cdot \mathbf{A}^{(2)}
		+ (\nabla \psi)^{(1)} \cdot \nabla_K \bm{D} \cdot \mathbf{A}^{(1)} \\
		&\qquad + (\nabla \psi)^{(2)} \cdot \nabla_K \bm{D} \cdot \hat{\mathbf{e}}  A^{(0)}
		+ \frac{1}{2} (\nabla \psi)^{(1)} (\nabla \psi)^{(1)} : \nabla_K \nabla_K \bm{D} \cdot \hat{\mathbf{e}} A^{(0)} \\
		&\qquad + \mathbf{w} \cdot \nabla \bm{\epsilon} \cdot \mathbf{A}^{(1)}
		+ \frac{1}{2} \mathbf{w} \mathbf{w} : \nabla \nabla \bm{\epsilon} \cdot \hat{\mathbf{e}} A^{(0)} \\
		&\qquad - \rmi \frac{c^2}{\Omega^2} \left[\frac{\hat{\mathbf{g}}}{g} \frac{\rmd}{\rmd \tau} \left(A^{(0)} \hat{\mathbf{e}} \right) + \nabla_w \mathbf{A}^{(1)} \right] \cdot (\nabla \psi)^{(0)} \\
		&\qquad + 2 \rmi \frac{c^2}{\Omega^2} (\nabla \psi)^{(0)} \cdot \left[\frac{\hat{\mathbf{g}}}{g} \frac{\rmd}{\rmd \tau} \left(A^{(0)} \hat{\mathbf{e}} \right) + \nabla_w \mathbf{A}^{(1)} \right] \\
		&\qquad - \rmi \frac{c^2}{\Omega^2} (\nabla \psi)^{(0)} \left[\frac{\hat{\mathbf{g}}}{g} \cdot \frac{\rmd}{\rmd \tau} \left(A^{(0)} \hat{\mathbf{e}} \right) + \nabla_w \cdot \mathbf{A}^{(1)} \right] \\
		&\qquad - \rmi \frac{c^2}{\Omega^2} A^{(0)} \bm{\Psi} \cdot \hat{\mathbf{e}}
		+ \rmi \frac{c^2}{\Omega^2} A^{(0)} \hat{\mathbf{e}}^{} \bm{\Psi} : \bm{1}
		= 0 .
	}
	\end{equation}
	To eliminate $\bm{D} \cdot \mathbf{A}^{(2)}$ from the equation, we contract the free index with $\hat{\mathbf{e}}^*$.
	We remark that some of the terms depend on $\mathbf{w}$ while the others do not. Since the equation must be valid regardless of the particular value of $\mathbf{w}$, we can separate it into two separate independent equations, one of which depends on $\mathbf{w}$, and the other having no $\mathbf{w}$ dependence. The equations that we derive from the piece of equation (\ref{eq:second_order}) that depends on $\mathbf{w}$ will determine the evolution of $\bm{\Psi}_{w}$, whereas the piece independent of $\mathbf{w}$ gives the equation for $A^{(0)}$. Note that the components of $\bm{\Psi}$ that are not in $\bm{\Psi}_{w}$ are determined by equation (\ref{eq:constraint}).

	\subsubsection{Pieces proportional to $\mathbf{w} \mathbf{w}$.}
	Using equations (\ref{eq:grad_psi_ordered}), (\ref{eq:firstorder_final}), (\ref{eq:grad_A1}), and (\ref{eq:grad_psi1}), the terms proportional to $\mathbf{w} \mathbf{w}$ in equation (\ref{eq:second_order}) give
	\begin{equation} \label{eq:ww_piece}
	\fl
	\eqalign{
		& \frac{1}{2g} \left[
			\mathbf{w} \cdot \frac{\rmd \bm{\Psi}_w }{\rmd \tau} \cdot \mathbf{w}
			+ 2 g K_g (\boldsymbol{\kappa} \cdot \mathbf{w})^2
			+ 2 (\boldsymbol{\kappa} \cdot \mathbf{w}) \frac{\rmd \mathbf{K}_w}{\rmd \tau} \cdot \mathbf{w}
		\right]
		\left(\hat{g}_\mu \hat{\mathbf{e}}^* \cdot \frac{\partial  \bm{D}}{\partial K_\mu} \cdot \hat{\mathbf{e}} \right) \\
		& \qquad + \textrm{w}_\mu \Psi_{\mu \nu} \hat{\mathbf{e}}^* \cdot \frac{\partial  \bm{D}}{\partial K_\nu} \cdot \left(
			\textrm{w}_\alpha \Psi_{\alpha \beta} \frac{\partial \hat{\mathbf{e}}} {\partial K_\beta}
			+ \textrm{w}_\beta \frac{\partial \hat{\mathbf{e}} }{\partial r_\beta}
		\right) \\
		& \qquad + \frac{1}{2}  (\textrm{w}_\mu \Psi_{\mu \nu}) (\textrm{w}_\alpha \Psi_{\alpha \beta}) \hat{\mathbf{e}}^* \cdot \frac{\partial^2 \bm{D}}{\partial K_\nu K_\beta} \cdot \hat{\mathbf{e}}\\
		& \qquad + \textrm{w}_\nu  \hat{\mathbf{e}}^* \cdot \frac{\partial \bm{D}}{\partial r_\nu} \cdot \left(
			\textrm{w}_\alpha \Psi_{\alpha \beta} \frac{\partial \hat{\mathbf{e}}} {\partial K_\beta}
			+ \textrm{w}_\beta \frac{\partial \hat{\mathbf{e}} }{\partial r_\beta}
		\right) \\
		& \qquad + \frac{1}{2} \textrm{w}_\nu \textrm{w}_\beta \hat{\mathbf{e}}^* \cdot \frac{\partial^2 \bm{D}}{\partial r_\nu r_\beta} \cdot \hat{\mathbf{e}}
		= 0 .
	}
	\end{equation}
	We now use equations (\ref{eq:dK_dtau_w}), (\ref{eq:Psi_definition}), and (\ref{eq:grad_grad_psi}) to write
	\begin{equation} \label{eq:dPsiw_dtau}
	\fl
	\eqalign{
		\mathbf{w} \cdot \frac{\rmd \bm{\Psi}_w }{\rmd \tau} \cdot \mathbf{w}
		= \mathbf{w} \cdot \frac{\rmd \bm{\Psi} }{\rmd \tau} \cdot \mathbf{w}
		- 2 g K_g (\boldsymbol{\kappa} \cdot \mathbf{w})^2
		- 2 (\boldsymbol{\kappa} \cdot \mathbf{w}) \frac{\rmd \mathbf{K}_w}{\rmd \tau} \cdot \mathbf{w} .
	}
	\end{equation}
	We then substitute equation (\ref{eq:dPsiw_dtau}) into equation (\ref{eq:ww_piece}) and use
	\begin{equation}
		\hat{g}_\mu \hat{\mathbf{e}}^* \cdot \frac{\partial  \bm{D}}{\partial K_\mu} \cdot \hat{\mathbf{e}}
		= \hat{\mathbf{g}} \cdot \nabla_K H
		= g 
	\end{equation}
	to get
	\begin{equation}
	\fl
	\eqalign{
		& \mathbf{w} \cdot \frac{\rmd \bm{\Psi} }{\rmd \tau} \cdot \mathbf{w} \\
		&  + 2 \textrm{w}_\mu \Psi_{\mu \nu} \hat{\mathbf{e}}^* \cdot \frac{\partial  \bm{D}}{\partial K_\nu} \cdot \left(
			\textrm{w}_\alpha \Psi_{\alpha \beta} \frac{\partial \hat{\mathbf{e}}} {\partial K_\beta}
			+ \textrm{w}_\beta \frac{\partial \hat{\mathbf{e}} }{\partial r_\beta}
		\right)
		+ (\textrm{w}_\mu \Psi_{\mu \nu}) (\textrm{w}_\alpha \Psi_{\alpha \beta}) \hat{\mathbf{e}}^* \cdot \frac{\partial^2 \bm{D}}{\partial K_\nu K_\beta} \cdot \hat{\mathbf{e}}\\
		&+ 2 \textrm{w}_\nu \hat{\mathbf{e}}^* \cdot \frac{\partial \bm{D}}{\partial r_\nu} \cdot \left(
			\textrm{w}_\alpha \Psi_{\alpha \beta} \frac{\partial \hat{\mathbf{e}}} {\partial K_\beta}
			+ \textrm{w}_\beta  \frac{\partial \hat{\mathbf{e}} }{\partial r_\beta}
		\right)
		+ \textrm{w}_\nu \textrm{w}_\beta \hat{\mathbf{e}}^* \cdot \frac{\partial^2 \bm{D}}{\partial r_\nu \partial r_\beta} \cdot \hat{\mathbf{e}}
		= 0 .
	}
	\end{equation}
	This expression can be rewritten in terms of derivatives of the dispersion relation $H$. Differentiating $\bm{D} \cdot \hat{\mathbf{e}} = H \hat{\mathbf{e}}$, we find
	\begin{equation} \label{eq:derivatives_of_H}
	\eqalign{
		& \partial_\mu \partial_\nu \bm{D} \cdot \hat{\mathbf{e}}
		+ \partial_\nu \bm{D} \cdot \partial_\mu \hat{\mathbf{e}}
		+ \partial_\mu \bm{D} \cdot \partial_\nu \hat{\mathbf{e}}
		+ \bm{D} \cdot \partial_\mu \partial_\nu \hat{\mathbf{e}} \\
		& \qquad = (\partial_\mu \partial_\nu H) \hat{\mathbf{e}}
		+(\partial_\nu H) \partial_\mu \hat{\mathbf{e}}
		+ (\partial_\mu H) \partial_\nu \hat{\mathbf{e}}
		+ H \partial_\mu \partial_\nu \hat{\mathbf{e}} .
	}
	\end{equation}
	Contracting the free index in this equation with $\hat{\mathbf{e}}^*$, we get
	\begin{equation} \label{eq:derivatives_of_H_contracted}
	\eqalign{
		& \hat{\mathbf{e}}^* \cdot \partial_\mu \partial_\nu \bm{D} \cdot \hat{\mathbf{e}}
		+ \hat{\mathbf{e}}^* \cdot \partial_\nu \bm{D} \cdot \partial_\mu \hat{\mathbf{e}}
		+ \hat{\mathbf{e}}^* \cdot \partial_\mu \bm{D} \cdot \partial_\nu \hat{\mathbf{e}} \\
		& \qquad = \partial_\mu \partial_\nu H
		+(\partial_\nu H) \partial_\mu \hat{\mathbf{e}} \cdot \hat{\mathbf{e}}^*
		+ (\partial_\mu H) \partial_\nu \hat{\mathbf{e}} \cdot \hat{\mathbf{e}}^* ,
	}
	\end{equation} 
	where we have used $H=0$ and $\hat{\mathbf{e}}^* \cdot \bm{D} = 0$. Using these results, employing the fact that the equation must hold true for arbitrary $\mathbf{w}$, and remembering that $\bm{\Psi}$ is symmetric, we find that
	\begin{equation}
	\fl
	\eqalign{
		& \left( 
			\frac{\rmd \bm{\Psi} }{\rmd \tau}
			+ \nabla \nabla H
			+ \bm{\Psi} \cdot \nabla_K \nabla H
			+ \nabla \nabla_K H \cdot \bm{\Psi}
			+ \bm{\Psi} \cdot \nabla_K \nabla_K H \cdot \bm{\Psi} 
		\right)_w \\
		& \qquad + \left[
			\left( \bm{\Psi} \cdot \nabla_K H \right) \left( \bm{\Psi} \cdot \nabla_K \hat{\mathbf{e}} \cdot \hat{\mathbf{e}}^* \right)
			+ \left( \bm{\Psi} \cdot \nabla_K \hat{\mathbf{e}} \cdot \hat{\mathbf{e}}^* \right) \left( \bm{\Psi} \cdot \nabla_K H \right)
		\right]_w\\
		& \qquad + \left[ \nabla H (\nabla \hat{\mathbf{e}} \cdot \hat{\mathbf{e}}^* )
		+ (\nabla \hat{\mathbf{e}} \cdot \hat{\mathbf{e}}^* ) \nabla H \right]_w \\
		& \qquad + \left[ \bm{\Psi} \cdot \nabla_K H (\nabla \hat{\mathbf{e}} \cdot \hat{\mathbf{e}}^* )
		+ (\bm{\Psi} \cdot \nabla_K \hat{\mathbf{e}} \cdot \hat{\mathbf{e}}^* ) \nabla H \right]_w \\
		& \qquad + \left[ \nabla H (\bm{\Psi} \cdot \nabla_K \hat{\mathbf{e}} \cdot \hat{\mathbf{e}}^* )
		+ (\nabla \hat{\mathbf{e}} \cdot \hat{\mathbf{e}}^* ) \bm{\Psi} \cdot \nabla_K H \right]_w
		= 0 .
	}
	\end{equation}
	Here we use the notation $\bm{C}_w = (\bm{1} - \hat{\mathbf{g}} \hat{\mathbf{g}}) \cdot \bm{C} \cdot (\bm{1} - \hat{\mathbf{g}} \hat{\mathbf{g}})$, where $\bm{C}$ is an arbitrary 3D matrix. We use equation (\ref{eq:constraint}) to get
	\begin{equation} \label{eq:dPsi_dtau_w}
	\fl
	\eqalign{
		\left(
		\frac{\rmd \bm{\Psi}}{\rmd \tau}
		+ \bm{\Psi} \cdot \nabla_K \nabla_K H \cdot \bm{\Psi}
		+ \bm{\Psi} \cdot \nabla_K \nabla H
		+ \nabla \nabla_K H \cdot \bm{\Psi}
		+ \nabla \nabla H
		\right)_w
		= 0 ,
	}
	\end{equation}
	which gives the components of $\rmd \bm{\Psi}^{} / \rmd \tau$ perpendicular to the beam. To obtain the components parallel to the beam, $\rmd \bm{\Psi}^{} / \rmd \tau \cdot \hat{\mathbf{g}}$, we differentiate equation (\ref{eq:constraint}) with respect to $\tau$,
	\begin{equation} \label{eq:Psi_dot_g} 
	\fl
		\frac{\rmd \bm{\Psi}}{\rmd \tau} \cdot \hat{\mathbf{g}}
		+ ( \bm{\Psi} \cdot \nabla_K \nabla_K H \cdot \bm{\Psi}
		+ \bm{\Psi} \cdot \nabla_K \nabla H
		+ \nabla \nabla_K H \cdot \bm{\Psi}
		+ \nabla \nabla H ) \cdot \hat{\mathbf{g}}
		= 0 ,
	\end{equation}
	where we have used $\rmd (\partial_\mu H)^{} / \rmd \tau = \mathbf{g} \cdot \nabla (\partial_\mu H) + \rmd \mathbf{K}^{} / \rmd \tau \cdot \nabla_K (\partial_\mu H) $, equations (\ref{eq:group_velocity}), (\ref{eq:dK_dtau}), and (\ref{eq:constraint}), as well as the symmetry of $\bm{\Psi}$. Recalling that $(\rmd \bm{\Psi}^{} / \rmd \tau)_w = (\bm{1} - \hat{\mathbf{g}} \hat{\mathbf{g}} ) \cdot (\rmd \bm{\Psi}^{} / \rmd \tau) \cdot (\bm{1} - \hat{\mathbf{g}} \hat{\mathbf{g}} )$, equations (\ref{eq:dPsi_dtau_w}) and (\ref{eq:Psi_dot_g}) give
	\begin{equation} \label{eq:Psi_evolution}
	\eqalign{
		\frac{\rmd \bm{\Psi}}{\rmd \tau}
		+ \bm{\Psi} \cdot \nabla_K \nabla_K H \cdot \bm{\Psi}
		+ \bm{\Psi} \cdot \nabla_K \nabla H
		+ \nabla \nabla_K H \cdot \bm{\Psi}
		+ \nabla \nabla H
		= 0 .
	}
	\end{equation}

	\subsubsection{Pieces independent of $\mathbf{w}$.}

	Using equations (\ref{eq:grad_psi_ordered}) and (\ref{eq:grad_A1}), the terms independent of $\mathbf{w}$ in equation (\ref{eq:second_order}) give
	\begin{equation} \label{eq:nonww_piece}
	\fl
	\eqalign{
		& \frac{1}{g A^{(0)}} \left[
			(\mathbf{K} \cdot \hat{\mathbf{e}}) \hat{\mathbf{e}}^*
			+ (\mathbf{K} \cdot \hat{\mathbf{e}}^*) \hat{\mathbf{e}}
			- 2 \mathbf{K}
		\right] \cdot \hat{\mathbf{g}} \frac{\rmd A^{(0)}}{\rmd \tau} \\
		& \qquad + \bm{\Psi} :(\hat{\mathbf{e}}^* \hat{\mathbf{e}} - \bm{1})
		+ \frac{1}{g} \left[
			(\hat{\mathbf{e}}^* \cdot \hat{\mathbf{g}}) \frac{\rmd \hat{\mathbf{e}}}{\rmd \tau} \cdot \mathbf{K}
			+ (\mathbf{K} \cdot \hat{\mathbf{e}}^*) \hat{\mathbf{g}} \cdot \frac{\rmd \hat{\mathbf{e}}}{\rmd \tau}
			- 2 (\mathbf{K} \cdot \hat{\mathbf{g}}) \frac{\rmd \hat{\mathbf{e}}}{\rmd \tau} \cdot \hat{\mathbf{e}}^*
		\right] \\
		& \qquad + \hat{\mathbf{e}}^* \cdot (\bm{1} - \hat{\mathbf{g}} \hat{\mathbf{g}}) \cdot (\bm{\Psi} \cdot \nabla_K \hat{\mathbf{e}} + \nabla \hat{\mathbf{e}} ) \cdot \mathbf{K} \\
		& \qquad + (\mathbf{K} \cdot \hat{\mathbf{e}}^*) (\bm{1} - \hat{\mathbf{g}} \hat{\mathbf{g}}) : (\bm{\Psi} \cdot \nabla_K \hat{\mathbf{e}} + \nabla \hat{\mathbf{e}} )  \\
		& \qquad -2 \mathbf{K} \cdot (\bm{1} - \hat{\mathbf{g}} \hat{\mathbf{g}}) \cdot (\bm{\Psi} \cdot \nabla_K \hat{\mathbf{e}} + \nabla \hat{\mathbf{e}} ) \cdot \hat{\mathbf{e}}^*
		= 0 .
	}
	\end{equation}
	Noting that equation (\ref{eq:constraint}) can be used to write
	\begin{equation} \label{eq:de_dtau}
	\eqalign{
		g \hat{\mathbf{g}} \cdot (\bm{\Psi} \cdot \nabla_K \hat{\mathbf{e}} + \nabla \hat{\mathbf{e}} )
		&= \frac{\rmd \mathbf{K}}{\rmd \tau} \cdot \nabla_K \hat{\mathbf{e}}
		+ \frac{\rmd \mathbf{q}}{\rmd \tau} \cdot \nabla \hat{\mathbf{e}}
		= \frac{\rmd \hat{\mathbf{e}}}{\rmd \tau} ,
	}
	\end{equation}
	we find that
	\begin{equation} \label{eq:to_simplify_the_nonww_piece} 
	\fl
	\eqalign{
		&\frac{1}{g} \left[
			(\hat{\mathbf{e}}^* \cdot \hat{\mathbf{g}}) \frac{\rmd \hat{\mathbf{e}}}{\rmd \tau} \cdot \mathbf{K}
			+ (\mathbf{K} \cdot \hat{\mathbf{e}}^*) \hat{\mathbf{g}} \cdot \frac{\rmd \hat{\mathbf{e}}}{\rmd \tau}
			- 2 (\mathbf{K} \cdot \hat{\mathbf{g}}) \frac{\rmd \hat{\mathbf{e}}}{\rmd \tau} \cdot \hat{\mathbf{e}}^*
		\right] \\
		& \qquad =  (\hat{\mathbf{e}}^* \cdot \hat{\mathbf{g}}) \hat{\mathbf{g}} \cdot (\bm{\Psi} \cdot \nabla_K \hat{\mathbf{e}} + \nabla \hat{\mathbf{e}}) \cdot \mathbf{K}
		+ (\mathbf{K} \cdot \hat{\mathbf{e}}^*)  \hat{\mathbf{g}} \hat{\mathbf{g}} : (\bm{\Psi} \cdot \nabla_K \hat{\mathbf{e}} + \nabla \hat{\mathbf{e}}) \\
		& \qquad - 2 (\mathbf{K} \cdot \hat{\mathbf{g}}) \hat{\mathbf{g}} \cdot (\bm{\Psi} \cdot \nabla_K \hat{\mathbf{e}} + \nabla \hat{\mathbf{e}}) \cdot \hat{\mathbf{e}}^*. }
	\end{equation}
	Using equations (\ref{eq:d_D_d_K}), (\ref{eq:group_velocity}), and (\ref{eq:to_simplify_the_nonww_piece}), equation (\ref{eq:nonww_piece}) becomes
	\begin{equation} \label{eq:nonww_piece_simplified}
	\fl
	\eqalign{
		& \frac{\Omega^2}{c^2} \frac{\rmd \ln A^{(0)}}{\rmd \tau}
		+\bm{\Psi} : (\hat{\mathbf{e}}^* \hat{\mathbf{e}} - \bm{1})
		+ \hat{\mathbf{e}}^* \cdot (\bm{\Psi} \cdot \nabla_K \hat{\mathbf{e}} + \nabla \hat{\mathbf{e}} ) \cdot \mathbf{K} \\
		& \qquad + (\mathbf{K} \cdot \hat{\mathbf{e}}^*) (\bm{\Psi} : \nabla_K \hat{\mathbf{e}} + \nabla \cdot \hat{\mathbf{e}} )
		-2 \mathbf{K} \cdot (\bm{\Psi} \cdot \nabla_K \hat{\mathbf{e}} + \nabla \hat{\mathbf{e}} ) \cdot \hat{\mathbf{e}}^*
		= 0  .
	}
	\end{equation}
	Noting that
	\begin{equation} \label{eq:estar_dot_gradk_D_expanded}
	\fl
	\eqalign{
		\hat{\mathbf{e}}^* \cdot \frac{\partial \bm{D}}{\partial K_\mu} \cdot \partial_\nu \hat{\mathbf{e}}
		= \frac{c^2}{\Omega^2} (\mathbf{K} \cdot \hat{\mathbf{e}}^*) \partial_\nu \hat{e}_\mu
		+ \frac{c^2}{\Omega^2} (\mathbf{K} \cdot \partial_\nu \hat{\mathbf{e}}) \hat{e}^*_\mu
		- 2 \frac{c^2}{\Omega^2} (\hat{\mathbf{e}}^* \cdot \partial_\nu \hat{\mathbf{e}}) K_\mu,
	}
	\end{equation}
	\begin{equation} \label{eq:gradk_gradk_D_expanded}
	\eqalign{
		\hat{\mathbf{e}}^* \cdot \frac{\partial^2 \bm{D}}{\partial K_\mu \partial K_\nu} \cdot \hat{\mathbf{e}}
		= \frac{c^2}{\Omega^2} \hat{e}^*_\mu \hat{e}_\nu 
		+ \frac{c^2}{\Omega^2} \hat{e}_\mu \hat{e}^*_\nu 
		- \frac{c^2}{\Omega^2} 2 \delta_{\mu \nu} ,
	}
	\end{equation}
	and that $\bm{\Psi}$ is symmetric, we can simplify equation (\ref{eq:nonww_piece_simplified}) further, getting
	\begin{equation} \label{eq:nonww_piece_further_simplified}
	\eqalign{
		& \frac{\rmd \ln A^{(0)}}{\rmd \tau}
		+ \Psi_{\mu \nu} \left(\frac{1}{2} \hat{\mathbf{e}}^* \cdot \frac{\partial^2 \bm{D}}{\partial K_\mu \partial K_\nu} \cdot \hat{\mathbf{e}}
		+ \hat{\mathbf{e}}^* \cdot \frac{\partial \bm{D}}{\partial K_\mu} \cdot \frac{\partial \hat{\mathbf{e}}}{\partial K_\nu} \right) \\
		& \qquad + \hat{\mathbf{e}}^* \cdot \frac{\partial \bm{D}}{\partial K_\mu} \cdot \frac{\partial \hat{\mathbf{e}}}{\partial r_\mu}
		= 0  .
	}
	\end{equation}
	Using equation (\ref{eq:derivatives_of_H_contracted}), equation (\ref{eq:nonww_piece_further_simplified}) simplifies to
	\begin{equation}
	\fl
	\eqalign{
		\frac{\rmd \ln A^{(0)}}{\rmd \tau}
		+ \frac{1}{2} \bm{\Psi} : \nabla_K \nabla_K H
		+ \nabla_K H \cdot \bm{\Psi} \cdot \nabla_K \hat{\mathbf{e}} \cdot \hat{\mathbf{e}}^*
		+ \hat{\mathbf{e}}^* \cdot \frac{\partial \bm{D}}{\partial K_\mu} \cdot \frac{\partial \hat{\mathbf{e}}}{\partial r_\mu}
		= 0  .
	}
	\end{equation}
	We then use equation (\ref{eq:constraint}), getting
	\begin{equation} \label{eq:amplitude_pre_conjugate}
	\fl
	\eqalign{
		\frac{\rmd \ln A^{(0)}}{\rmd \tau}
		+ \frac{1}{2} \bm{\Psi} : \nabla_K \nabla_K H
		+ \frac{\rmd \mathbf{K}}{\rmd \tau} \cdot \nabla_K \hat{\mathbf{e}} \cdot \hat{\mathbf{e}}^*
		+ \hat{\mathbf{e}}^* \cdot \frac{\partial \bm{D}}{\partial K_\mu} \cdot \frac{\partial \hat{\mathbf{e}}}{\partial r_\mu}
		= 0 .
	}
	\end{equation}

	To solve for $A^{(0)}$ using equation (\ref{eq:amplitude_pre_conjugate}), we write $A^{(0)}$ as $A^{(0)} = |A^{(0)}| \exp (\rmi \phi^{(0)})$. The real part of equation (\ref{eq:amplitude_pre_conjugate}) gives the equation for $|A^{(0)}|$, and its imaginary part the equation for $\phi^{(0)}$. The real part of equation (\ref{eq:amplitude_pre_conjugate}) is
	\begin{equation}
	\fl
	\eqalign{
		\frac{\rmd \ln |A^{(0)}|}{\rmd \tau}
		+ \frac{1}{2} \textrm{Re}(\bm{\Psi}) : \nabla_K \nabla_K H
		+ \frac{1}{2} \left( \hat{\mathbf{e}}^* \cdot \frac{\partial \bm{D}}{\partial K_\mu} \cdot \frac{\partial \hat{\mathbf{e}}}{\partial r_\mu}
		+ \frac{\partial \hat{\mathbf{e}}^*}{\partial r_\mu} \cdot \frac{\partial \bm{D}}{\partial K_\mu} \cdot \hat{\mathbf{e}} \right)
		= 0  ,
	}
	\end{equation}
	where we have used $\hat{\mathbf{e}} \cdot \partial \hat{\mathbf{e}}^*  / \partial K_\mu + \hat{\mathbf{e}}^* \cdot \partial \hat{\mathbf{e}}^{}  / \partial K_\mu = \partial (\hat{\mathbf{e}}^* \cdot \hat{\mathbf{e}} ) / \partial K_\mu = 0$, the fact that $\partial \bm{D}^{}  / \partial K_\mu$ is Hermitian, and that only $\bm{\Psi}$ and $\hat{\mathbf{e}}$ are complex. We then differentiate $H = \hat{\mathbf{e}}^* \cdot \bm{D} \cdot \hat{\mathbf{e}}$ and use $\partial^2 \bm{D} / \partial K_\mu \partial r_\nu = 0$ and equation (\ref{eq:deriv_eigenvalue_equation}) to obtain
	\begin{equation}
	\eqalign{
		\frac{\partial^2 H}{\partial K_\mu \partial r_\mu}
		= \frac{\partial \hat{\mathbf{e}}^* }{\partial r_\mu} \cdot \frac{\partial \bm{D} }{\partial K_\mu} \cdot \hat{\mathbf{e}}
		+ \hat{\mathbf{e}}^* \cdot \frac{\partial \bm{D} }{\partial K_\mu} \cdot \frac{\partial \hat{\mathbf{e}} }{\partial r_\mu} ,
	}
	\end{equation}
	to get
	\begin{equation} \label{eq:d_dtau_abs_A}
	\eqalign{
		\frac{\rmd \ln |A^{(0)}|}{\rmd \tau}
		+ \frac{1}{2} \textrm{Re}(\bm{\Psi}) : \nabla_K \nabla_K H
		+ \frac{1}{2} \nabla \cdot \nabla_K H
		= 0 .
	}
	\end{equation}
	The magnitude $|A^{(0)}|$ is related to $\det[\textrm{Im}(\bm{\Psi}_w)]$, where we have defined the determinant to be
	\begin{eqnarray}
		\det (\bm{\Psi}_w) =
		\begin{array}{|cc|}
			\Psi_{xx} & \Psi_{xy} \\
			\Psi_{yx} & \Psi_{yy} \\
		\end{array} \ .
	\end{eqnarray}
	To prove that $|A^{(0)}|$ is related to $\det[\textrm{Im}(\bm{\Psi}_w)]$, we evaluate
	\begin{equation} \label{eq:d_dtau_det_ImPsi}
		\frac{\rmd}{\rmd \tau} \det[\textrm{Im}(\bm{\Psi}_w)] 
		= \det[\textrm{Im}(\bm{\Psi}_w)] [\textrm{Im}(\bm{\Psi}_w)]^{-1} : \frac{\rmd \ \textrm{Im}(\bm{\Psi}_w) }{\rmd \tau},
	\end{equation}
	where we define
	\begin{eqnarray}
		\bm{\Psi}_w^{-1} =
		\left(
		\begin{array}{cc}
			&
			\left(
			\begin{array}{cc}
				\Psi_{xx} & \Psi_{xy} \\
				\Psi_{yx} & \Psi_{yy}
			\end{array}
			\right) ^{-1}

			\begin{array}{c}
				0 \\
				0 \\
			\end{array} \\

			& \begin{array}{cc}
				\ \ \ 0 \ \ \ & 0~~
			\end{array}

			\begin{array}{c}
				\ \ \ \ \ \ 0
			\end{array}

		\end{array} \right) .
	\end{eqnarray}
	We use equation (\ref{eq:Psi_evolution}) to find $\rmd \ \textrm{Im}(\bm{\Psi}) / \rmd \tau $, which is
	\begin{equation} \label{eq:d_dtau_ImPsi}
	\fl
	\eqalign{
		\frac{\rmd \ \textrm{Im}(\bm{\Psi})}{\rmd \tau}
		=
		&- \textrm{Im}(\bm{\Psi}) \cdot \nabla_K \nabla_K H \cdot \textrm{Re}(\bm{\Psi})
		- \textrm{Re}(\bm{\Psi}) \cdot \nabla_K \nabla_K H \cdot \textrm{Im}(\bm{\Psi}) \\
		& - \textrm{Im}(\bm{\Psi}) \cdot \nabla_K \nabla H
		- \nabla \nabla_K H \cdot \textrm{Im}(\bm{\Psi}) .
	}
	\end{equation}
	Substituting equation (\ref{eq:d_dtau_ImPsi}) into equation (\ref{eq:d_dtau_det_ImPsi}), and using $\textrm{Im}(\bm{\Psi}_w) \cdot [\textrm{Im}(\bm{\Psi}_w)]^{-1} = \bm{1} - \hat{\mathbf{g}} \hat{\mathbf{g}}$, we get
	\begin{equation}
	\fl
	\frac{\rmd}{\rmd \tau} \ln( \det[\textrm{Im}(\bm{\Psi}_w)] )
	=
	-2 (\bm{1} - \hat{\mathbf{g}} \hat{\mathbf{g}}) : [
	\textrm{Re}(\bm{\Psi}) \cdot \nabla_K \nabla_K H
	+ \nabla_K \nabla H
	] .
	\end{equation}
	Using this result and equation (\ref{eq:d_dtau_abs_A}), we find
	\begin{equation}\label{eq:abs_A_semifinal}
	\fl
	\eqalign{
		\frac{\rmd \ln |A^{(0)}|}{\rmd \tau}
		= \frac{1}{4} \frac{\rmd}{\rmd \tau} \ln( \det[\textrm{Im}(\bm{\Psi}_w)] )
		- \frac{1}{2} \hat{\mathbf{g}} \cdot \textrm{Re}(\bm{\Psi}) \cdot \nabla_K \nabla_K H \cdot \hat{\mathbf{g}}
		- \frac{1}{2} \hat{\mathbf{g}} \cdot \nabla_K \nabla H  \cdot \hat{\mathbf{g}}.
	}
	\end{equation}
	Using the fact that $\hat{\mathbf{g}}$ is purely real and taking the real part of equation (\ref{eq:constraint}), we find that
	\begin{equation}
		\hat{\mathbf{g}} \cdot \textrm{Re}(\bm{\Psi}) = \frac{1}{g} \frac{\rmd \mathbf{K}}{\rmd \tau} .
	\end{equation}
	Hence,
	\begin{equation}
	\eqalign{
		&\hat{\mathbf{g}} \cdot \textrm{Re}(\bm{\Psi}) \cdot \nabla_K \nabla_K H \cdot \hat{\mathbf{g}}
		+ \hat{\mathbf{g}} \cdot \nabla_K \nabla H  \cdot \hat{\mathbf{g}} \\
		&\qquad = \frac{\hat{\mathbf{g}}}{g} \cdot \nabla_K \nabla_K H \cdot \frac{\rmd \mathbf{K}}{\rmd \tau}
		+ \frac{\hat{\mathbf{g}}}{g} \cdot \nabla \nabla_K H  \cdot \frac{\rmd \mathbf{q}}{\rmd \tau} ,	
	}
	\end{equation}
	where we have used the definition of group velocity in equation (\ref{eq:group_velocity}). Substituting this result into equation (\ref{eq:abs_A_semifinal}), and noting that
	\begin{equation}
		\frac{\rmd \mathbf{g}}{\rmd \tau}
		= \frac{\rmd }{\rmd \tau} (\nabla_K H)
		= \frac{\rmd \mathbf{K}}{\rmd \tau} \cdot \nabla_K \nabla_K H
		+ \frac{\rmd \mathbf{q}}{\rmd \tau} \cdot \nabla \nabla_K H ,
	\end{equation}
	we find
	\begin{equation}
	\frac{\rmd \ln |A^{(0)}|}{\rmd \tau}
	=
	\frac{1}{4} \frac{\rmd}{\rmd \tau} \ln( \det[\textrm{Im}(\bm{\Psi}_w)] )
	- \frac{1}{2} \frac{\hat{\mathbf{g}}}{g} \cdot \frac{\rmd \mathbf{g}}{\rmd \tau}  .
	\end{equation} \label{eq:abs_amplitude_evolution}
	Using $g^{-1} \hat{\mathbf{g}} \cdot \rmd \mathbf{g} / \rmd \tau = \rmd \ln g / \rmd \tau$ and simplifying, we get
	\begin{equation}
		|A^{(0)}| 
		= C \frac{[\det (\textrm{Im} (\bm{\Psi}_w)) ]^{\frac{1}{4}} }{g^{\frac{1}{2}} },
	\end{equation} 
	where $C$ is a constant of integration. 
	
	The imaginary part of equation (\ref{eq:amplitude_pre_conjugate}) gives us
	\begin{equation} \label{eq:beam_tracing_phi}
	\fl
	\eqalign{
		\frac{\rmd \phi^{(0)}}{\rmd \tau}
		&= 
		- \textrm{Im}(\bm{\Psi}) : \nabla_K \nabla_K H \\
		& 
		+ \rmi \frac{\rmd K}{\rmd \tau} \cdot \nabla_K \hat{\mathbf{e}} \cdot \hat{\mathbf{e}}^*
		- \frac{1}{2 \rmi} \left( 
			\hat{\mathbf{e}}^* \cdot \frac{\partial \bm{D}}{\partial K_\mu} \cdot \frac{\partial \hat{\mathbf{e}} }{\partial r_\mu}  
			- \frac{\partial \hat{\mathbf{e}^*} }{\partial r_\mu} \cdot \frac{\partial \bm{D}}{\partial K_\mu} \cdot \hat{\mathbf{e}}	
		\right) ,
	}
	\end{equation} 
	where we have once again used $\hat{\mathbf{e}} \cdot \partial \hat{\mathbf{e}}^*  / \partial K_\mu + \hat{\mathbf{e}}^* \cdot \partial \hat{\mathbf{e}}^{}  / \partial K_\mu = \partial (\hat{\mathbf{e}}^* \cdot \hat{\mathbf{e}} ) / \partial K_\mu = 0$. Using equations (\ref{eq:group_velocity}) and (\ref{eq:partial_H}), we can rewrite this equation as
	\begin{equation}
	\fl
	\eqalign{
		\frac{\rmd \phi^{(0)}}{\rmd \tau}
		&= 
		- \textrm{Im}(\bm{\Psi}) : \nabla_K \nabla_K H \\
		& 
		+ \rmi \frac{\rmd \hat{\mathbf{e}}}{\rmd \tau} \cdot \hat{\mathbf{e}}^* 
		- \frac{1}{2 \rmi} \left( 
			\hat{\mathbf{e}}^* \cdot \frac{\partial \bm{D}}{\partial K_\mu} \cdot \left( \bm{1} - \hat{\mathbf{e}} \hat{\mathbf{e}}^* \right) \cdot \frac{\partial \hat{\mathbf{e}} }{\partial r_\mu}  
			- \frac{\partial \hat{\mathbf{e}^*} }{\partial r_\mu} \cdot \left( \bm{1} - \hat{\mathbf{e}} \hat{\mathbf{e}}^* \right) \cdot \frac{\partial \bm{D}}{\partial K_\mu} \cdot \hat{\mathbf{e}}	
		\right) .
	}
	\end{equation}
	For clarity, we have re-expressed the second and third terms of equation (\ref{eq:beam_tracing_phi}) as two new terms, one parallel to $\hat{\mathbf{e}}$, and the other perpendicular to it. Using equation (\ref{eq:shift_derivative}), this simplifies to
	\begin{equation}
	\fl
	\eqalign{
		\frac{\rmd \phi^{(0)}}{\rmd \tau}
		&= 
		- \textrm{Im}(\bm{\Psi}) : \nabla_K \nabla_K H \\
		& 
		+ \rmi \frac{\rmd \hat{\mathbf{e}}}{\rmd \tau} \cdot \hat{\mathbf{e}}^* 
		- \frac{1}{2 \rmi} \left( 
			\frac{\partial \hat{\mathbf{e}}^*}{\partial K_\mu} \cdot \bm{D} \cdot \frac{\partial \hat{\mathbf{e}} }{\partial r_\mu} 
			- \frac{\partial \hat{\mathbf{e}}^*}{\partial r_\mu} \cdot \bm{D} \cdot \frac{\partial \hat{\mathbf{e}} }{\partial K_\mu} 
		\right) .
	}
	\end{equation}	
	We separate $\phi^{(0)}$ into two parts, the Gouy phase $\phi_G$ and the phase $\phi_P$ associated with how the polarisation changes when the probe beam passes through a plasma. The former is a beam effect, whereas the latter is a result of ray tracing. We define the phases $\phi_G$ and $\phi_P$ by their evolution equations,
	\begin{equation}
	\eqalign{
		\frac{\rmd \phi_G}{\rmd \tau}
		= - \textrm{Im}(\bm{\Psi}) : \nabla_K \nabla_K H ,
	}
	\end{equation}
	and
	\begin{equation}
	\eqalign{
		\frac{\rmd \phi_P}{\rmd \tau}
		=   
		\rmi \frac{\rmd \hat{\mathbf{e}}}{\rmd \tau} \cdot \hat{\mathbf{e}}^* 
		- \frac{1}{2 \rmi} \left( 
			\frac{\partial \hat{\mathbf{e}}^*}{\partial K_\mu} \cdot \bm{D} \cdot \frac{\partial \hat{\mathbf{e}} }{\partial r_\mu} 
			- \frac{\partial \hat{\mathbf{e}}^*}{\partial r_\mu} \cdot \bm{D} \cdot \frac{\partial \hat{\mathbf{e}} }{\partial K_\mu} 
		\right) .
	}
	\end{equation}
	The $\rmd \hat{\mathbf{e}} / \rmd \tau \cdot \hat{\mathbf{e}}^*$ piece corrects for our choice of the vectors $\partial_\mu \hat{\mathbf{e}} \cdot \hat{\mathbf{e}}^*$, as discussed at the end of \ref{appendix_beam_tracing_first_order}. If we had chosen the vectors $\partial_\mu \hat{\mathbf{e}} \cdot \hat{\mathbf{e}}^* = 0$, then we would get $\rmd \hat{\mathbf{e}} / \rmd \tau \cdot \hat{\mathbf{e}}^* = 0$. The other piece of $\rmd \phi_P / \rmd \tau$ is physical, and accounts for how the polarisation changes as $\mathbf{B}$ and $n_e$ vary along the beam path.

	\section{Eigenvalues of $\bm{D}$} \label{appendix_Cardano}
	We find the eigenvalues of $\bm{D}$, given in equation (\ref{eq:D_explicit}) by solving the eigenvalue equation 
	\begin{eqnarray}
		\det \left( \bm{D} - H \bm{1} \right) 
		= 
		\left|
		\begin{array}{ccc}
			D_{11} - H & - \rmi D_{12} & D_{1b}\\
			\rmi D_{12} & D_{22} - H & 0\\
			D_{1b} & 0 & D_{bb} - H\\
		\end{array}
		\right|	
		=
		0 ,
	\end{eqnarray}
	getting
	\begin{equation}
		H^3 
		+ h_2 H^2 
		+ h_1 H 
		+ h_0		
		= 0 ,
	\end{equation}
	where the coefficients $h_2$, $h_1$, and $h_0$ are
	\begin{equation}
		h_2 =
		- D_{11} 
		- D_{22} 
		- D_{bb} ,
	\end{equation}
	\begin{equation}
		h_1 =
		D_{11} D_{bb}
		+ D_{11} D_{22} 
		+ D_{22} D_{bb}
		- D_{12}^2
		- D_{1b}^2 ,
	\end{equation}	
	and
	\begin{equation}
		h_0 =
		D_{22} D_{1b}^2
		+ D_{bb} D_{12}^2 
		- D_{11} D_{22} D_{bb} . 
	\end{equation}		
	We use Cardano's formula to solve for $H$, getting three possible solutions
	\begin{equation} \label{eq:Cardano_H_1}
		H_1 =
		\frac{h_t}{3 \sqrt[3]{2}}
		-
		\sqrt[3]{2} \frac{ 3 h_1 - h_2^2 }{3 h_t}
		-\frac{h_2}{3} ,
	\end{equation}
	\begin{equation} \label{eq:Cardano_H_2}
		H_2 =
		- \frac{1 - i \sqrt{3}}{6 \sqrt[3]{2}} h_t
		+ \left( 1 + i \sqrt{3} \right) \frac{ 3 h_1 - h_2^2 }{3 \sqrt[3]{4} h_t}
		- \frac{h_2}{3} ,
	\end{equation}
	and
	\begin{equation} \label{eq:Cardano_H_3}
		H_3 =
		- \frac{1 + i \sqrt{3}}{6 \sqrt[3]{2}} h_t
		+ \left( 1 - i \sqrt{3} \right) \frac{ 3 h_1 - h_2^2 }{3 \sqrt[3]{4} h_t}
		- \frac{h_2}{3} ,
	\end{equation}
	where we have used the shorthand
	\begin{equation}
	\fl
		h_t = \Bigl[
			- 2 h_2^3
			+ 9 h_2 h_1
			-27 h_0
			+ 3 \sqrt{3} \Bigl( 
				4 h_2^3 h_0 
				- h_2^2 h_1^2 
				- 18 h_2 h_1 h_0 
				+ 4 h_1^3 
				+ 27 h_0^2
			\Bigr)^{\frac{1}{2}}
		\Bigr]^{\frac{1}{3}} .
	\end{equation}
	We select the solution that is zero along the entire path of the ray, using it to calculate $g$ in post-processing.

	\section{Discontinuity of $\bm{\Psi}$ at the vacuum-plasma boundary} \label{appendix_vacuum_plasma_interface}
	We are going to look at a special case of the vacuum-plasma boundary, where the density $n_e$ is continuous, but its gradient $\nabla n_e$ is discontinuous. Hence, the ray part of beam tracing is well-behaved, but there will be a discontinuity in the beam part. We need to find the matching conditions at the boundary for $\bm{\Psi}$. This is a special case of what was presented in earlier work \cite{Poli:beam_tracing_BC:2001}. 
	
	We take the density to be zero on the plasma-vacuum interface. We also assume a general 2D curvature of the interface surface, which is relevant to our concerns because density is specified on flux surfaces, curved in 2D.
%
%
%

	We re-express the exponential piece in $\mathbf{E}_b$ at the vacuum-plasma boundary as
	\begin{equation}
	\eqalign{
		E_b \propto
		\exp \Big[
			&\rmi s^{vp}
			+ \rmi K_R^{vp} \Delta R
			+ \rmi K_Z^{vp} \Delta Z
			+ \rmi K_\zeta^{vp} \Delta \zeta \\
			&+ \frac{\rmi}{2} \big(
				\Psi_{RR} \left( \Delta R \right)^2
				+ 2 \Psi_{RZ} \Delta R \Delta Z
				+ \Psi_{ZZ} \left( \Delta Z \right)^2 \\
				&+ \Psi_{\zeta \zeta} \left( \Delta \zeta \right)^2
				+ 2 \Psi_{R \zeta} \Delta R \Delta \zeta
				+ 2 \Psi_{\zeta Z} \Delta \zeta \Delta Z
			\big)
		\Big] ,
	}
	\end{equation}
	where the superscript $^{vp}$ indicates that the variables are continuous across this boundary. We need to impose continuity of $\mathbf{E}_b$ across this boundary. Since we are dealing with the special case where the density is zero at the boundary, there is no charge or current, so all components of $\mathbf{E}_b$ are continuous. To proceed, we have to figure out how to parameterise the vacuum-plasma boundary.
	Since the density is specified on a flux surface, the discontinuity in $\bm{\Psi}$ occurs on a flux surface, which we take to also be the vacuum-plasma boundary. Two points on the same flux surface must have the same flux label. Hence, they are related by
	\begin{equation} \label{eq:flux_surface_expansion}
	\fl
	\eqalign{
		\psi_{p} (R,Z) 
		&= \psi_{p} (R + \Delta R, Z + \Delta Z) \\ 
		&\simeq \psi_{p} (R,Z)
		+ \frac{\partial \psi_{p}}{\partial R} \Delta R
		+ \frac{\partial \psi_{p}}{\partial Z} \Delta Z \\
		&+ \frac{1}{2} \left(
			\frac{\partial^2 \psi_{p}}{\partial R^2} \left( \Delta R \right)^2
			+ 2 \frac{\partial^2 \psi_{p}}{\partial R \partial Z} \Delta R \Delta Z
			+ \frac{\partial^2 \psi_{p}}{\partial Z^2} \left( \Delta Z \right)^2
		\right) ,
	}
	\end{equation}
	where we have expanded to second order. We write $\Delta R = \Delta R^{(0)} + \Delta R^{(1)} + \ldots$ and $\Delta Z = \Delta Z^{(0)} + \Delta Z^{(1)} + \ldots$. We parameterise poloidal displacement with $\xi$, such that
	\begin{equation} \label{eq:delta_R_delta_Z_zeroth_order}
		\left( \begin{array}{c}
		\Delta R^{(0)} \\
		\Delta Z^{(0)} \\
		\end{array} \right)
		=
		\left( \begin{array}{c}
		-\partial \psi_{p} / \partial Z \\
		\partial \psi_{p} / \partial R \\
		\end{array} \right)		
		\xi .
	\end{equation}
	This choice ensures that equation (\ref{eq:flux_surface_expansion}) is satisfied to first order. To second order, we find
	\begin{equation}
	\fl
	\eqalign{
		&\frac{\partial \psi_{p}}{\partial R} \Delta R^{(1)}
		+ \frac{\partial \psi_{p}}{\partial Z} \Delta Z^{(1)} \\
		&+ \frac{1}{2} \left(
			\frac{\partial^2 \psi_{p}}{\partial R^2} \left( \Delta R^{(0)} \right)^2
			+ 2 \frac{\partial^2 \psi_{p}}{\partial R \partial Z} \Delta R^{(0)} \Delta Z^{(0)}
			+ \frac{\partial^2 \psi_{p}}{\partial Z^2} \left( \Delta Z^{(0)} \right)^2
		\right)
		= 0	.
	}
	\end{equation}
	The next order corrections are thus
	\begin{equation}
	\eqalign{
		\left( \begin{array}{c}
			\Delta R^{(1)} \\
			\Delta Z^{(1)} \\
		\end{array} \right)
		&=
		\left( \begin{array}{c}
			-\partial \psi_{p} / \partial Z \\
			\partial \psi_{p} / \partial R \\
		\end{array} \right)		
		\xi^{(1)}
		+ 
		\eta
		\left( \begin{array}{c}
			\partial \psi_{p} / \partial R \\
			\partial \psi_{p} / \partial Z \\
		\end{array} \right) ,
	}
	\end{equation}
	where $\eta$ is given by 
	\begin{equation}
	\fl
	\eqalign{
		\eta
		=
		- \frac{1}{2} &\left[
			\frac{\partial^2 \psi_{p}}{\partial R^2} \left( \frac{\partial \psi_p }{\partial Z} \right)^2
			+ 2 \frac{\partial^2 \psi_{p}}{\partial R \partial Z} \frac{\partial \psi_p }{\partial R} \frac{\partial \psi_p }{\partial Z}
			+ \frac{\partial^2 \psi_{p}}{\partial Z^2} \left( \frac{\partial \psi_p }{\partial R} \right)^2		
		\right] \\		 
		&\times \left[
			\left( \frac{\partial \psi_{p}}{\partial R} \right)^2
			+ \left( \frac{\partial \psi_{p}}{\partial Z} \right)^2
		\right]^{-1}  
		\xi^2 .
	}		
	\end{equation}
	We redefine $\xi$ such that it incorporates the $\xi^{(1)}$ term, getting
	\begin{equation} \label{eq:delta_R_delta_Z_first_order}
	\eqalign{
		\left( \begin{array}{c}
		\Delta R^{(1)} \\
		\Delta Z^{(1)} \\
		\end{array} \right)
		&=
		\eta
		\left( \begin{array}{c}
		\partial \psi_{p} / \partial R \\
		\partial \psi_{p} / \partial Z \\
		\end{array} \right)	.
	}
	\end{equation}
	
	Imposing continuity across the plasma-vacuum boundary, we find the formula 
	\begin{equation}
	\fl
	\eqalign{
		&s^{vp}
		+ K_R^{vp} \Delta R
		+ K_Z^{vp} \Delta Z
		+ K_\zeta^{vp} \Delta \zeta 
		+\frac{1}{2} \left(
			\Psi_{RR}^{v} \left( \Delta R \right)^2
			+ \Psi_{ZZ}^{v} \left( \Delta Z \right)^2 
			+ \Psi_{\zeta \zeta}^{v} \left( \Delta \zeta \right)^2
		\right) \\
		&+ \Psi_{RZ}^{v} \Delta R \Delta Z
		+ \Psi_{R \zeta}^{v} \Delta R \Delta \zeta
		+ \Psi_{\zeta Z}^{v} \Delta \zeta \Delta Z	\\
		=
		&s^{vp}
		+ K_R^{vp} \Delta R
		+ K_Z^{vp} \Delta Z
		+ K_\zeta^{vp} \Delta \zeta 
		+\frac{1}{2} \left(
			\Psi_{RR}^{p} \left( \Delta R \right)^2
			+ \Psi_{ZZ}^{p} \left( \Delta Z \right)^2 
			+ \Psi_{\zeta \zeta}^{p} \left( \Delta \zeta \right)^2
		\right) \\
		&+ \Psi_{RZ}^{p} \Delta R \Delta Z
		+ \Psi_{R \zeta}^{p} \Delta R \Delta \zeta
		+ \Psi_{\zeta Z}^{p} \Delta \zeta \Delta Z .		
	}
	\end{equation}
	We now consider displacement in the toroidal direction only, $\xi = 0$,
	\begin{equation}
		K_\zeta^{vp}
		+ \frac{1}{2} \Psi_{\zeta \zeta}^{v} \left( \Delta \zeta \right)^2
		=
		K_\zeta^{vp}
		+ \frac{1}{2} \Psi_{\zeta \zeta}^{p} \left( \Delta \zeta \right)^2 ,
	\end{equation}
	hence
	\begin{equation} \label{eq:BC_1}
		\Psi_{\zeta \zeta}^{v}
		=
		\Psi_{\zeta \zeta}^{p} .
	\end{equation}
	It may seem surprising that the curvature of the flux surface in the toroidal direction does not appear in this equation. However, this curvature is already accounted for in our definition of $\Psi_{\zeta \zeta}$ that gives rise to the $K_R R$ term in equation (\ref{eq:Psi_zeta_zeta}). 
	Moving forward, we consider purely poloidal displacement, that is, $\Delta \zeta = 0$,
	\begin{equation}
	\fl
	\eqalign{
		&K_R^{vp} \Delta R
		+ K_Z^{vp} \Delta Z
		+ \frac{1}{2} \left(
			\Psi_{RR}^{v} \left( \Delta R \right)^2
			+2 \Psi_{RZ}^{v} \Delta R \Delta Z
			+ \Psi_{ZZ}^{v} \left( \Delta Z \right)^2
		\right) \\
		& \qquad = K_R^{vp} \Delta R
		+ K_Z^{vp} \Delta Z
		+ \frac{1}{2} \left(
			\Psi_{RR}^{p} \left( \Delta R \right)^2
			+2 \Psi_{RZ}^{p} \Delta R \Delta Z
			+ \Psi_{ZZ}^{p} \left( \Delta Z \right)^2
		\right) .
	}
	\end{equation}
	Hence, to leading order, we use equations (\ref{eq:delta_R_delta_Z_zeroth_order}) and (\ref{eq:delta_R_delta_Z_first_order}) to get
	\begin{equation} \label{eq:BC_2}
	\eqalign{
		&\Psi_{RR}^{v} \left( \frac{\partial \psi_{p}}{\partial Z} \right)^2
		-2 \Psi_{RZ}^{v} \frac{\partial \psi_{p}}{\partial R} \frac{\partial \psi_{p}}{\partial Z}
		+ \Psi_{ZZ}^{v} \left( \frac{\partial \psi_{p}}{\partial R} \right)^2 \\
		& \qquad = \Psi_{RR}^{p} \left( \frac{\partial \psi_{p}}{\partial Z} \right)^2
		-2 \Psi_{RZ}^{p} \frac{\partial \psi_{p}}{\partial R} \frac{\partial \psi_{p}}{\partial Z}
		+ \Psi_{ZZ}^{p} \left( \frac{\partial \psi_{p}}{\partial R} \right)^2  .
	}
	\end{equation}
	Note that $\Delta R^{(2)}$ and $\Delta Z^{(2)}$ do not contribute because of the continuity of $\mathbf{K}$ across the boundary \cite{Poli:beam_tracing_BC:2001}. Finally, we consider a displacement that is simultaneously in the toroidal and poloidal directions. Following similar steps as earlier, and using the results we already have, we find that
	\begin{equation}
		\Psi_{R \zeta}^{v} \Delta R \Delta \zeta
		+ \Psi_{\zeta Z}^{v} \Delta \zeta \Delta Z
		=
		\Psi_{R \zeta}^{p} \Delta R \Delta \zeta
		+ \Psi_{\zeta Z}^{p} \Delta \zeta \Delta Z	.
	\end{equation}
	Hence,
	\begin{equation} \label{eq:BC_3}
		- \Psi_{R \zeta}^{v} \frac{\partial \psi_{p}}{\partial Z}
		+ \Psi_{\zeta Z}^{v} \frac{\partial \psi_{p}}{\partial R}
		=
		- \Psi_{R \zeta}^{p} \frac{\partial \psi_{p}}{\partial Z}
		+ \Psi_{\zeta Z}^{p} \frac{\partial \psi_{p}}{\partial R} .
	\end{equation}
	Equations (\ref{eq:BC_1}), (\ref{eq:BC_2}), and (\ref{eq:BC_3}) give us 3 conditions for the transition, but we have 6 variables we want to find (either $\bm{\Psi}^{p}$ or $\bm{\Psi}^{v}$). The other 3 conditions can be found from equation (\ref{eq:constraint}), giving us
	\begin{equation}
		\bm{\Psi}^{p} \cdot \nabla_K H + \nabla H = 0.
	\end{equation}
	We solve these six linear equations numerically in Scotty.
	
	If the equilibrium density were also discontinuous, then we would have to solve three more equations. Two are linear equations, which state that the components of $\mathbf{K}$ parallel to the flux surface have to be continuous. The third component of the wavevector can be determined from the dispersion relation, $H = 0$, which is nonlinear. A discontinuous $\mathbf{K}$ also means that the six equations for $\bm{\Psi}$ are changed slightly, because $\mathbf{K}^{v} \neq \mathbf{K}^{p}$. This is not currently implemented in Scotty.
	
	\section{Derivation of $u_1$ and $u_2$} \label{appendix_derivation_u1_u2}

	We seek to find two coordinates whose gradients are perpendicular to $\hat{\mathbf{b}}$ as well as to each other. Unfortunately, due to magnetic shear, it is in general not possible to find such a set of coordinates throughout the whole volume of a plasma, but it is possible to do it along a line. Hence, we first seek to find $u_1$ and $u_2$ such that $\nabla u_1$ and $\nabla u_2$ are perpendicular to $\hat{\mathbf{b}}$ and to each other along $\mathbf{w} = 0$, that is, along the central ray. We then find the higher order corrections to $u_1$ and $u_2$ at finite $\mathbf{w}$ by imposing that $\nabla u_1$ and $\nabla u_2$ be perpendicular to $\hat{\mathbf{b}}$ everywhere. 
	
	We know that $u_i$ must satisfy
	\begin{equation}
		\hat{\mathbf{b}} (\mathbf{q}(\tau) + \mathbf{w}) \cdot \nabla u_i
		\simeq \left[
		\hat{\mathbf{b}} (\mathbf{q}(\tau))
		+ \mathbf{w} \cdot \nabla \hat{\mathbf{b}} (\mathbf{q}(\tau))
		\right]
		\cdot \nabla u_i
		= 0 .
	\end{equation}	
	Using the chain rule, we get
	\begin{equation}
	\fl
	\eqalign{
		\underbrace{
			\left(
			\hat{\mathbf{b}}
			+ \mathbf{w} \cdot \nabla \hat{\mathbf{b}}
			\right)
			\cdot
			\nabla \tau \frac{\partial u_i}{\partial \tau}
		}_{\sim \frac{1}{L} u_i}
		+
		\underbrace{
			\left(
			\hat{\mathbf{b}}
			+ \mathbf{w} \cdot \nabla \hat{\mathbf{b}}
			\right)
			\cdot
			\nabla w_x \frac{\partial u_i}{\partial w_x}
		}_{\sim \frac{1}{W} u_i}
		+
		\underbrace{
			\left(
			\hat{\mathbf{b}}
			+ \mathbf{w} \cdot \nabla \hat{\mathbf{b}}
			\right)
			\cdot
			\nabla w_y \frac{\partial u_i}{\partial w_y}
		}_{\sim \frac{1}{W} \frac{W}{L} \sim \frac{1}{L} u_i}
		= 0 .
	} 
	\end{equation}
	The size of the term proportional to $\partial u_i / \partial w_y$ is determined by the fact that $\nabla w_y$ points mostly along $\hat{\mathbf{y}}$ in Figure \ref{fig:basis}, and this direction is perpendicular to $\hat{\mathbf{b}}$. Using reciprocal vectors, we follow the same procedure used to derive equation (\ref{eq:grad_tau}) and get
	\begin{equation} \label{eq:grad_wx}
		\nabla w_x
		= \hat{\mathbf{x}} + \frac{ w_y \hat{\mathbf{g}} }{ g (1 - \boldsymbol{\kappa} \cdot \mathbf{w}) } \left( \frac{\rmd \hat{\mathbf{x}}}{\rmd \tau} \cdot \hat{\mathbf{y}} \right) ,
	\end{equation}
	and
	\begin{equation} \label{eq:grad_wy}
		\nabla w_y
		= \hat{\mathbf{y}} - \frac{ w_x \hat{\mathbf{g}} }{ g (1 - \boldsymbol{\kappa} \cdot \mathbf{w}) } \left( \frac{\rmd \hat{\mathbf{x}}}{\rmd \tau} \cdot \hat{\mathbf{y}} \right) .
	\end{equation}
	
	Having obtained these results, we then expand $u_i$ in $W / L$ to obtain
	\begin{equation}
		u_i = u_i^{(0)} + u_i^{(1)} + u_i^{(2)} + \ldots ,
	\end{equation}
	where $u_i^{(n)} \sim (W / L)^n \ u_i^{(0)}$. Since the term with $\partial u_i / \partial w_x$ is much larger than the other two terms, $\partial u_i^{(0)} / \partial w_x = 0$. Hence, $u_i^{(0)} = u_i^{(0)} (\tau,w_y)$. Since we want $\nabla u_1 (\mathbf{w}=0) = \hat{\mathbf{u}}_{1}$ and $\nabla u_2 (\mathbf{w}=0) = \hat{\mathbf{u}}_{2}$, where these directions are shown in Figure \ref{fig:basis}, we choose
	\begin{equation} \label{eq:u1_order0}
		u_1^{(0)} = u_1^{(0)} (\tau) = \int_0^\tau g(\tau') \cos \theta(\tau') \rmd \tau ,
	\end{equation}
	and
	\begin{equation} \label{eq:u2_order0}
		u_2^{(0)} = u_2^{(0)} (w_y) = w_y .
	\end{equation}
	This turns out to be convenient because one direction is only a function of $\tau$, while the other is only a function of $w_y$. 
	
	We now find the higher order corrections. The first order ($W / L$) equation for $u_1$ is
	\begin{equation}
	\eqalign{
		- \sin \theta \cos \theta
		- \frac{\partial u_1^{(1)}}{\partial w_x} \cos \theta
		= 0 ,
	}
	\end{equation}
	which gives us
	\begin{equation} \label{eq:u1_order1}
		u_1^{(1)} = - w_x \sin \theta .
	\end{equation}
	The first order ($W / L$) equation for $u_2$ is
	\begin{equation}
	\eqalign{
		- \frac{\partial u_2^{(1)}}{\partial w_x} \cos \theta
		+ \left( 
		w_x \frac{\sin \theta}{g} \frac{\rmd \hat{\mathbf{x}}}{\rmd \tau} \cdot \hat{\mathbf{y}} 
		+ \mathbf{w} \cdot \nabla \hat{\mathbf{b}} \cdot \hat{\mathbf{y}} 
		\right)
		\frac{\partial u_2^{(0)}}{\partial w_y}
		= 0 ,
	}
	\end{equation}
	which gives us
	\begin{equation} \label{eq:u2_order1}		
		u_2^{(1)} =  
		\frac{w_x^2}{2 \cos \theta} \left( 
			\frac{\sin \theta}{g} \frac{\rmd \hat{\mathbf{x}}}{\rmd \tau} \cdot \hat{\mathbf{y}} 
			+ \hat{\mathbf{x}} \cdot \nabla \hat{\mathbf{b}} \cdot \hat{\mathbf{y}} 
		\right)
		+ \frac{w_x w_y}{\cos \theta} \hat{\mathbf{y}} \cdot \nabla \hat{\mathbf{b}} \cdot \hat{\mathbf{y}}.
	\end{equation}
	Here we have used equations (\ref{eq:grad_tau}), (\ref{eq:grad_wx}), and (\ref{eq:grad_wy}). Since $k_{\perp,1} u_1^{(1)} \sim W / \lambda \gg 1$, we need to find the next order correction $u_1^{(2)}$, whereas $k_{\perp,2} u_2^{(1)} \sim (1 / \lambda) (W^2 / L) \sim 1$ indicates that expanding $u_2$ to this order is already sufficient. To get the second order contribution $u_1^{(2)}$, we use
	\begin{equation}
	\fl
	\eqalign{
		& - \frac{\sin \theta}{g} \frac{\partial u_1^{(1)}}{\partial \tau}
		+ \left(
			- \frac{\sin \theta}{g} \boldsymbol{\kappa} \cdot \mathbf{w}
			+ \frac{\mathbf{w} \cdot \nabla \hat{\mathbf{b}} \cdot \hat{\mathbf{g}}}{g}
		\right)
		\frac{\partial u_1^{(0)}}{\partial \tau} \\
		& \qquad - \cos \theta \frac{\partial u_1^{(2)}}{\partial w_x}
		+ \left(
			- \frac{\sin \theta}{g} \frac{\rmd \hat{\mathbf{x}}}{\rmd \tau} \cdot \hat{\mathbf{y}} w_y
			+ \mathbf{w} \cdot \nabla \hat{\mathbf{b}} \cdot \hat{\mathbf{x}}
		\right)
		\frac{\partial u_1^{(1)}}{\partial w_x}
		= 0 ,
	}
	\end{equation}  
	and find that
	\begin{equation} \label{eq:u1_order2}
	\fl
	\eqalign{
		u_1^{(2)} &=
		\frac{w_x^2}{2} \left(
			\frac{\sin \theta}{g} \frac{\rmd \theta}{\rmd \tau}
			- \boldsymbol{\kappa} \cdot \hat{\mathbf{x}} \sin \theta 
			+ \hat{\mathbf{x}} \cdot \nabla \hat{\mathbf{b}} \cdot \hat{\mathbf{g}}
			- \hat{\mathbf{x}} \cdot \nabla \hat{\mathbf{b}} \cdot \hat{\mathbf{x}} \tan \theta
		\right) \\
		&+ w_x w_y \left(
			- \boldsymbol{\kappa} \cdot \hat{\mathbf{y}} \sin \theta
			+ \hat{\mathbf{y}} \cdot \nabla \hat{\mathbf{b}} \cdot \hat{\mathbf{g}}
			+ \frac{\sin \theta \tan \theta}{g} \frac{\rmd \hat{\mathbf{x}}}{\rmd \tau} \cdot \hat{\mathbf{y}}
			- \hat{\mathbf{y}} \cdot \nabla \hat{\mathbf{b}} \cdot \hat{\mathbf{x}} \tan \theta
		\right) .
	}
	\end{equation}

	\section{Relationship between $\theta$ and $\theta_m$} \label{appendix_theta_thetam}
	
	To shed light on how $\theta$ and $\theta_m$ relate to each other (when they are small), we seek to make explicit the piece
	\begin{equation}
		\frac{\partial H}{\partial (\mathbf{K} \cdot \hat{\mathbf{b}})^2 } \left( \frac{\partial H}{\partial K^2 } \right)^{-1} ,
	\end{equation}
	in equation (\ref{eq:theta_theta_m}). This algebra-heavy process is detailed in \ref{appendix_theta_thetam_derivation}, while the results and their significance are discussed in \ref{appendix_theta_thetam_results}. 
	
	\subsection{Derivation of $\theta / \theta_{m}$} \label{appendix_theta_thetam_derivation}
	
	The derivation is somewhat onerous because we cannot neglect $\theta_m$ from the onset; we need to keep $K_\parallel$ for part of the derivation to be able to calculate $\partial H / \partial K^2_\parallel$. Fortunately, since we are finding the ratio of two derivatives of $H$, we are free to choose a $\bar{H}$ that satisfies equation (\ref{eq:H_bar}) because
	\begin{equation}
		\frac{\partial H}{\partial K_\parallel^2 } \left( \frac{\partial H}{\partial K^2 } \right)^{-1}
		=
		\frac{\partial \bar{H}}{\partial K_\parallel^2 } \left( \frac{\partial \bar{H}}{\partial K^2 } \right)^{-1} .
	\end{equation}
	For this calculation, we choose a convenient $\bar{H}$ to minimise the unpleasantness of the resulting algebra: $\bar{H} = \det (\bm{D})$,
	\begin{equation}
	\eqalign{
		\det (\bm{D})
		&=
		\left(
			\epsilon_{11} - N^2_\parallel
		\right)
		\left(
			\epsilon_{11} - N^2
		\right)	
		\left(
			\epsilon_{bb} - N^2 + N^2_\parallel
		\right) \\
		&
		- N^2_\parallel \left(
			N^2 -  N^2_\parallel
		\right) \left(
			\epsilon_{11} - N^2
		\right)
		- \epsilon_{12}^2 \left(
			\epsilon_{bb} - N^2 + N^2_\parallel
		\right) .
	}
	\end{equation}
	Here we have used equation (\ref{eq:D_explicit}), as well as the notation $N = K c / \Omega$ and $N_\parallel = K_\parallel c / \Omega$. The derivatives that we need can then be calculated for $N_\parallel^2 = N^2 \theta_{m}^2 \ll N^2$. We find
	\begin{equation}
		\frac{\partial \det (\bm{D})}{\partial K^2}
		= \frac{c^2}{\Omega^2} \left[
			\epsilon_{11} \left(
				N^2 - \epsilon_{bb}
			\right)
			+ \epsilon_{11} \left(
				N^2 - \epsilon_{11}
			\right)
			+ \epsilon_{12}^2 
		\right] ,
	\end{equation}
	and
	\begin{equation}
	\eqalign{
		\frac{\partial \det (\bm{D})}{\partial K_\parallel^2}
		= \frac{c^2}{\Omega^2} \Bigl[
			&\epsilon_{11} \left(
				\epsilon_{11} - N^2
			\right)
			- \left(
				\epsilon_{11} - N^2
			\right) \left(
				\epsilon_{bb} - N^2
			\right) \\
			&- N^2 \left(
				\epsilon_{11} - N^2
			\right)
			- \epsilon_{12}^2 
		\Bigr] ,
	}
	\end{equation}
	Now that we have evaluated the derivatives of $\bar{H}$, we find the piece that we care about
	\begin{equation} \label{eq:H_deriv_ratio}
		\frac{\partial H}{\partial K_\parallel^2 } \left( \frac{\partial H}{\partial K^2 } \right)^{-1}
		=
		\frac{
			  \epsilon_{11}^{2} 
			- \epsilon_{12}^{2} 
			- \epsilon_{11} \epsilon_{bb} 
			+ N^2 \left( 
				\epsilon_{bb} - \epsilon_{11} 
			\right)
		}{
			- \epsilon_{11}^{2} 
			+ \epsilon_{12}^{2}			
			- \epsilon_{11} \epsilon_{bb} 		
			+ 2 N^2 \epsilon_{11}		
		} .
	\end{equation}	
	For the O mode, $N^2 = \epsilon_{bb}$, we have
	\begin{equation} \label{eq:H_deriv_ratio_O_mode}
	\eqalign{
		\frac{\partial H}{\partial K_\parallel^2 } \left( \frac{\partial H}{\partial K^2 } \right)^{-1}
		&=
		- \frac{
			\left( \epsilon_{11} - \epsilon_{bb} \right)^2
			- \epsilon_{12}^2
		}{
			\epsilon_{11} \left( \epsilon_{11} - \epsilon_{bb} \right)
			- \epsilon_{12}^2
		} \\
		&=		
		- \frac{\Omega_{pe}^{2}}{\Omega^{2}} . 
	}
	\end{equation} 
	Similarly, for the X mode, $\epsilon_{11} (\epsilon_{11} - N^2) = \epsilon_{12}^2$, we have
	\begin{equation} \label{eq:H_deriv_ratio_X_mode}
	\eqalign{
		\frac{\partial H}{\partial K_\parallel^2 } \left( \frac{\partial H}{\partial K^2 } \right)^{-1}
		&=
		- \frac{
			\epsilon_{12}^2 \epsilon_{bb}
		}{
			\epsilon_{11} \left[
				\epsilon_{11} \left( \epsilon_{11} - \epsilon_{bb} \right)
				- \epsilon_{12}^2
			\right]
		} \\
		&=
		\frac{\Omega_{pe}^{2} \left(\Omega^{2} - \Omega_{pe}^{2}\right)}{\Omega^{2} \left(\Omega^{2} - \Omega_{ce}^{2} - \Omega_{pe}^{2}\right)} . 
	}
	\end{equation}
	
	\subsection{Properties of $\theta / \theta_m$} \label{appendix_theta_thetam_results}
\begin{figure}
		\centering
		
		\begin{subfigure}{.45\textwidth}
		\centering
		\begin{tikzpicture}
		
		\begin{groupplot}[
		group style={
			group size=1 by 2,
			vertical sep=0pt
		},
		]
		\nextgroupplot[
		legend cell align={left},
		legend style={fill opacity=0.8, draw opacity=1, text opacity=1},
		scaled x ticks=manual:{}{\pgfmathparse{#1}},
		tick align=outside,
		tick pos=left,
		title={O-mode},
		x grid style={white!69.0196078431373!black},
		xmin=-0.4, xmax=0.41,
		xtick style={color=black},
		xticklabels={},
		y grid style={white!69.0196078431373!black},
		ylabel={angle / $^{\circ}$},
		ymin=-2.4, ymax=2.4,
		ytick style={color=black},
		width=6cm,
		height=6cm,
		scale only axis
		]
		\addplot [ultra thick, blue]
		table {%
		-0.352570424333031 -0.0176103168874513
		-0.340321159237281 -0.107976843349496
		-0.328469754734922 -0.188682071184891
		-0.317014938310637 -0.262916894174749
		-0.305951373767537 -0.332547815037504
		-0.295270262516097 -0.398346058348292
		-0.284959452235854 -0.460327554989039
		-0.275004354516252 -0.518093692557119
		-0.265387174212452 -0.571071261734096
		-0.256088337130649 -0.61561024340736
		-0.247088509953242 -0.649552912502543
		-0.238368636984674 -0.675280991554529
		-0.22990997708687 -0.694422483419938
		-0.221694142993455 -0.707903144599361
		-0.213703142177311 -0.71660878618148
		-0.205919418236827 -0.721309129420767
		-0.198325891581765 -0.722487770838118
		-0.190905998058727 -0.720358585934882
		-0.183643724076422 -0.714896803800111
		-0.176523935456617 -0.705995389685177
		-0.169535418881836 -0.693904126799726
		-0.162667909999033 -0.678978384687393
		-0.155911471170953 -0.661572895783199
		-0.149256802619912 -0.642038024407899
		-0.142695230724462 -0.62070434731323
		-0.136218693446852 -0.597873560008922
		-0.129819723135348 -0.57381663917917
		-0.123491426991022 -0.548782763356359
		-0.117227465517183 -0.523003584470832
		-0.111022029281015 -0.496684977966794
		-0.104869814312117 -0.469999996919742
		-0.0987659964432693 -0.443085056938563
		-0.0927062048671312 -0.416047164207178
		-0.0866864951412505 -0.388973676339905
		-0.0807033218253898 -0.361940260213929
		-0.0747533912066941 -0.335019890755297
		-0.0688333611969818 -0.308287127553412
		-0.0629402531392075 -0.281798971337476
		-0.0570712455573491 -0.255599232439117
		-0.0512236483645154 -0.229720816249499
		-0.0453948942586195 -0.204187306991345
		-0.0395825299952612 -0.17901425317255
		-0.033784207567538 -0.154210646486391
		-0.0279976753179607 -0.129780160014365
		-0.0222207690030891 -0.105722015831667
		-0.0164514028270064 -0.0820318403308151
		-0.0106875604554579 -0.0587023057842868
		-0.00492728601844827 -0.0357236234362291
		0 -0.0163320319449293
		0.000831344945212364 -0.0130840070965912
		0.00659015429680082 0.00922968169742483
		0.0123509906214001 0.0312309674844614
		0.018115684968791 0.0529341895319545
		0.0238861812752366 0.0743593465375456
		0.0296643660793645 0.0955255667295322
		0.0354521396041003 0.116451847088781
		0.0412514381891381 0.137157391680844
		0.0470642435584872 0.15766160962412
		0.052892592061175 0.17798400554859
		0.058738583884165 0.19814421543945
		0.0646043922337484 0.218162000120103
		0.0704922724784749 0.23805685921212
		0.076404571243156 0.257847536352192
		0.0823437354396893 0.277551567849849
		0.0883123212164772 0.297184495300009
		0.0943130028041621 0.316758583050677
		0.100348581231362 0.336282394038258
		0.106421992880196 0.355760941395331
		0.112536352643057 0.375197193641687
		0.118695185663749 0.394594440015832
		0.124902044464765 0.413949824439345
		0.131160685683606 0.433258976481171
		0.13747515061653 0.45250986418327
		0.143849789820126 0.471676016756809
		0.15028928738231 0.490710799596907
		0.156798684762134 0.509543282576426
		0.163383404068072 0.528080977147198
		0.170049270618111 0.546207113192116
		0.176802534603124 0.563774812306787
		0.183649891659085 0.580598167741411
		0.190598502144749 0.596440959159962
		0.197656008920432 0.61100223356627
		0.204830550339556 0.623876199877142
		0.212129933573405 0.63454987933086
		0.219563852708665 0.642354056108754
		0.227143721351604 0.646502271726636
		0.234881476660194 0.646172354444743
		0.242789596764062 0.640533610309069
		0.250881115380802 0.628781031025134
		0.259169633580632 0.61014255585848
		0.267669328716497 0.5837849400042
		0.276394960600393 0.5488844797976
		0.285361875066005 0.505078788352368
		0.294586005110029 0.452629705339223
		0.304083869847837 0.390312133820352
		0.31387257155187 0.312586862754731
		0.323969791062703 0.216906525553761
		0.334393275168124 0.10060350514462
		0.345158542573667 -0.0392167982435996
		0.35627972699155 -0.205459953394903
		0.36776879303264 -0.400915056313383
		0.379635139549282 -0.627907290173542
		};
		\addlegendentry{$\theta$}
		\addplot [ultra thick, red]
		table {%
		-0.352570424333031 0.0176213804622029
		-0.340321159237281 0.115262851570347
		-0.328469754734922 0.215381000851642
		-0.317014938310637 0.321450012412408
		-0.305951373767537 0.435892832875228
		-0.295270262516097 0.559937337118695
		-0.284959452235854 0.693651113775745
		-0.275004354516252 0.836062931534736
		-0.265387174212452 0.985371194069473
		-0.256088337130649 1.13364075535305
		-0.247088509953242 1.27387629276365
		-0.238368636984674 1.40706027321365
		-0.22990997708687 1.53337249233915
		-0.221694142993455 1.65196791277548
		-0.213703142177311 1.76223413769789
		-0.205919418236827 1.86366172733962
		-0.198325891581765 1.9554249397423
		-0.190905998058727 2.03636577484755
		-0.183643724076422 2.10503778539286
		-0.176523935456617 2.16007792125377
		-0.169535418881836 2.20087842266538
		-0.162667909999033 2.2273328931726
		-0.155911471170953 2.23964693648693
		-0.149256802619912 2.23826495497196
		-0.142695230724462 2.22380928398117
		-0.136218693446852 2.19702864516693
		-0.129819723135348 2.15876424627735
		-0.123491426991022 2.10995232604384
		-0.117227465517183 2.05161146819913
		-0.111022029281015 1.98478178558742
		-0.104869814312117 1.91046527044645
		-0.0987659964432693 1.82957808660096
		-0.0927062048671312 1.74295029987546
		-0.0866864951412505 1.6513433645889
		-0.0807033218253898 1.55546781815752
		-0.0747533912066941 1.4559824232552
		-0.0688333611969818 1.35348377206473
		-0.0629402531392075 1.24857400240945
		-0.0570712455573491 1.14180816064534
		-0.0512236483645154 1.03369100499147
		-0.0453948942586195 0.92467764240461
		-0.0395825299952612 0.815174611491174
		-0.033784207567538 0.705543456744975
		-0.0279976753179607 0.596104679796952
		-0.0222207690030891 0.487141194349927
		-0.0164514028270064 0.378902679220737
		-0.0106875604554579 0.271609663645212
		-0.00492728601844827 0.165457419593799
		0 0.0756679700888061
		0.000831344945212364 0.0606200039344443
		0.00659015429680082 -0.0427447499918031
		0.0123509906214001 -0.144488818876287
		0.018115684968791 -0.244474746680403
		0.0238861812752366 -0.342599874406688
		0.0296643660793645 -0.438763261850351
		0.0354521396041003 -0.532868280850613
		0.0412514381891381 -0.624823397147346
		0.0470642435584872 -0.71454097236254
		0.052892592061175 -0.801935414534264
		0.058738583884165 -0.886921926695243
		0.0646043922337484 -0.969415088034637
		0.0704922724784749 -1.04932587822922
		0.076404571243156 -1.12655851399102
		0.0823437354396893 -1.20100787493392
		0.0883123212164772 -1.27255598718122
		0.0943130028041621 -1.3410670784387
		0.100348581231362 -1.40638710018904
		0.106421992880196 -1.4683463048562
		0.112536352643057 -1.52676801898268
		0.118695185663749 -1.58147851548984
		0.124902044464765 -1.63227783346713
		0.131160685683606 -1.67896627695529
		0.13747515061653 -1.72132311384071
		0.143849789820126 -1.75908528144131
		0.15028928738231 -1.79193388753353
		0.156798684762134 -1.81949025216497
		0.163383404068072 -1.84133892586044
		0.170049270618111 -1.857032343598
		0.176802534603124 -1.86608795723281
		0.183649891659085 -1.86797982306567
		0.190598502144749 -1.86212889199709
		0.197656008920432 -1.84789129475709
		0.204830550339556 -1.82448249877091
		0.212129933573405 -1.79120171830949
		0.219563852708665 -1.7469511350593
		0.227143721351604 -1.69053372234211
		0.234881476660194 -1.62115546211571
		0.242789596764062 -1.5384756217985
		0.250881115380802 -1.44264565024635
		0.259169633580632 -1.33426220183376
		0.267669328716497 -1.21410508731956
		0.276394960600393 -1.08325719252353
		0.285361875066005 -0.943891364590549
		0.294586005110029 -0.799281927181904
		0.304083869847837 -0.649953605077989
		0.31387257155187 -0.489934654137011
		0.323969791062703 -0.319441115440664
		0.334393275168124 -0.139011319231136
		0.345158542573667 0.0507909155154372
		0.35627972699155 0.249283967516295
		0.36776879303264 0.455748828342264
		0.379635139549282 0.669372391066776
		};
		\addlegendentry{$\theta_m$}
		
		\nextgroupplot[
		legend cell align={left},
		legend style={fill opacity=0.8, draw opacity=1, text opacity=1, anchor = south, at={(axis cs:0,-1)} },
		tick align=outside,
		tick pos=left,
		x grid style={white!69.0196078431373!black},
		xlabel={($l - l_c$) / m},
		xmin=-0.4, xmax=0.41,
		xtick style={color=black},
		y grid style={white!69.0196078431373!black},
		ylabel={ratio},		
		ymin=-1.05, ymax=0.05,
		ytick style={color=black},
		width=6cm,
		height=3cm,
		scale only axis,
		clip mode=individual
		]
		\addplot [semithick, black, mark=*, mark size=3, mark options={solid}, only marks]
		table {%
		-0.352570424333031 -0.999372150509135
		-0.340321159237281 -0.936787888538367
		-0.328469754734922 -0.876038603399649
		-0.317014938310637 -0.817909111907069
		-0.305951373767537 -0.762911867221947
		-0.295270262516097 -0.711411852615662
		-0.284959452235854 -0.663629807329713
		-0.275004354516252 -0.619682649493944
		-0.265387174212452 -0.579549377098833
		-0.256088337130649 -0.543038207210221
		-0.247088509953242 -0.509902661814479
		-0.238368636984674 -0.479923287161128
		-0.22990997708687 -0.452872662636984
		-0.221694142993455 -0.428521122671208
		-0.213703142177311 -0.406647885687668
		-0.205919418236827 -0.387038655588231
		-0.198325891581765 -0.369478652007647
		-0.190905998058727 -0.353747148391752
		-0.183643724076422 -0.339612337964134
		-0.176523935456617 -0.326837926881543
		-0.169535418881836 -0.315285078745682
		-0.162667909999033 -0.304839203321897
		-0.155911471170953 -0.295391601687421
		-0.149256802619912 -0.286846301632749
		-0.142695230724462 -0.279117616687891
		-0.136218693446852 -0.272128249817834
		-0.129819723135348 -0.265807922365158
		-0.123491426991022 -0.260092494310204
		-0.117227465517183 -0.254923309104875
		-0.111022029281015 -0.250246642514302
		-0.104869814312117 -0.246013368675323
		-0.0987659964432693 -0.242178817172946
		-0.0927062048671312 -0.238702827175799
		-0.0866864951412505 -0.235549846676944
		-0.0807033218253898 -0.232689005834048
		-0.0747533912066941 -0.230098856555067
		-0.0688333611969818 -0.227773050491121
		-0.0629402531392075 -0.225696651374826
		-0.0570712455573491 -0.22385479562053
		-0.0512236483645154 -0.222233544782945
		-0.0453948942586195 -0.220819989180618
		-0.0395825299952612 -0.219602341202806
		-0.033784207567538 -0.218570018631938
		-0.0279976753179607 -0.217713707697399
		-0.0222207690030891 -0.217025406715499
		-0.0164514028270064 -0.216498443609648
		-0.0106875604554579 -0.216127456572997
		-0.00492728601844827 -0.215908259200048
		0 -0.215838113877795
		0.000831344945212364 -0.215836460696052
		0.00659015429680082 -0.215925504282859
		0.0123509906214001 -0.216147987971317
		0.018115684968791 -0.216522116295121
		0.0238861812752366 -0.217044290125093
		0.0296643660793645 -0.217715508647379
		0.0354521396041003 -0.218537772417021
		0.0412514381891381 -0.219513853525717
		0.0470642435584872 -0.220647402629456
		0.052892592061175 -0.22194306713834
		0.058738583884165 -0.223406603755704
		0.0646043922337484 -0.225044981053882
		0.0704922724784749 -0.226866471275684
		0.076404571243156 -0.228880731138167
		0.0823437354396893 -0.231098874239372
		0.0883123212164772 -0.233533532743253
		0.0943130028041621 -0.236198910661094
		0.100348581231362 -0.239110835127153
		0.106421992880196 -0.242286809466363
		0.112536352643057 -0.245746039330645
		0.118695185663749 -0.249509832824767
		0.124902044464765 -0.253602552183211
		0.131160685683606 -0.258051029629292
		0.13747515061653 -0.262884905538511
		0.143849789820126 -0.26813709473502
		0.15028928738231 -0.273844254529019
		0.156798684762134 -0.280047272564463
		0.163383404068072 -0.286791839204958
		0.170049270618111 -0.294129025310265
		0.176802534603124 -0.302115883724365
		0.183649891659085 -0.310816080865665
		0.190598502144749 -0.320300577324856
		0.197656008920432 -0.33064836405682
		0.204830550339556 -0.341946935800934
		0.212129933573405 -0.354259306947149
		0.219563852708665 -0.36770007083624
		0.227143721351604 -0.382424948513275
		0.234881476660194 -0.398587531883863
		0.242789596764062 -0.416343035426375
		0.250881115380802 -0.435852720255844
		0.259169633580632 -0.457288346338466
		0.267669328716497 -0.480835593311821
		0.276394960600393 -0.506698209424239
		0.285361875066005 -0.535102668908796
		0.294586005110029 -0.566295433371174
		0.304083869847837 -0.600523069294336
		0.31387257155187 -0.638017458277845
		0.323969791062703 -0.679018808378946
		0.334393275168124 -0.723707290176457
		0.345158542573667 -0.772122294816288
		0.35627972699155 -0.824200430705485
		0.36776879303264 -0.87968422820014
		0.379635139549282 -0.938053762828265
		};
		\addlegendentry{ $\theta / \theta_m$}
		\addplot [ultra thick, green]
		table {%
		-0.352570424333031 -0.999501544283161
		-0.340321159237281 -0.936736405119276
		-0.328469754734922 -0.875863721563059
		-0.317014938310637 -0.817507231394555
		-0.305951373767537 -0.762163780542436
		-0.295270262516097 -0.71019781083676
		-0.284959452235854 -0.661840610766239
		-0.275004354516252 -0.617201862380434
		-0.265387174212452 -0.576278694450098
		-0.256088337130649 -0.538980567385154
		-0.247088509953242 -0.505156355716479
		-0.238368636984674 -0.47460959115121
		-0.22990997708687 -0.447113477604894
		-0.221694142993455 -0.422424547371029
		-0.213703142177311 -0.40029407035042
		-0.205919418236827 -0.38047690747664
		-0.198325891581765 -0.362737891891393
		-0.190905998058727 -0.346856043791543
		-0.183643724076422 -0.332627019117636
		-0.176523935456617 -0.319865203893669
		-0.169535418881836 -0.308413606215785
		-0.162667909999033 -0.298132002746687
		-0.155911471170953 -0.288893730365422
		-0.149256802619912 -0.28058584124673
		-0.142695230724462 -0.273108025301212
		-0.136218693446852 -0.266371478515056
		-0.129819723135348 -0.260297777788875
		-0.123491426991022 -0.254817801792279
		-0.117227465517183 -0.249870721969037
		-0.111022029281015 -0.245403079182634
		-0.104869814312117 -0.241367950871711
		-0.0987659964432693 -0.237724201986299
		-0.0927062048671312 -0.234435816504001
		-0.0866864951412505 -0.231471306127027
		-0.0807033218253898 -0.228803191024088
		-0.0747533912066941 -0.2264075852366
		-0.0688333611969818 -0.224263726333671
		-0.0629402531392075 -0.222353237761197
		-0.0570712455573491 -0.220659934297233
		-0.0512236483645154 -0.219169611323592
		-0.0453948942586195 -0.217869853400525
		-0.0395825299952612 -0.216749866908232
		-0.033784207567538 -0.21580033442169
		-0.0279976753179607 -0.215013288682859
		-0.0222207690030891 -0.214382004243652
		-0.0164514028270064 -0.213900905062973
		-0.0106875604554579 -0.213565486544651
		-0.00492728601844827 -0.213372250696021
		0 -0.213317838656658
		0.000831344945212364 -0.213318653267516
		0.00659015429680082 -0.213403062024358
		0.0123509906214001 -0.213624726056317
		0.018115684968791 -0.213983806039349
		0.0238861812752366 -0.214481476249967
		0.0296643660793645 -0.215119546121682
		0.0354521396041003 -0.215900664863199
		0.0412514381891381 -0.216828368170723
		0.0470642435584872 -0.217907104985165
		0.052892592061175 -0.21914227361599
		0.058738583884165 -0.220540267528663
		0.0646043922337484 -0.222108531207619
		0.0704922724784749 -0.223855626621714
		0.076404571243156 -0.225791310934427
		0.0823437354396893 -0.227926626217214
		0.0883123212164772 -0.230274002040988
		0.0943130028041621 -0.232847371937151
		0.100348581231362 -0.235662304834735
		0.106421992880196 -0.238736152691875
		0.112536352643057 -0.242088379586937
		0.118695185663749 -0.245741674784761
		0.124902044464765 -0.249720286172765
		0.131160685683606 -0.254051201316912
		0.13747515061653 -0.258764742883517
		0.143849789820126 -0.263894937420771
		0.15028928738231 -0.269479931102907
		0.156798684762134 -0.275562455321921
		0.163383404068072 -0.28219033962983
		0.170049270618111 -0.289417074066846
		0.176802534603124 -0.29730242507859
		0.183649891659085 -0.305913104319534
		0.190598502144749 -0.315323486278929
		0.197656008920432 -0.325616366307136
		0.204830550339556 -0.336883735589398
		0.212129933573405 -0.349225253284428
		0.219563852708665 -0.362755993425358
		0.227143721351604 -0.377607229469793
		0.234881476660194 -0.393923780879226
		0.242789596764062 -0.411864563491037
		0.250881115380802 -0.431602775903049
		0.259169633580632 -0.453325537622571
		0.267669328716497 -0.477232741145765
		0.276394960600393 -0.503534821607539
		0.285361875066005 -0.532449090570439
		0.294586005110029 -0.564194237626819
		0.304083869847837 -0.598982604996387
		0.31387257155187 -0.63700988483309
		0.323969791062703 -0.678441982328385
		0.334393275168124 -0.723396639431601
		0.345158542573667 -0.771912251309005
		0.35627972699155 -0.823928300009802
		0.36776879303264 -0.879261911028535
		0.379635139549282 -0.937590215843165
		};
		\addlegendentry{$r_O $}
		\end{groupplot}
		
		\end{tikzpicture}
			
		\end{subfigure}
		\qquad
		\begin{subfigure}{.45\textwidth}
		\centering
		\begin{tikzpicture}
		
		\begin{groupplot}[
				group style={
					group size=1 by 2,
					vertical sep=0pt
				},
				]
		\nextgroupplot[
		legend cell align={left},
		legend style={fill opacity=0.8, draw opacity=1, text opacity=1, at={(0.97,0.03)}, anchor=south east},
		scaled x ticks=manual:{}{\pgfmathparse{#1}},
		tick align=outside,
		tick pos=left,
		title={X-mode},
		x grid style={white!69.0196078431373!black},
		xmin=-0.25659857790235, xmax=0.225753081177974,
		xtick style={color=black},
		xticklabels={},
		y grid style={white!69.0196078431373!black},
		ylabel={angle / $^{\circ}$},
		ymin=-1.2, ymax=1.2,
		ytick style={color=black},
		width=6cm,
		height=6cm,
		scale only axis
		]
		\addplot [ultra thick, blue]
		table {%
		-0.234673502489608 0.874630929760089
		-0.227673632743631 0.86877635170175
		-0.220804845649404 0.86057395265187
		-0.214067797964014 0.849156081414709
		-0.207462695198663 0.833676349822357
		-0.200989319321115 0.813350270257272
		-0.194647113906197 0.787494572540022
		-0.188435187382464 0.75555472907799
		-0.182352316566019 0.717125822485689
		-0.176396950990916 0.671965475265852
		-0.170567218236238 0.619998841838205
		-0.164861087175742 0.561345083964878
		-0.15927619374263 0.496295876237735
		-0.153809912839053 0.425294993464939
		-0.148459564583947 0.348951825471245
		-0.143222414226678 0.268762662888547
		-0.138095672258192 0.189665326638972
		-0.133076494747171 0.112312861212898
		-0.128161983930599 0.0367730676976083
		-0.12334918908977 -0.037050921948575
		-0.118635107744001 -0.109064677222315
		-0.11401668719498 -0.178957203834634
		-0.109490826454649 -0.246275015429909
		-0.105054378588645 -0.310586775911267
		-0.100704153720616 -0.371528196244393
		-0.096437115274321 -0.428808330618343
		-0.0922503762163327 -0.482124750111451
		-0.0881409898750145 -0.531124812172009
		-0.0841060312743504 -0.575441261327136
		-0.080142595061213 -0.614704871422389
		-0.0762477933818524 -0.648554406037994
		-0.0724187537135119 -0.676639550484698
		-0.0686526166582495 -0.698639223817778
		-0.0649465337072617 -0.714275252696629
		-0.0612976649852623 -0.723322581748415
		-0.0577031769857397 -0.725640719080156
		-0.0541602403091568 -0.721186827964474
		-0.0506660274173584 -0.710009143504976
		-0.0472177104185365 -0.6922357720511
		-0.0438124588980487 -0.668065000718684
		-0.0404474378111471 -0.637755338545295
		-0.0371198054541634 -0.601617731318565
		-0.0338267115309066 -0.560004995757755
		-0.0305653108664209 -0.513328782507713
		-0.0273327488495939 -0.462059491001746
		-0.0241261347480264 -0.406684373918808
		-0.0209425586071458 -0.347719127817353
		-0.0177790917052044 -0.285701855839806
		-0.0146327872116972 -0.221186830721499
		-0.0115006810832822 -0.154738059245106
		-0.00837979318309875 -0.0869229786496011
		-0.00526712860420572 -0.0183062707048469
		-0.00215967861416491 0.050556027465556
		0 0.098303930000592
		0.000945652710336004 0.119121773953276
		0.00405175617128939 0.186868975447721
		0.00716174805633438 0.253300418814181
		0.0102786814084741 0.317948098587133
		0.0134056158016942 0.380377000307012
		0.0165456153311557 0.440188075816579
		0.0197017467242884 0.497020368090561
		0.022877080233018 0.550552202457933
		0.0260746882169646 0.60050167512141
		0.0292976447058656 0.646626155404167
		0.0325490255952428 0.688721104217195
		0.0358319084664504 0.726618239082661
		0.039149374835218 0.760182704659369
		0.0425045115764655 0.789309513087391
		0.0459004132538771 0.81391601897767
		0.049340185342625 0.833934605527126
		0.052826948185406 0.849303165801
		0.0563638416616025 0.859956503714908
		0.0599540306622225 0.865814863255358
		0.0636007113746453 0.866772617132792
		0.0673071180973269 0.86270154164019
		0.0710765301753148 0.853484167172967
		0.0749122783705316 0.839031099655475
		0.0788177523846231 0.819294103387887
		0.0827964070830047 0.794289388433047
		0.0868517695258439 0.764106772743884
		0.0909874477156181 0.728921142830067
		0.0952071367761354 0.689011016122334
		0.0995146255907414 0.644776044208182
		0.10391380581418 0.59669621607319
		0.108408680970405 0.545273589493808
		0.113003370834368 0.491015820045624
		0.117702122432847 0.434507625778836
		0.122509305568548 0.376636688297515
		0.127429402171262 0.318708745986084
		0.132467008100852 0.262312987055964
		0.137626835258604 0.208710418530448
		0.142913713480807 0.155737393360996
		0.148332465669281 0.0996694106736709
		0.153888153074875 0.0409110927593977
		0.159585943611869 -0.0203432237435839
		0.165430756244211 -0.0838769499176788
		0.171427250503216 -0.149464121906732
		0.177579818195765 -0.21687995092791
		0.183892576496236 -0.285909556665007
		0.190369362315784 -0.356355716123044
		0.197013671210967 -0.42804643841376
		0.203828005765232 -0.500841311195831
		};
		\addlegendentry{$\theta$}
		\addplot [ultra thick, red]
		table {%
		-0.234673502489608 -0.874199564354674
		-0.227673632743631 -0.838140082947912
		-0.220804845649404 -0.802524397923615
		-0.214067797964014 -0.766605332169096
		-0.207462695198663 -0.729721112466596
		-0.200989319321115 -0.691310817181218
		-0.194647113906197 -0.650931023976021
		-0.188435187382464 -0.60826037733457
		-0.182352316566019 -0.563098988188313
		-0.176396950990916 -0.515363355603313
		-0.170567218236238 -0.465077519256923
		-0.164861087175742 -0.412384011164568
		-0.15927619374263 -0.357515939521286
		-0.153809912839053 -0.300778214973242
		-0.148459564583947 -0.242557241988566
		-0.143222414226678 -0.183812320625726
		-0.138095672258192 -0.127758087732129
		-0.133076494747171 -0.0745823887730072
		-0.128161983930599 -0.0240951527304451
		-0.12334918908977 0.0239747237912473
		-0.118635107744001 0.0697485162156136
		-0.11401668719498 0.113191924218967
		-0.109490826454649 0.154170530111901
		-0.105054378588645 0.192556932527649
		-0.100704153720616 0.228257757520799
		-0.096437115274321 0.261217018740308
		-0.0922503762163327 0.291363229150208
		-0.0881409898750145 0.318586276895806
		-0.0841060312743504 0.342761061321718
		-0.080142595061213 0.363755761721181
		-0.0762477933818524 0.381438115606421
		-0.0724187537135119 0.395677327365204
		-0.0686526166582495 0.406354818864965
		-0.0649465337072617 0.413371950481222
		-0.0612976649852623 0.416655553457303
		-0.0577031769857397 0.416175520327853
		-0.0541602403091568 0.411951566425902
		-0.0506660274173584 0.404048813296899
		-0.0472177104185365 0.392571025557261
		-0.0438124588980487 0.377654975408317
		-0.0404474378111471 0.359464864339214
		-0.0371198054541634 0.338188136017762
		-0.0338267115309066 0.314029848847498
		-0.0305653108664209 0.287222762444135
		-0.0273327488495939 0.258027444137706
		-0.0241261347480264 0.226708860710757
		-0.0209425586071458 0.193543602537031
		-0.0177790917052044 0.158817020445225
		-0.0146327872116972 0.122820278491454
		-0.0115006810832822 0.085847312291613
		-0.00837979318309875 0.0481918716228418
		-0.00526712860420572 0.0101446294504004
		-0.00215967861416491 -0.0280095497965094
		0 -0.0544603646078471
		0.000945652710336004 -0.0659940006648043
		0.00405175617128939 -0.10354322989055
		0.00716174805633438 -0.140405019349221
		0.0102786814084741 -0.176342455303434
		0.0134056158016942 -0.211135644229954
		0.0165456153311557 -0.244583027071674
		0.0197017467242884 -0.276502265596113
		0.022877080233018 -0.306730659029584
		0.0260746882169646 -0.335125220945566
		0.0292976447058656 -0.361562250380605
		0.0325490255952428 -0.385936558137253
		0.0358319084664504 -0.408160353354449
		0.039149374835218 -0.42816159920618
		0.0425045115764655 -0.445881963739036
		0.0459004132538771 -0.461272519354806
		0.049340185342625 -0.474289496485942
		0.052826948185406 -0.484888688634157
		0.0563638416616025 -0.493020232342966
		0.0599540306622225 -0.498621548689108
		0.0636007113746453 -0.501610092047633
		0.0673071180973269 -0.501884222619577
		0.0710765301753148 -0.499341490248066
		0.0749122783705316 -0.493887976529595
		0.0788177523846231 -0.485445375017651
		0.0827964070830047 -0.473964413759392
		0.0868517695258439 -0.459430367260982
		0.0909874477156181 -0.441870325005118
		0.0952071367761354 -0.421365324094812
		0.0995146255907414 -0.398062757797985
		0.10391380581418 -0.372153565748989
		0.108408680970405 -0.343837478358408
		0.113003370834368 -0.313312993471653
		0.117702122432847 -0.28082332038777
		0.122509305568548 -0.246806893407172
		0.127429402171262 -0.21199002683965
		0.132467008100852 -0.177322066486083
		0.137626835258604 -0.143580588407206
		0.142913713480807 -0.109193953214956
		0.148332465669281 -0.0713396438746532
		0.153888153074875 -0.029946909432583
		0.159585943611869 0.0152590224373168
		0.165430756244211 0.0646070326147759
		0.171427250503216 0.11850106989069
		0.177579818195765 0.177442752798074
		0.183892576496236 0.242055931589866
		0.190369362315784 0.313113580390584
		0.197013671210967 0.391566573283001
		0.203828005765232 0.478568139231048
		};
		\addlegendentry{$\theta_m$}
		
		\nextgroupplot[
		legend cell align={left},
		legend style={fill opacity=0.8, draw opacity=1, text opacity=1, at={(0.5,0.91)}, anchor=north},
		tick align=outside,
		tick pos=left,
		x grid style={white!69.0196078431373!black},
		xlabel={($l - l_c$) / m},
		xmin=-0.25659857790235, xmax=0.225753081177974,
		xtick style={color=black},
		y grid style={white!69.0196078431373!black},
		ylabel={ratio},
		ymin=-2.05, ymax=-0.95,
		ytick style={color=black},
		width=6cm,
		height=3cm,
		scale only axis,
		clip mode=individual
		]
		\addplot [ultra thick, black, mark=*, mark size=3, mark options={solid}, only marks]
		table {%
		-0.234673502489608 -1.00049344042597
		-0.227673632743631 -1.03655268299075
		-0.220804845649404 -1.07233369462467
		-0.214067797964014 -1.1076835051643
		-0.207462695198663 -1.14245885939133
		-0.200989319321115 -1.17653340587619
		-0.194647113906197 -1.20979726504637
		-0.188435187382464 -1.24215674278978
		-0.182352316566019 -1.27353420540309
		-0.176396950990916 -1.30386739367453
		-0.170567218236238 -1.33310860268801
		-0.164861087175742 -1.36121932171824
		-0.15927619374263 -1.38817831983177
		-0.153809912839053 -1.41398203823629
		-0.148459564583947 -1.43863701042451
		-0.143222414226678 -1.46215804236428
		-0.138095672258192 -1.48456610462614
		-0.133076494747171 -1.50588983620147
		-0.128161983930599 -1.52616039039023
		-0.12334918908977 -1.54541600859241
		-0.118635107744001 -1.5636845504378
		-0.11401668719498 -1.5810068171334
		-0.109490826454649 -1.59741952791597
		-0.105054378588645 -1.61296075832881
		-0.100704153720616 -1.62766952711581
		-0.096437115274321 -1.64157884002438
		-0.0922503762163327 -1.65472064377382
		-0.0881409898750145 -1.66713022716203
		-0.0841060312743504 -1.67884082021506
		-0.080142595061213 -1.68988353205402
		-0.0762477933818524 -1.70028735855856
		-0.0724187537135119 -1.71007915715163
		-0.0686526166582495 -1.71928371803053
		-0.0649465337072617 -1.72792385130417
		-0.0612976649852623 -1.73602049881843
		-0.0577031769857397 -1.74359298814239
		-0.0541602403091568 -1.75065926856766
		-0.0506660274173584 -1.75723605697922
		-0.0472177104185365 -1.76333892973497
		-0.0438124588980487 -1.76898238927311
		-0.0404474378111471 -1.77417990411288
		-0.0371198054541634 -1.77894392867456
		-0.0338267115309066 -1.78328588130394
		-0.0305653108664209 -1.78721483680304
		-0.0273327488495939 -1.7907377742158
		-0.0241261347480264 -1.79386183955849
		-0.0209425586071458 -1.79659323924604
		-0.0177790917052044 -1.79893726150304
		-0.0146327872116972 -1.80089829984297
		-0.0115006810832822 -1.80247995090958
		-0.00837979318309875 -1.80368547065106
		-0.00526712860420572 -1.80452827718851
		-0.00215967861416491 -1.80495680340626
		0 -1.80505456965721
		0.000945652710336004 -1.80503943924112
		0.00405175617128939 -1.80474354185446
		0.00716174805633438 -1.80406954101949
		0.0102786814084741 -1.80301503707678
		0.0134056158016942 -1.80157643061316
		0.0165456153311557 -1.79974907125335
		0.0197017467242884 -1.79752728976391
		0.022877080233018 -1.79490437701838
		0.0260746882169646 -1.79187252283513
		0.0292976447058656 -1.78842275354655
		0.0325490255952428 -1.78454486805124
		0.0358319084664504 -1.780227386396
		0.039149374835218 -1.77545745827922
		0.0425045115764655 -1.77022076979403
		0.0459004132538771 -1.76450142773759
		0.049340185342625 -1.75828183357597
		0.052826948185406 -1.75154254101767
		0.0563638416616025 -1.74426209575246
		0.0599540306622225 -1.73641685870098
		0.0636007113746453 -1.7279808179188
		0.0673071180973269 -1.71892540701386
		0.0710765301753148 -1.70921940964483
		0.0749122783705316 -1.69882876184008
		0.0788177523846231 -1.68771636429351
		0.0827964070830047 -1.67584182561914
		0.0868517695258439 -1.66316122571373
		0.0909874477156181 -1.64962682846291
		0.0952071367761354 -1.63518680043852
		0.0995146255907414 -1.6197848996851
		0.10391380581418 -1.60336020124459
		0.108408680970405 -1.58584687189169
		0.113003370834368 -1.56717349831215
		0.117702122432847 -1.54726332976497
		0.122509305568548 -1.52603796068271
		0.127429402171262 -1.50341386685684
		0.132467008100852 -1.47930256089449
		0.137626835258604 -1.45361166746669
		0.142913713480807 -1.42624558206457
		0.148332465669281 -1.39711113288979
		0.153888153074875 -1.36612069607708
		0.159585943611869 -1.33319312080133
		0.165430756244211 -1.29826346332606
		0.171427250503216 -1.2612892191151
		0.177579818195765 -1.22225307885475
		0.183892576496236 -1.18117145399868
		0.190369362315784 -1.13810367368454
		0.197013671210967 -1.09316388992273
		0.203828005765232 -1.04654127623408
		};
		\addlegendentry{$ \theta / \theta_m$}
		\addplot [ultra thick, green]
		table {%
		-0.234673502489608 -1.00051105866107
		-0.227673632743631 -1.03713920201891
		-0.220804845649404 -1.07331586344712
		-0.214067797964014 -1.10891551633013
		-0.207462695198663 -1.14382467324344
		-0.200989319321115 -1.17794267324686
		-0.194647113906197 -1.21118178553116
		-0.188435187382464 -1.24346762283316
		-0.182352316566019 -1.27473927600688
		-0.176396950990916 -1.30494916939636
		-0.170567218236238 -1.33406265261913
		-0.164861087175742 -1.36205630366192
		-0.15927619374263 -1.38891820114551
		-0.153809912839053 -1.4146465646755
		-0.148459564583947 -1.43924742862614
		-0.143222414226678 -1.46273502782153
		-0.138095672258192 -1.48513638917668
		-0.133076494747171 -1.50647620153821
		-0.128161983930599 -1.52678182191469
		-0.12334918908977 -1.54608311487433
		-0.118635107744001 -1.56441244725667
		-0.11401668719498 -1.58180414922745
		-0.109490826454649 -1.59829381738948
		-0.105054378588645 -1.61391742622676
		-0.100704153720616 -1.62871076392769
		-0.096437115274321 -1.64270796365328
		-0.0922503762163327 -1.65594130215155
		-0.0881409898750145 -1.66844242971227
		-0.0841060312743504 -1.68024178812175
		-0.080142595061213 -1.69136850706771
		-0.0762477933818524 -1.70185033818234
		-0.0724187537135119 -1.71171363412552
		-0.0686526166582495 -1.72098329768817
		-0.0649465337072617 -1.7296827548903
		-0.0612976649852623 -1.73783394608704
		-0.0577031769857397 -1.74545725137368
		-0.0541602403091568 -1.75257147472983
		-0.0506660274173584 -1.75919389500488
		-0.0472177104185365 -1.76534033530586
		-0.0438124588980487 -1.77102522871875
		-0.0404474378111471 -1.77626168449884
		-0.0371198054541634 -1.78106154501223
		-0.0338267115309066 -1.78543544958932
		-0.0305653108664209 -1.78939268413453
		-0.0273327488495939 -1.79294134760531
		-0.0241261347480264 -1.79608872036438
		-0.0209425586071458 -1.79884108508322
		-0.0177790917052044 -1.80120375226717
		-0.0146327872116972 -1.80318108255586
		-0.0115006810832822 -1.8047765056374
		-0.00837979318309875 -1.80599253477462
		-0.00526712860420572 -1.80683077726419
		-0.00215967861416491 -1.80729194020667
		0 -1.80739028500622
		0.000945652710336004 -1.80737583618919
		0.00405175617128939 -1.8070813904636
		0.00716174805633438 -1.80640662432447
		0.0102786814084741 -1.80534864255997
		0.0134056158016942 -1.80390361702764
		0.0165456153311557 -1.8020667649468
		0.0197017467242884 -1.79983232101246
		0.022877080233018 -1.79719348508514
		0.0260746882169646 -1.79414238374155
		0.0292976447058656 -1.79067001349804
		0.0325490255952428 -1.78676617797873
		0.0358319084664504 -1.78241942136754
		0.039149374835218 -1.77761693450163
		0.0425045115764655 -1.77234446025364
		0.0459004132538771 -1.7665861713508
		0.049340185342625 -1.76032453512497
		0.052826948185406 -1.75354015709212
		0.0563638416616025 -1.74621161379418
		0.0599540306622225 -1.73831526030068
		0.0636007113746453 -1.72982502283814
		0.0673071180973269 -1.72071222710982
		0.0710765301753148 -1.71094551223446
		0.0749122783705316 -1.70049067378191
		0.0788177523846231 -1.68931046392872
		0.0827964070830047 -1.67736441529023
		0.0868517695258439 -1.66460859560782
		0.0909874477156181 -1.65099534670823
		0.0952071367761354 -1.63647306154746
		0.0995146255907414 -1.62098595980952
		0.10391380581418 -1.60447370126393
		0.108408680970405 -1.58687102666785
		0.113003370834368 -1.56810766065916
		0.117702122432847 -1.54810850907361
		0.122509305568548 -1.5267944734667
		0.127429402171262 -1.50408306057463
		0.132467008100852 -1.47988860960043
		0.137626835258604 -1.45412199873127
		0.142913713480807 -1.42668660169369
		0.148332465669281 -1.39748672389961
		0.153888153074875 -1.36643878175842
		0.159585943611869 -1.3334664852423
		0.165430756244211 -1.29850798923393
		0.171427250503216 -1.26152083935078
		0.177579818195765 -1.22248744909027
		0.183892576496236 -1.18142091452067
		0.190369362315784 -1.13837086043608
		0.197013671210967 -1.09342935934081
		0.203828005765232 -1.04674076243109
		};
		\addlegendentry{$r_X$}
		\end{groupplot}
		
		\end{tikzpicture}

		\end{subfigure}
								
	\caption{The variation of $\theta$ and $\theta_m$ along the ray (top), and how they relate to each other (bottom). Here $l-l_c$ is the distance along the ray from the cut-off. The standard MAST test scenario for our paper was used for the O-mode. For the X-mode, the toroidal launch angle was changed to $-7^{\circ}$ such that $\theta_m = 0$ at the cut-off. In the lower subplots, we perform a sanity check; as expected, we have agreement between the ratio of $\theta$ to $\theta_m$ as calculated directly from these angles (black points) and from the right-hand sides of equations (\ref{eq:ratio_O_mode}) and (\ref{eq:ratio_X_mode}) (green lines). } 
		\label{fig:theta_thetam_MAST}
	\end{figure}
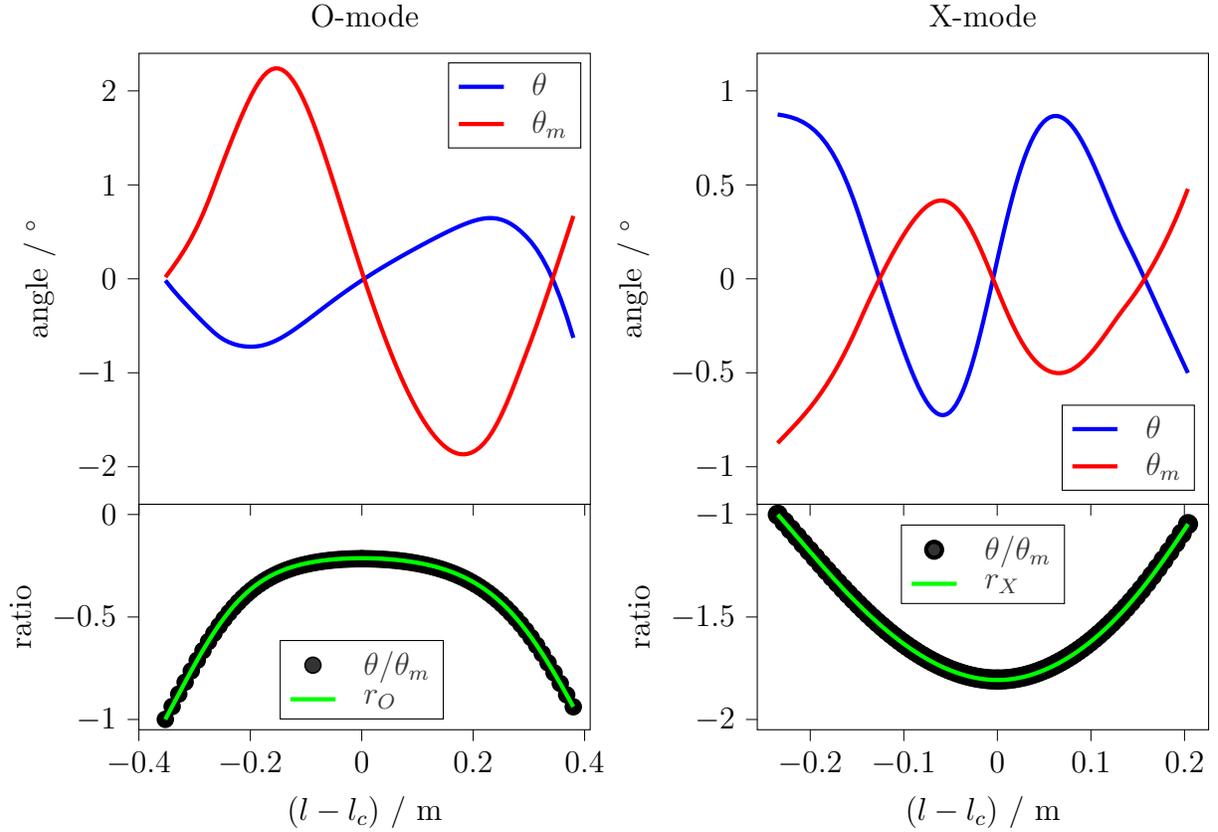

	Now that we have the ratio between $\theta_{m}$ and $\theta$, we discuss what it means for the O and X modes. We have
	\begin{equation} \label{eq:ratio_O_mode}
		\frac{\theta}{\theta_m}
		= 
		- 1
		+ \frac{\Omega_{pe}^{2}}{\Omega^{2}}
		,
	\end{equation}
	for the O-mode and
	\begin{equation} \label{eq:ratio_X_mode}
		\frac{\theta}{\theta_m}
		= 
		- 1
		- \frac{\Omega_{pe}^{2} \left(\Omega^{2} - \Omega_{pe}^{2}\right)}{\Omega^{2} \left(\Omega^{2} - \Omega_{ce}^{2} - \Omega_{pe}^{2}\right)} ,
	\end{equation}	
	for the X-mode, Figure \ref{fig:theta_thetam_MAST}. For the O-mode, since we launch a beam into the plasma from vacuum, $\Omega > \Omega_{pe}$ always, and thus $\theta / \theta_m \leq 0$ and $| \theta / \theta_m | \leq 1$. However, for the X-mode, the cut-off frequency is $\Omega_R = (\Omega_{ce}^2/4 + \Omega_{pe}^2)^{1 / 2} + \Omega_{ce}/2$, which is above the upper hybrid frequency $\Omega_{UH}^2 = \Omega_{ce}^{2} + \Omega_{pe}^{2}$. Consequently, we still have $\theta / \theta_m \leq 0$, but now $| \theta / \theta_m | > 1$. 
	
\begin{figure}
		\centering
		
		\begin{subfigure}{.49\textwidth}
		\centering
		\begin{tikzpicture}[>=latex]
		\tikzmath{
					\thet = -25; \thetmO = 15; \thetmX = 35; 
				 }
		
		\draw[style=help lines] (0,0) (3,2);
		
		\coordinate (g_hat) at (90:2.5);
		\coordinate (x_hat) at (0:2.5);
		\coordinate (b_hat) at (180+\thet:2.5);
		\coordinate (u1_hat) at (90 + \thet:2.5);
		\coordinate (K_vec_O) at (90 + \thet + \thetmO:3.5);
		\coordinate (K_vec_X) at (90 + \thet + \thetmX:3.5);		
		\coordinate (vec6) at (180:2.5);
		
		\draw[->,thick,black] (0,0) -- (g_hat) node[above] {$\hat{\mathbf{g}}$};
		\draw[->,thick,black] (0,0) -- (x_hat) node[right] {$\hat{\mathbf{x}}$};
		\draw[->,thick,DarkBlue] (0,0) -- (b_hat) node [left] {$\hat{\mathbf{b}}$};
		\draw[->,thick,DarkBlue] (0,0) -- (u1_hat) node [above] {$\hat{\mathbf{u}}_{1}$};
		\draw[->,thick,DarkRed] (0,0) -- (K_vec_O) node [above] {$\mathbf{K}_{O}$};
		\draw[->,thick,DarkRed] (0,0) -- (K_vec_X) node [above] {$\mathbf{K}_{X}$};		
		\draw[dashed,thick,black] (0,0) -- (vec6) ;
		
		\draw [->,red,thick,domain=90+\thet:90+\thet+\thetmO] plot ({1.25*cos(\x)}, {1.25*sin(\x)});
		\draw [->,red,thick,domain=90+\thet:90+\thet+\thetmX] plot ({2.0*cos(\x)}, {2.0*sin(\x)});		
		\draw [->,blue,thick,domain=180:180+\thet] plot ({1.5*cos(\x)}, {1.5*sin(\x)}); 
		\node[above left,blue] at (-1.5,0.0) {$\theta$};
		
		\draw (0,.2)-|(.2,0);
		\draw [DarkBlue] ({0.4*cos(180+\thet)}, {0.4*sin(180+\thet)}) -- ({0.4*cos(180+\thet)-0.4*cos(90-\thet)}, {0.4*sin(180+\thet)+0.4*sin(90-\thet)}) -- ({0.4*cos(90+\thet)}, {0.4*sin(90+\thet)});
		
		\draw[DarkBlue,thick] (0,-0.5) circle (0.2cm) node[below right] {$\hat{\mathbf{u}}_{2} = \hat{\mathbf{y}}$};
		\draw[DarkBlue,thick] (-0.14,-0.64) -- (0.14,-0.36);
		\draw[DarkBlue,thick] (0.14,-0.64) -- (-0.14,-0.36);
		
		\end{tikzpicture} \\
		\caption{$K_X$ and $K_O$ when $\theta < 0$.}
		\end{subfigure}
		\begin{subfigure}{.49\textwidth}
		\centering
		\begin{tikzpicture}[>=latex]
		\tikzmath{
					\thet = 25; \thetmO = -15; \thetmX = -35; 
				 }
		
		\draw[style=help lines] (0,0) (3,2);
		
		\coordinate (g_hat) at (90:2.5);
		\coordinate (x_hat) at (0:2.5);
		\coordinate (b_hat) at (180+\thet:2.5);
		\coordinate (u1_hat) at (90 + \thet:2.5);
		\coordinate (K_vec_O) at (90 + \thet + \thetmO:3.5);
		\coordinate (K_vec_X) at (90 + \thet + \thetmX:3.5);		
		\coordinate (vec6) at (180:2.5);
		
		\draw[->,thick,black] (0,0) -- (g_hat) node[above] {$\hat{\mathbf{g}}$};
		\draw[->,thick,black] (0,0) -- (x_hat) node[right] {$\hat{\mathbf{x}}$};
		\draw[->,thick,DarkBlue] (0,0) -- (b_hat) node [left] {$\hat{\mathbf{b}}$};
		\draw[->,thick,DarkBlue] (0,0) -- (u1_hat) node [above] {$\hat{\mathbf{u}}_{1}$};
		\draw[->,thick,DarkRed] (0,0) -- (K_vec_O) node [above] {$\mathbf{K}_{O}$};
		\draw[->,thick,DarkRed] (0,0) -- (K_vec_X) node [above] {$\mathbf{K}_{X}$};		
		\draw[dashed,thick,black] (0,0) -- (vec6) ;
		
		\draw [->,red,thick,domain=90+\thet:90+\thet+\thetmO] plot ({1.25*cos(\x)}, {1.25*sin(\x)});
		\draw [->,red,thick,domain=90+\thet:90+\thet+\thetmX] plot ({2.0*cos(\x)}, {2.0*sin(\x)});		
		\draw [->,blue,thick,domain=180:180+\thet] plot ({1.5*cos(\x)}, {1.5*sin(\x)}); 
		\node[blue,below left] at (-1.5,0.0) {$\theta$};
		
		\draw (0,.2)-|(.2,0);
		\draw [DarkBlue] ({0.4*cos(180+\thet)}, {0.4*sin(180+\thet)}) -- ({0.4*cos(180+\thet)-0.4*cos(90-\thet)}, {0.4*sin(180+\thet)+0.4*sin(90-\thet)}) -- ({0.4*cos(90+\thet)}, {0.4*sin(90+\thet)});
		
		\draw[DarkBlue,thick] (0,-0.5) circle (0.2cm) node[below right] {$\hat{\mathbf{u}}_{2} = \hat{\mathbf{y}}$};
		\draw[DarkBlue,thick] (-0.14,-0.64) -- (0.14,-0.36);
		\draw[DarkBlue,thick] (0.14,-0.64) -- (-0.14,-0.36);
		
		\end{tikzpicture} \\
		\caption{$K_X$ and $K_O$ when $\theta > 0$.}
		\end{subfigure}
								
		\caption{The wavevectors of the X-mode and O-mode are denoted by $K_X$ and $K_O$ respectively. This notation is only used in this section. We have shown that for the X-mode, $| \theta | > | \theta_m |$, and the O-mode, $|\theta| < |\theta_m|$.  The unlabelled bright red arrows show $\theta_m$. Note that $\theta$ and $\theta_m$ always have different signs, for reasons explained in this section. The angles used in this figure were chosen for illustration purposes: the real values of $\theta$ and $\theta_m$ for which our expansion is valid are too small for legible schematics.} 
		\label{fig:theta_thetam_illustration}
	\end{figure}
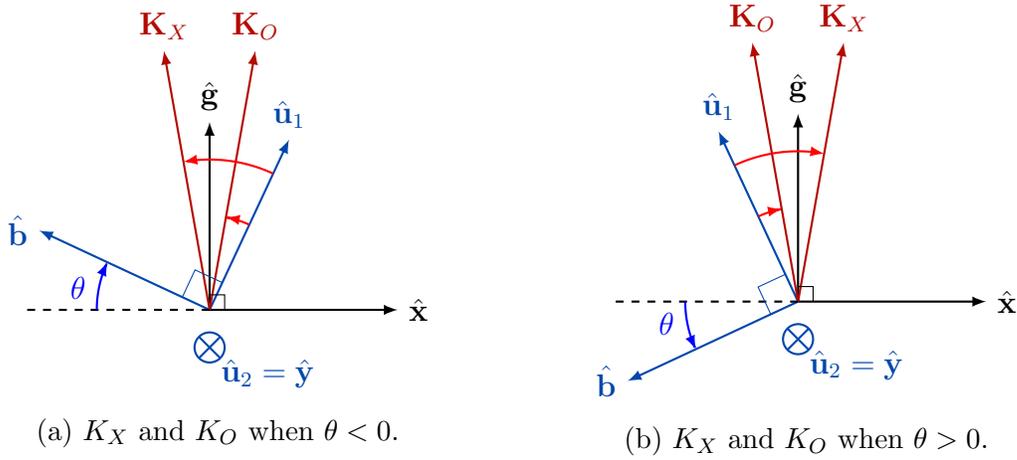

	Since $\theta / \theta_m$ is always negative, $\theta$ and $\theta_m$ have different signs no matter the mode. For the O-mode, the magnitude of this ratio is always less than one, so $| \theta | < | \theta_m |$, and for the X-mode, the magnitude of the same ratio is always more than one, so $| \theta | > | \theta_m |$. This is illustrated in Figure \ref{fig:theta_thetam_illustration} for both positive and negative $\theta$.

	\section{Region near the cut-off} \label{appendix_airy_function} 
	We use $\tau_c$ to denote the location where $\rmd K / \rmd \tau = 0$. The subscript $_c$ means that the function is evaluated at $\tau = \tau_c$, the nominal cut-off location.

	We call $k_{c,1}$ the particular $k_{\perp,1}$ which scatters at $\tau_c$,
	\begin{equation} \label{eq:definition_ks}
		k_{c,1} = - 2 K_{c} = -2 K (\tau_c) .
	\end{equation}
	We Taylor expand the large phase piece $2 \rmi s + \rmi k_{\perp, 1} \int_0^\tau  g (\tau') \ \rmd \tau'$ of equation (\ref{eq:A_r_after_wx_wy_integration}) around $k_{c,1}$ and $\tau_c$ to determine what happens at the scattering location $\tau_c$,
	\begin{equation} \label{eq:phase_2D_expansion}
	\fl
	\eqalign{
		2 s + k_{\perp, 1} \int_0^\tau g (\tau') \ \rmd \tau
		&= 2 s_c + k_{c,1} \int_0^{\tau_c} g (\tau') \ \rmd \tau
		+ (k_{\perp, 1} - k_{c,1}) \int_0^{\tau_c} g (\tau') \ \rmd \tau' \\
		& + g_c (k_{\perp, 1} - k_{c,1}) (\tau - \tau_c)
		+ \frac{1}{3} g_c \frac{\rmd^2 K_g}{\rmd \tau^2} \Bigr|_{\tau_c} (\tau - \tau_c)^3 .
	}
	\end{equation}
	We order
	\begin{equation} \label{eq:ordering_taus-tau}
		g_c \frac{\rmd^2 K_g}{\rmd \tau^2} \Bigr|_{\tau_c} (\tau - \tau_c)^3
		\sim \frac{L}{\lambda} \left( \frac{\tau - \tau_c}{\tau_L} \right)^3
		\sim 1 ,
	\end{equation}
	and
	\begin{equation} \label{eq:ordering_ks-k}
		g_c (k_{\perp, 1} - k_{c,1}) (\tau - \tau_c)
		\sim (k_{\perp, 1} - k_{c,1}) L \frac{\tau - \tau_c}{\tau_L}
		\sim 1 ,
	\end{equation}
	where $g \tau_L \sim L$. With these orderings, we can safely neglect the terms of the Taylor expansion of the phase that we have not included in equation (\ref{eq:phase_2D_expansion}). Equations (\ref{eq:ordering_taus-tau}) and (\ref{eq:ordering_ks-k}) give $(\tau - \tau_c) / \tau_L \sim (\lambda / L)^{1 / 3} \ll 1$ and $(k_{\perp,1} - k_{c,1}) / k_{c,1} \sim (\lambda / L)^{2 / 3} \ll 1$. As a result, functions that do not oscillate quickly with $\tau$ are evaluated only at $\tau_c$. Then, using the integration variable
	\begin{equation}
		\xi = \left( g_c \frac{\rmd^2 K_g}{\rmd \tau^2} \Bigr|_{\tau_c} \right)^{\frac{1}{3}} (\tau - \tau_c),
	\end{equation}
	equation (\ref{eq:A_r_after_wx_wy_integration}) becomes
	\begin{equation} \label{eq:A_r_Airy}
	\fl
	\eqalign{
		\tilde{A}_r
		&= - \frac{\Omega A_{ant} g_{ant} \pi}{c} \int \left[
		\frac{\det \left[\textrm{Im} \left( \bm{\Psi}_{w,c} \right) \right]}{\det\left( \bm{M}_{w,c} \right) }
		\right]^{\frac{1}{2}} \exp (2 \rmi \phi_{G,c} ) \\
		& \times \frac{ \delta \tilde{n}_{e,c} }{n_{e,c}} \ \hat{\mathbf{e}}^*_c
		\cdot (\bm{\epsilon}_{eq,c} -\bm{1}) \cdot \hat{\mathbf{e}}_c
		\ \exp \left[
			2 \rmi s_t 
			+ \rmi k_{\perp,1} \int_0^{\tau_t} g (\tau') \left( 1 - \frac{\theta^2 (\tau')}{2} \right) \ \rmd \tau' 
		\right] \\
		& \times \left( g_t \frac{\rmd^2 K_g}{\rmd \tau^2} \Bigr|_{\tau_t} \right)^{- \frac{1}{3}} \textrm{Ai} \left[
			(k_{\perp, 1} - k_{t,1}) g_t^{ \frac{2}{3}} \left( \frac{\rmd^2 K_g}{\rmd \tau^2} \Bigr|_{\tau_t} \right)^{- \frac{1}{3}}
		\right] \\
		& \times \exp\left[
		- \frac{\rmi}{4} (2 \mathbf{K}_{w,c} + \mathbf{k}_{\perp,w,c} )
		\cdot \bm{M}_{w,c}^{-1} \cdot
		(2 \mathbf{K}_{w,c} + \mathbf{k}_{\perp,w,c} )
		\right]
		\ \rmd k_{\perp,1} \ \rmd k_{\perp,2} .
	}
	\end{equation}
	where $\textrm{Ai}(z) = (2 \pi)^{-1} \int_{- \infty}^{\infty} \exp ( z \xi + \xi^3 / 3 ) \rmd \xi$ is the Airy function. We have thus shown that there is indeed no divergence at the cut-off.


	\section*{References}
	\bibliographystyle{iopart-num}
	\bibliography{references}

\providecommand{\newblock}{}
\begin{thebibliography}{10}
\expandafter\ifx\csname url\endcsname\relax
  \def\url#1{{\tt #1}}\fi
\expandafter\ifx\csname urlprefix\endcsname\relax\def\urlprefix{URL }\fi
\providecommand{\eprint}[2][]{\url{#2}}

\bibitem{Hillesheim:DBS:2009}
Hillesheim J~C, Peebles W~A, Rhodes T~L, Schmitz L, Carter T~A, Gourdain P~A
  and Wang G 2009 {\em Review of Scientific Instruments\/} {\bf 80} ISSN
  00346748

\bibitem{Hirsch:DBS:2001}
Hirsch M, Holzhauer E, Baldzuhn J, Kurzan B and Scott B 2001 {\em Plasma
  Physics and Controlled Fusion\/} {\bf 43} 1641--1660 ISSN 07413335

\bibitem{Hillesheim:DBS_rotation:2015}
Hillesheim J~C, Parra F~I, Barnes M, Crocker N~A, Meyer H, Peebles W~A,
  Scannell R and Thornton A 2015 {\em Nuclear Fusion\/} {\bf 55} 032003 ISSN
  0029-5515 (\textit{Preprint} \eprint{1407.2121v2})

\bibitem{Conway:DBS_Flow:2005}
Conway G~D, Scott B, Schirmer J, Reich M, Kendl A and Team t~A~U 2005 {\em
  Plasma Physics and Controlled Fusion\/} {\bf 47} 1165--1185 ISSN 0741-3335

\bibitem{Conway:flows:2010}
Conway G~D, Poli E, Happel T {\em et~al.\/} 2010 {\em Plasma and Fusion
  Research\/} {\bf 5} S2005--S2005

\bibitem{Schmitz:LH:2017}
Schmitz L 2017 {\em Nuclear Fusion\/} {\bf 57} 025003

\bibitem{Tynan:drift_turbulence:2009}
Tynan G~R, Fujisawa A and McKee G 2009 {\em Plasma Physics and Controlled
  Fusion\/} {\bf 51} 113001

\bibitem{Hennequin:DBS:2004}
Hennequin P, Honor{\'e} C, Truc A, Qu{\'e}m{\'e}neur A, Lemoine N, Chareau J~M
  and Sabot R 2004 {\em Review of Scientific Instruments\/} {\bf 75} 3881--3883

\bibitem{Happel:DBS:2009}
Happel T, Estrada T, Blanco E, Tribaldos V, Cappa A and Bustos A 2009 {\em
  Review of Scientific Instruments\/} {\bf 80} 073502

\bibitem{Zhou:DBS:2013}
Zhou C, Liu A~D, Zhang X~H, Hu J~Q, Wang M~Y, Li H, Lan T, Xie J~L, Sun X, Ding
  W~X {\em et~al.\/} 2013 {\em Review of Scientific Instruments\/} {\bf 84}
  103511

\bibitem{Happel:DBS_synthetic:2017}
Happel T, G{\"o}rler T, Hennequin P, Lechte C, Bernert M, Conway G~D, Freethy
  S~J, Honor{\'e} C, Pinz{\'o}n J, Stroth U {\em et~al.\/} 2017 {\em Plasma
  Physics and Controlled Fusion\/} {\bf 59} 054009

\bibitem{Shi:DBS:2016}
Shi Z, Zhong W, Jiang M, Yang Z, Zhang B, Shi P, Chen W, Wen J, Chen C, Fu B
  {\em et~al.\/} 2016 {\em Review of Scientific Instruments\/} {\bf 87} 113501

\bibitem{Rhodes:DBS:2016}
Rhodes T~L, Barada K, Peebles W~A and Crocker N~A 2016 {\em Review of
  Scientific Instruments\/} {\bf 87} 11E726

\bibitem{Hu:DBS:2017}
Hu J, Zhou C, Liu A, Wang M, Doyle E, Peebles W, Wang G, Zhang X, Zhang J, Feng
  X {\em et~al.\/} 2017 {\em Review of Scientific Instruments\/} {\bf 88}
  073504

\bibitem{Tokuzawa:DBS_LHD:2018}
Tokuzawa T, Tsuchiya H, Tsujimura T, Emoto M, Nakanishi H, Inagaki S, Ida K,
  Yamada H, Ejiri A, Watanabe K~Y {\em et~al.\/} 2018 {\em Review of Scientific
  Instruments\/} {\bf 89} 10H118

\bibitem{MolinaCabrera:DBS_TCV:2019}
Molina~Cabrera P, Coda S, Porte L, Smolders A and Team T 2019 {\em Review of
  Scientific Instruments\/} {\bf 90} 123501

\bibitem{Wen:DBS:2021}
Wen J, Shi Z, Zhong W, Yang Z, Yang Z, Wang B, Jiang M, Shi P, Hillesheim J,
  Freethy S {\em et~al.\/} 2021 {\em Review of Scientific Instruments\/} {\bf
  92} 063513

\bibitem{Tokuzawa:DBS_LHD:2021}
Tokuzawa T, Tanaka K, Tsujimura T, Kubo S, Emoto M, Inagaki S, Ida K, Yoshinuma
  M, Watanabe K, Tsuchiya H {\em et~al.\/} 2021 {\em Review of Scientific
  Instruments\/} {\bf 92} 043536

\bibitem{Carralero:DBS_JT60SA:2021}
Carralero D, Happel T, Estrada T, Tokuzawa T, Mart{\'\i}nez J, de~la Luna E,
  Cappa A and Garc{\'\i}a J 2021 {\em Fusion Engineering and Design\/} {\bf
  173} 112803

\bibitem{Yashin:DBS:2021}
Yashin A, Bulanin V, Petrov A and Ponomarenko A 2021 {\em Applied Sciences\/}
  {\bf 11} 8975

\bibitem{Volpe:microwave:2017}
Volpe F~A 2017 {\em Journal of Instrumentation\/} {\bf 12} C01094

\bibitem{Costley:new_diagnostics:2010}
Costley A~E 2010 {\em IEEE Transactions on Plasma Science\/} {\bf 38}
  2934--2943

\bibitem{Piliya:reciprocity:2002}
Piliya A~D and Popov A~Y 2002 {\em Plasma Physics and Controlled Fusion\/} {\bf
  44} 467

\bibitem{Gusakov:scattering_slab:2004}
Gusakov E~Z and Surkov A~V 2004 {\em Plasma Physics and Controlled Fusion\/}
  {\bf 46} 1143

\bibitem{Bulanin:spatial_spectral_resolution:2006}
Bulanin V~V and Yafanov M~V 2006 {\em Plasma Physics Reports\/} {\bf 32} 47--55

\bibitem{Gusakov:1D_RCDR:2011}
Gusakov E~Z and Kosolapova N~V 2011 {\em Plasma Physics and Controlled
  Fusion\/} {\bf 53} 045012
  \urlprefix\url{https://doi.org/10.1088/0741-3335/53/4/045012}

\bibitem{Krutkin:DBS_synthetic:2019}
Krutkin O, Altukhov A~B, Gurchenko A~D, Gusakov E~Z, Irzak M~A, Esipov L~A,
  Sidorov A, Ch{\^o}n{\'e} L, Kiviniemi T~P, Leerink S {\em et~al.\/} 2019 {\em
  Nuclear Fusion\/}

\bibitem{Catto:GK:1978}
Catto P~J 1978 {\em Plasma Physics\/} {\bf 20} 719

\bibitem{Frieman:GK:1982}
Frieman E and Chen L 1982 {\em The Physics of Fluids\/} {\bf 25} 502--508

\bibitem{Hillesheim:DBS_MAST:2015}
Hillesheim J~C, Crocker N~A, Peebles W~A, Meyer H, Meakins A, Field A~R, Dunai
  D, Carr M, Hawkes N, Team M {\em et~al.\/} 2015 {\em Nuclear Fusion\/} {\bf
  55} 073024

\bibitem{Coelho:Pitch:2009}
Coelho R, Alves D, Hawkes N, Brix M and Contributors J~E 2009 {\em Review of
  Scientific Instruments\/} {\bf 80} 063504

\bibitem{Ko:Pitch:2016}
Ko J 2016 {\em Review of Scientific Instruments\/} {\bf 87} 11E541

\bibitem{Damba:mismatch:2021}
Damba J, Hong R, Pratt Q and Rhodes T 2021 {\em Bulletin of the American
  Physical Society\/}

\bibitem{Stix:Waves:1992}
Stix T~H 1992 {\em Waves in plasmas\/} (Springer Science \& Business Media)

\bibitem{Hirsch:2D_fullwave:2001}
Hirsch M, Holzhauer E, Baldzuhn J, Kurzan B and Scott B 2001 {\em Plasma
  Physics and Controlled Fusion\/} {\bf 43} 1641

\bibitem{Silva:2D_fullwave:2004}
Silva F~d, Heuraux S, Lemoine N, Honor{\'e} C, Hennequin P, Manso M and Sabot R
  2004 {\em Review of Scientific Instruments\/} {\bf 75} 3816--3818

\bibitem{Hillesheim:2D_fullwave:2012}
Hillesheim J, Holland C, Schmitz L, Kubota S, Rhodes T and Carter T 2012 {\em
  Review of Scientific Instruments\/} {\bf 83} 10E331

\bibitem{Williams:3D_fullwave:2014}
Williams T, K{\"o}hn A, O'Brien M and Vann R 2014 {\em Plasma Physics and
  Controlled Fusion\/} {\bf 56} 075010

\bibitem{Peysson:C3PO:2012}
Peysson Y, Decker J and Morini L 2012 {\em Plasma Physics and Controlled
  Fusion\/} {\bf 54} 045003

\bibitem{Marushchenko:TRAVIS:2007}
Marushchenko N~B, Erckmann V, Hartfuss H~J, Hirsch M, Laqua H~P, Maassberg H
  and Turkin Y 2007 {\em Plasma and Fusion Research\/} {\bf 2} S1129

\bibitem{Farina:GRAY:2007}
Farina D 2007 {\em Fusion Science and Technology\/} {\bf 52} 154--160

\bibitem{Prater:benchmarking_codes:2008}
Prater R, Farina D, Gribov Y, Harvey R~W, Ram A, Lin-Liu Y~R, Poli E, Smirnov
  A~P, Volpe F, Westerhof E {\em et~al.\/} 2008 {\em Nuclear Fusion\/} {\bf 48}
  035006

\bibitem{Honore:quasioptics:2006}
Honor{\'e} C, Hennequin P, Truc A and Qu{\'e}m{\'e}neur A 2006 {\em Nuclear
  fusion\/} {\bf 46} S809

\bibitem{Casperson:beam_tracing:1973}
Casperson L~W 1973 {\em Applied optics\/} {\bf 12} 2434--2441

\bibitem{Cerveny:beam_tracing:1982}
{\v{C}}erven{\`y} V, Popov M~M and P{\v{s}}en{\v{c}}{\'\i}k I 1982 {\em
  Geophysical Journal International\/} {\bf 70} 109--128

\bibitem{Kravtsov:beam_tracing:2007}
Kravtsov Y~A and Berczynski P 2007 {\em Studia Geophysica et Geodaetica\/} {\bf
  51} 1--36

\bibitem{Peeters:Beam_Tracing:1996}
Peeters A~G 1996 {\em Physics of Plasmas\/} {\bf 3} 4386--4395

\bibitem{Pereverzev:Beam_tracing:1992}
Pereverzev G 1992 {\em Nuclear fusion\/} {\bf 32} 1091

\bibitem{Pereverzev:Beam_tracing:1993}
Pereverzev G~V 1993

\bibitem{Pereverzev:Beam_Tracing:1996}
Pereverzev G~V 1996 Paraxial wkb solution of a scalar wave equation {\em
  Reviews of Plasma Physics: Volume 19\/} ed Kadomtsev B~B (Springer) pp 1--48

\bibitem{Pereverzev:Beam_tracing:1998}
Pereverzev G~V 1998 {\em Physics of Plasmas\/} {\bf 5} 3529--3541

\bibitem{Poli:paraxial:1999}
Poli E, Pereverzev G~V and Peeters A~G 1999 {\em Physics of Plasmas\/} {\bf 6}
  5--11

\bibitem{Poli:beam_tracing_BC:2001}
Poli E, Peeters A~G and Pereverzev G~V 2001 {\em Physics of Plasmas\/} {\bf 8}
  4325--4330

\bibitem{Poli:Torbeam:2001}
Poli E, Peeters A~G and Pereverzev G~V 2001 {\em Computer Physics
  Communications\/} {\bf 136} 90--104

\bibitem{Ramponi:ITER_EC:2008}
Ramponi G, Farina D, Henderson M~A, Poli E, Sauter O, Saibene G, Zohm H and
  Zucca C 2008 {\em Nuclear Fusion\/} {\bf 48} 054012

\bibitem{Bertelli:ECCD:2010}
Bertelli N, Balakin A, Westerhof E and Buyanova M 2010 {\em Nuclear Fusion\/}
  {\bf 50} 115008

\bibitem{Rodrigues:LHCD:2002}
Rodrigues P and Bizarro J~P 2002 {\em IEEE transactions on plasma science\/}
  {\bf 30} 68--69

\bibitem{Thomas:SAMI:2016}
Thomas D~A 2016 {\em Phased array imaging of two dimensional Doppler microwave
  backscattering from spherical tokamak edge plasmas\/} Ph.D. thesis University
  of York

\bibitem{Stegmeir:reflectometry:2011}
Stegmeir A, Conway G~D, Poli E and Strumberger E 2011 {\em Fusion Engineering
  and Design\/} {\bf 86} 2928--2942

\bibitem{Hillesheim:zonal_flow:2016}
Hillesheim J~C, Delabie E, Meyer H, Maggi C~F, Meneses L, Poli E, Contributors
  J, Consortium E {\em et~al.\/} 2016 {\em Physical Review Letters\/} {\bf 116}
  065002

\bibitem{Maj:Beam_Tracing:2009}
Maj O, Pereverzev G~V and Poli E 2009 {\em Physics of Plasmas\/} {\bf 16}
  062105

\bibitem{Fidone:ModeConversion:1971}
Fidone I and Granata G 1971 {\em Nuclear Fusion\/} {\bf 11} 133

\bibitem{Boyd:ModeConversion:1985}
Boyd D~A 1985 {\em Proceedings of the Fifth Joint Workshop on Electron
  Cyclotron Emission and Electron Cyclotron Resonant Heating\/}  77

\bibitem{Donne:ModeConversion:1997}
Donn{\'e} A, De~Baar M and Cavazzana R 1997 {\em Review of scientific
  instruments\/} {\bf 68} 473--476

\bibitem{Minami:ModeConversion:1998}
Minami K, Ejiri A, Tanaka K and Watanabe T 1998 {\em Japanese journal of
  applied physics\/} {\bf 37} 6601

\bibitem{Nagasaki:ModeConversion:1999}
Nagasaki K, Ejiri A, Mizuuchi T, Obiki T, Okada H, Sano F, Zushi H, Besshou S
  and Kondo K 1999 {\em Physics of Plasmas\/} {\bf 6} 556--564

\bibitem{Tokuzawa:ModeConversion:2003}
Tokuzawa T, Kawahata K, Tanaka K, Nagayama Y, Group L~E, Kaneba T and Ejiri A
  2003 {\em Review of scientific instruments\/} {\bf 74} 1506--1509

\bibitem{Tokuzawa:ModeConversion:2010}
Tokuzawa T, Kawahata K, Nagayama Y, Inagaki S, De~Vries P, Mase A, Kogi Y,
  Yokota Y, Hojo H, Tanaka K {\em et~al.\/} 2010 {\em Fusion Science and
  Technology\/} {\bf 58} 364--374

\bibitem{Booker:propagation:1936}
Booker H~G 1936 {\em Proceedings of the Royal Society of London. Series
  A-Mathematical and Physical Sciences\/} {\bf 155} 235--257

\bibitem{Booker:propagation:1938}
Booker H~G 1938 {\em Philosophical Transactions of the Royal Society of London.
  Series A, Mathematical and Physical Sciences\/} {\bf 237} 411--451

\bibitem{Lao:EFIT:1985}
Lao L, John H~S, Stambaugh R, Kellman A and Pfeiffer W 1985 {\em Nuclear
  fusion\/} {\bf 25} 1611

\bibitem{Appel:EFIT:2006}
Appel L, Huysmans G, Lao L, McCarthy P, Muir D, Solano E, Storrs J, Taylor D,
  Zwingmann W {\em et~al.\/} 2006 A unified approach to equilibrium
  reconstruction {\em Proceedings-33rd EPS conference on Controlled Fusion and
  Plasma Physics, pp. P--2.160\/}

\bibitem{SciPy:algorithms:2020}
Virtanen P, Gommers R, Oliphant T~E, Haberland M, Reddy T, Cournapeau D,
  Burovski E, Peterson P, Weckesser W, Bright J, {van der Walt} S~J, Brett M,
  Wilson J, Millman K~J, Mayorov N, Nelson A~R~J, Jones E, Kern R, Larson E,
  Carey C~J, Polat {\.I}, Feng Y, Moore E~W, {VanderPlas} J, Laxalde D,
  Perktold J, Cimrman R, Henriksen I, Quintero E~A, Harris C~R, Archibald A~M,
  Ribeiro A~H, Pedregosa F, {van Mulbregt} P and {SciPy 10 Contributors} 2020
  {\em Nature Methods\/} {\bf 17} 261--272

\bibitem{Dormand:RK:1980}
Dormand J~R and Prince P~J 1980 {\em Journal of computational and applied
  mathematics\/} {\bf 6} 19--26

\bibitem{Scannell:Thomson:2010}
Scannell R, Walsh M, Dunstan M, Figueiredo J, Naylor G, O’Gorman T, Shibaev
  S, Gibson K and Wilson H 2010 {\em Review of Scientific Instruments\/} {\bf
  81} 10D520

\bibitem{Slusher:scattering:1980}
Slusher R~E and Surko C~M 1980 {\em The Physics of Fluids\/} {\bf 23} 472--490

\bibitem{Hutchinson:diagnostics:2002}
Hutchinson I~H 2002 {\em Principles of Plasma Diagnostics\/} (IOP Publishing)

\bibitem{Sheffield:scattering:2011}
Froula D, Glenzer S~H, Luhmann~Jr N~C and Sheffield J 2011 {\em Plasma
  scattering of electromagnetic radiation: theory and measurement techniques\/}
  (Academic press)

\bibitem{Brooker:Optics_Textbook:2003}
Brooker G and Brooker G 2003 {\em Modern classical optics\/} vol~8 (Oxford
  University Press)

\bibitem{Goldsmith:Quasioptics_Textbook:1998}
Goldsmith P~F 1998 {\em Quasioptical systems\/} (Chapman \& Hall New York)

\bibitem{Villeneuve:reciprocity:1958}
Villeneuve A and Harrington R 1958 {\em IRE Transactions on Microwave Theory
  and Techniques\/} {\bf 6} 308--310

\bibitem{Gusakov:Reflectometry:1997}
Gusakov E~Z and Tyntarev M~A 1997 {\em Fusion Engineering and Design\/}
  501--505

\bibitem{Holzhauer:DBS:1998}
Holzhauer E, Hirsch M, Grossmann T, Branas B and Serra F 1998 {\em Plasma
  physics and controlled fusion\/} {\bf 40} 1869

\bibitem{Conway:DBS:2004}
Conway G, Schirmer J, Klenge S, Suttrop W, Holzhauer E {\em et~al.\/} 2004 {\em
  Plasma Physics and Controlled Fusion\/} {\bf 46} 951

\bibitem{Pinzon:DBS_tilt_2019}
Pinz{\'o}n J, Estrada T, Happel T, Hennequin P, Blanco E, Stroth U, Teams T~I
  {\em et~al.\/} 2019 {\em Plasma Physics and Controlled Fusion\/} {\bf 61}
  105009

\bibitem{Lechte:2D_fullwave:2012}
Lechte C 2009 {\em IEEE Transactions on Plasma Science\/} {\bf 37} 1099--1103

\bibitem{Lechte:2D_fullwave:2017}
Lechte C, Conway G, G{\"o}rler T, Tr{\"o}ster-Schmid C {\em et~al.\/} 2017 {\em
  Plasma Physics and Controlled Fusion\/} {\bf 59} 075006

\bibitem{RuizRuiz:RCDR:2022}
Ruiz~Ruiz J, Parra F~I, Hall-Chen V~H, Christen N, Barnes B, Candy J, Garcia J,
  Giroud C, Guttenfelder W, Hillesheim J~C, Holland C, Howard N~T, Ren Y, White
  A~E and JET~contributors In press {\em Plasma Physics and Controlled
  Fusion\/}

\bibitem{Schekochihin:spectrum:2008}
Schekochihin A, Cowley S, Dorland W, Hammett G, Howes G~G, Plunk G, Quataert E
  and Tatsuno T 2008 {\em Plasma Physics and Controlled Fusion\/} {\bf 50}
  124024

\bibitem{Schekochihin:spectrum:2009}
Schekochihin A, Cowley S, Dorland W, Hammett G, Howes G~G, Quataert E and
  Tatsuno T 2009 {\em The Astrophysical Journal Supplement Series\/} {\bf 182}
  310

\bibitem{Barnes:spectrum:2011}
Barnes M, Parra F and Schekochihin A 2011 {\em Physical Review Letters\/} {\bf
  107} 115003

\end{thebibliography}

\end{document}